\RequirePackage[hyphens]{url}
\documentclass[manyauthors,nocleardouble]{cernphprep} 
\usepackage[english,german]{babel}
\usepackage[latin1]{inputenc}
\usepackage{ifpdf}
\ifpdf
  \usepackage[pdftex]{graphicx}
  \usepackage{epstopdf}
\else
  \usepackage{graphicx}
\fi
\DeclareGraphicsExtensions{.png,.pdf,.jpg,.mps,.eps}
\graphicspath{{./fig/}}
\usepackage[dvipsnames,usenames]{xcolor}
\usepackage{latexsym}
\usepackage[displaymath,mathlines]{lineno}
\usepackage{eurosym}
\usepackage{xspace} 
\usepackage{subfig}
\usepackage{fix2col}
\usepackage[sort&compress,square,comma,numbers,merge]{natbib}
\usepackage{amsmath,amssymb}
\usepackage{amsfonts}
\usepackage{upgreek} 
\usepackage{hepnames}
\usepackage{units}
\usepackage{import}
\usepackage{multirow}
\usepackage{tabularx} 
\usepackage{booktabs}
\usepackage{tikz}
\usepackage[plainpages=false,colorlinks,backref=section]{hyperref}
\expandafter\def\expandafter\UrlBreaks\expandafter{\UrlBreaks\do\a\do\b\do\c\do\d\do\e\do\f\do\g\do\h\do\i\do\j\do\k\do\l\do\m\do\n\do\o\do\p\do\q\do\r\do\s\do\t\do\u\do\v\do\w\do\x\do\y\do\z\do\&}
\hypersetup{citecolor=ForestGreen}
\renewcommand*{\backref}[1]{}
\renewcommand*{\backrefalt}[4]{%
  \ifcase #1 %
  (Not cited.)%
  \or
  (Cited in Sec.~#2.)%
  \else
  (Cited in Secs.~#2.)%
  \fi}

\input ./sty/mymathphys.sty
\input ./sty/myunits.sty
\usepackage[textsize=tiny]{todonotes}
\usepackage[section]{placeins}
\usepackage{svn-multi} 
\svnid{$Id: loi2017.tex 4408 2018-10-11 13:27:56Z avauth $}
\svnid{$Id: newcom.tex 4377 2018-09-11 11:51:52Z avauth $} 
\newcommand{\figref}[1]{Fig.~\ref{#1}}
\newcommand{\Figref}[1]{Figure~\ref{#1}}
\newcommand{\secref}[1]{section~\ref{#1}}

\newcommand*{\dif}[1][]{\ifthenelse{\equal{#1}{}}{\mathop{}\!\mathrm{d}}{\mathop{}\!\mathrm{d}^{#1}}}

\newcommand{\pluLeft}{{\stackrel{+}{\leftarrow}}}
\newcommand{\minRight}{{\stackrel{-}{\rightarrow}}}

\newcommand*{\JP}{\ensuremath{J^P}\xspace}

\newcommand{\Xminus}{\relax\ifmmode\mathrm{X^-}%
                \else$\mathrm{X^-}$\fi}%
\newcommand{\X}{\relax\ifmmode\mathrm{X}%
                \else$\mathrm{X}$\fi}%

\def\hermes{{\sc Hermes}}
\def\compass{{\sc Compass}\xspace}

\begin{document}
\selectlanguage{english}


\hypersetup{pageanchor=false}   
\svnid{$Id: title.tex 4608 2019-01-24 18:41:46Z vandrieu $}

\begin{titlepage}
\PHnumber{2019--003}
\EXPnumber{250}
\PHdate{\today}
\title{             Letter of Intent:

{\bf A New QCD facility at the M2 beam line of the CERN SPS\footnote{email: NQF-M2@cern.ch} }}

\renewcommand{\thefootnote}{\fnsymbol{footnote}}

\vspace*{-3\baselineskip}\centerline{\large \bf COMPASS++\footnote{COmmon Muon Proton Apparatus for Structure and Spectroscopy}/AMBER\footnote{Apparatus for Meson and Baryon Experimental Research}}

\vspace{1.5\baselineskip}

\label{app:collab}
\renewcommand\labelenumi{\textsuperscript{\theenumi}~}
\renewcommand\theenumi{\arabic{enumi}}
\begin{flushleft}
B.~Adams\Irefnn{chicago}{charlton}, 
C.A.~Aidala\Irefn{randall},
R.~Akhunzyanov\Irefn{dubna},
G.D.~Alexeev\Irefn{dubna},
M.G.~Alexeev\Irefn{turin_u},
A.~Amoroso\Irefnn{turin_u}{turin_i},
V.~Andrieux\Irefn{illinois},
N.V.~Anfimov\Irefn{dubna},
V.~Anosov\Irefn{dubna},
A.~Antoshkin\Irefn{dubna}
% ...
\end{flushleft}

%
%
\begin{Authlist}
\item \Idef{randall}{Randall Laboratory of Physics, University of Michigan, Ann Arbor, MI 48109-1040, USA}
\item \Idef{astana}{L.N.Gumilyov Eurasian National University, Astana, Kazakhstan}
\item \Idef{georgiasu}{Georgia State University, Atlanta, GA 30302, USA}
\item \Idef{aveiro}{University of Aveiro, i3N, Dept.\ of Physics, 3810-193 Aveiro, Portugal}
\item \Idef{beijing}{Tsinghua University, Beijing, 100084, P. R. China}
\item \Idef{bochum}{Universit\"at Bochum, Institut f\"ur Experimentalphysik, 44780 Bochum, Germany}
\item \Idef{bologna}{INFN Sezione di Bologna, Bologna, Italy}
\item \Idef{bonniskp}{Universit\"at Bonn, Helmholtz-Institut f\"ur  Strahlen- und Kernphysik, 53115 Bonn, Germany}
\item \Idef{bonnpi}{Universit\"at Bonn, Physikalisches Institut, 53115 Bonn, Germany}
\item \Idef{brno}{Institute of Scientific Instruments, AS CR, 61264 Brno, Czech Republic}
\item \Idef{calcutta}{Matrivani Institute of Experimental Research \& Education, Calcutta-700 030, India}
\item \Idef{charlton}{Incom Inc., Charlton, MA 01507, USA}
\item \Idef{chicago}{Enrico Fermi Institute, University of Chicago, Chicago, IL 60637, USA}
\item \Idef{dubna}{Joint Institute for Nuclear Research, 141980 Dubna, Moscow region, Russia}
\item \Idef{darmstadt}{GSI Helmholtzzentrum f\"ur Schwerionenforschung, 64291 Darmstadt, Germany}
\item \Idef{freiburg}{Universit\"at Freiburg, Physikalisches Institut, 79104 Freiburg, Germany}
\item \Idef{munichtu}{Technische Universit\"at M\"unchen, Physik Dept., 85748 Garching, Germany}
\item \Idef{gatchina}{Petersburg Nuclear Physics Institute, 188300 Gatchina, Russia}
\item \Idef{cern}{CERN, 1211 Geneva 23, Switzerland}
\item \Idef{saclay}{IRFU, CEA, Universit\'e Paris-Saclay, 91191 Gif-sur-Yvette, France}
\item \Idef{lanzhou}{Institute of Modern Physics, Chinese Academy of Science, Lanzhou 730000, China}
\item \Idef{liberec}{Technical University in Liberec, 46117 Liberec, Czech Republic}
\item \Idef{lisbon}{LIP, 1649-003 Lisbon, Portugal}
\item \Idef{losalamos}{Physics Division, Los Alamos National Laboratory, Los Alamos, NM 87545, USA}
\item \Idef{mainz}{Universit\"at Mainz, Institut f\"ur Kernphysik, 55099 Mainz, Germany}
\item \Idef{minsk}{Institute for Nuclear Problems of Belarusian State University, Minsk, Belarus}
\item \Idef{miyazaki}{University of Miyazaki, Miyazaki 889-2192, Japan}
\item \Idef{moscowlpi}{Lebedev Physical Institute, 119991 Moscow, Russia}
\item \Idef{moscowmephi}{National Research Nuclear University MEPhI, Moscow, Russia}
\item \Idef{nagoya}{Nagoya University, 464 Nagoya, Japan}
\item \Idef{praguecu}{Charles University in Prague, Faculty of Mathematics and Physics, 18000 Prague, Czech Republic}
\item \Idef{praguectu}{Czech Technical University in Prague, 16636 Prague, Czech Republic}
\item \Idef{protvino}{State Scientific Center Institute for High Energy Physics of National Research Center `Kurchatov Institute', 142281 Protvino, Russia}
\item \Idef{nysb}{Stony Brook University,  SUNY, Stony Brook, NY 11794-3800, USA}
\item \Idef{taipei}{Academia Sinica, Institute of Physics, Taipei 11529, Taiwan}
\item \Idef{telaviv}{Tel Aviv University, School of Physics and Astronomy, 69978 Tel Aviv, Israel}
\item \Idef{trento}{Universit\`{a} degli Studi di Trento and INFN-TIFPA, Trento, Italy}
\item \Idef{triest_u}{University of Trieste, Dept.\ of Physics, 34127 Trieste, Italy}
\item \Idef{triest_i}{Trieste Section of INFN, 34127 Trieste, Italy}
\item \Idef{tomsk}{Tomsk Polytechnic University,634050 Tomsk, Russia}
\item \Idef{turin_u}{University of Turin, Dept.\ of Physics, 10125 Turin, Italy}
\item \Idef{turin_i}{Torino Section of INFN, 10125 Turin, Italy}
\item \Idef{torino}{Universit\`{a} di Torino and INFN-Torino, Turin, Italy}
\item \Idef{illinois}{University of Illinois at Urbana-Champaign, Dept.\ of Physics, Urbana, IL 61801-3080, USA}
\item \Idef{nybnl}{RIKEN BNL Research Center, Brookhaven  National  Laboratory,  Upton, NY 11973-5000, USA}
\item \Idef{warsaw}{National Centre for Nuclear Research, 00-681 Warsaw, Poland}
\item \Idef{warsawu}{University of Warsaw, Faculty of Physics, 02-093 Warsaw, Poland}
\item \Idef{warsawtu}{Warsaw University of Technology, Institute of Radioelectronics, 00-665 Warsaw, Poland }
\item \Idef{yamagata}{Yamagata University, Yamagata 992-8510, Japan }
\item \Idef{yerevan}{A.Alikhanyan National Science Laboratory, Yerevan Physics Institute Foundation, Alikhanian Br. Street, 0036, Yerevan, Armenia}
\end{Authlist}

%
%
\renewcommand\theenumi{\alph{enumi}}
\begin{Authlist}
\item \Adef{a}{Also at Instituto Superior T\'ecnico, Universidade de Lisboa, Lisbon, Portugal}
\item \Adef{bbnlonly}{Also at Physics Dept., Brookhaven National Laboratory, Upton, NY 11973, USA}
\item \Adef{ictp}{Also at Abdus Salam ICTP, 34151 Trieste, Italy}
\item \Adef{r}{Supported by the DFG cluster of excellence `Origin and Structure of the Universe' (www.universe-cluster.de) (Germany)}
\item \Adef{d}{Also at Chubu University, Kasugai, Aichi 487-8501, Japan}
\item \Adef{x}{Also at Dept.\ of Physics, National Central University, 300 Jhongda Road, Jhongli 32001, Taiwan}
\item \Adef{e}{Also at KEK, 1-1 Oho, Tsukuba, Ibaraki 305-0801, Japan}
\item \Adef{g}{Also at Moscow Institute of Physics and Technology, Moscow Region, 141700, Russia}
\item \Adef{h}{Present address: RWTH Aachen University, III.\ Physikalisches Institut, 52056 Aachen, Germany}
\item \Adef{yerevan}{Also at Yerevan Physics Institute, Alikhanian Br. Street, Yerevan, Armenia, 0036}
\item \Adef{y}{Also at Dept.\ of Physics, National Kaohsiung Normal University, Kaohsiung County 824, Taiwan}
\item \Adef{regen}{Also at Institut f\"ur Theoretische Physik, Universit\"at T\"ubingen, 72076 T\"ubingen, Germany}
\item \Adef{turin_p}{Also at University of Eastern Piedmont, 15100 Alessandria, Italy}
%
%
%
\end{Authlist}
%
%


\clearpage

\tableofcontents

\end{titlepage}

\hypersetup{pageanchor=true}


\svnid{$Id: exec.tex 4519 2018-11-28 18:16:10Z avauth $} 
\section*{Executive Summary}
\addcontentsline{toc}{section}{\protect\numberline{}Executive Summary}%

In this Letter of Intent, we propose a broad experimental programme for the ``New QCD facility at the M2 beam line of the CERN SPS''. This unrivalled installation will provide the site for a great variety of measurements to address fundamental issues of Quantum Chromodynamics, which are expected to lead to significant improvements in the understanding of QCD as our present theory of strong interactions. The proposed measurements cover the range from lowest-$Q^2$ physics as the determination of the proton radius by elastic muon-proton scattering, over average-$Q^2$-reactions to study hadron spectroscopy, to high-$Q^2$ hadron-structure investigations using the Drell-Yan process and Deeply Virtual Compton Scattering.

In this Executive Summary, we briefly describe a number of unique first-generation experiments that are already possible using the existing M2 beam line with muons or unseparated hadrons. We also describe a number of unique second-generation experiments that rely on high-energy high-intensity radio-frequency (RF) separated hadron beams, which could be made possible by a major upgrade of the M2 beam line. Such an upgrade, which is presently under study by CERN EN-EA in the context of the Physics beyond Colliders Initiative, would make the CERN SPS M2 beam line absolutely unique in the world for many years to come. For every proposed experiment, individual instrumentation using modern detector architecture will be constructed and installed in the experimental hall EHN2, where the upgraded multi-purpose two-stage magnetic spectrometer will serve as experimental backbone of the new facility.  

\svnid{$Id: exec-main.tex 4408 2018-10-11 13:27:56Z avauth $} 

\textbf{Proton radius measurement using muon-proton elastic scattering} \\
The aim of this experiment is an independent precision determination of the electric mean-square charge radius of the proton using elastic muon-proton scattering. Such a measurement appears timely, since in spite of many years of intense activity the proton-radius puzzle remains unsolved up to now.  Presently, a discrepancy as large as 5 standard deviations exists between the two most recent precision measurements: $r^{rms}_{CREMA}=0.841 \pm 0.001$ from line-splitting measurements in laser spectroscopy of muonic hydrogen and  $r^{rms}_{MAMI}=0.879 \pm 0.008$ from elastic electron-proton scattering.

We propose to perform a one-year measurement using high-energy muons of the CERN M2 beam line, which will provide a new and completely independent result on the proton radius with a statistical accuracy of 0.01 and considerably smaller systematic uncertainty. Using muons instead of electrons is highly advantageous, as several experimental systematic effects and also theoretical (radiative) corrections are considerably smaller. The measurement will employ a time-projection chamber filled with pure hydrogen up to pressures of 20\,bar, which serves at the same time as a target and as detector gas. It has been developed by the PNPI group for several other experiments and will be adapted in a common effort to the proton-radius measurement.


\textbf{Hard exclusive reactions using a muon beam and a transversely polarised target} \\
The main goal of this experiment is to extend our fundamental knowledge on the angular-momentum structure of the nucleon, which can be accomplished by measurements of Generalized Parton Distributions (GPDs). These functions describe the correlations between longitudinal momentum fractions and transverse spatial positions of partons in the nucleon, commonly known as ``3-dimensional'' picture of the nucleon. Experimental data on these GPDs are required over the largest possible domain of the Bjorken-$x$ variable, as the evaluation of the average total angular momentum $J^f$ carried by quarks of flavour $f$ through the Ji sum rule requires an integration over the sum of the GPDs $H^f$ and $E^f$. While some experimental knowledge on the GPDs $H^f$ exists already for $u$ and $d$ quarks, no experimental data exists on the ``elusive'' GPDs $E^f$. 

Access to the GPDs $E^f$ is possible by measuring certain cross section asymmetries in Deeply Virtual Compton Scattering, using a polarised charged-lepton beam and a transversely polarised target. A measurement in the so-far uncharted Bjorken-$x$ domain between 0.005 and 0.05, where phenomenological models predict a particularly large sensitivity to $E^f$, can only be performed using the high-energy self-polarised muon beam of the CERN SPS M2 beam line. A 2-year measurement will yield an unprecedented statistical accuracy of 0.03 for the cross section asymmetry when using four bins per variable, with the systematic uncertainty expected to be smaller. A new 3-layer Silicon detector has to be constructed and installed inside of the existing polarised target, in order to accomplish identification of the target-recoil proton by the dE/dx method and to measure its momentum and trajectory. The proposed measurement is complementary in kinematics to a JLab experiment planned for 2022, which will use a transversely-polarised HD ice target.


\textbf{Drell-Yan and charmonium production} \\
The main objective of the proposed Drell-Yan and J/$\psi$ production experiments is to make a major step forward in the determination of the nearly unknown pion and kaon parton distribution functions (PDFs). The planned measurements will provide key benchmarks for testing the most recent predictions of fundamental, non-perturbative QCD calculations, such as lattice QCD and Dyson-Schwinger Equations formalism. At medium and large values of Bjorken-$x$, a quantitative comparison between the pion and the kaon valence distributions is of utmost importance.  At smaller values of Bjorken-$x$, improved knowledge of the onset of the sea and gluon distributions in the meson will help in explaining the differences between the gluon contents of pions, kaons and nucleons, and hopefully provide clues to understand the mechanism that generates the hadron masses.  Furthermore, a comparison between cross sections for positive and negative meson beams is expected to provide stringent experimental constraints on the J/$\psi$ production mechanism, as well as an alternative way of accessing both quark and gluon distributions in the incoming meson. In parallel to meson structure measurements, the availability of heavier nuclear targets in the setup will allow the study of cold nuclear effects such as nuclear PDFs and partonic energy loss.  

The M2 secondary hadron beam line at the CERN SPS provides an exclusive opportunity for such measurements. For both Drell-Yan and J/$\psi$ production processes, the expected statistical accuracies correspond to a few $10^4$ and nearly $10^6$ events for the Drell-Yan and J/$\psi$ production processes, respectively. The first part of this programme can be completed using the presently available positive ($\pi^+$,  $K^+$, $p$) and negative ($\pi^-$, $K^-$, $\bar{p}$) hadron beams, in combination with an optimized experimental setup. A detailed study of the presently uncharted kaon structure, which is envisaged in a later stage of the programme, requires the existence of a high-intensity kaon beam that can only be produced using radio-frequency beam-separation cavities. The proposed studies need a high-performance beam particle identification. They would also greatly benefit from a novel and compact lepton absorber/detector immersed in a magnetic field.  Such detector could largely improve the performances of the setup, while ensuring a large geometric acceptance and allowing for $e^+/e^-$ detection in parallel with the muon pair.   


\textbf{Measurement of antiproton production cross sections for Dark Matter Search} \\
The purpose of this experiment is the measurement of the antiproton production cross section 
on proton and $^{4}He$  targets for projectile energies from several ten to a few hundred $\mathrm{GeV}$. In combination with similar measurements by LHCb in the $\mathrm{TeV}$ range, our data will provide a fundamental data set that will greatly improve the accuracy of the predicted natural flux of antiprotons in the galactic cosmic rays. This is of great importance as the indirect detection of Dark Matter (DM) is based on the search for products of DM annihilation or decay, which are expected to appear as distortions in the spectra of rare Cosmic Ray components like positrons or antiprotons. The new data set is expected to vastly improve the sensitivity of the very accurate AMS antiproton flux measurements to DM signals, which is presently spoiled by the poor knowledge  of the antiproton production cross section. 

The existing M2 beam line with its momentum range between $20-30$ and $280\, \mathrm{GeV/c}$
is an ideal tool to perform this measurement. The double-differential antiproton production cross section will be measured in $p+p$ and $p+^4He$ collisions using the existing spectrometer in EHN2 equipped with liquid hydrogen and liquid helium targets. Measuring for several beam momenta the cross section in 20 bins each for antiproton momentum and pseudorapidity, a statistical uncertainty of 1\% will be reached, with an anticipated point-to-point systematic uncertainty of less than 5\%. The main challenges for these measurements  are the precise determination of the efficiencies of the trigger systems and of the particle identification efficiency and purity.


\textbf{Spectroscopy with low-energy antiprotons} \\
The main objective of this experiment is a significant improvement in the understanding of the nature of the so-called $X$, $Y$, $Z$ states, which remains so far unclear in spite of worldwide efforts in the last 15 years. The discovery of narrow resonance-like signals at $e^+e^-$ colliders, which have masses above the open-charm threshold and decay to charmonium, had triggered a wealth of theoretical interpretations ranging from conventional excited charmonium to multi-quark states and hybrid mesons. Of particular interest is the production of states in association with a recoil particle, which opens the possibility of observing states with spin-exotic quantum numbers such as hybrids or glueballs. The annihilation of low-energy antiprotons with momenta below $20\,\mathrm{GeV}$ provides a unique tool to study these states. 

An early experiment at the M2 beamline of the SPS using low-energy antiprotons and a liquid hydrogen target can reach luminosities of the order of $10^{30}\,\mathrm{cm}^{-2}\mathrm{s}^{-1}$. It is expected to make important contributions to the production and spectroscopy of these long sought-after states even before the start of PANDA at FAIR, where such a measurement is in the planning state. Our measurements will at the same time provide vital input for PANDA, e.g.\ by measuring production cross sections for $X$, $Y$, $Z$ states in $\overline{p}p$ reactions, for which theoretical estimates range from  $0.1\,\nb$ to $10\,\nb$. Owing to the low beam momentum, the experiment requires a dedicated target spectrometer including charged-particle tracking and electromagnetic calorimetry, in addition to the forward spectrometer in EHN2.   


\textbf{Spectroscopy of kaons} \\
The goal of the kaon spectroscopy programme is to map out the complete spectrum of excited kaons with unprecedented precision using novel analysis methods. A more precise knowledge of the kaon spectrum will have a broad impact not only on low-energy QCD, but also on many processes in modern hadron and particle physics where excited kaons appear, e.g.\ in the study of CP violation in heavy-meson decays, which are studied at LHCb and Belle II. 

The high-intensity RF-separated kaon beam of the upgraded M2 beam line is an unrivalled place to accumulate a large data set for kaon spectroscopy as most of the planned or proposed measurements in the strange-meson sector elsewhere can either not compete with the measurement proposed here or are complementary to it. Taking data over one year with a beam energy of at least $50\,\GeV$, about $20\times 10^6$ events can be collected, which allows us to keep systematic effects under control. The main experimental challenges are high-precision vertex reconstruction, photon detection with electromagnetic calorimeters for reconstructing neutral hadrons, and final-state particle identification, i.e.\ kaons have to be distinguished from pions with high efficiency over a broad kinematic range.


\textbf{Study of the gluon distribution in the kaon via prompt-photon production} \\ 
The purpose of this experiment is the study of the gluon content of charged kaons. This 
is of fundamental importance for understanding the internal structure of light mesons and may also shed light onto the mechanism that generates their mass. In contrast to the well-known gluon distribution of the nucleon, our knowledge on the gluon distribution in light mesons is rather limited. It can be considerably improved by studying prompt-photon production in hadronic collisions using a high-energy meson beam.

The high-intensity RF-separated kaon beam of the upgraded M2 beam line is an unrivalled site to investigate the gluon content of the kaon. Using a beam of positive kaons with an energy of at least $100\,\GeV$ over 1-2 years, the gluon PDF in the charged kaon will be measured using the dominant hard gluon Compton scattering process in the range of $x_g$ above 0.05 and for $Q^2 \sim p^2_T>7$ (GeV/c)$^2$. A separate data set will have to be taken with a beam of negative kaons in order to allow for a separation of the subdominant quark-antiquark annihilation process. For systematic studies and also to improve the knowledge on the gluon structure of the pion, it is foreseen to also collect data with incoming pions. This can be done either in parallel or separately in a preceding running period. The basic requirements for the proposed measurement are a sufficiently high transparency of the experimental setup for the produced photons, a wide kinematic range of photon detection by the system of electromagnetic calorimeters, efficient beam hadron identification, and a dedicated calorimeter-based trigger on high-$p_T$ photons. We are not aware of plans to study the gluon structure of charged kaons at other laboratories.

\filbreak 

\textbf{Low-energy tests of QCD using Primakoff reactions} \\
The main objective of this experiment is the first measurement of the kaon polarisability. Electric and magnetic polarisabilities, which describe the rigidity against deformation by an external electromagnetic field, constitute fundamental quantities of low-energy QCD. Apart from a quite vague experimental upper limit, presently only model predictions exist for the electric polarisability of the kaon, e.g.\ by chiral perturbation theory ($\chi PT$) or the quark-confinement model. It can be determined using the so-called Primakoff reaction, $K^- Z \rightarrow K^- \gamma \, Z$, using a high-energy high-intensity kaon beam, on a nuclear target Z with Z protons. Such an experimental approach allows the parallel determination of another fundamental quantity of low-energy QCD, i.e.\ the lifetime of the neutral pion, which is related to the chiral-anomaly hypothesis.

The RF-separated beam of the upgraded M2 beam line is a unique place for these measurements. One year of data taking with a $100\,\GeV$ beam delivering $5 \times 10^6$ kaons per second will allow to determine the electric kaon polarisability with a statistical accuracy of $0.03 \times 10^{-4} \mathrm{fm}^3$, with the total systematic uncertainty estimated to be smaller. This accuracy is even smaller than that of the prediction of $\chi PT$ in the one-loop approximation and will clearly allow us to distinguish between models for the kaon polarisability. The neutral pion lifetime will be measured in parallel using those beam pions that remain in the RF-separated kaon beam. Here, the resulting statistical accuracy of about 1\% will be smaller by more than a factor of two compared to the accuracy of the theoretical prediction and that of existing measurements. The proposed measurement requires advanced beam particle identification, excellent tracking capabilities at small angles and precise electromagnetic calorimeters. All these expectations are estimated to be met by the upgraded spectrometer that is planned to be available in EHN2 at the time of running. The proposed measurements have presently no competitors in the world.


\textbf{Production of vector mesons and excited kaons off nuclei} \\
The main goal of this measurement is to study pion-induced exclusive vector-meson production off various nuclei in order to determine the nuclear dependence of the cross section for longitudinally polarised vector mesons ($\rho^0(770)$, $K^{0*}(892)$). The absorption strength for a vector meson produced in a nuclear environment is predicted to strongly depend on its polarisation and also on the beam energy. However, neither existing data nor phenomenological models allow presently for a satisfactory description of nuclear transparency. Of particular interest is a clarification of a possible colour screening effect, as predicted by the colour transparency model. When using incoming kaons, the absorption of polarised $K^*$ mesons can be studied, which is a topic of increasing interest.

About one year of running using pions and kaons produced by the M2 beam line is required to reach a sufficient accuracy of the cross section measurement. The beam energy should be higher than about $50\,\mathrm{GeV}$ to avoid effects related to meson decays in the nuclear medium and lower than about $100\,\mathrm{GeV}$ as the cross section for the charge-exchange reaction decreases with energy. In this energy range, the cross section is sufficiently large, so that parallel running with other physics programmes can be considered. The proposed measurements are kinematically fully complementary to the studies that are planned to be performed at JLab.
 
 
\textbf{Tentative schedule of the CERN M2 beam line} \\
A tentative schedule of the CERN M2 beam line is outlined in table~\ref{tab:schedule} according to the information given on \url{https://lhc-commissioning.web.cern.ch/lhc-commissioning/schedule/LHC-long-term.htm}.

\begin{table}[h]
\newcolumntype{L}[1]{>{\hsize=#1\hsize\raggedright\arraybackslash}X}%
\begin{tabularx}{\textwidth}{L{0.25}|L{2.6}|L{0.55}|L{0.6}}
\textbf{Year} & \textbf{Activity} & \textbf{Duration} & \textbf{Beam} \\
\hline
2019 \newline 2020 & Long Shutdown 2 & 2 years & - \\
\hline
2021 & COMPASS-II transversity with polarised deuteron target & 1 year & muon \\
\hline
2022 & proton radius & 1 year & muon \\
\hline
2023 \newline 2024 & Drell-Yan for $\pi$ and $K$ PDFs and charmonium production mechanism \newline Antiproton cross section for Dark Matter Search & $\lesssim$ 2 years \newline \newline 2 month  & $p$, $K^+$, $\pi^+$ \newline $\bar{p}$, $K^-$, $\pi^-$  \newline $p$ \\
\hline
2025  & Long Shutdown 3 (for SPS) &   &  \\
\hline
\end{tabularx}
\caption{Tentative schedule of the CERN M2 beam line for the time between LS2 and LS3.}
\label{tab:schedule}
\end{table}

\svnid{$Id: intro.tex 4588 2019-01-10 02:03:19Z vandrieu $} 
\section{Introduction}
\label{sec:intro}

The structure of hadrons as building blocks of visible matter is still at the core of interest in the quest for understanding nature. In the late seventies, the quarks suggested by Gell-Mann and Zweig in 1964 as building blocks of nucleons and mesons were identified with Feynman's partons. This paved the way for the huge success of the Quark-Parton Model (QPM). In the same decade, Quantum Chromodynamics (QCD) became widely accepted as the theory of strong interactions. However, in spite of the huge effort spent by generations of experimentalists and theoreticians in the past four decades towards a deeper understanding of the force binding quarks and gluons to nucleons, the major fraction of visible matter in the universe, QCD can by far not compete with the overwhelming accuracy provided by the Standard Model of electroweak interactions that was confirmed recently by the spectacular discovery of the Higgs boson at the CERN LHC. 

Over the last four decades, measurements at the external beam lines of the CERN SPS received worldwide attention by making a great variety of measurements possible that resulted in a steady consolidation of QCD as our theory for the description of visible matter. As of today, these beam lines bear great potential for future significant improvements of our understanding of hadronic matter. Especially the world-unique SPS M2 beam line, when operated with high-energy muons, can be used to determine the average charged proton radius in a muon-proton elastic scattering experiment. Such a measurement constitutes a highly-welcomed complementary approach in this area of world-wide activity. When operated with pions, the M2 beam line can be used to shed new light to the light-meson structure using the Drell-Yan process, in particular by determining the widely unknown parton distribution functions (PDFs) of the pion. Operating with antiprotons, completely new results in the search for XYZ exotic states are expected. Operating with protons, the largely unknown antiproton production cross section can be measured, which constitutes important input for the upcoming activities in the Search for Dark Matter. 

A completely new horizon in hadron physics would be opened by upgrading the M2 beam line to provide radio-frequency (RF) separated high-intensity and high-energy kaon and antiproton beams. Kaon beams open access to the virgin field of high-precision spectroscopy of the strange-meson spectrum and a first measurement of the kaon polarisability through the Primakoff process. Measurements of prompt photons allow first determinations of the gluon PDF in the kaon. Studying the Drell-Yan process with incoming kaons opens access to the unknown kaon PDFs. Using an antiproton beam will provide a complementary approach to study the transverse-momentum-dependent (TMD) PDFs of the nucleon, which are presently under intense theoretical discussion using (mostly) the existing COMPASS data on semi-inclusive deep inelastic scattering. 

Last but not least, novel data from hadron physics experiments will be required to confirm results on the structure of mesons and nucleons in terms of PDFs, which are expected to emerge from near-future computations in the framework of QCD on the lattice.

In this Letter of Intent (LoI), we propose to establish a world-unique QCD facility by using the external M2 beam line of the CERN SPS in conjunction with a universal spectrometer that will use upgraded installations of the COMPASS experiment in the experimental hall EHN2. As sketched above, unique measurements are already possible using the existing beam line. A major upgrade to provide RF-separated beams in a second phase would make this facility absolutely unique in the world for many years to come.

The main features of the envisaged physics programme and the anticipated hardware upgrades are summarized in Table~\ref{tab:intro:programs}.

\begin{table}[h]
  \begin{center}
      \resizebox{\textwidth}{!}{\setlength{\tabcolsep}{5pt}%
      \begin{tabular}{c|c|c|c|c|c|c|c|c}
        & Physics & Beam & Beam & Trigger & Beam & & Earliest & Hardware \\
        Program & Goals& Energy&Intensity&Rate&Type&Target&start time,&additions\\ 
        &  & [GeV]&[s$^{-1}$]&[kHz]&&&duration\\\hline \hline 

        muon-proton & Precision&& & & &  high-&& active TPC,\\ 
        elastic&proton-radius&100&$4\cdot 10^6$&100&\color{blue}{$\mu^\pm$}&pressure&2022&SciFi  trigger, \\
        scattering&measurement&&&&&H2&1 year&silicon veto,\\\hline

        Hard&&&&&&&&recoil silicon,\\
        exclusive &GPD $E$ & 160 & $2\cdot 10^7$ & 10 & \color{blue}{$\mu^\pm$} & NH$_3^\uparrow$ & 2022& modified polarised  \\
        reactions&&&&&&&2 years&target magnet \\\hline
        
        Input for Dark&$\overline{p}$ production&20-280&$5\cdot 10^5$&25& \color{ForestGreen}{$p$}&LH2,&2022&liquid helium\\
        Matter Search & cross section&  & & & & LHe & 1 month & target \\\hline

        &&&&&&&&target spectrometer:\\
        $\overline{p}$-induced & Heavy quark& 12, 20 & $5\cdot 10^7$ &25& \color{ForestGreen}{$\overline{p}$} & LH2 & 2022& tracking, \\
        spectroscopy&exotics&&&&&&2 years& calorimetry\\\hline

        Drell-Yan & Pion PDFs& 190 & $7\cdot 10^7$ & 25 & \color{ForestGreen}{$\pi^{\pm}$}  &  C/W & 2022 &\\
        &&&&&&&1-2 years&\\\hline\hline

        &&&&&&&&\\
        Drell-Yan & Kaon PDFs \& & $\sim$100 & $10^8$& 25-50 & \color{red}{$K^{\pm}$, $\overline{p}$} & NH$_3^\uparrow$, &2026 &''active absorber'', \\ 
        (RF)&Nucleon TMDs& &&&&C/W&2-3 years&vertex detector \\ \hline

        &Kaon polarisa-&&&&&&\footnotesize{non-exclusive}& \\
        Primakoff & bility \& pion& $\sim$100 & $5\cdot 10^6$ & $>10$ & \color{red}{$K^-$} & Ni &2026&  \\
        (RF)&life time&&&&&&1 year&\\\hline

        Prompt&&&&&&&\footnotesize{non-exclusive}&\\
        Photons & Meson gluon& $\geq 100$ & $5\cdot 10^6$ & 10-100 & \color{red}{$K^{\pm}$} & LH2, & 2026 &  hodoscope\\
        (RF)&PDFs &&&&\color{ForestGreen}{$\pi^{\pm}$}&Ni&1-2 years&\\\hline

        $K$-induced&High-precision&&&&&&&\\
        Spectroscopy &strange-meson& 50-100 & $5\cdot 10^6$ & 25 & \color{red}{$K^-$} &LH2 & 2026 & recoil TOF, \\
        (RF)&spectrum&&&&&&1 year&forward PID\\\hline

        &Spin Density&&&&&&&\\
        Vector mesons &Matrix& 50-100 & $5\cdot 10^6$ & 10-100 & \color{red}{$K^{\pm},\pi^{\pm}$} & from H  & 2026 &  \\
        (RF)&Elements&&&&&to Pb&1 year&\\

      \end{tabular}%
    }

  \end{center}
  \caption{Requirements for future programmes at the M2 beam line after 2021. {\color{blue}Muon beams} are in blue, {\color{ForestGreen}conventional hadron beams} in green, and {\color{red}RF-separated hadron beams} in red. }
  \label{tab:intro:programs}
\end{table}

\newpage

This LoI is structured as follows:
\begin{itemize}
\item future hadron physics with the existing SPS M2 muon beam;
\item future hadron physics with existing SPS M2 hadron beams;
\item future hadron physics with new RF-separated kaon and antiproton beams;
\item instrumental upgrades of the COMPASS spectrometer to provide the equipment necessary to carry out the proposed measurements.
\end{itemize}


\svnid{$Id: stdmuo.tex 4376 2018-09-11 09:50:37Z avauth $} 

\section{Hadron physics using the muon beam}
\label{sec:stdmuo}
\svnid{$Id: stdmuo_mup.tex 4400 2018-10-05 08:46:26Z avauth $} 
\subsection{Proton radius measurement using muon-proton elastic scattering}
\label{sec:stdmuo.mup}

\svnid{$Id: stdmuo_mup_update.tex 4588 2019-01-10 02:03:19Z vandrieu $} 

The physics of the proton as the charged nuclear building block
of matter is at the core of interest in the quest for understanding 
nature. As a consequence of its inner structure, the electromagnetic form 
factors $G_E$ and $G_M$ encode the response of the proton to outer
electric and magnetic fields, respectively. As will be detailed below, the squares $G_E^2$ and $G_M^2$ can be measured in 
non-polarised elastic lepton scattering off the proton. These measurements have been pioneered in the 1950's by R.~Hofstadter~\cite{Hofstadter:1955ae}, and are still actively pursued.
The gross feature of the form factors is a dependence on the squared
momentum transfer $Q^2$ given by
\begin{equation}
G_E(Q^2)=G_M(Q^2)/\mu_p=\frac{1}{(1+Q^2/a^2)^2},
\label{Eq:DipoleFF}
\end{equation}
which is called dipole approximation. It can be motivated by a substructure of 
the proton consisting of three constituent quarks. Here, $\mu_p\approx 2.79$ is the proton anomalous magnetic moment. The constant $a$ was 
determined in electron scattering to be about $a^2\approx 0.71\,\text{GeV}^2/c^2$. 
The functional behavior shown in Eq.~(\ref{Eq:DipoleFF}) with $a^2=0.71\,\text{GeV}^2/c^2$ is used as the standard 
reference dipole form factor $G_D(Q^2)$.

\begin{figure}
\begin{center}
\includegraphics[width=1.1\textwidth]{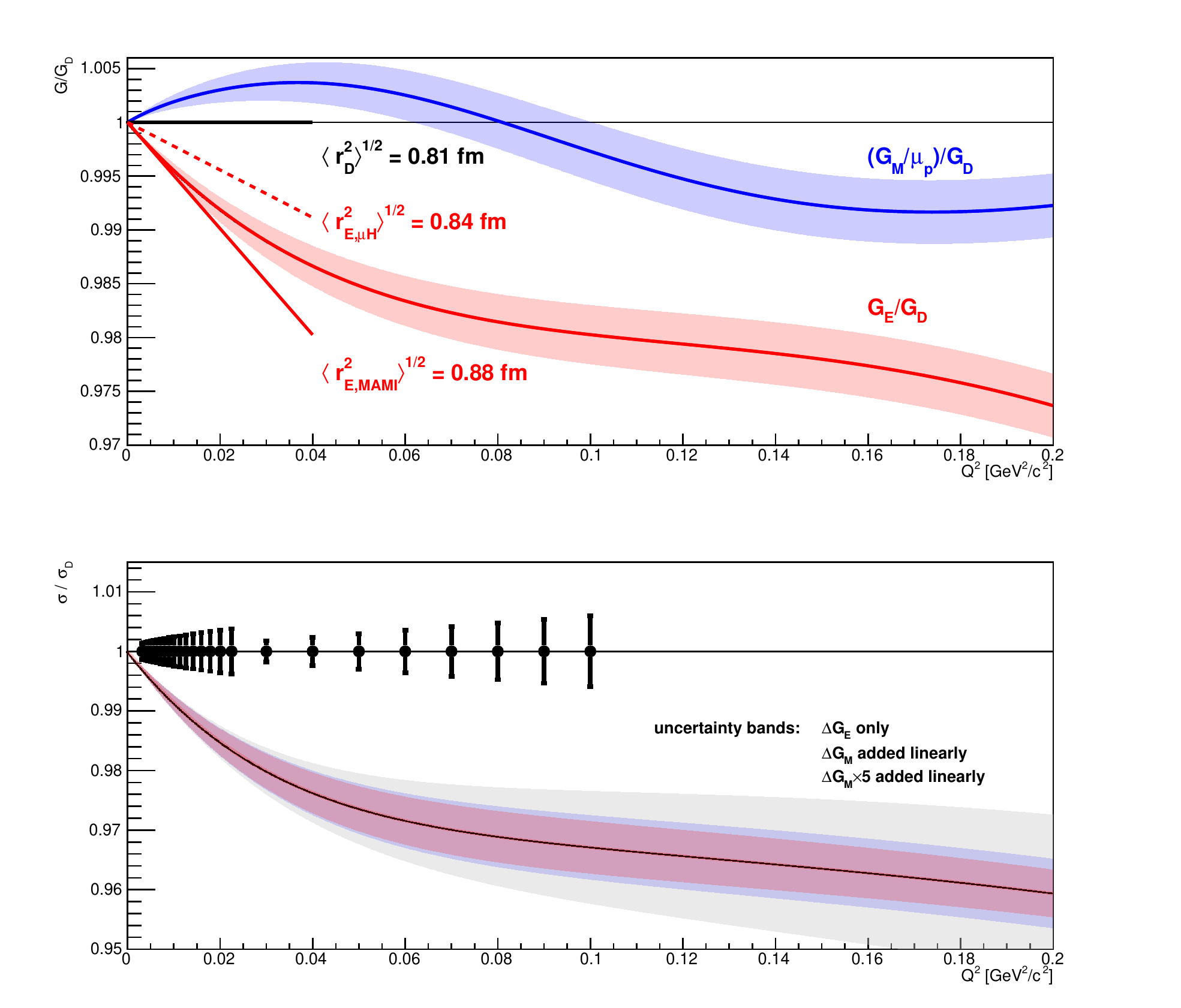}
\caption{Top panel: proton form factors $G_E$ and $G_M$ as measured at MAMI, 
shown as ratio to the dipole form factor $G_D$ as given in the text.
Bottom panel: ratio of the cross section measured at MAMI over the prediction for the cross section using the standard dipole form factor. The innermost (red) uncertainty band 
corresponds to the effect of the uncertainty of $G_E$ only, while for the 
(blue) middle band the uncertainty 
from $G_M$ was added linearly, and for the outer (gray) band the 
contribution from
$\Delta G_M$ was increased by a factor of five. The dots with
error bars, arbitarily placed at 1, represent the anticipated statistical 
precision of the proposed measurement, down to $Q^2=0.003\,\text{GeV}^2/c^2$,
where the statistical uncertainties are expected to dominate the
systematic point-to-point uncertainty. There will be data from the
proposed experiment down to $Q^2\le0.001\,\text{GeV}^2/c^2$, with the statistical uncertainty
further shrinking according to the increasing cross section with
$Q^2\rightarrow$0, {\it cf.{}} Eq.~(\ref{Eq::elCS}), 
which are omitted here for conciseness.
See text for a discussion of the 
uncertainty contributions in different $Q^2$ regions.}
\label{fig::FFandCS}
\end{center}
\end{figure}

The respective charge and magnetic-moment distributions in space are obtained 
by Fourier transformation of the form factors, and specifically the electric 
mean-square charge radius is related to the form factor by
\begin{equation}
  \langle r_E^2 \rangle = -6\hbar^2 \left.\frac{d G_E(Q^2)}{d Q^2}\right|_{Q^2\rightarrow 0} 
  \stackrel{dipole}{=} \frac{12\hbar^2}{a^2} \approx (0.81\,\text{fm})^2 \equiv \langle r_D^2\rangle.
\label{Eq:reDef}
\end{equation}
More refined fits to the measured shape of the form factors are often given
as polynomials or other analytic functions of $Q^2$ multiplying the dipole 
approximation shown in Eq.~(\ref{Eq:DipoleFF}).
The so far most elaborate measurement of the proton form factors by elastic 
electron
scattering was carried out at the Mainz university accelerator 
MAMI~\cite{PhysRevLett.105.242001,Bernauer:2012hi}, and 
a parameterisation of the results at small values $Q^2<0.2\,\text{GeV}^2/c^2$ is 
shown in the upper panel of   Fig.~\ref{fig::FFandCS}.
Unlike earlier electron scattering data, $G_M^2$ shows a more shallow 
slope than that of $G_D^2$, while $G_E^2$ shows a steeper one resulting in $r_{E,\text{MAMI}}^{rms}=\sqrt{\langle r_{E,\text{MAMI}}^2\rangle}=(0.879\pm0.008)\,\text{fm}$.
In both cases total uncertainties are shown. They represent statistical and systematic uncertainties added in quadrature, where the systematic one is obtained by summing up its components linearly.
This value for the proton radius is at variance with $r_{E,\mu H}^{rms}=(0.841\pm0.001)\,\text{fm}$ that was found in laser spectroscopy of muonic 
hydrogen~\cite{Pohl:2010zza,Pohl:2014bpa}, which is a different way to measure the proton radius.
This striking discrepancy of 
about five standard deviations has triggered many efforts to clarify its 
origin~\cite{Antognini:2015moa,Gasparian:2017cgp,Pohl:2016glp,crema::2017,Gilman:2013eiv,Henderson:2016dea,suda:2016}. 

\subsubsection{Experiments targeting the proton radius puzzle: the M2 beam line case}
We propose to measure
{\it elastic muon-proton scattering} 
with a high-energetic muon beam impinging on a hydrogen
gas target, over a momentum-transfer range that is particularly sensitive to
the proton charge radius.
This means, on the one hand, to measure the cross section
while coming as close as possible to $Q^2=0$ as required in Eq.~(\ref{Eq:reDef}),
and on the other hand
to cover a sufficient range in momentum transfer in order to constrain the
slope of the cross section within the desired level of precision. As illustrated
in Fig.~\ref{fig::FFandCS}, this range is approximately
$0.001<Q^2/(\text{GeV}^2/c^2)<0.02$. At values of $Q^2$ smaller than $0.001\,\text{GeV}^2/c^2$, the deviation
from a point-like proton is on the level of $10^{-3}$ and thus smaller than
unavoidable systematic effects, as the variation of the detector 
efficiencies with $Q^2$ cannot be controlled more accurately with the
currently available methods. At values of $Q^2$ larger than $0.02\,\text{GeV}^2/c^2$, the non-linearity
of the $Q^2$ dependence becomes the predominant source of uncertainty, and the data
cannot be used to determine the proton radius unless more elaborate 
theory input is used. 

In order to reach the precision required at small momentum transfer, 
it is important to observe the recoil proton. 
Due to its small energy, this requires an active target, \textit{i.e.} the target
being detector at the same time. This can be realized by a Time Projection
Chamber (TPC) operated with pure hydrogen gas. Such a target has been developed
by PNPI~\cite{Vorobyov:1974ds,Ilieva:2012vhs},
and it is in the testing phase for a similar experiment using
electron scattering at Mainz.

Several experiments are currently ongoing or proposed for 
refining the knowledge of the proton radius by {\it electron-proton} elastic 
scattering~\cite{Antognini:2015moa,Gasparian:2017cgp,vorobyov:2016,Mihovilovic:2016rkr}.
This includes the mentioned TPC experiment at 
MAMI~\cite{vorobyov:2016}, but also the 
inital-state radiation experiment of
the A1 collaboration~\cite{Mihovilovic:2016rkr}. 
All experiments using electron-proton scattering are
challenged by the required QED radiative corrections, which are as large
as 20\% due to the small electron mass. Currently, it is unclear how 
the precision of these corrections can be controlled on the desired
level of less than $1\%$. Hence independent of the outcome
of any measurement performed using an electron beam, those using a muon beam will test
systematic effects related to radiative corrections, which are 
substantially smaller for muons due to their much larger mass.

Despite this obvious benefit, there are still significant systematic
effects expected in
measurements using {\it muon-proton} elastic scattering at 
{\it low muon-beam energies}, as
{\it e.g.{}} discussed for the proposed MUSE experiment in Ref.~\cite{Gilman:2013eiv}. 
Apart from corrections for
the pion component in the beam and for muon decays, there is a substantial
correction for the Coulomb distortion of the low-velocity muon wave function.
The latter is estimated to be on the level of one percent for larger
scattering angles, however with an unclear relation to the other 
radiative corrections, which introduces a systematic uncertainty for
which an experimental test is required (see, {\it e.g.}, Eq.~(2.24) and Appendix~C in \cite{Friedrich:2000zha}). Such a test could best be realized
using muon scattering at very high energies, where the Coulomb correction reduces
to a negligible level.

The highest precision on proton radius measurements is claimed to be achievable by
the investigation of atomic level 
splittings~\cite{Pohl:2010zza,Pohl:2014bpa,Pohl:2016glp,crema::2017}, which are very accurately
measured by laser spectroscopy. From 1S transitions in
{\it muonic hydrogen}, the above-mentioned value of $0.841\,\text{fm}$ was
determined by correcting the measured frequency for all known QED
effects and attributing the remaining effect to the finite size of the proton. In contrast, scattering experiments do not determine the proton radius directly, but by measuring the dependence of the form factor over an extended range in $Q^2$. This additional information allows for checking {\it e.g.{}} the
assumption made on the linear behaviour of the form factor in the studied
$Q^2$ range.

Altogether, we regard the proposed muon-proton scattering using a high-energy
muon beam for the determination of the proton radius to be an
important and unique cornerstone in the quest for solving the proton
radius puzzle. In view of the highly competitive
and dynamic ongoing research in the field, it appears very timely to realise the measurement 
at the CERN M2 beam line  as soon as the scheduling and the required preparatory steps
will allow.

\subsubsection{Formalism of elastic lepton-proton scattering}
The cross section for elastic muon-proton scattering, $\mu p \to  \mu^{\prime} p$, is to first order given by
\begin{eqnarray}
  \frac{d\sigma}{d Q^2}&\hspace{-2mm}=\hspace{-2mm}&
  \frac{\pi\alpha^2}{Q^4\,m_p^2\, \vec{p}_\mu^{\,2}}
\left[\left(G_E^2+\tau G_M^2\right)
  \frac{4E_\mu^2 m_p^2-Q^2(s-m_\mu^2)}{1+\tau}
-G_M^2\frac{2m_\mu^2Q^2-Q^4}{2}\right], \label{Eq::elCS}
\end{eqnarray}
where $Q^2=-t=-(p_\mu-p_{\mu'})^2$, $\tau=Q^2/(4m_p^2)$ and $s=(p_\mu+p_p)^2$. 
The squared centre-of-momentum energy $s$ is given in the laboratory system 
by $s=2E_\mu m_p + m_p^2+m_\mu^2$ with $E_\mu$ being the energy and $\vec{p}_\mu$ 
the three-momentum of the incoming muon when colliding with a proton at rest. 

As $G_M^2$ enters in two terms, the first being for small $Q^2$ proportional to $E_\mu^2$ and the second to $m_\mu^2$ measuring at different energies around $E_\mu\approx m_\mu$ enables us to separate the contributions from $G_E^2$ and $G_M^2$ (``Rosenbluth separation''). This is not possible at the high energies proposed here.
In this case, the last term in Eq.\ref{Eq::elCS} can be safely neglected, and the measured cross section is directly proportional to the 
linear
combination $(G_E^2+\tau G_M^2)$.
At the small values of $Q^2$ 
({\it i.e.{}} small $\tau$) of interest here, 
the contribution from $G_M$ is small
and can be corrected for by assuming some shape for $G_M$, so that only $G_E$ is left as a free parameter. Even with large variations of $G_M$, 
{\it i.e.} a factor of five larger
than the one given in the MAMI analysis, the uncertainty on $G_E$
and thus on the proton charge radius stays well below 0.1\%, which is about a 
factor of 10 smaller than the precision of 1\% that the measurement aims at.

\subsubsection{Measurement at the CERN M2 beam line }
We propose to measure elastic muon-proton scattering using 
a 100\,GeV muon beam and a pressurized hydrogen gas target.
For a precise measurement of the
proton radius, the relevant momentum transfer region is $0.001<Q^2/(\text{GeV}^2/c^2)<0.02$, which requires to operate the target
as a TPC in order to detect the proton recoil track. The pressure of the gas
will be optimized for having on the one hand sufficiently low stopping power such
that a proton recoil track is detectable, and on the other hand it still 
stops in the TPC volume.
The respective gas system has been developed and is in the test
phase at MAMI. The details of the readout electronics, which are to be adapted to the
\compass\ environment, are currently under study. At $Q^2$ values larger than $0.02\,\text{GeV}^2/c^2$, recoil protons can no longer be stopped in the gas. In order to study the form factor also in this region, a hydrogen target is foreseen in conjunction with multiple layers of scintillating fibers that are arranged as hollow cylinders. These scintillating fibers will be located inside the gas volume, surrounding the part of the volume that is traversed by the beam.

The scattered muon will be measured with the \compass\ spectrometer 
using its standard
muon setup. In order to allow for the detection of the elastically scattered muon track, the beam killer
components have to be excluded from the trigger. The central parts of 
the tracking detectors will be activated, and silicon telescopes surrounding
the TPC will be used for measuring the muon scattering angle with high accuracy. 
In addition, the electromagnetic calorimeters serve to detect the (rare)
radiative events, by which soft photons with energies up to $2\,\text{GeV}$ are emitted.

Since triggering only on the proton recoil would imply
$Q^2$-dependent efficiency variations that could not be controlled
using the data themselves, a trigger component from the muon trajectory
has to be added. As presently the beam rate is too high to record all events, the beam
trigger will have to be extended by a new component that allows to veto muons
with a scattering angle below about 5\,$\mu$rad.
This suppresses those muons that have
experienced multiple (small-angle) scattering only, which amounts to about
$99\%$ of the incoming muons.
In contrast, muons are efficiently selected if their scattering angle
is larger than $100\,\mu\text{rad}$, which corresponds to momentum transfers
larger than $10^{-4}\,\text{GeV}^2/c^2$. A possible detector system to realise this trigger would be the extension of the silicon telescopes with new
scintillating fiber tracking detectors. However, solutions
with thinner detectors, such as silicon pixel detectors with a sufficiently fast readout, would be desirable for minimizing
 multiple scattering as a source of systematic uncertainty.  Proton recoil measurement, muon measurement, and the
trigger of this experiment are detailed in Sec.~\ref{sec:inst-mup}. 
 
A triggerless readout (Sec.~\ref{sec:inst:daqfeet})
could solve current issues of rate capability and would allow for implementing the
above described event selection in an elegant and efficient manner. It is therefore highly advisable to pursue its development in the context of this proposal.

The statistical uncertainties that can be achieved in the sketched experiment
are shown in Fig.~\ref{fig::FFandCS}, using a suitable segmentation of the
data in $Q^2$ bins. The dataset is sufficient to constrain the proton radius
to a precision better than $0.01\,\text{fm}$.


\svnid{$Id: stdmuo_dvcs.tex 4588 2019-01-10 02:03:19Z vandrieu $} 
\subsection{Hard exclusive reactions using muon beam and transversely polarised target}
\label{sec:stdmuo.dvcs}
\subsubsection{Motivations for a measurement of  the GPD E}

Our understanding of the nucleon structure has been improving recently thanks 
to the ongoing intense effort of research on Generalized Parton Distributions 
(GPDs)~\cite{Mueller:1998fv,Ji:1996ek,Ji:1996nm,Radyushkin:1996nd,Radyushkin:1997ki,
Collins:1996fb,Burkardt:2000za,Burkardt:2002hr,Burkardt:2004bv}. These functions describe the correlations between longitudinal 
momentum and transverse spatial position of the partons inside the nucleon. Among other features, 
they give access to the contribution of the orbital angular momentum of the quarks and gluons to the spin 
of the nucleon.
The nucleon GPDs are accessible via measurements of deeply virtual Compton scattering (DVCS) 
$\ell N\to \ell' N' \gamma$  or deeply virtual meson production (DVMP) $\ell N\to \ell' N' M$.
In the one-photon-exchange approximation, a virtual photon $\gamma^*$ is exchanged between incoming lepton 
and target nucleon, producing a real photon $\gamma$ or a meson M.
We denote by $k$, $k^\prime$, $q$, $p$ and $p^\prime$ the four-momenta of the incoming  and scattered 
 leptons $\ell$ and $\ell^\prime$, the virtual photon $\gamma^*$, and  the initial  
and final state  nucleons $N$ and $N^\prime$ respectively.
In order to apply the GPD formalism~\cite{Collins:1996fb},
the photon virtuality $Q^2=-q^2=-(k-k^\prime)^2$ must be sufficiently large 
to assure a hard interaction  at the parton level, i.e. $Q^2>>M^2$ where $M$ is the proton mass.
It also must be much larger than the squared four-momentum 
transfer to the target $t= (p - p^\prime)^2$, i.e. $|t|/Q^2<<1$. Also, the squared hadronic center-of-mass energy 
$W=(p + q)^2$ must be large, i.e. above the nucleon resonance domain, for fixed values of the Bjorken variable  
$x_B=\frac{Q^2}{2M\nu}$, 
where  $\nu = E_{\ell}-E_{\ell'}$ is the energy of the photon in the laboratory system.



One of the major goals of the ongoing worldwide GPD physics programmes is the precise mapping 
of the GPDs $H^f$ and $E^f$, which enter in the ``Ji sum rule''\cite{Ji:1996ek,Ji:1996nm} that provides access to the average total angular momentum of partons in the nucleon, as a function of $Q^2$:
\begin{equation}
J^f(Q^2)\ =\ \frac{1}{2}\lim_{t \to 0} \int_{-1}^1 dx\ x \left[ H^f(x,\xi,t, Q^2) + E^f(x,\xi,t, Q^2) \right].
\label{eq:ji_sum_rule}
\end{equation}
Here
\begin{equation}
\frac{1}{2}\ =\ \sum_{q=u,d,s} J^q(Q^2)\ +\ J^g(Q^2)
\end{equation}
with $f$ denoting either a quark flavor ($u,d,s$) or a gluon ($g$).
The GPDs in Eq.~\eqref{eq:ji_sum_rule}  depend upon the four variables 
$x$, $\xi$, $t$ and $Q^2$, where 
$x$ is the average longitudinal momentum fraction of the struck parton in DVCS or DVMP, and 
$\xi$  is half of the difference of longitudinal momentum fractions between initial and final parton.
The skewness $\xi$ is related to $x_B$ by $\xi\simeq\frac{x_B}{2-x_B}$. 
In an experiment like DVCS one measures Compton Form factors (CFF)~\cite{Belitsky:2001ns}, which are 
integrals over $x$ of GPDs 
convoluted with the operator describing the elementary interaction between 
the parton and the virtual photon. 

While some information on  the GPDs $H^f$ is already provided by the existing data, the GPDs $E^f$ are basically unknown.
 The most promising DVCS observables that are sensitive to $E^f$ are the transverse target-spin asymmetry in 
the case of a proton target, and the longitudinal beam-spin asymmetry in the case of a neutron target.
 Two pioneering experiments were done at JLab using a deuteron target~\cite{Mazouz:2007aa} 
 and at HERMES using a transversely polarised proton target~\cite{Airapetian:2008aa}.
The two new measurements~\cite{Jlab12_CLAS_n,Jlab12_CLAS_HDice}  planned with CLAS12 at Jlab  
represent a flagship goal of the Jlab physics
  program~\cite{Dudek:2012vr} with the new 12~GeV beam.

The \compass\ experiment has already performed in 2016 and 2017 measurements of exclusive photon 
and meson production with an unpolarised proton target and high-energy polarised $\mu^+$ and $\mu^-$ beams, 
mainly covering the kinematic domain of sea quarks. The kinematics of these  measurements at smaller 
$x_B$ are complementary to earlier or planned measurements at lower energies. 
The \compass\ measurement is mostly sensitive to the GPDs $H^f$ and 
will provide a separate measurement of the real and imaginary parts of the corresponding CFF 
$\mathcal{H}$ by combining cross sections measured with beams of opposite charge and polarisation.

By making use of the existing transversely polarised proton target, 
the GPDs $E^f$ can be accessed using the existing \compass\ setup  by a measurement of the 
transverse-target spin-dependent DVCS cross sections. 
Such a measurement would be complementary to the CLAS12 data, 
and would provide an extension of the kinematical coverage to the small $x_B$ domain. 




\subsubsection{Measurements of Deeply Virtual Compton Scattering}
\label{subsubsec:physics.gpd.observables.DVCS}
\begin{figure}
	\begin{center}
		\includegraphics[width=0.5\textwidth]{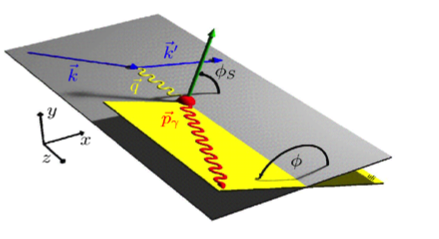}
	\end{center}
	\caption{Definition of the relevant angles in DVCS on a transversely polarised target.}
	\label{fig:future.compass_dvcs_transv_angles}
\end{figure}

Since at \compass\ both beam and target are polarised, the relevant observables for accessing the GPDs $E^f$ are represented by the ``transverse'' (refering to transversely polarised target) beam charge \& spin 
difference  of the $\mu p^\uparrow \to\ \mu \gamma p$ cross section (denoted $d \sigma$),
\begin{equation}
\label{eq:DCST}
{\cal{D}}_{CS,T} \equiv 
\Bigl(d\sigma^\pluLeft(\phi, \phi_S) 
- d\sigma^\pluLeft (\phi, \phi_S + \pi)\Bigr) 
- \Bigl(d\sigma^\minRight (\phi, \phi_S)
- d\sigma^\minRight (\phi, \phi_S + \pi)\Bigr),
\end{equation}
and the transverse beam charge \& spin sum,  
\begin{equation}
\label{eq:SCST}
{\cal{S}}_{CS,T} \equiv 
\Bigl(d\sigma^\pluLeft(\phi,\phi_S) 
- d\sigma^\pluLeft(\phi,\phi_S+\pi) \Bigr)
+ \Bigl( d\sigma^\minRight(\phi,\phi_S) 
- d\sigma^\minRight(\phi,\phi_S+\pi) \Bigr).
\end{equation}
Here, the subscript  $CS$ indicates that both lepton charge ($+$ or $-$) and lepton spin ($\leftarrow$ or $\rightarrow$) are reversed simultaneously
when changing from $\mu^+$ to $\mu^-$, while $T$ denotes the transversely polarised target.
The quantities between parenthesis in these two equations represent the differences of cross sections 
with the two opposite target spin orientations (denoted by $\phi_S$ and $\phi_S + \pi$, 
with $\phi$ and $\phi_S$ denoting the azimuthal angle of the produced photon and of the target spin vector, respectively, 
see Fig.~\ref{fig:future.compass_dvcs_transv_angles}).

Two experimental asymmetries can be derived from these expressions:
\begin{equation}
\label{eq:ACST}
{\cal{A}}_{CS,T}^D = \frac {{\cal{D}}_{CS,T}} {{\Sigma}_{unpol} }   
\;\; {\mbox{and}} \;\; 
{\cal{A}}_{CS,T}^S = \frac {{\cal{S}}_{CS,T}} {{\Sigma}_{unpol} },
\end{equation}
where ${\Sigma}_{unpol}$ represents the lepton-charge-average unpolarised cross section.

\begin{figure}
	\begin{center}
		\includegraphics[width=\textwidth]{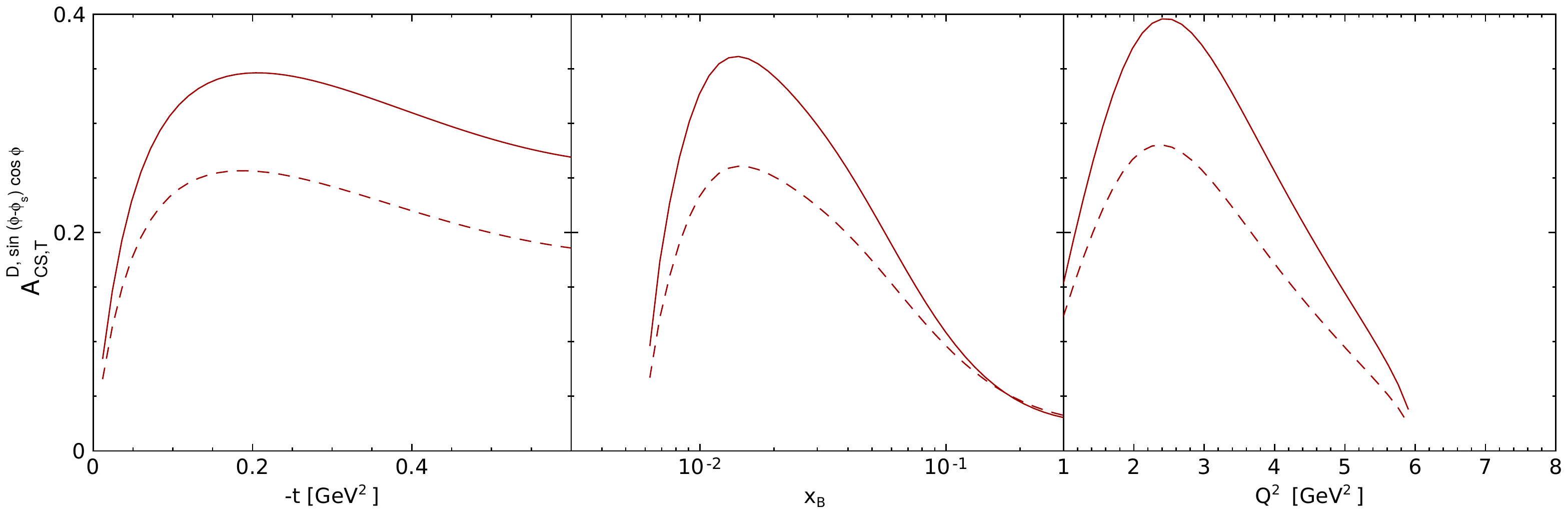}
	\end{center}
	\caption{Estimation of the amplitude of the $\left[\sin(\phi-\phi_S)\cos(\phi)\right]$ 
	modulation in the \compass\ kinematics, based on predictions from the  
	 GK~\cite{Goloskokov:2011rd} model at leading order in $\alpha _S$ (solid lines) 
	and with the additional assumption of $E^f=0$ (dashed lines). 
	The estimates were obtained using the PARTONS~\cite{Berthou:2015oaw} framework.}
	\label{fig:future.compass_partons_predictions}
\end{figure}

The above defined  difference and sum of cross sections can be decomposed into angular harmonics~\cite{Belitsky:2001ns}
of the type $\left[\sin(\phi-\phi_S)\sin(n\phi)\right]$, $\left[\sin(\phi-\phi_S)\cos(n\phi)\right]$, 
$\left[\cos(\phi-\phi_S)\sin(n\phi)\right]$ and\\ $\left[\cos(\phi-\phi_S)\cos(n\phi)\right]$, 
with coefficients that represent linear or bi-linear combinations of the CFFs
$\mathcal{H}$, $\mathcal{\widetilde{H}}$ and $\mathcal{E}$, which are  related to the GPDs $H$, $\tilde{H}$, $E$ respectively.
For example, the leading twist-2 coefficient in ${\cal{D}}_{CS,T}$ is 
associated to the $\left[\sin(\phi-\phi_S)\cos(\phi)\right]$ modulation in the interference 
term~\footnote{As the DVCS process has  a final state identical to that of the Bethe-Heitler process, 
the measured $\mu p$ cross section contains additional interference terms.}, 
and it receives contributions from the imaginary parts of $\mathcal{H}$ and $\mathcal{E}$ at 
the same level~\cite{Belitsky:2001ns}:
\begin{equation}
\label{eq:abcsdt2}
c_{1T_-}^I \propto  \frac{t}{4M^2} {\rm{Im}}\left[(2-x_B) F_1 {\mathcal{E}} 
- 4 \frac{1-x_B}{2-x_B} F_2 \mathcal{H}\right].
\end{equation}
The coefficients can be extracted from a Fourier analysis of the measured cross sections or asymmetries. 
The size of the asymmetry ${\cal{A}}_{CS,T}^{D,\sin(\phi-\phi_S)\cos(\phi)}$
 associated to the $c_{1T_-}^I$ term 
has been recently estimated  
in the context of the PARTONS framework~\cite{Berthou:2015oaw}, 
using the GK~\cite{Goloskokov:2011rd} model prediction 
as a function of $-t$, $x_B$ and $Q^2$ in the typical kinematic \compass\ domain. 
As presented in Fig.~\ref{fig:future.compass_partons_predictions}, the predicted asymmetries are sizeable 
and show a clear sensitivity to the contribution of the GPDs $E^f$, particularly
in the small-$x_B$ range covered by the \compass\ setup.
\begin{figure}
	\begin{center}
		\includegraphics[width=\textwidth]{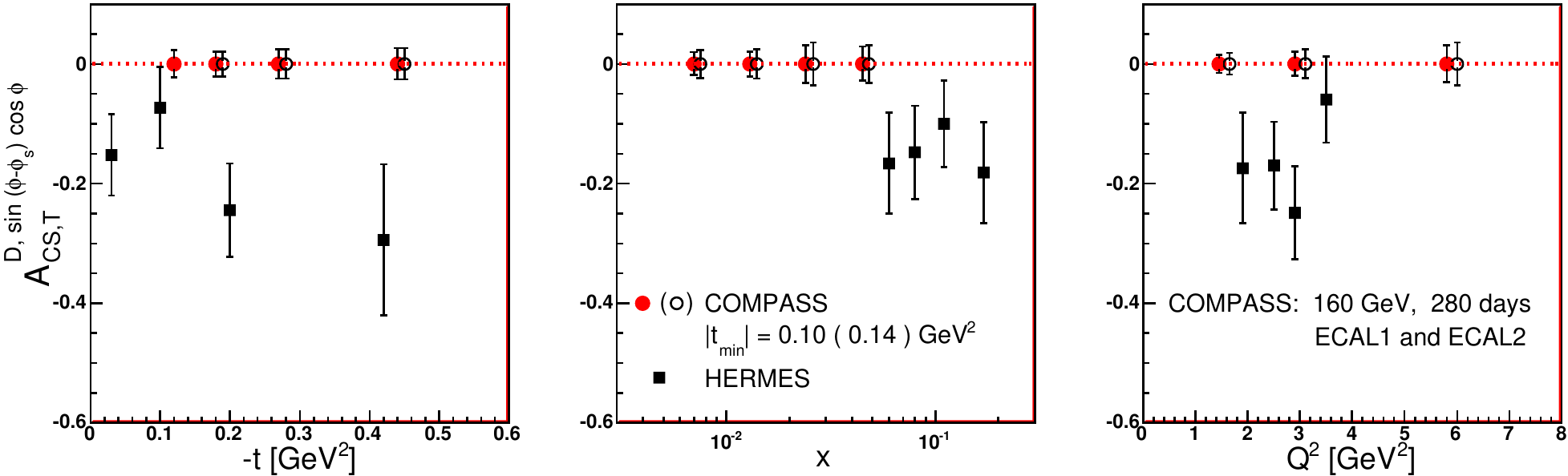}
	\end{center}
	\caption{Expected statistical accuracy of 
		$A_{CS,T}^{D,\sin(\phi-\phi_s)\cos\phi}$ as a function of $-t$, $x_B$ 
		and $Q^2$ from a 280 days measurement  with the \compass\ spectrometer, 
		using a 160~GeV muon beam and a transversely polarised NH$_3$ target. 
		Solid and open circles correspond to a minimum accessible $|t_{min}|$ of 0.10~GeV$^2$ and 0.14~GeV$^2$, 
		respectively. Also shown is
		the asymmetry $A_{U,T}^{\sin(\phi-\phi_s)\cos\phi  }$ measured at 
		\hermes~\cite{Airapetian:2008aa} with its statistical errors. 
		Figure from ref.~\cite{COMPASS:2010}.}
	\label{fig:future.compass_dvcs_transv_proj}
\end{figure}

The expected statistical accuracy for a  measurement of the $\left[\sin(\phi-\phi_S)\cos(\phi)\right]$
modulation using the \compass\ spectrometer and a transversely polarised NH$_3$ target is shown in
Fig.~\ref{fig:future.compass_dvcs_transv_proj}. The red points and the open circles  represent the
projections for a 280 days measurement at the nominal muon beam intensity, 
and for a minimum accessible value of $|t|$ of 0.10~GeV$^2$ and 0.14~GeV$^2$, respectively. 
For comparison, the black squares show the asymmetry $A_{U,T}^{\sin(\phi-\phi_s)\cos\phi  }$ measured by 
\hermes~\cite{Airapetian:2008aa} with its statistical errors. 
The new data could therefore provide a measurement of the $\left[\sin(\phi-\phi_S)\cos(\phi)\right]$ modulation 
with a statistical accuracy of approximately 0.03 in the so far uncharted region of
 $5\cdot 10^{-3} \lesssim x_B \lesssim 5\cdot 10^{-2}$.

The technical realisation of detecting recoil particles with polarised solid-state targets is 
detailed in Section~\ref{sec:inst-dvcs}. As described there, the minimum detectable $|t|$ value of 0.10~GeV$^2$ appears reasonable
when using the 3-layer silicon detector with a layer thickness of 300 $\mu$m.

\subsubsection{Measurements of Deeply Virtual Meson Production}
\label{subsubsec:physics.gpd.observables.DVMP}

The ``transverse target spin asymmetry'' measured
in deeply virtual exclusive vector meson production on a transversely polarised target was shown to be sensitive to the GPDs $E^f$~\cite{Goloskokov:2007nt}.

In perturbative QCD (pQCD), there exists a general proof of
factorisation~\cite{Collins:1996fb} for exclusive meson production by longitudinal
virtual photons. In this case, the amplitude for hard exclusive meson
leptoproduction can be factorised into a hard-scattering part and soft
parts, the latter depending on the structure of the nucleon described by
GPDs and on the structure of the meson described by its
distribution amplitude (DA).  No rigorous
proof of factorisation exists for transverse virtual photons. 
However, phenomenological pQCD-inspired models have been proposed
\cite{Martin:1996bp, Goloskokov:2005sd, Goloskokov:2007nt, Goloskokov:2008ib}
that go beyond the collinear factorisation by postulating the so called
`$k_\perp$ factorisation', where $k_\perp$ denotes the parton transverse
momentum. In the model of Refs.~\cite{Goloskokov:2005sd, Goloskokov:2007nt,
Goloskokov:2008ib, Goloskokov:2013mba, Goloskokov:2014ika}, by using $k_\perp$ factorisation, cross sections and spin-density matrix elements (SDMEs) for
DVMP by both longitudinal and transverse virtual photons can be described
simultaneously.

At leading twist, the chiral-even GPDs $H^f$ and $E^f$ introduced in Sect.~\ref{subsubsec:physics.gpd.observables.DVCS}, are sufficient to describe exclusive vector
meson production on a spin-1/2 target.  Depending on the quark content and the
quantum numbers of the meson, there exists sensitivity to various types
of GPDs and different quark flavours. Because of this property DVMP can be regarded as a quark flavour filter, which
motivates the study of a wide spectrum of mesons.

In exclusive vector meson production on transversely polarised targets,
the observable that is sensitive to the GPDs $E^f$ is the azimuthal asymmetry
$A_{UT}^{\rm{sin}(\phi-\phi_S)}$ (the definitions of angles are as in 
Fig.~\ref{fig:future.compass_dvcs_transv_angles} 
with the replacement of the real photon by the meson). Here, the indices
$U$ and $T$ refer to the beam spin-independent and transverse target
spin-dependent cross section, and ${\sin(\phi-\phi_S)}$ indicates the
type of azimuthal modulation of the cross section. The GPDs $E^f$ appear
at leading twist only in this azimuthal asymmetry for vector meson
production by longitudinal photons. In the latter case, the relation between asymmetry and GPDs 
is the following:
\begin{equation}
A_{UT}^{\rm{sin}(\phi-\phi_S)} \propto \mathrm{Im} \mathcal{E}^*_M \mathcal{H}_M .
\label{eq:sinphi-phis}
\end{equation}
Here, the quantities $\mathcal{H}_M$ and $\mathcal{E}_M$ denote
 integrals over GPDs $H^f$
and $E^f$, respectively, in a form appropriate for the production of the meson $M$. For vector mesons, both 
quark and gluon GPDs  contribute in general at leading order of $\alpha _s$.

The published  \compass\ results on $A_{UT}^{\rm{sin}(\phi-\phi_S)}$ for exclusive $\rho ^0$ and $\omega $ 
production on transversely polarised targets~\cite{Adolph:2012ht, Adolph:2013zaa, Adolph:2016ehf} are compatible with zero within 
total experimental uncertainties, which amount to 0.011 for $\rho ^0$ and 0.08 for $\omega $ produced on 
transversely polarised protons. The following kinematic constraints were used: 1\,(GeV/$c$)$^2$
$<$ $Q^2$ $<$ 10\,(GeV/$c$)$^2$, $0.1 < y < 0.9$ and 
$0.05~(\mathrm{GeV}/\mathit{c})^2 < p_{T}^{2} < {\rm 0.5}~(\mathrm{GeV}/\mathit{c})^2$, 
where $p_T$ is the transverse momentum of the produced meson relative to the direction of the
virtual photon. The extracted transverse target-spin azimuthal asymmetries were corrected for the semi-inclusive 
(SIDIS) background in the exclusive sample, which was on the level of about 20\%. Note that these 
results correspond to the unseparated contributions of transverse and longitudinal virtual photons.

For an experiment with the muon beam, the only feasible method to separate
transverse and longitudinal contributions would be the one proposed in Ref.~\cite{Diehl:2005pc} to exploit the 
angular
distributions of vector meson decays and use the assumption of $s$-channel helicity conservation, which was found experimentally to be valid at the 10\% level.
 Apart from a precise control of acceptance 
corrections for angular distributions, a necessary prerequisite for this 
method is a sufficient suppression of the SIDIS background 
contribution to the meson decay angular distributions. This was not feasible for the past measurements with the 
\compass\ polarised target, because of a sizable contribution of the SIDIS background. On the contrary, a significant
suppression of this background (to the percent level) is expected for the future polarised solid-state target when combined with the detection of recoil particles. 

Projections for statistical uncertainties for the future measurement are based on experimental conditions of 
Refs.~\cite{Adolph:2012ht, Adolph:2013zaa, Adolph:2016ehf}, 
as well as assuming 280 days of data taking with the $\mu ^+$ and $\mu ^-$ beam intensities being the same 
and equal to that for $\mu ^-$ beam. The kinematic range for $Q^2$ and $y$ is expected to be the same as 
for the previous 
\compass\ measurements, while that for $p_{T}$ will be $0.10 \;\; (0.14) ~(\mathrm{GeV}/\mathit{c})^2 < p_{T}^{2} <
{\rm 0.7}~(\mathrm{GeV}/\mathit{c})^2$. Due to the smaller value of the integrated muon flux and a higher minimal value of the measured $p_{T}$, for 
the proposed measurement
the statistical uncertainty for the $A_{UT}^{\rm{sin}(\phi-\phi_S)}$ asymmetry averaged over the entire kinematic region 
will increase by a factor of 1.7 (2.0) for $\rho ^0$ and 1.4 (1.6) for $\omega $ compared to the published results. Above, each pair of numbers (without or with
parentheses) corresponds to 
the respective lower limit of the measured $p_{T}$ indicated above.     
While for the unseparated asymmetry
the new data may significantly reduce the uncertainty when combined with published 
results, it is more important that they will provide a unique possibility  
for a muon-beam experiment to separate the contributions from 
transverse and longitudinal virtual photons. After such a separation,   a measurement of
$A_{UT}^{\rm{sin}(\phi-\i_S)}$ for longitudinal photons could open access to the GPDs $E^f$ at leading twist.

\svnid{$Id: stdhad.tex 4376 2018-09-11 09:50:37Z avauth $} 
\section{Hadron physics using conventional hadron beams}
\label{sec:stdhad}
\svnid{$Id: stdhad_dryan_1.tex 4588 2019-01-10 02:03:19Z vandrieu $} 
\subsection{Drell-Yan and charmonium production using conventional hadron beams}
\label{sec:stdhad.dryan}

\subsubsection{Introduction: Meson structure and the origin of nuclear mass}
\label{ssec:stdhad.dryan.intro}

The quark-gluon structure of light mesons and the physical origin of their small masses are still largely unknown.
While there are ample data available on the proton, the experimental determination of meson structure remains the long-awaited and critical input to theoretical efforts that seek to explain the emergence of massive composite hadrons, including the large mass difference between pion and proton. 

Two Standard Model mechanisms contribute to the generation of mass. Spontaneous electroweak symmetry breaking gives rise to the Higgs mechanism providing fundamental particles with their current masses. Strong-interaction chiral symmetry breaking leads to the large masses of composite hadrons. In chiral QCD with massless quarks, hadron masses in the Lagrangian emerge through the trace anomaly of the energy-momentum tensor. For the proton, the binding energy and the mass of dressed quarks add to about $m_p=1\,\GeV$. Very differently for the pion, the Goldstone Boson of the interaction, the binding energy and the dressed quark masses cancel to $m_\pi = 0\,\GeV$ \cite{C.Roberts}. In lattice QCD, the recently proposed Large Momentum Effective Theory (LaMET) \cite{X.Ji} will make it possible to calculate quark and gluon distribution functions for hadrons quantitatively, see for example \cite{H.W.Lin}  and \cite{Abdel-Rehim}. Such calculations greatly benefit from the arrival of Peta-scale supercomputers.
Recently, there has been increasing interest in theoretical calculations of the parton structure of mesons, including the Nambu-Jona-Lasinio model \cite{Nam:2012vm, Hutauruk:2016sug}, the chiral constituent quark model~\cite{Watanabe:2016lto}, the light-front constituent model~~\cite{Pasquini:2014ppa}, and from QCD Dyson-Schwinger equations~\cite{Nguyen:2011jy, Chen:2016sno}.

Detailed experimental information on the quark and gluon structure of the proton is available from the analysis of numerous lepton-nucleon deep inelastic scattering experiments combined with several data sets of jet, hadron, and Drell-Yan cross sections observed in proton-proton and proton-antiproton collider experiments over a broad range of the scattering kinematics. Global analyses have been carried out using NNLO in perturbative QCD and have resulted in precise knowledge of quark and gluon distribution functions of the proton.

In contrast, the quark and gluon structure of mesons is only poorly constrained from early Drell-Yan cross-section measurements for pions \cite{Badier:1983mj,Betev:1985pg,Conway:1989fs,Bordner:1996,Chekanov:2002yh,Aaron:2010ab}  and completely unconstrained for kaons \cite{GRV}. The sparse experimental information on meson structure limits the ability to test theoretical progress directed at determining quark and gluon distributions from ab initio lattice QCD calculations. Furthermore, it limits our understanding of the dynamical generation of hadron masses in QCD. Important experimental activities are underway to study the pion structure through final-state neutron-tagged DIS at JLab, while the feasibility of pion-structure measurements at a future Electron-Ion Collider is being evaluated. But the need to relate the experimental neutron-tagged DIS to the physics of DIS off-shell pions translates into model uncertainties that still need to be assessed.

The current high-intensity pion-dominated hadron beam from the CERN M2 beamline provides a unique opportunity for measurements of the pion structure and the nucleon structure through pion-induced Drell-Yan production on polarised and unpolarised proton, deuteron and nuclear targets. A significant improvement in the statistical precision can be achieved using novel analysis methods that access the Drell-Yan signal also at lower invariant masses.
At a later stage, an RF-separated kaon beam at CERN would lead to the first measurement of the kaon structure, and an RF-separated antiproton beam would allow for precision measurements of the spin-dependent transverse-momentum  dependent (TMD) PDFs of the nucleon. These will be described in Section~\ref{sec:rfsep.dryan}.

We propose to perform detailed studies of the pion structure using the existing CERN M2 hadron beams. The following physics goals are foreseen: 
\begin{itemize}
\item determine pion valence and sea-quark distributions;
\item study the charmonium production mechanism in order to extract the gluon distribution in the pion;
\item investigate flavour-dependent effects in nuclear targets.
\end{itemize} 

The first two topics aim at a detailed picture of the pion structure. The third one will contribute to the determination of nuclear PDFs in the large Bjorken-$x$ region and, for the first time, check the flavour (in)dependence of the EMC effect. In parallel, we propose to perform precise measurements of the Drell-Yan and J/$\psi$ angular distributions using an isoscalar target, which shall complement those presently being performed  at the COMPASS experiment with an ammonia target. As will be shown, these goals can be achieved simultaneously in  two years of dedicated data taking using $190\,\GeV$ pion beams with positive and negative polarity.
 
 
\begin{figure}[tb]
\centering
\includegraphics[width=.48\textwidth]{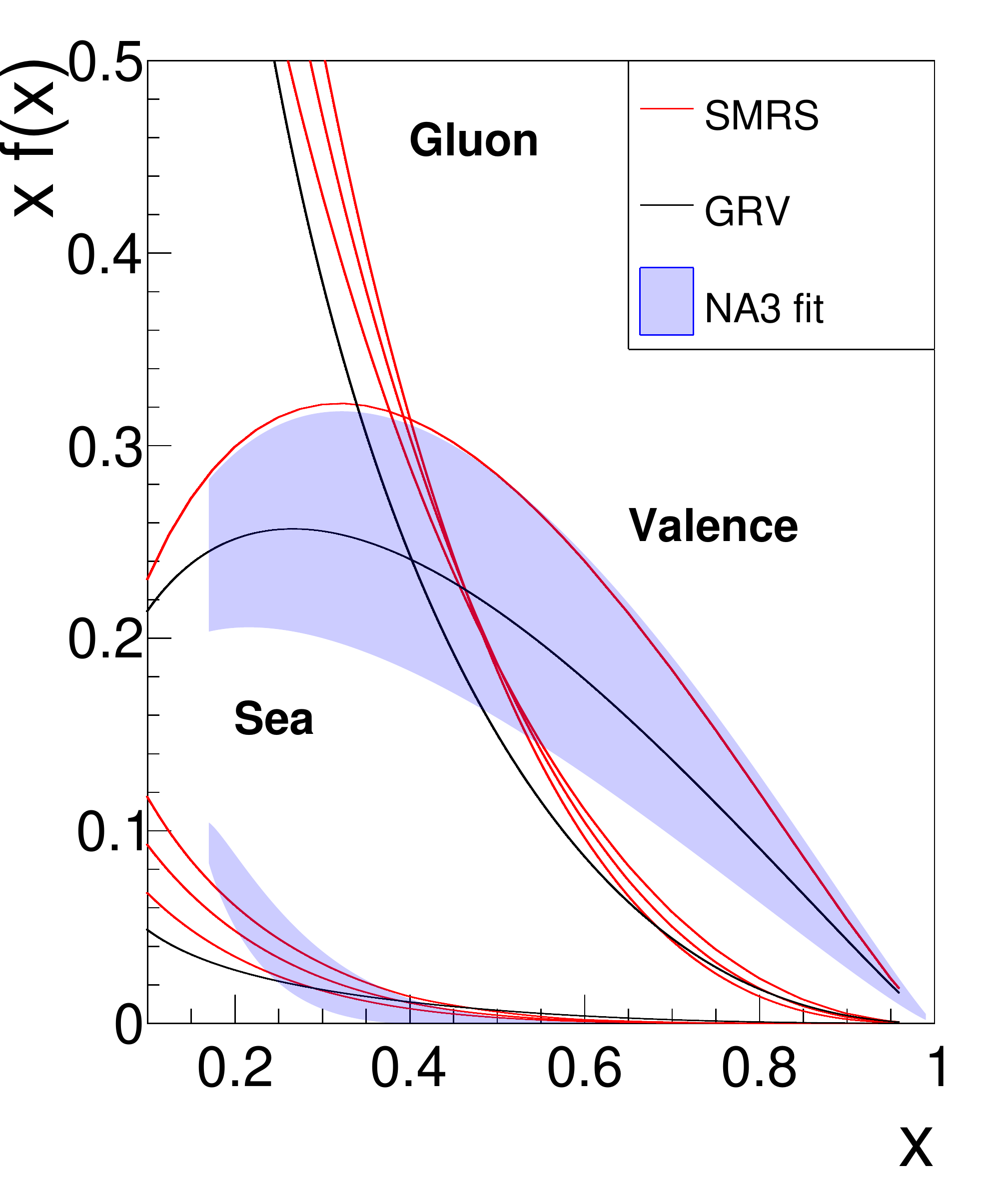}
\caption{Valence, sea and gluon distributions for the pion from global fits of SMRS~\cite{Sutton:1991ay}  and GRV/S~\cite{Gluck:1991ey,Gluck:1999xe}, shown together with the NA3 extraction~\cite{Badier:1983mj}.
The three sea curves labelled SMRS correspond to three different hypotheses for the sea content. As a result, there are also three curves  for the gluon contribution. The shaded area accounts for the experimental uncertainty, evaluated using the published NA3 data.}
\label{fig:SeaValPion}
\end{figure}

\subsubsection{Separation of valence and sea-quark contributions in the pion}
\label{sssec:seaval}
Pion-induced Drell-Yan data were collected by the NA3~\cite{Badier:1983mj}, NA10~\cite{Betev:1985pg}, and WA39~\cite{corden:1980} collaborations at CERN and by the E615~\cite{Conway:1989fs} collaboration at Fermilab. Drell-Yan data using both $\pi^+$ and $\pi^-$ beams were taken by NA3 and WA39. 
Figure~\ref{fig:SeaValPion} shows the pion valence distribution from two global analyses, SMRS~\cite{Sutton:1991ay} and GRV/S~\cite{Gluck:1991ey,Gluck:1999xe}. These extractions rely on the $\pi^-$ Drell-Yan data from E615 and NA10, and do not include uncertainty estimates. In the analysis of GRV/S the sea content is derived from momentum conservation, while the gluon contribution is constrained by the direct-photon measurements of WA70~\cite{Bonesini:1987mq} and NA24~\cite{deMarzo:1987}.                                                     
Here, $x_\pi$ is the momentum fraction carried by the active beam parton.                   Sutton et al.~\cite{Sutton:1991ay} provide their own parameterisation for the sea, assuming that the sea contribution amounts to 10\%, 15\%, and 20\%, which then leads to three different results for the gluon contribution. The valence and sea distributions from NA3 are also shown, together with the respective error bands, as calculated using the published fit coefficients and correlation matrix.

We propose to determine the shape of the sea-quark distribution in the pion for values of $x_{\pi}$ larger than 0.1 and better constrain this contribution by collecting data with pion beams of positive and negative charge using an isoscalar target, as proposed by \cite{Londergan:1995wp}. Assuming charge conjugation, SU(2)$_{f}$ symmetry for valence quarks and SU(3)$_{f}$ symmetry for sea quarks, it is possible to build the two linear combinations:
\begin{equation}
  \Sigma_{val}^{\pi D}=-\sigma^{\pi^+D}+\sigma^{\pi^-D}
  \label{eq:SigmaVal}
\end{equation}
\begin{equation}
  \Sigma_{sea}^{\pi D}=4\sigma^{\pi^+D}-\sigma^{\pi^-D}
\label{eq:SigmaSea}
\end{equation}

The first combination contains only valence-valence terms, while the second one comprises only sea-valence and valence-sea terms~\cite{Londergan:1995wp}. If nuclear effects are neglected, the ratio $\Sigma_{sea}/\Sigma_{val}$ can be evaluated for each of  the measured $x_N$ values. Here, $x_N$ is the momentum fraction carried by the active target parton. The use of a light isoscalar carbon target instead of the non-isoscalar platinum and tungsten targets used by NA3 and NA10, respectively, reduces nuclear effects.

\begin{figure}[tb]
\centering
\includegraphics[width=.80\textwidth]{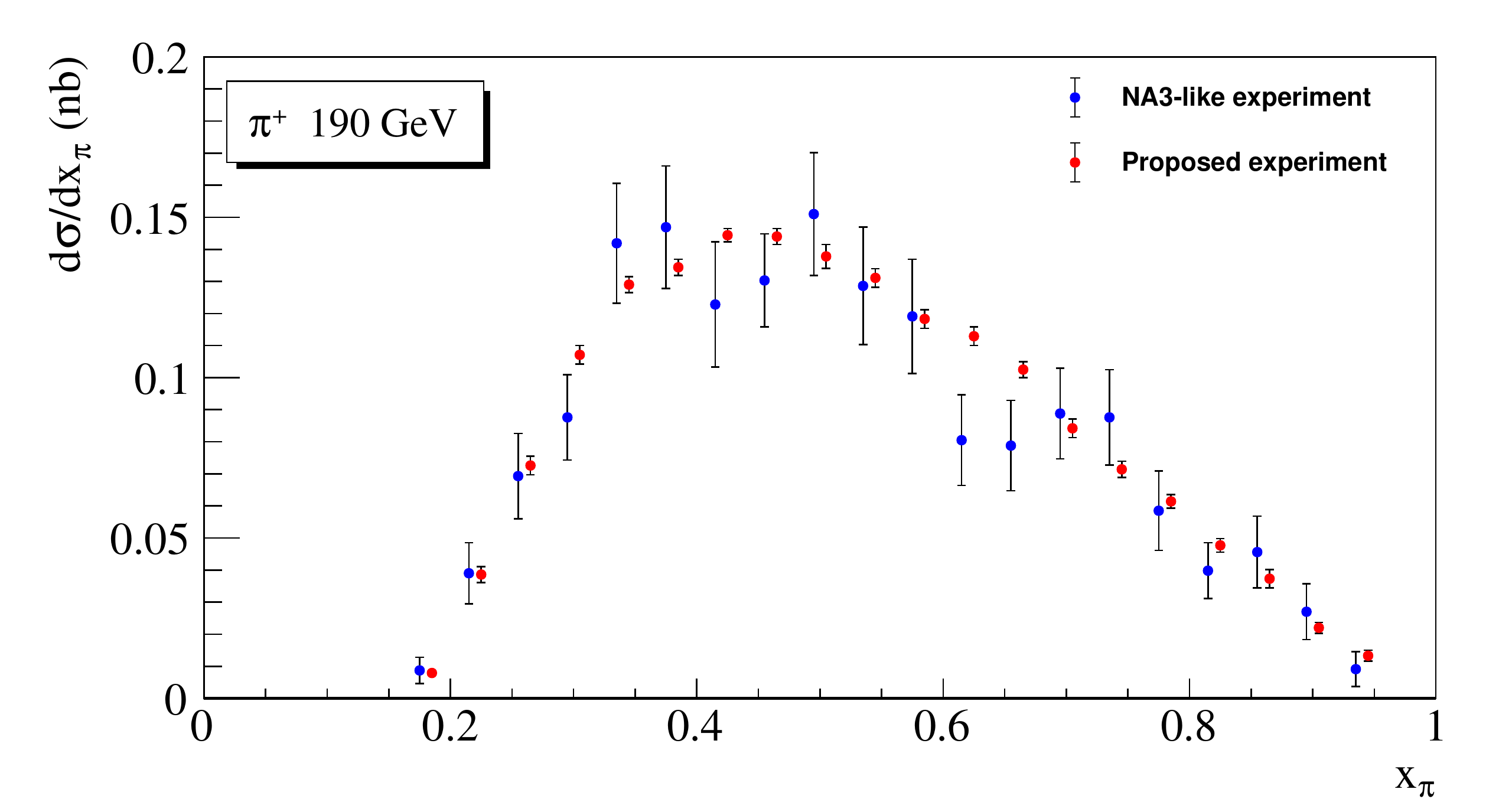} 
\caption{Expected statistical accuracy of the Drell-Yan cross sections as a function of $x_\pi$ for a carbon target and a $\pi^+$ beam. 
The accuracy achievable with a NA3-like experiment, estimated from the reported~\cite{Badier:1983mj} number of events, is shown for comparison. In order to avoid overlapping symbols, the data points were slightly shifted horizontally.}
\label{fig:pion-positive}
\end{figure}

In order to have minimal systematic uncertainties when evaluating Eqs.~\ref{eq:SigmaVal} and \ref{eq:SigmaSea}, precise cross-section determinations are required. We aim at absolute cross-section measurements at the level of a 3\% systematic uncertainty. A cross-check of the relative normalisation can be performed by comparing the J/$\psi$ cross sections for $\pi^-$ and $\pi^+$, as for isoscalar targets the J/$\psi$ cross  section does not depend on the charge of the incident meson. The cross-section ratio for  $\pi^-$ and $\pi^+$-induced J/$\psi$ production on a Platinum target was measured to be $1.016\pm0.006$) by NA3~\cite{Badier:1983dg}.

The relative contribution from the sea quarks increases as $x_{\pi}$ decreases. Therefore, a good separation between sea and valence quarks requires as low as possible $x_{\pi}$ values, and an as high as possible incident pion momentum. For a reasonable geometrical acceptance down to $x_F = -0.2$ and an incident momentum of $190\,\GeV/c$, values as low as $x_{\pi}=0.10$ can be reached.


Figure~\ref{fig:pion-positive} shows the achievable cross-section accuracy of the proposed experiment,  as simulated with PYTHIA at leading order, corrected by a K-factor of $K=2$ that is consistent with observations made by past  experiments. Represented is the option of 255 days of data taking using a $\pi^+$ beam with   momentum of $190\,\GeV/c$ and four consecutive carbon targets of 25 cm length each. For the sake of comparison, the statistical accuracy that would be achieved by a NA3-like experiment~\cite{Badier:1983mj} is also shown. Since the NA3 cross sections are not available, the comparison is made assuming the number of $\pi^+$-induced Drell-Yan events on a Pt target, as reported in their publication.
 
The $\pi^+$ data should be complemented with  25 days data taking with a $\pi^-$ beam.  The difference in data-collection time between the  two beam charges is explained by the Drell-Yan cross-section difference itself and by the different  hadron compositions for positive and negative beams. Altogether, the two effects lead to a ratio of 10:1  between the $\pi^+$ and $\pi^-$ running times. Beam intensities  of $7\times10^7$ particles per second, with two pulses of 4.8 s in each SPS  super-cycle of 52~s are assumed. The fraction of pions in the positive and negative hadron beams is 24\% and 97\%, respectively. A beam-particle identification efficiency for the CEDAR detectors of 90\% is assumed.  The product of other efficiencies, acceptance and live time is estimated to be 0.13~. 

\begin{figure}[tb]
\centering
\includegraphics[width=.80\textwidth]{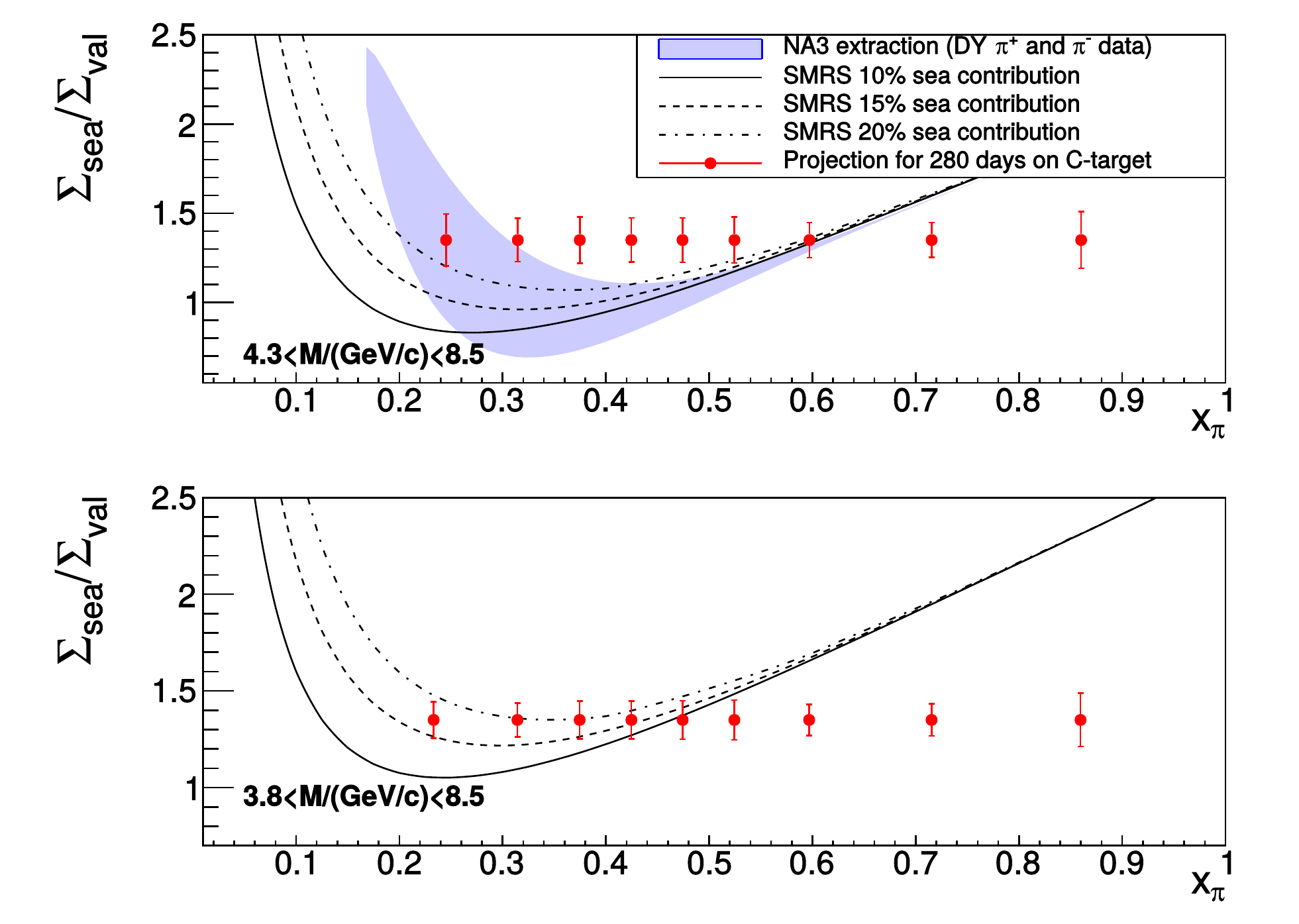}
\caption{The ratio $\Sigma_{sea}/\Sigma_{val}$ as a function of $x_\pi$, using three different sea-quark distributions from
\cite{Sutton:1991ay}. The top and bottom panels show the statistical accuracy
that would be achieved in two of the three mass ranges discussed in the text. The shaded area in the upper panel is the uncertainty derived from the uncertainties quoted in the NA3 paper~\cite{Badier:1983mj}.}
\label{fig:pion-separation}
\end{figure}

Figure~\ref{fig:pion-separation} shows accuracy estimates for the ratio $\Sigma_{sea}/\Sigma_{val}$ as a function of  $x_\pi$, in three possible dimuon mass ranges. The top panel represents the background-free Drell-Yan mass range. The bottom panels incorporate the assumption that machine-learning techniques will succeed in isolating the Drell-Yan contribution from competing processes. The curves labelled SMRS represent the predictions~\cite{Sutton:1991ay}  for three possible contributions of the sea quarks to the pion momentum, ranging from 10\% to 20\%. The three different assumptions for the pion sea yield increasingly different predictions for $x_{\pi}$ values below 0.5. The shaded area in the top panel represents the uncertainty band on the sea distribution as estimated from the sea/valence separation of NA3 that is based solely on their own data.

Recently developed techniques of data analysis are planned to be employed in order to disentangle the different physics contributions to the dimuon mass spectrum. Machine-learning algorithms allow to combine data in clusters according to their behaviour in terms of a set of physics variables. Models are then used to attribute a physics origin to each given set. The clustered data are used to train neural networks by classifying each event according to its probability to originate from a particular physics process. The whole method can be validated using Monte-Carlo samples. It is presently being developed  for use in the COMPASS experiment to process the collected Drell-Yan data. With such an approach  the analysis of the Drell-Yan events could be extended beyond the traditionally considered "safe range" of $4.3 < M < 8.5\,\GeV$, which is free from contamination, to the region $3.8 < M < 8.5\,\GeV$,  which is still dominated by the Drell-Yan process.  Eventually, it may become possible to apply this method in the very challenging lower-mass range $2.0 < M < 3.8\,\GeV$, where charmonia provide the  dominant contribution together with a non-negligible fraction of semi-leptonic open-charm decays into muons, all mixed with the Drell-Yan contribution.

In Table~\ref{table:pipluspiminus} the achievable statistics of the proposed experiment for a running period of two years, i.e. $2 \times 140$ days, is compared to the Drell-Yan statistics of earlier experiments. In the experimental  conditions assumed above, the sea contribution to the pion momentum could be evaluated with an accuracy of 5\% or better.

\begin{table}[h]
\centering
{\scriptsize
\begin{tabular}{lcccccc}\toprule
 Experiment                   & Target type                  & Beam energy (GeV)      & Beam type   & Beam intensity (part/sec) & DY mass (GeV/c$^2$) & DY events \\\midrule\midrule
 \multirow{2}{*}{E615}        & \multirow{2}{*}{20\,cm W}      &  \multirow{2}{*}{252}  & $\pi^+$     & $17.6 \times 10^7$                & \multirow{2}{*}{4.05 -- 8.55} & 5000 \\
                              &                              &                        & $\pi^-$     & $18.6 \times 10^7$                &                               & 30000 \\\midrule
 \multirow{4}{*}{NA3}         & \multirow{2}{*}{30\,cm H$_2$}  &  \multirow{2}{*}{200}   & $\pi^+$     & $2.0 \times 10^7$                 & \multirow{2}{*}{4.1 -- 8.5} & 40 \\
                              &                              &                         & $\pi^-$     & $3.0 \times 10^7$                 &                             & 121 \\\cmidrule{2-7}
                              & \multirow{2}{*}{6\,cm Pt}      &  \multirow{2}{*}{200}  & $\pi^+$     & $2.0 \times 10^7$                 & \multirow{2}{*}{4.2 -- 8.5} & 1767 \\
                              &                              &                        & $\pi^-$     & $3.0 \times 10^7$                 &                            & 4961 \\\midrule
 \multirow{5}{*}{NA10}        & \multirow{2}{*}{120\,cm D$_2$} &  286                   & \multirow{2}{*}{$\pi^-$}  & \multirow{2}{*}{$65 \times 10^7$} & 4.2 -- 8.5 & 7800 \\
                              &                              &  140                   &                          &                                   & 4.35 -- 8.5 & 3200 \\\cmidrule{2-7}
                              & \multirow{3}{*}{12\,cm W}      &  286                   & \multirow{3}{*}{$\pi^-$} & \multirow{3}{*}{$65 \times 10^7$} & 4.2 -- 8.5 & 49600 \\
                              &                              &  194                   &                          &                                   & 4.07 -- 8.5 & 155000 \\
                              &                              &  140                   &                          &                                   & 4.35 -- 8.5 & 29300 \\\midrule
 COMPASS 2015                 & \multirow{2}{*}{110\,cm NH$_3$}&\multirow{2}{*}{ 190} & \multirow{2}{*}{$\pi^-$}& \multirow{2}{*}{$7.0 \times 10^7$}& \multirow{2}{*}{4.3 -- 8.5} & 35000 \\
 COMPASS 2018                 &                               &                         &    &                                   &                            & 52000 \\\midrule\midrule

\multirow{12}{*}{\color{red}This exp}  &\multirow{6}{*}{100\,cm C} &\multirow{3}{*}{190}& \multirow{3}{*}{$\pi^+$} & \multirow{3}{*}{$1.7 \times 10^7$} & 4.3 -- 8.5 & 23000 \\
                                       &                         &                         &     &                                    & 3.8 -- 8.5 & 37000 \\
                                       &                         &                         &     &                                    & 2.0 -- 8.5 & 170000 \\\cmidrule{3-7}
                                       &                         &\multirow{3}{*}{ 190}& \multirow{3}{*}{$\pi^-$} & \multirow{3}{*}{$6.8 \times 10^7$} & 4.3 -- 8.5 & 22000 \\
                                       &                         &                         &     &                                    & 3.8 -- 8.5 & 34000 \\
                                       &                         &                         &     &                                    & 2.0 -- 8.5 & 161000 \\\cmidrule{2-7}
                                       &\multirow{6}{*}{24\,cm W}  &\multirow{3}{*}{190}& \multirow{3}{*}{$\pi^+$ } & \multirow{3}{*}{$0.2 \times 10^7$} & 4.3 -- 8.5 & 7000 \\
                                       &                         &                          &    &                                    & 3.8 -- 8.5 & 11000 \\
                                       &                         &                          &    &                                    & 2.0 -- 8.5 & 51000 \\\cmidrule{3-7}
                                       &                         &\multirow{3}{*}{ 190}& \multirow{3}{*}{$\pi^-$} & \multirow{3}{*}{$1.0 \times 10^7$} & 4.3 -- 8.5 & 6000 \\
                                       &                         &                         &     &                                    & 3.8 -- 8.5 & 9000 \\
                                       &                         &                         &     &                                    & 2.0 -- 8.5 & 48000 \\
\bottomrule
\end{tabular}}
\caption{Statistics collected by earlier experiments (top rows), compared with the achievable statistics of the new experiment (bottom rows).}
\label{table:pipluspiminus}
\end{table}

\subsubsection{J/$\psi$ production mechanism and the gluon distribution in the pion}
\label{ssec:stdbeam.dryan.Jpsi}
Charmonium production provides a particularly attractive alternative for investigating the meson structure. The cross sections are large, typically a factor of 20 to 30 higher in comparison with the Drell-Yan process. While at collider energies the main contribution to the cross section comes from gluon-gluon fusion, at the relatively low fixed-target  energies both quark-antiquark annihilation and gluon-gluon fusion processes do contribute.  The different quark and gluon densities then lead to different $x_F$ dependences.  Separating the two contributions should, with minimised model uncertainties, allow access to the parton distributions  of the beam particle.

Analyses aiming at a determination of the gluon distribution in the pion, $g_{\pi}(x)$, were  performed by some of the pioneering dimuon production  experiments, NA3~\cite{Badier:1983dg} and WA11~\cite{McEwen:1982fe} at CERN. Assuming that the J/$\psi$ production cross section at sufficiently low $x_F$ proceeds through gluon-gluon fusion, the authors provide phenomenological fits to the data using the simple parameterisation
\begin{equation}
g_{\pi}(x)=A(1-x)^{\beta},
\end{equation}
where the parameter $\beta$ describes the slope of the $g_{\pi}(x)$ distribution as a function  of $x_{\pi}$, and A is a normalisation factor. A similar analysis was later done by the E537 collaboration at Fermilab~\cite{Akerlof:1990ij}, using Be and W targets. In all these experiments, the determination of $g_{\pi}(x)$ depends on the assumptions for  the fraction of J/$\psi$ (or $\Upsilon$) events produced by gluon-gluon fusion and on the  knowledge of the nucleon or nuclear gluon distribution.  Not surprisingly, the corresponding uncertainties  lead to a large spread of the derived $\beta$ values.

The large number of J/$\psi$ dimuon events, which will be collected simultaneously with the proposed pion-induced Drell-Yan data, should greatly help in improving the situation. About one million events would be expected for the combined $\pi^+$ and $\pi^-$ data on a $^{12}$C target and about half of this number on a $^{184}$W target, allowing for a precise determination of the  corresponding $x_F$ and $p_T$ distributions. 


The extraction of $g_{\pi}(x)$ from the data relies on a good understanding of  the J/$\psi$ production mechanism. Two basic models are used to describe J/$\psi$ production. The simpler Colour Evaporation Model (CEM) was shown~\cite{Gavai:1994in}  to successfully describe cross sections and momentum distributions.  
The more recent and more rigorous Non-Relativistic QCD  model~\cite{Bodwin:1994jh} (NRQCD) explicitly uses colour and spin to calculate the various charmonium states. It separates the short-distance  perturbative and the long-distance non-perturbative effects. 
\begin{figure}[tb]
\centering
  \includegraphics[width=.70\textwidth]{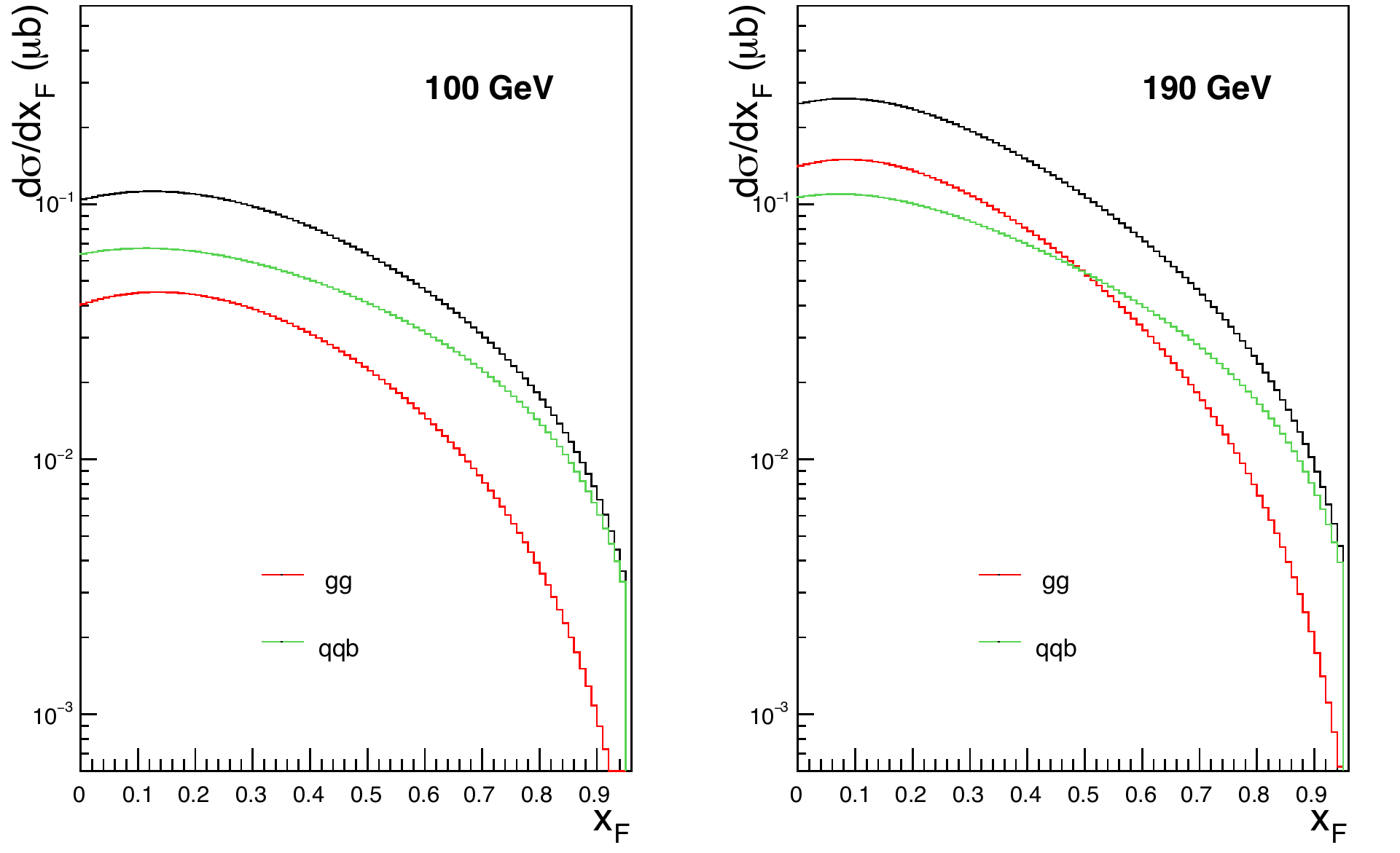}
  \caption{Pion-induced J/$\psi$ cross sections for a $^{12}$C target as a function of $x_F$ for incident energies of $100\,\GeV$ (Left) and $190\,\GeV$ (Right). The red, green and black curves show the $gg$, $q\bar q$, and  total cross sections, respectively.}
  \label{fig:Jpsi-qqgg}
\end{figure}

In both models, the cross section is a sum of two main contributions: $q\bar q$ annihilation and $gg$ fusion. A detailed study of their $x_F$ dependence~\cite{Vogt:1999dw} shows that the two models qualitatively agree: the $gg$ term  dominates at low $x_F$, whereas the $q\bar q$ term becomes important at large $x_F$. Furthermore, when lowering the incident energy, the $gg$ contribution to the cross section decreases. 
Figure~\ref{fig:Jpsi-qqgg} shows the different $gg$ and $q\bar q$ cross-section contributions at  $100\,\GeV$ and $190\,\GeV$, as obtained using the LO CEM model.  For both energies the same scaling factor of 0.389 is used, as fitted to the J/$\psi$ production data~\cite{Badier:1983dg} on a Pt target.
At $100\,\GeV$ the $q\bar q$ contribution dominates the cross section over the entire range of $x_F$, whereas at $190\,\GeV$ it is larger for $x_F > 0.5$ only.

The strong sensitivity of the J/$\psi$ cross section to the relative amount of  $gg$ and $q\bar q$ contributions as a function of the energy could then be used  to constrain the J/$\psi$ production models at fixed-target energies.

\svnid{$Id: stdhad_dryan_2.tex 4411 2018-10-12 10:18:43Z splatchk $} 
\subsubsection{Nuclear dependence studies: Flavour-dependent valence modifications}
\label{sec:stdhad.dryan.nudep}

The parton distributions in a bound nucleon differ from those in a free nucleon. More than thirty years ago, a measurement made by the European Muon Collaboration (EMC) showed ~\cite{Aubert:1983xm} that medium modifications  can play a significant role for nuclear observables. Since then, an impressive amount of DIS data taken in several  laboratories around the world has been accumulated~\cite{Norton:2003cb,Rith:2014tma}. One of the main  findings of these studies is that quarks play an important role in the determination of the properties of nuclei. On  the theoretical side, many models have been proposed, but a satisfactory explanation of the EMC effect is still missing ~\cite{Hen:2016kwk}. The situation has recently become more perplexing, after a JLab experiment on light  nuclei~\cite{Seely:2009gt} provided evidence that the nuclear dependence is not always a function of the atomic number or  the mean nuclear density.

Inclusive DIS experiments are only sensitive to the charge-weighted sum of the quark and antiquark distributions.  However, nuclear effects could be different for up and down quarks. Here, Drell-Yan experiments can play a major role,  as with different pion beam charges one or the other valence quark distribution is preferentially accessed.

\begin{figure}[tb]
\centering
  \includegraphics[width=.7\textwidth]{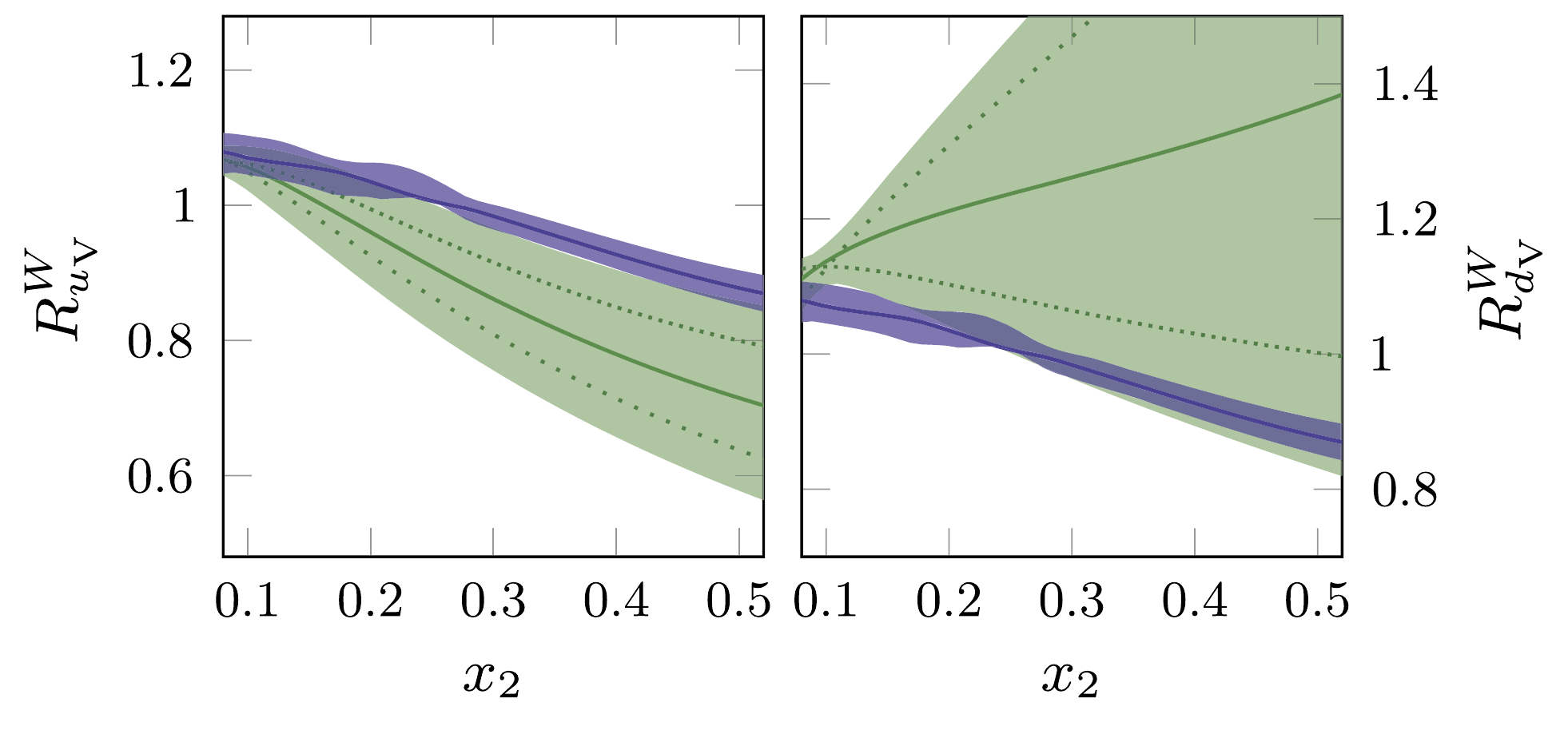}
  \caption{The modification of $u_v$ (left) and $d_v$ (right) distributions in tungsten, as obtained by the  nCTEQ15 global fit in green, and by the EPS09 global fit in blue. From Paakkinen et al. \cite{Paakinen:2017}.}
  \label{fig:paakinen}
\end{figure}

The possibility of flavour-dependent PDF modifications in nuclei was raised by several authors, most recently by ~\cite{Paakinen:2017}. The inclusion of pion-induced Drell-Yan data, together with the independent constraints on up  and down quarks, may have strong impact on global fits of nuclear PDFs. This is illustrated in  Fig.~\ref{fig:paakinen}, where the results for the modification of the valence distribution in tungsten found by the  nCTEQ15 group~\cite{nCTEQ15:2016} releases such quark flavour constraints, but includes no data effectively  constraining it. This leads to the over-estimated green error bands. On the contrary, the EPS09~\cite{EPS09:2009} extraction  shown by the blue band in the figure, which imposes the same nuclear modifications for up and down quarks, severely  underestimates the uncertainties. The potential impact of pion-induced Drell-Yan data becomes evident. 

A new measurement would aim at a precise pion-induced DY measurement in order to evaluate the  EMC effect on the valence quarks. In order to allow for such nuclear-dependence studies, a tungsten target is  proposed to be placed 40 cm downstream of the carbon target discussed above. A comparison between the Drell-Yan  data collected with both positive and negative pion beams on tungsten should allow for a flavour-dependent study of  the nuclear effects.

A recent calculation~\cite{Cloet:2009qs} based on the Nambu-Jona-Lasinio (NJL) model was used to evaluate the nuclear  quark distributions inside a large-A nucleus. A remarkable success of the NJL model is that it accounts ~\cite{Cloet:2009qs} for a large fraction of the so-called NuTeV anomaly of the weak mixing angle.  An important feature of this calculation is that for nuclei with N$>$Z, the isovector mean field affects the lighter quarks differently as compared to the heavier ones, which leads to the prediction of different nuclear modifications for $u$ and $d$ quarks. 

%

\begin{figure}[tb]
\centering
  \includegraphics[width=.7\textwidth]{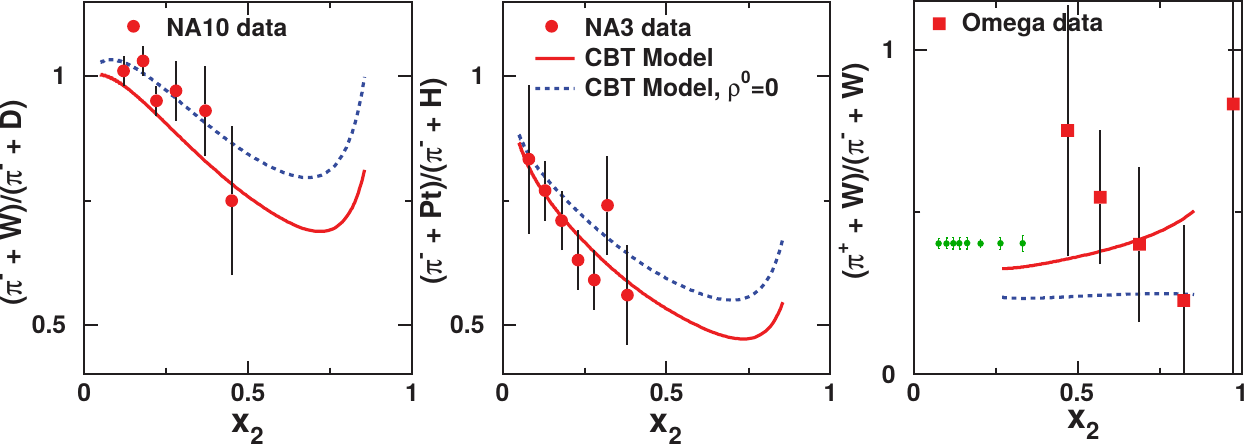}\\
  \includegraphics[width=.7\textwidth]{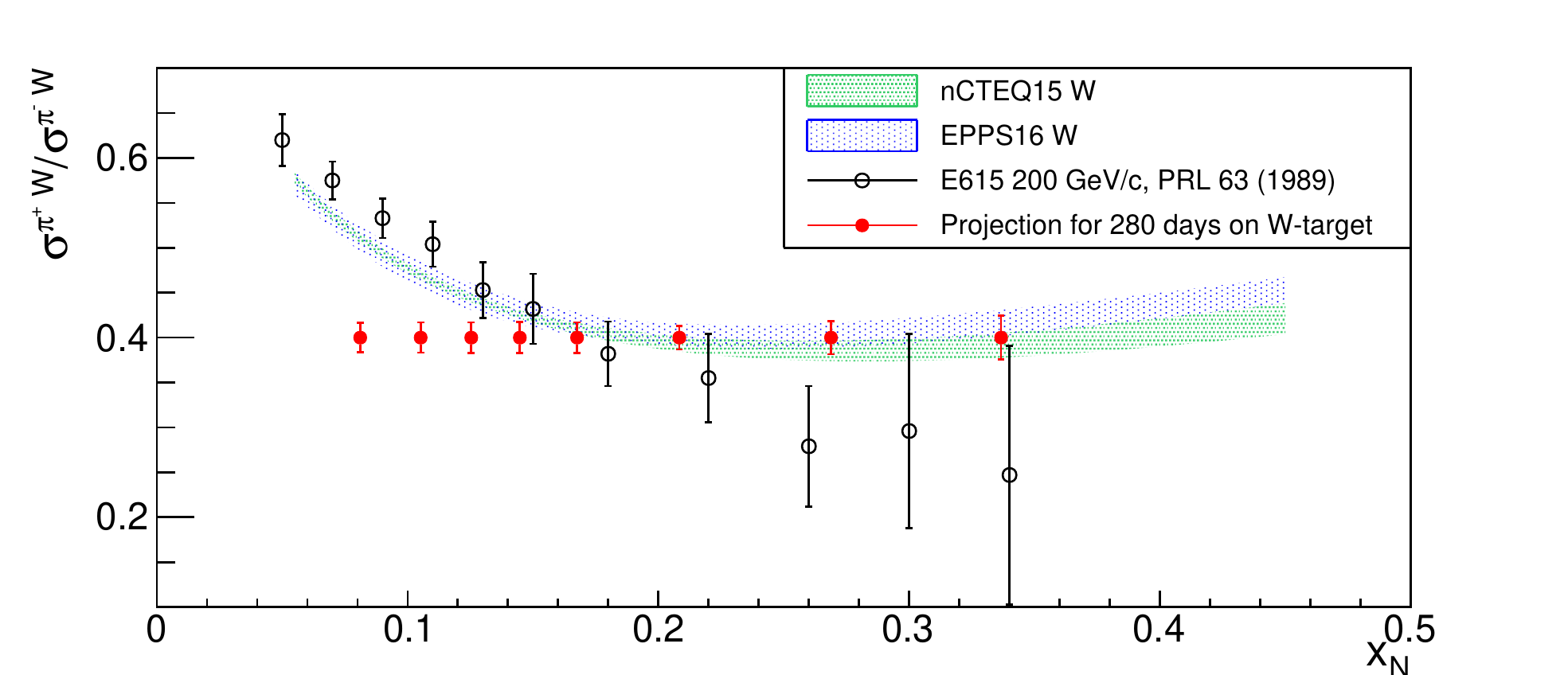}\\
  \includegraphics[width=.7\textwidth]{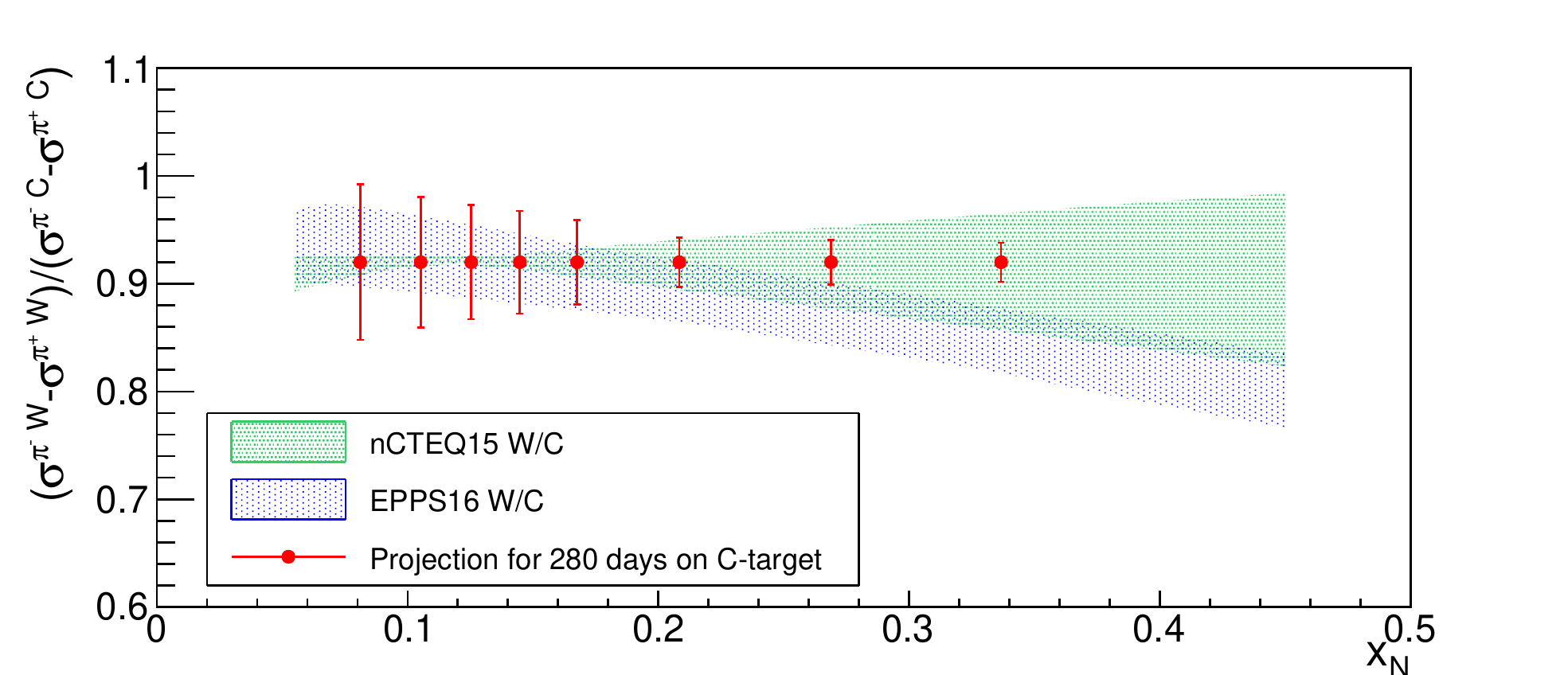}\\
  \caption{Top row: The Dutta et al. CBT model calculations~\cite{Dutta:2010pg} are compared to 
NA10~\cite{Bordalo:1987cs}, NA3~\cite{Badier:1981ci} and Omega (WA39)~\cite{corden:1980} ratios. Note that  the projected statistical uncertainties for the proposed experiment are also shown as small green dots together  with Omega data in the rightmost panel. Middle row: Drell-Yan cross-section ratio for positive-over-negative pion  beam polarity, shown \textit{vs.} $x_N$. The expected statistical uncertainties from the proposed experiment (shown as  full red dots) are compared to E615 results and two sets of nuclear PDFs. Bottom row: Drell-Yan ratio of the cross-section beam-charge differences for tungsten over carbon, shown \textit{vs.} $x_N$. The expected accuracy of the  proposed experiment is shown together with two set of nuclear PDFs.}
  \label{fig:flav-dep}
\end{figure}

Using the Cloet-Bentz-Thomas (CBT) model~\cite{Cloet:2006bq,Cloet:2009qs}, Dutta et al.~\cite{Dutta:2010pg}  have explored the sensitivity of a future Drell-Yan experiment to the flavour-dependent EMC effect. The data  available from the NA10 experiment~\cite{Bordalo:1987cs} seem to be in agreement with the flavour-dependent PDFs,  although a better accuracy is necessary to confirm the effect. This is shown in the top row of  Fig.~\ref{fig:flav-dep}, where the right-most panel presents the expected accuracy of the proposed measurement of  the cross-section ratio $\sigma^{\pi^+ W}/\sigma^{\pi^- W}$.


The same experimental conditions as described in Sec.\ref{sssec:seaval} are considered. PYTHIA simulations at  leading order with a K-factor of K=2 are performed for a "nuclear" target with a given composition of protons  and neutrons. The projected statistical uncertainties on the Drell-Yan cross-section ratio of positive-over-negative pion beam polarity on tungsten are shown in the middle panel of Fig. \ref{fig:flav-dep}. The results are compared to  the previous measurement performed by E615~\cite{Heinrich:1989jg} and to a leading-order calculation using two  recent nuclear PDFs. The bottom panel of Fig.~\ref{fig:flav-dep} shows another observable,  $\frac{\sigma^{\pi^- W} - \sigma^{\pi^+ W}}{\sigma^{\pi^- C} - \sigma^{\pi^+ C}}$, which was introduced by \cite{Paakinen:DIS}, where the sensitivity to the nuclear valence asymmetry is enhanced,  as it can be inferred from the larger error bands. This new observable makes the best usage of the statistics to be  collected by the proposed experiment.

In parallel to the collection of Drell-Yan events, the proposed new measurement will also accumulate a large  statistics of J/$\psi$ events. The comparison of the pion-induced J/$\psi$ production cross section for a heavy  target relative to that for a light isoscalar target could therefore be used to attempt accessing the nuclear  gluon distribution, assuming that a separation of the $q\bar q$ and $gg$-fusion processes can be performed.

\subsubsection{Drell-Yan and $J/\psi$ angular distributions}
\label{sec:stdhad.dryan.angular}

In parallel to the pion-structure measurements discussed above, studies of the Drell-Yan and J/$\psi$ angular distributions will be performed. The data collected with both pion beam polarities will be used.  A light isoscalar target, like deuterium or carbon, will provide results complementary to those that will be  obtained using the COMPASS data already taken with an ammonia target.

\subsubsection{Run plan: physics goals and required beam time}
\label{ssec:stdhad.dryan.timeline}

CERN is presently the only place in the world where high-energy and high-intensity hadron beams of both  polarities are available. The beam intensity in the target region is presently limited by radio-protection  constraints. An improved radiation shielding in the target region would allow for a wider opening of the beam  line collimators, thereby allowing for an increase of the beam intensity to be used in the proposed measurements. The  estimates presented above are conservative ones, which do not take into account these possible improvements.  Once implemented, these modifications could reduce the two years of data-taking to one year, still reaching  identical statistics.

The present COMPASS apparatus efficiently complements the uniqueness of the CERN M2 beam line. With two large spectrometers, the setup has a large geometrical acceptance of nearly 40\%. This acceptance compares extremely favourably to the acceptance of previous Drell-Yan experiments, which were limited to about 10\%. In addition, the azimuthal acceptance is quite uniform.

Presently, the Drell-Yan data analysis concentrates on the high mass region,  $4.3$\,GeV$ < M_{\mu^+\mu^-} < 8.5$\,GeV. This high-mass region avoids background from decays of heavy  vector mesons and for lower masses also combinatorial background. New analysis tools based on machine-learning  techniques are expected to extend the data analysis to the region below $4.3$\,GeV and increase the available  Drell-Yan statistics by a large factor. Estimates based on such future improvements are shown in  Table~\ref{table:pipluspiminus}.

The proposed Drell-Yan physics programme relies on the development of new and improved instrumentation. In  order to improve the present beam identification, tracking efficiency and energy resolution, we would need to:

\begin{itemize}
\item improve the read-out of the two CEDAR detectors to reach a beam-PID efficiency higher than 90\% and high purity;
\item foresee a dedicated detector for a precise luminosity measurement, which reaches a precision of the order of 3\%;
\item install beam trackers to achieve a precise beam reconstruction;
\item build a dedicated vertex-detection system for improved vertex resolution;
\item design a high-purity and high-efficiency dimuon trigger with target pointing capability.
\end{itemize} 

We propose to run for 2 equivalent years with both positive and negative hadron beams with a share of 10:1,  as explained previously. The target setup will include from upstream to  downstream: (1) a segmented carbon target, (2) a small tungsten target and (3) the tungsten beam plug,  the first part of which will also be used as target. New vertex detectors and beam counters will be  placed downstream of each sub-target. The requirement of good statistics to be taken with a positive pion  beam puts an additional constraint on the choice of the incident beam momentum. For a nominal momentum  of 190 GeV/$c$, the fraction of the positive pions in the beam is 24\%. This fraction could be further  increased by installing a passive polyethylene absorber along the beam path. Due to different interaction  lengths, protons in the beam are more absorbed  than pions. With a 2\,m long absorber, the NA3 experiment reached a $\pi^+$ fraction of 36\% at 200\,GeV/$c$.  For an incident momentum of 190\,GeV/$c$, this translates into a pion fraction of about 40\%.  If necessary  from counting-rate considerations, this fraction could be further increased by up to 50\% or more by choosing  a slightly lower incident momentum, e.g.\ 160\,GeV/$c$.

In all proposed measurements with negative hadron beams a good separation between pions and kaons is mandatory.  For the positive hadron beam, the challenge is to identify the 24\% pions out of the most abundant protons.  An excellent beam particle tagging system, with an efficiency at the level of 90\% or higher, is hence  mandatory for the success of the program. This may be achieved by the ongoing upgrade of the present CEDARs  (differential Cherenkov counters) that are used in the COMPASS experiment, or by new threshold Cherenkov detectors.

Assuming a 10:1 time share between beams with positive and negative polarity, the number of Drell-Yan events  that could be collected in two "years". i.e. two times 140 effective physics data-taking days, would lead to  a statistical accuracy better by an order of magnitude than that of NA3.

\subsubsection{Worldwide competition}

High-energy pion beams are exclusively available at CERN. Secondary meson beam lines are also under  construction at the J-PARC facility in Japan. However, their energy of up to 15\,GeV remains too low for extensive studies of the pion structure.  

The only alternative way of accessing the form factors and the parton distribution functions of the pion is  a model-dependent one and relies on the validity of the pion-cloud model. The investigation of the pion  structure through leading-neutron DIS electro-production was initiated~\cite{Aaron:2010ab} at HERA. While  these experiments cover the $x$ region below $x=0.01$, the resulting extraction of the pion sea suffers from  large model uncertainties, which are mainly coming from the unknown normalisation of the pion flux. An  experiment at JLab~\cite{Camsonne:2014pr} proposes to make similar measurements in the large $x$ region, while  normalising the pion flux using the available Drell-Yan data.  

Concerning the proposed J/$\psi$ studies, there are currently no other laboratories where pion-induced charmonium  production can be investigated.

Studies of the nuclear sea-quark distribution have just been completed by the SeaQuest experiment ~\cite{Arrington:2006} at Fermilab. Using the proton-induced Drell-Yan process at incident momentum of  120\,GeV/$c$, this experiment probes the antiquark distributions in nuclei. If combined with future Drell-Yan  data for the valence quarks, as detailed in this proposal, the two experiments are complementary. 

The JLab EMC PVDIS experiment~\cite{Riordan:2014pr} proposes to investigate possible effects of flavour-dependent  nuclear-medium modifications using parity-violating deep inelastic scattering on a ${}^{48}$Ca target,  as suggested in Ref.~\cite{Cloet:2012td}. Such modifications could impact the extracted nuclear PDFs.

\svnid{$Id: stdhad_spect.tex 4404 2018-10-08 18:17:46Z avauth $} 
\subsection{Spectroscopy with low-energy antiprotons}
\label{sec:stdhad.spect}

\subsubsection{Physics case}
\label{sec:lepbar.physics}

Although conceptually rather simple, the strong interaction between quarks and gluons is still far from being understood. At distance scales much smaller than the size of a nucleon, perturbative methods are routinely being used to make precision calculations of strong-interaction effects. The perturbative approach, however, fails  dramatically at distances approaching the nucleon size, when the strong  coupling constant $\alpha_\mathrm{s}$ is of order   unity and where pions and other light hadrons become the relevant degrees of freedom.    In this regime, the spectroscopy of hadrons is a powerful tool towards a better understanding of the strong interaction between quarks and gluons. 

\begin{figure}[bp]
  \begin{center}
    \hfill
    \includegraphics[width=0.8\columnwidth]{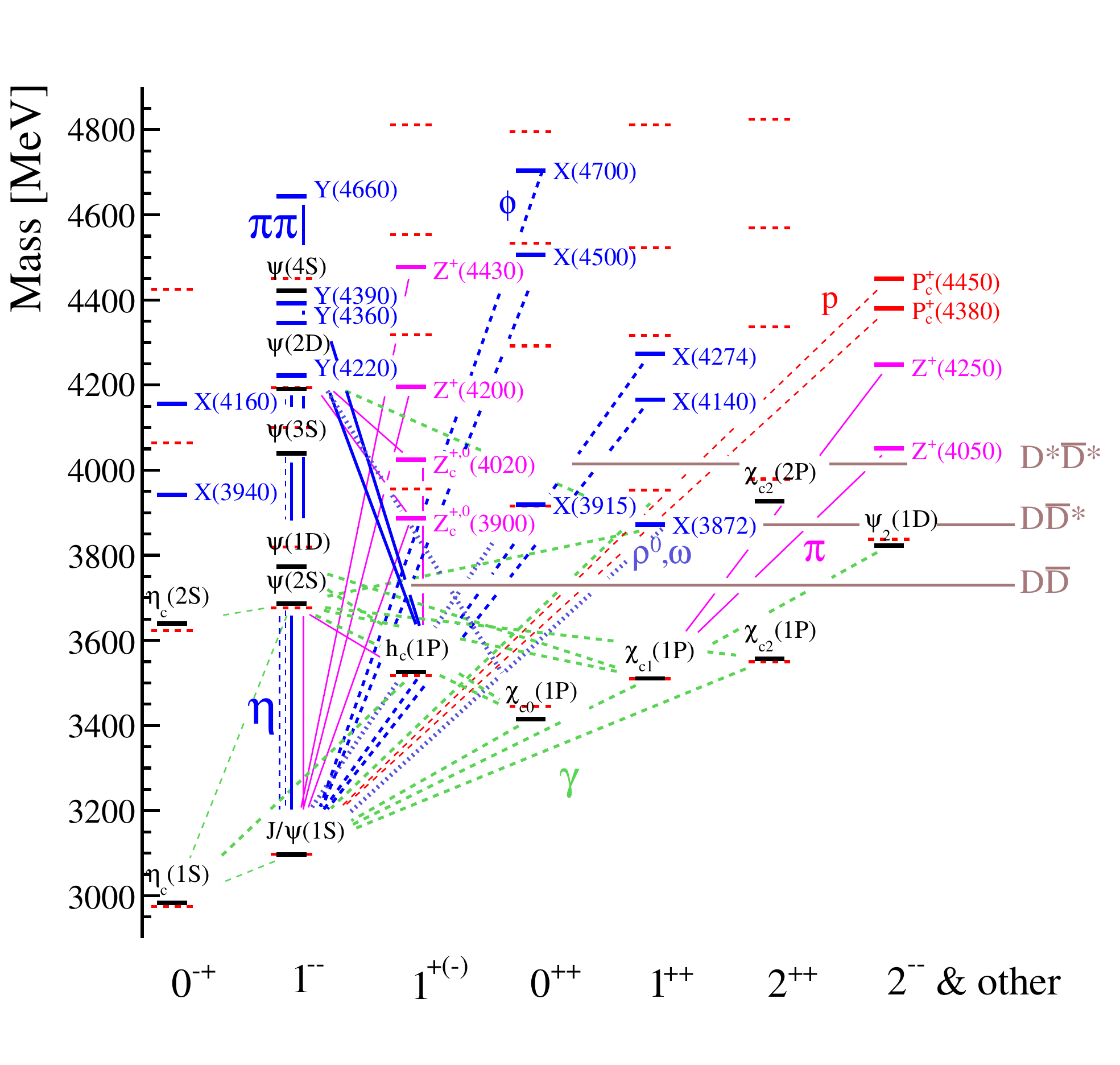}
    \caption{Current status of the charmonium-like spectrum~\cite{Olsen:2017bmm}. Horizontal lines indicate (red) expected   states, (black) experimentally established states, (brown) open   flavor thresholds, (blue, magenta) candidates for chamornium-like states. The lines connecting the states denote known transitions.} 
  \label{fig:charmonium_exotics}
  \end{center}
\end{figure}

The observation of many charmonium and bottomonium-like $X$, $Y$, $Z$ states, which do  not match the scheme expected from model calculations, has triggered a tremendous  interest in this exciting field of physics in recent years (see e.g.\ \cite{Olsen:2017bmm} for a recent review). 
\Figref{fig:charmonium_exotics} summarizes the current status of the charmonium-like spectrum~\cite{Olsen:2017bmm}. All states indicated by blue and magenta horizontal lines are candidates for states beyond the $q\overline{q}$ configuration of mesons, which have been sought-after for many years.  Similarly to the $X$, $Y$, $Z$ states, \compass has observed an unexpected,    resonance-like signal in the light-quark sector, the $a_1(1420)$~\cite{Adolph:2015pws},  which is currently being discussed as a possible  tetraquark state~\cite{Wang:2014bua,Chen:2015fwa,Gutsche:2017oro}, a molecular state~\cite{Gutsche:2017oro}, or a dynamically  generated signal~\cite{Basdevant:2015wma,Ketzer:2015tqa}. 

\begin{figure}[tbp]
  \begin{center}
    \includegraphics[width=0.8\columnwidth]{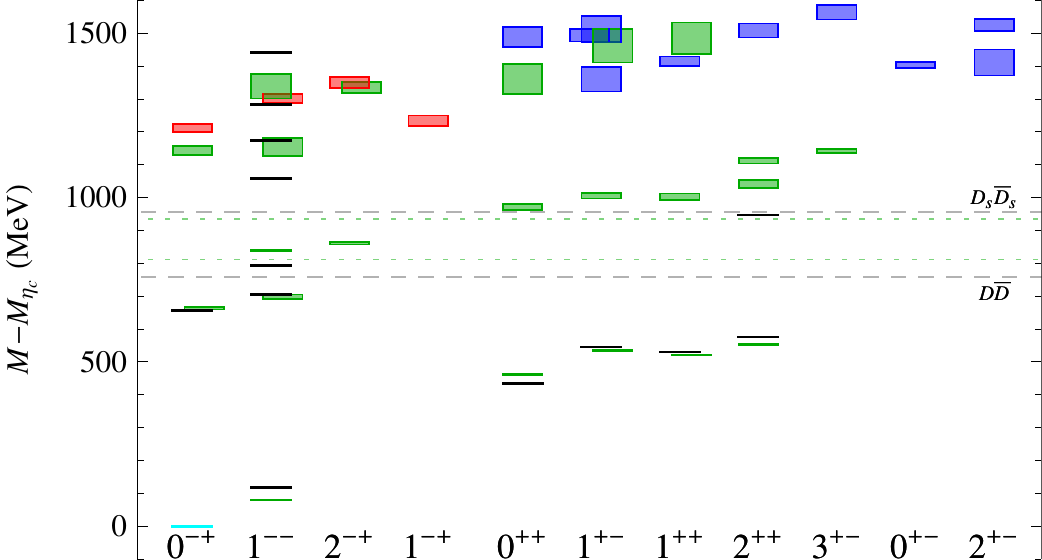}
    \caption{Lattice QCD spectrum for charmonium hybrid candidates~\cite{Liu:2012ze}. Red (dark blue) boxes are states suggested to  be members of the lightest (first excited) hybrid supermultiplet.} 
    \label{fig:charmonium_hybrids_lattice}
  \end{center}
\end{figure}

While QCD allows for and predicts full multiplets of such states, only  very few (or even none) have been unambiguously established experimentally. Recently, Lattice Gauge Theory started to make predictions for non-exotic and exotic charmonium-like states, albeit with an unphysical pion mass and still ignoring decays. Nevertheless, such calculations are useful as a guidance towards a future understanding of the spectrum. As an example,  \figref{fig:charmonium_hybrids_lattice} shows the spectrum of hybrid  candidates obtained by the  Hadron Spectrum Collaboration for a pion mass of $400\,\MeV$. The pattern of quantum numbers follows the same structure as in the light-meson sector~\cite{Ketzer:2012vn}, with a low-lying supermultiplet of hybrid mesons with quantum numbers $0^{-+}$, $1^{--}$, $2^{-+}$ and $1^{-+}$, the latter being a spin-exotic multiplet. 

\begin{figure}[tbp]
   \begin{center}
    \null\hfill
    \subfloat[]{
      \includegraphics[width=0.2\columnwidth]{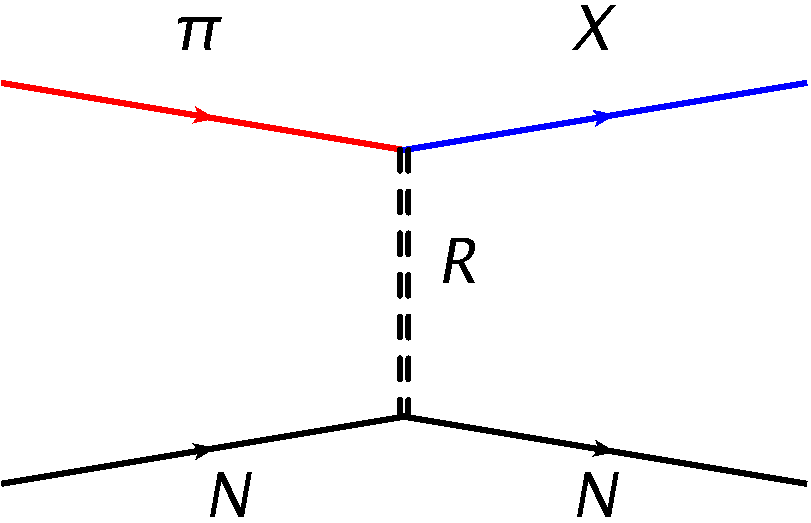}
      \label{fig:exp.production.diff}}
    \hfill
    \subfloat[]{
      \includegraphics[width=0.3\columnwidth]{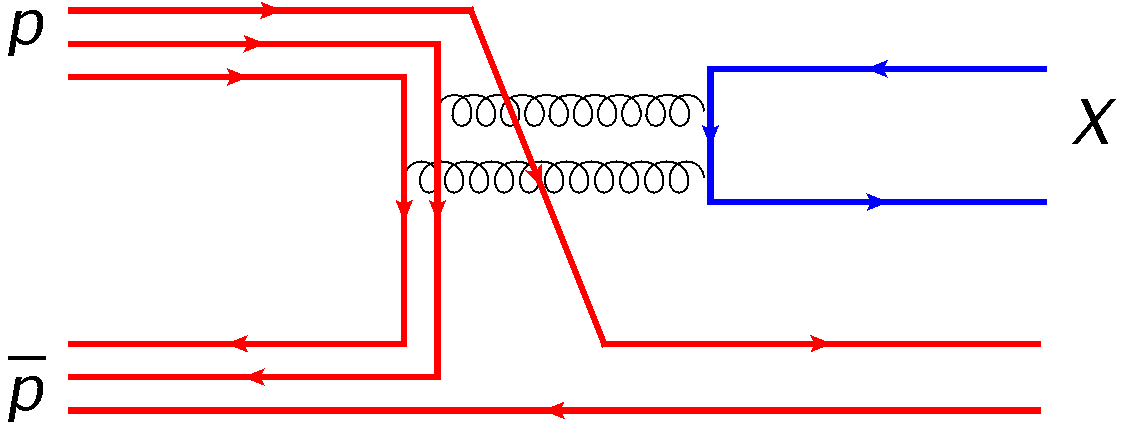}
      \label{fig:exp.production.pbarp}}
    \hfill\null
    \caption{Production mechanisms of mesons: (a) diffractive production in peripheral scattering of high-energy hadrons off a proton or nuclear target, (b) proton-antiproton annihilation of a low-energy antiproton beam on a proton target, where $X$ can be a $q\overline{q}$ state, a hybrid with gluonic degrees of freedom or a glueball without valence quark content.}
    \label{fig:exp.production}
  \end{center}
\end{figure}

In recent years, \compass has studied the spectrum of light-quark mesons with  unprecedented statistical precision,  thereby spurring  the development of novel analysis techniques in order to minimise the model bias when interpreting the data. Using a high-energy pion or proton beam scattering off a liquid hydrogen or solid nuclear targets, excited states were produced in diffractive reactions, which are dominated by the exchange of a Pomeron in the $t$-channel (\figref{fig:exp.production.diff}).    

Whereas diffractive reactions of beam pions or kaons dominantly produce final states containing light quarks, experiments employing the annihilation of antiprotons  of comparatively low energy between $12\,\GeV$ and $20\,\GeV$ provide a different and complementary access to excited states of mesons and baryons, which covers not only the light and strange quark sector, but also the charmonium and possibly the bottomonium region. In the past, a wealth of data was collected by experiments employing antiproton-proton annihilation, e.g.\ Crystal Barrel at LEAR~\cite{Aker:1992ny}  and experiments E760 and E835 at Fermilab~\cite{Garzoglio:2004kw,Bettoni:2009hbu}.  The PANDA experiment a FAIR will use a dedicated antiproton storage ring to study, among other physics topics, the spectrum of mesons in the charmonium region~\cite{Lutz:2009ff}. 
     
Antiproton annihilation (\figref{fig:exp.production.pbarp}) can proceed either in flight or at rest. For annihilation in flight, also  high-spin states can be populated (up to $L\sim 15$ at $\sqrt{s}\sim 6\,\GeV$). New states are generated either  resonantly in $s$-channel formation or in associated production together with a recoiling particle.  The quantum numbers of the multi-meson system are restricted  only by conservation laws of the reaction. The final state  will contain contributions from all possible intermediate states with different quantum numbers. A partial-wave analysis is usually required to disentangle the different contributions.  In antiproton annihilation, formation reactions provide direct  access to all states with non-exotic quantum numbers. This is in contrast to  e.g.\  $e^+e^-$ annihilation, where only states with quantum numbers $1^{--}$ are directly formed.  In production reactions with an additional recoil  particle in the final state, also  states with spin-exotic quantum numbers may be produced. This is the mode with the highest discovery potential for new states, including states with explicit gluonic degrees of freedom, i.e.\ hybrids or  glueballs in the charmonium sector. Consequently, this is where an experiment at the M2 beam line of the SPS can make important contributions, long before the start of PANDA, which is currently envisaged for 2025, albeit still with large uncertainties and low luminosity at best.  

The M2 beam line can provide, with minimal modifications compared to the present setup, a rather clean beam of antiprotons with momenta between $12$ and $20\,\mathrm{GeV}$. According to preliminary calculations,  the intensity of antiprotons at the target is between $1.1\cdot 10^{7}$ and  $1.8\cdot 10^{7}$ per pulse of $10^{13}$ protons on the production target at momenta of $12$ and $20$~GeV, respectively, and is limited by radiation protection issues. Employing a $40$~cm long liquid hydrogen target, as for the \compass measurements with a pion beam, a luminosity of the order of   $10^{30}\,\mathrm{cm}^{-2}\mathrm{s}^{-1}$ can be achieved. 

This opens the possibility to use antiproton annihilations as a tool to study the spectrum of quarkonia and  
possibly exotic states. 
According to model and lattice calculations, the lightest charmonium hybrid is predicted at a mass around $4.3\,\mathrm{GeV}$ with spin-exotic quantum numbers. Also the lowest-mass glueball with spin-exotic quantum numbers, predicted at a mass between $4$ and  $5\,\mathrm{GeV}$ is within the kinematic reach of this experiment.  A production survey of these states could thus be performed at the SPS of CERN, including the production of high-spin states. Other measurements are possible including that of $\overline{p}p$ production cross sections for $X,Y,Z$ states.  
The production cross sections of exotic charmonia are largely unknown, which makes them  one of the major uncertainties for the simulation of signal-to-background ratios in PANDA. Thus, besides improving our general understanding of these production mechanisms by the  measurements at the M2 beam line described below, a better knowledge of these quantities would pave the way for PANDA to strengthen and focus its physics perspectives for precision studies. 

The setup for these measurements will make use of the existing forward  spectrometer, augmented by a powerful target spectrometer to maximise the acceptance for exclusive measurements of multi-particle final states. We are currently investigating several options in this direction, including e.g.\ the use of parts of the WASA spectrometer. 

\subsubsection{Beam line}
\label{sec:lepbar.beam}
\begin{figure}[tbp]
  \begin{center}
    \includegraphics[width=0.8\columnwidth]{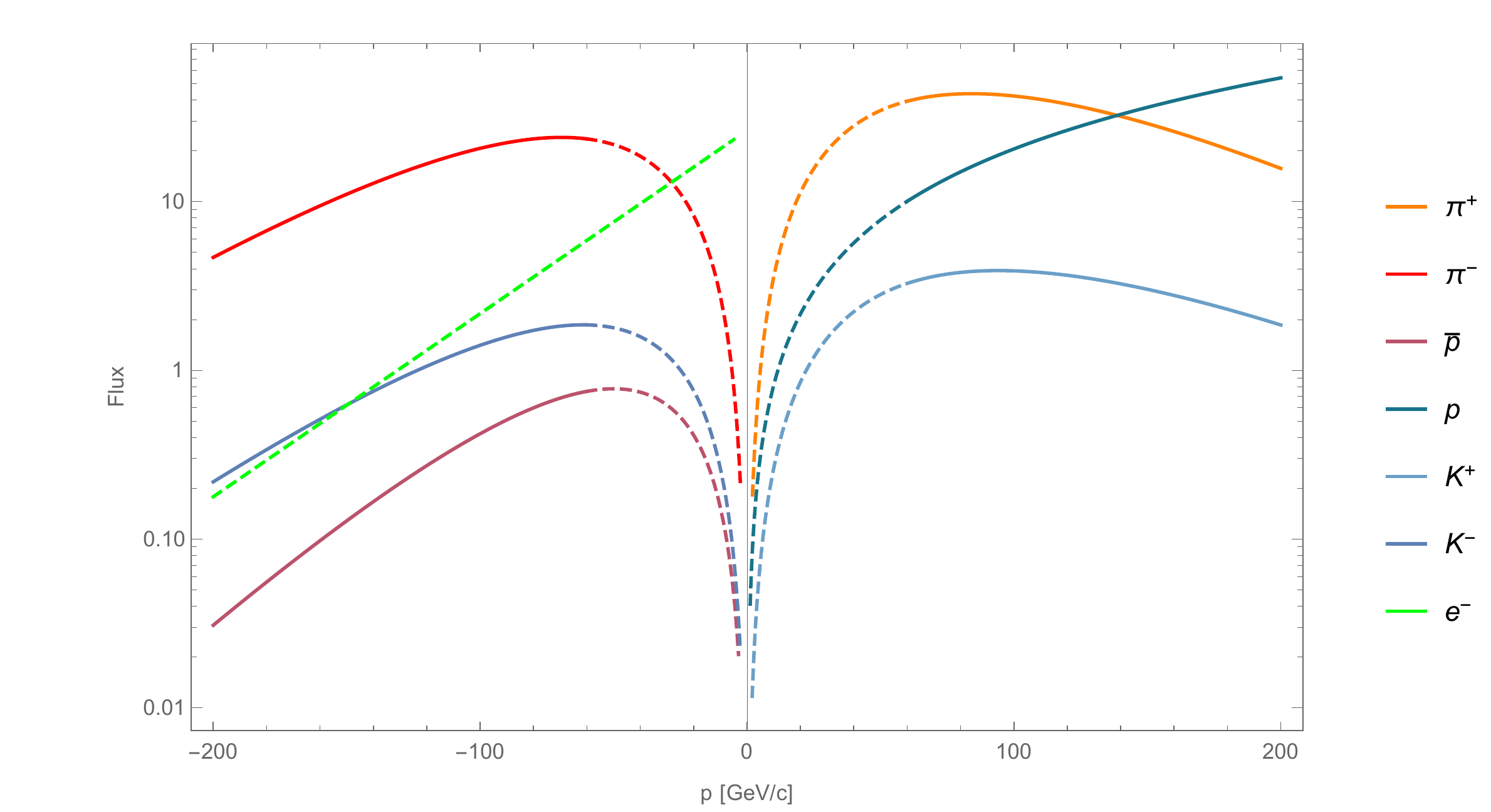}
    \caption{Atherton parameterisation for the production of different
      particle species given in flux per solid angle [steradian], per interacting proton, and per dp [GeV/c] as a function of secondary momenta for a 0\,mrad production angle\,\cite{Atherton:133786}. Negative values for momenta are used to indicate negatively charged particles.}
    \label{fig:Atherton}
  \end{center}
\end{figure}

Starting from the current layout of the M2 beam line, a study in the framework of the Physics Beyond Colliders Initiative has been launched by EN-EA in order to check principal limitations and feasibility of low-energy antiproton beams. \\ 
In a first step, the production of antiprotons at several desired energies has been estimated using the so-called Atherton parameterisation\,\cite{Atherton:133786}, based on results of production measurements on Beryllium targets in the North Area. In Fig.\,\ref{fig:Atherton}, the flux of secondary particles at 0\,mrad production angle versus the secondary momentum is shown. For the two study cases at $12$ ($20$)\,GeV$/c$, the flux is about 0.41 (0.20) antiprotons per interacting proton per steradian per GeV$/c$ momentum bite. This corresponds to about 4.4\% to 4.8\% of the total negative hadron flux. Based on the experience of operating the West Area in the 1990s, the main background contribution to the beam was identified as electrons. As depicted in Fig.\,\ref{fig:Atherton}, the electrons at lower energies contribute with more than 90\% to the total flux. Hence a suppression of the electron background has to be included, most probably by the insertion of a thin lead sheet at a focal position in the beam optics in order to keep the contribution by multiple scattering to the beam divergence at the CEDAR counters low.

Given a 99\% suppression of electrons and including the decay of hadrons along the M2 line, this would result in a fraction of 18.2\% (11.3\%) of antiprotons at the \compass target location for $12$ ($20$)\,GeV$/c$ beams. With a typical solid angle of $\pi \cdot 10^{-5}$, a target efficiency of 40\% for the 500\,mm T6 target head, a flux of $10^{13}$ protons on T6, and assuming a 2\% momentum bite for the new low-energy optics, the resulting antiproton flux would be $10^{8}$ ($5\cdot10^{7}$) for $12$ ($20$)\,GeV$/c$ beams. As the intensity in EHN2 is limited by radiation protection to about $10^{8}$ particles per $4.8$\,s spill, the total antiproton flux thus is limited by the purity of the beam. For the calculated purity of antiprotons of 18\% (11\%), an upper limit of the antiproton flux at the \compass target is estimated to be $1.8$ $(1.1) \cdot 10^{7}$ antiprotons per spill.
 
For an efficient transport of low-energy antiprotons, several optimisations of the M2 beam line could be envisaged. Besides a study of dedicated low-energy optics, a completion of the vacuum in the line would be necessary. So far, the M2 beam line is optimised for muon transport, which implies that several elements specific to muon beams were not designed for operation in vacuum, such as the magnetic collimators (``scrapers''), collimator 5, and 9.9\,m of Beryllium absorbers inside bend 4. As a consequence, a total of about 80\,m of the beam line remain without vacuum. Depending on the operational conditions, two solutions would be preferred. For a full year of operation without muon beams, the above mentioned elements could be removed from the beam line and/or be exchanged by standard magnets and absorbers, which are compatible with the vacuum requirements. In this case, the removal of scrapers will have the consequence of a large muon component in the beam of the order of 3-5\,\% and an increased muon halo due to the M2 geometry. If this background cannot be tolerated or an intermediate operation of muon beams is envisaged, another solution could be a fitting of vacuum tanks inside the scrapers. In addition to the optics change and vacuum optimisation, the CEDAR counters would have to be exchanged by so-called West Area CEDARs that are optimised for beam momenta below $100$\,GeV$/c$. In this configuration, other optical elements and another gas (N$_{2}$) are used.

\subsubsection{Measurements}
\label{sec:lepbar.exp}

Using an antiproton beam with a momentum between $12\,\GeV/c$ and $20\,\GeV/c$, we plan to perform spectroscopy of heavy-quark mesons by measuring exclusive production of multi-particle final states. With the available centre-of-mass energies  at the M2 beam line, we cover the full range of charmonium-like states up to masses of about $6\,\GeV$. 
In principle, higher beam momenta would allow us even to touch the bottomonium region, although then the intensity of antiprotons will be smaller. 

With a production survey at fixed antiproton beam momentum, we plan to study high-spin charmonia and charmonium-like states as well as exotics like hybrids and glueballs. Of particular interest at present is the study of the $Z_c$ multiplets (charged and neutral), which until now have only been observed in $e^+e^-$ reactions. 

The cross sections for the production of charmonium-like states in antiproton annihilations are largely unknown.  Experimental results on inclusive $J/\psi$ production in $p\bar{p}$ annihilation  were obtained e.g. at the CERN SPS with a $\bar{p}$ momentum of $39.5\,\GeV/c$, where a cross section of $(12\pm 5)\,\nb$ was extracted~\cite{Corden:1980ht}.  Theoretical estimates range from  $0.1\,\nb$ to $10\,\nb$ (see e.g.\ an estimate of $\bar{p}p\to\pi^0 J/\psi$ in~\cite{Wiele:2013vla}). It is thus important to measure these cross sections, first in order to test production models and second to provide input for simulations of the physics performance that can be achieved with future precision experiments like PANDA. 

Based on the luminosity estimated above and using an inclusive $J/\psi$ cross section of $12\,\nb$~\cite{Corden:1980ht}, we will produce of the order of $120,000$ inclusive $J/\psi$ per year of running, corresponding to $\sim 7,000$ $J/\psi$ decaying into $\mu^+\mu^-$. This number may be increased by a factor of 5 by including e.g.\ the $e^+e^-$ decay channel and by increasing the target length to $100\,\Cm$.  

\begin{figure}[!htb]
  \begin{center}
    \hfill
    \subfloat[]{
      \includegraphics[width=0.45\columnwidth]{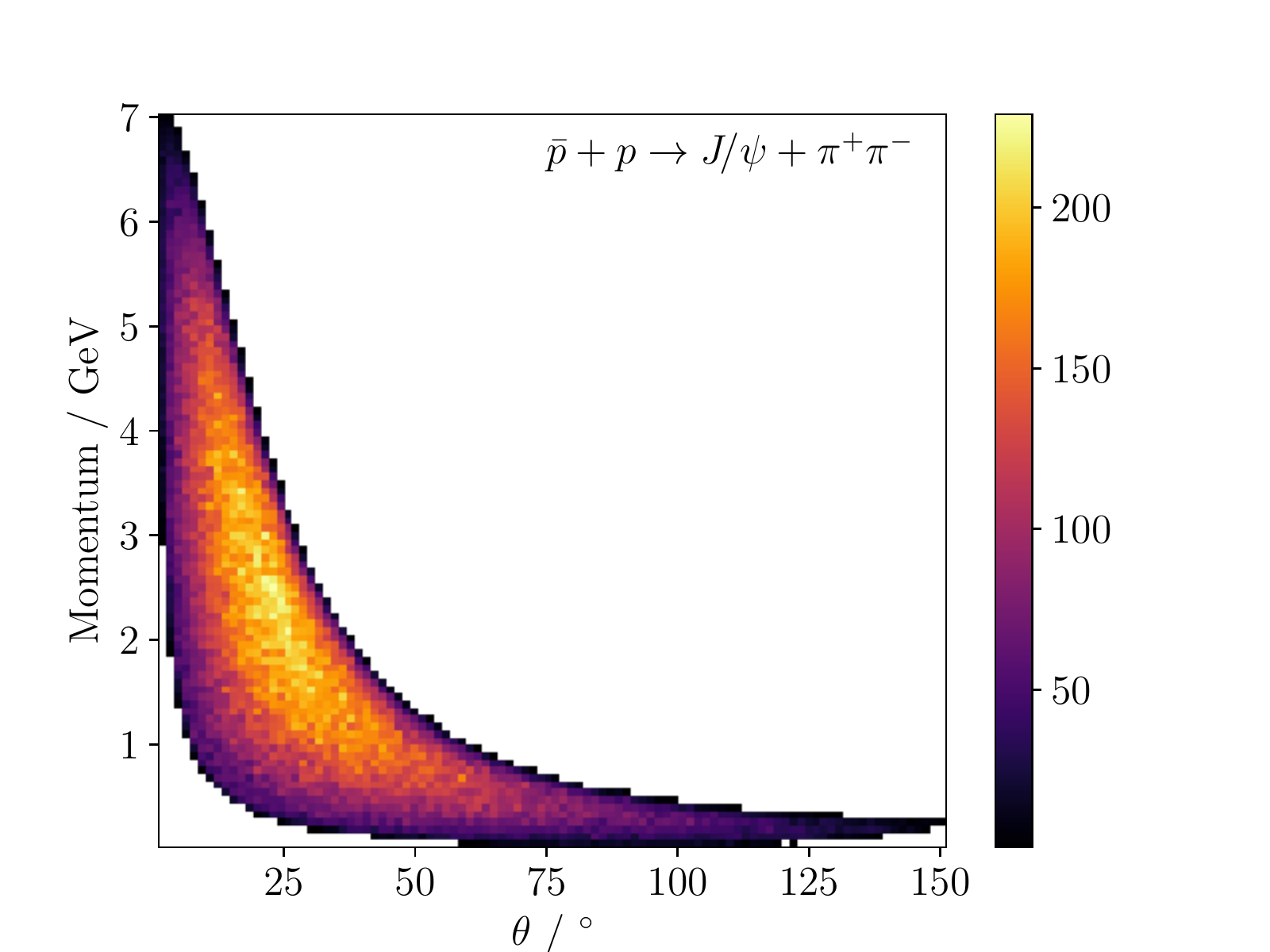}
      \label{fig:piZc.pions}}
    \hfill
    \subfloat[]{
      \includegraphics[width=0.45\columnwidth]{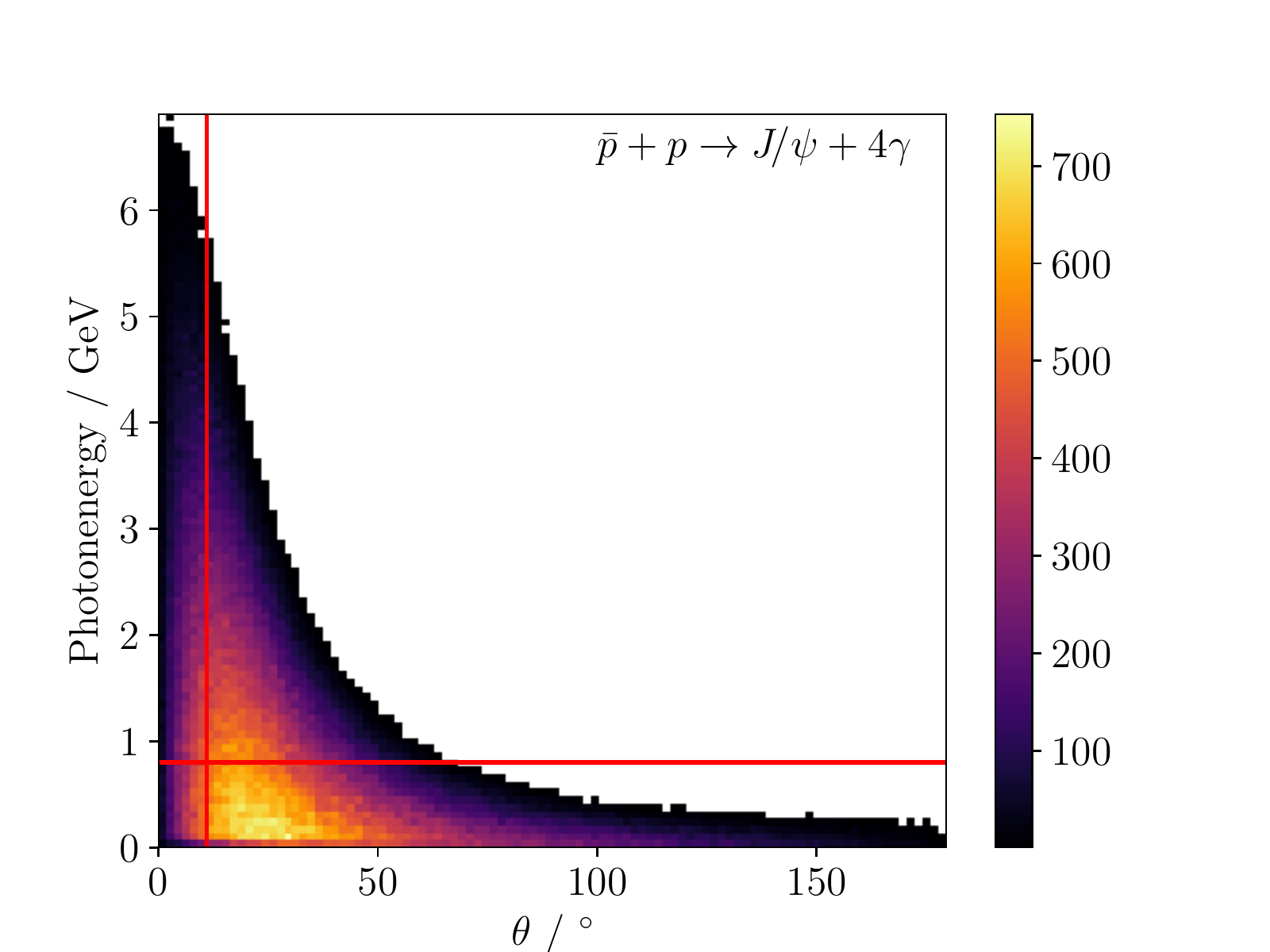}
      \label{fig:piZc.photons}}
    \hfill
    \subfloat[]{
      \includegraphics[width=0.45\columnwidth]{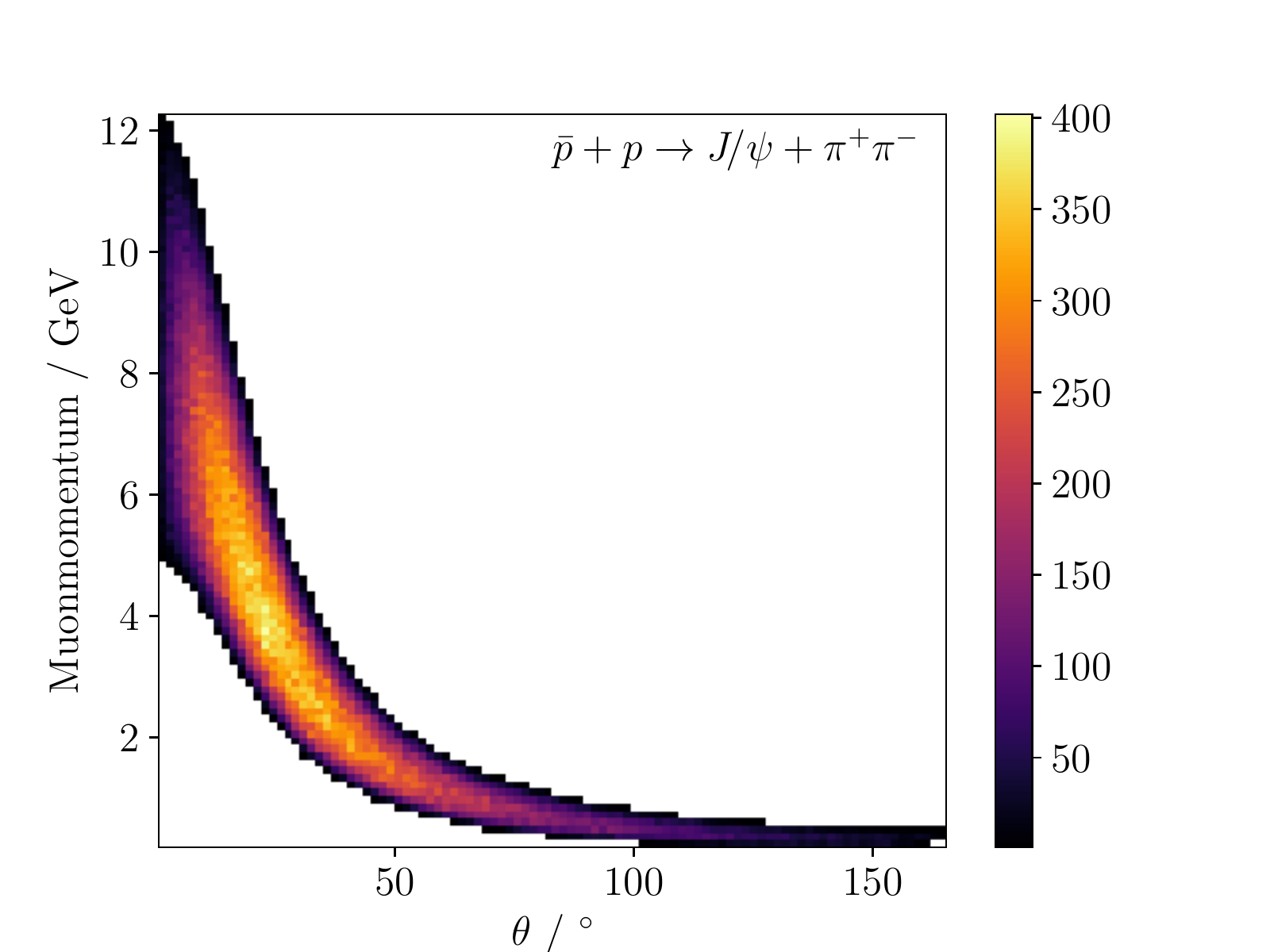}
      \label{fig:piZc.muons}}
    \hfill
    \caption{Kinematic distributions from phase-space simulations of  $\bar{p}p$ annihilation using an antiproton beam momentum of $12\,\GeV/c$: (a) charged-pion momentum vs.\ polar angle from the  reaction $p\bar{p}\to \pi^- Z_c^+(4430)$, with $Z_c^+ \to \pi^+ J/\psi$; (b) photon energy vs. polar angle from the reaction $p\bar{p}\to \pi^0 Z_c^0(4430)$, with $Z_c^0 \to \pi^0 J/\psi$,   and (c) muon momentum vs. polar angle from the decay of  $J/\psi$.}
    \label{fig:piZc}
  \end{center}
\end{figure}

\subsubsection{Experimental requirements}
\label{sec:lepbar.setup}

\begin{figure}[!htb]
  \begin{center}
    \hfill
    \subfloat[]{
      \includegraphics[width=0.45\columnwidth]{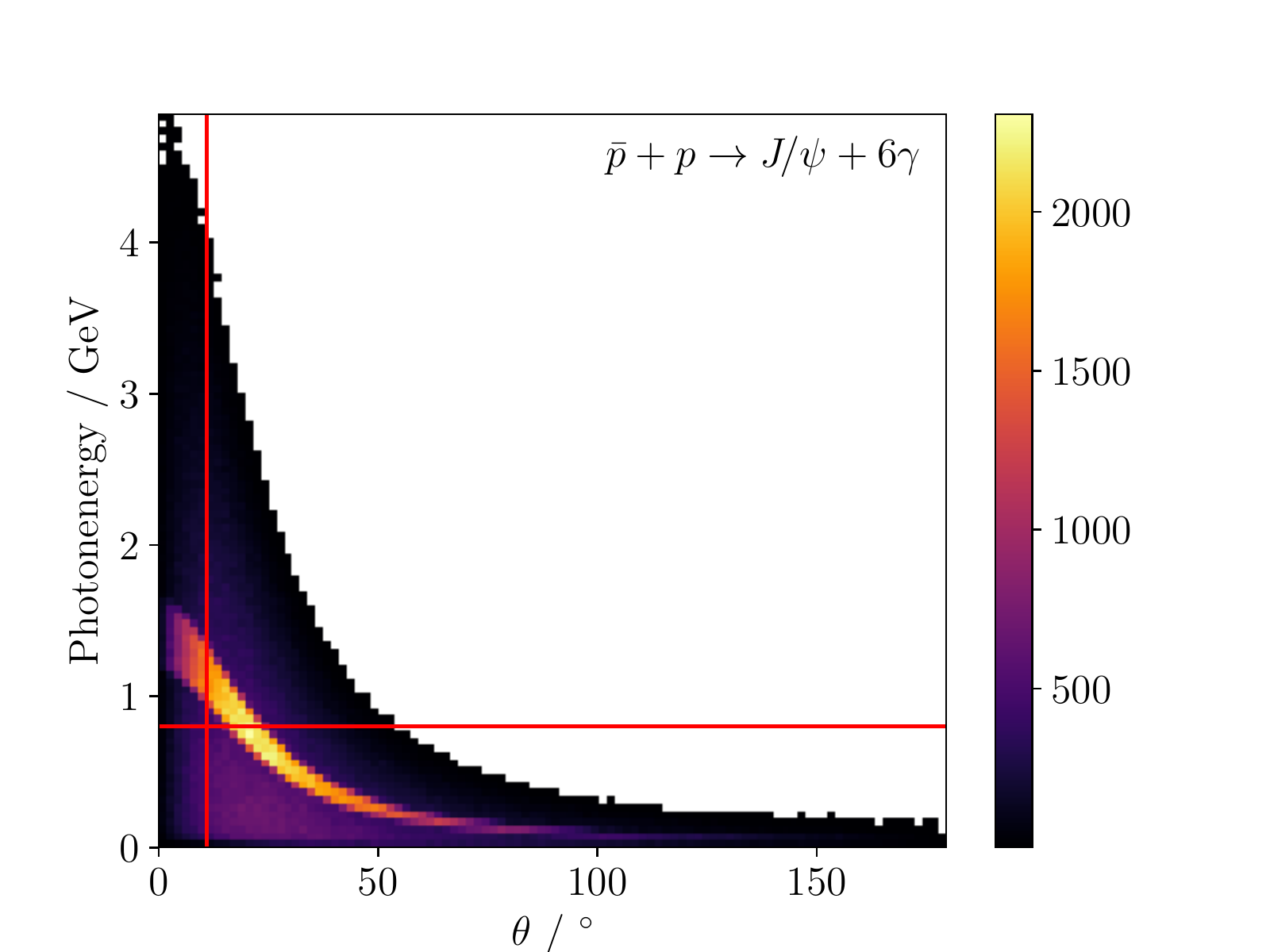}
      \label{fig:etah.photons}}
    \hfill
    \subfloat[]{
      \includegraphics[width=0.45\columnwidth]{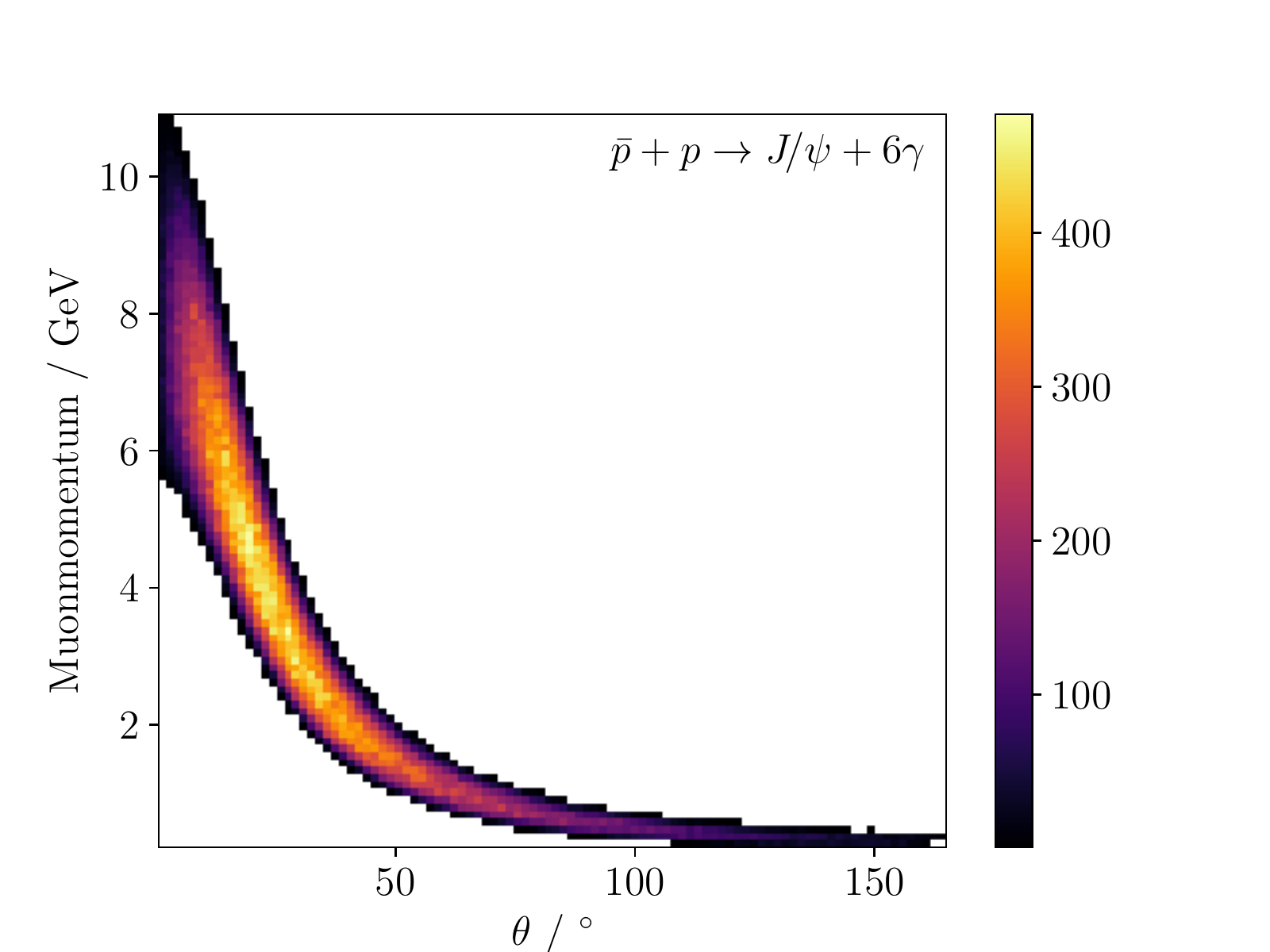}
      \label{fig:etah.muons}}
    \hfill
    \caption{Kinematic distributions from phase-space simulations of $\bar{p}p$ annihilation using an antiproton beam momentum of $12\,\GeV/c$: (a)  photon energy vs. polar angle from the reaction $p\bar{p}\to \eta h(4300)$, with $h(4300) \to \pi^0\pi^0 J/\psi$ and $\eta \to \gamma\gamma$, where the red lines indicate the acceptance of the WASA calorimeter discussed in \secref{sec:inst-lowespec};   (b) muon momenta vs. polar angle from the decay of  $J/\psi$.}
  \label{fig:etah}
  \end{center}
\end{figure}

Since the beam at the M2 beam line of the SPS will contain not only antiprotons, but also pions and electrons, it is important that each incoming beam particle is identified and tagged by CEDAR Cherenkov detectors. Since we need to push the intensity to the limit allowed by radioprotection, these detectors have to work efficiently at intensities of $\EE*{8}$ particles per spill.  

As target we envisage a $40-100\,\Cm$ long cylinder containing liquid hydrogen, similar  to what was used for the measurements with a pion beam  at \compass in the years 2008-2012. In addition, the use of foils or wires could be envisaged as nuclear targets. 

In order to study the required energy and angular acceptance, we performed phase-space simulations of the reactions 
\begin{enumerate}
\item\label{itm:Zc+} $p\bar{p}\to \pi^- Z_c^+(4430)$, with $Z_c^+ \to \pi^+ J/\psi$, 
\item\label{itm:Zc0} $p\bar{p}\to \pi^0 Z_c^0(4430)$, with $Z_c^0 \to \pi^0 J/\psi$, 
\item\label{itm:hybrid} $p\bar{p}\to \eta h(4300)$, with $h \to \pi^0\pi^0 J/\psi$
  (fictitious $c\bar{c}$ hybrid at $4.3\,\GeV$) and $\eta \to \gamma\gamma$, 
\end{enumerate}
all with $J/\psi\to\mu^+\mu^-$, using an antiproton beam momentum of $12\,\GeV/c$. \Figref{fig:piZc} shows the distributions of momenta or energies of charged pions, muons, and photons versus polar angle  from the production of $Z_c$ (reactions \ref{itm:Zc+} and \ref{itm:Zc0}) in the laboratory frame. The corresponding phase-space distributions in the laboratory frame for reaction \ref{itm:hybrid} are shown in \figref{fig:etah}. 

Because of the reduced beam energy compared to the earlier measurements performed by \compass, for which they used beam energies above $100\,\GeV$ that imply a correspondingly larger boost of final-state particles in forward direction, it is clear that in order to perform exclusive measurements an additional coverage with charged-particle tracking and calorimetry surrounding the target is needed.  In particular, the detection of photons from the decay of $\pi^0$ and $\eta$ will be of importance for the reduction of combinatorial background and thus the identification of states. 

The trigger should include dimuon and possibly dielectron production from the decay of $J/\psi$. While muon identification requires dedicated muon chambers, electron identification could be achieved by a transition-radiation tracker. 

The considerations about the experimental setup are detailed in Sec.~\ref{sec:inst-lowespec}.

\svnid{$Id: stdhad_xsect.tex 4404 2018-10-08 18:17:46Z avauth $} 
\subsection{Measurement of antiproton production cross sections for Dark Matter Search}
\label{sec:stdhad.xsect}

\subsubsection{Physics case}
\label{sec:stdhad.xsect.phys}
Multiple and concurring evidence indicate that the vast majority of the matter content of the universe is non baryonic and electrically neutral. This constituent of the universe is usually called  Dark Matter (DM), for its lack of electromagnetic interactions.

The DM surrounds the galaxies and the large structures of the universe, and it is the major constituent of the gravitational fabric of the universe. 
Origin and nature of Dark Matter constitute the most intriguing puzzle that is completely unresolved. The most appealing hypothesis is that DM consists of weakly-interacting massive particles (WIMPs), which are presently supposed to be cold thermal relics of the Big Bang. 

The indirect detection of DM is based on the search for the products of DM annihilation or decay. They should appear as distortions in the gamma-ray spectra or as anomalies in the rare Cosmic Ray components. In particular cosmic-ray antimatter components, like antiprotons, antideuterons and positrons, promise to provide sensitivity to DM annihilation  on top of the standard astrophysical production  

\begin{equation}
  \chi + \chi \rightarrow q\bar{q} ,W^{-} W^{+},... \rightarrow \bar{p},\bar{d}, e^{+}, \gamma, \nu ,
\label{eq:astroprod}
\end{equation}
where $\chi$ is a generic symbol for a DM candidate particle. 

The search for DM annihilation products motivated the development of new challenging experiments, either  ground-based or in space, which produced spectacular results; among them the AMS-02~\cite{Ag:2016am} experiment on the International Space Station.
In the following, we will briefly discuss how crucial the measurements of antiproton-production cross sections are for the indirect search of DM, in particular cross sections of antiprotons and antideuterons, which can be produced by secondary beams at accelerators.

\paragraph{Antiproton production cross section}
The dominant part of the antiprotons in our galaxy originates from inelastic scattering of incoming Cosmic Rays (CRs) off Interstellar Medium (ISM) nuclei at rest and represents the background when searching for small contributions from exotic sources.

After the breakthrough from the satellite-borne PAMELA detector~\cite{Adriani:2013}, the $\bar{p}$ flux and the $\bar{p}/p$ ratio have been measured with the unprecedented accuracy of a few percent by AMS-02
over an energy range from below $1\,$GeV up to a few hundreds of GeV, which shows that above about 60\,GeV this ratio is independent of the energy.

\begin{figure}[tb]
        \begin{center}
                \includegraphics[width=0.5\columnwidth]{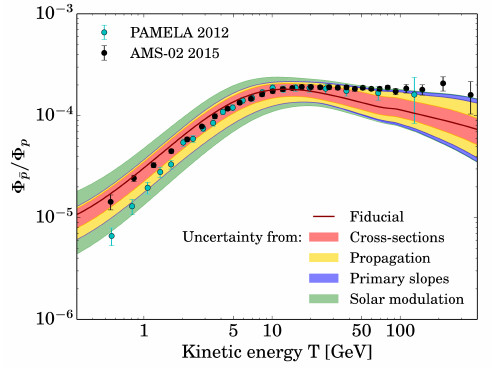}
        \end{center}
        \caption{The combined total uncertainty on the predicted secondary $\bar{p}$ / $p$ ratio, superimposed to the PAMELA~\cite{Adriani:2013} and the AMS-02~\cite{Ag:2016am} data.}        
        \label{fig:dm.pbarp_flux}
\end{figure}

The antiproton (secondary) component generated by cosmic rays is expected to decrease more rapidly than the primary proton spectrum, however the predictions are affected by several uncertainties. As shown in figure \ref{fig:dm.pbarp_flux} \cite{Giesen:2015ufa} 
, one can identify three sources of uncertainties: the primary slopes, the propagation in the Galaxy \cite{Johannesson:2016rlh} and the antiproton-production cross section. While AMS-02 measurements will contribute to reduce the first two, new dedicated measurements must be performed to reduce the latter uncertainty.

In order to be able to profit from the AMS-02 high-precision data, a similar accuracy in the computation of the $\bar{p}$ source term for all the production channels has to be achieved. 
Nuclei heavier than protons and helium give a very small or negligible contribution either as projectiles or targets, and thus play a marginal role in the production of secondary antiprotons. The dominant reactions are those involving  protons and helium ($p+p, p+{}^{4}He, {}^{4}He+p, {}^{4}He+{}^{4}He$).
Accurate measurements of the $\bar{p}$ production cross section in $p+p$ collisions and $p+{}^{4}He$ collisions are thus of fundamental importance in a wide energy range from $10\,\GeV$ to a few TeV in order to reduce the uncertainty on the secondary $\bar{p}$ production cross section and eventually clarify if there is an evidence of exotic components coming from DM annihilation or decay in the AMS-02 data.

While  on $p+p$ collisions a few experimental datasets are available, the very first  dataset on $p+{}^{4}He$ collisions was collected at the end of 2015 by the LHCb experiment at $4\,$TeV and $7\,$TeV.
A new \compass-like fixed-target experiment at the M2 beam line of the CERN SPS would contribute to this fundamental DM search by performing a complementary measurement with a proton beam of a few hundred GeV/c impinging on a liquid helium target.

\paragraph{Antideuteron production}
When compared to the indirect DM search using antiprotons and positrons, which suffer from relatively large and uncertain standard-astrophysical background, a search with low-energy antideuterons benefits from strongly suppressed background.

The dominant channel for secondary $\bar{d}$ production in our galaxy is the one that involves $p+H_2$ collisions followed by cosmic protons colliding on IS helium ($p+{}^{4}He$). 
The $\bar{d}$ flux from a wide range of DM models exceeds the background flux by more than
two orders of magnitude in the energy range below $0.25\,$GeV/nucleon, and by more than an order of magnitude up to $1\,$GeV/nucleon. Thus the measurement of low-energy $\bar{d}$ production offers a potential breakthrough in an unexplored phase space for indirect DM search. Also, many dark matter models predict the antideuteron flux to be within the reach of currently operating or planned experiments, like BESS, AMS-02, and GAPS.
Nevertheless, the largest uncertainties in the flux estimation, both for primary and secondary (background) $\bar{d}$,  are due to the hadronisation and coalescence models that are used to describe antideuteron formation, and due to the propagation models. 
 Understanding antideuteron production is thus one of the crucial points for the interpretation of the cosmic-ray data, which impacts both the antideuteron background expectation as well as the formation of antideuterons in the aftermath of dark matter annihilations or decays. 
The predicted antideuteron fluxes depend on the only free parameter of the coalescence model, i.e. the coalescence momentum $p_{0}$, which is defined as the radius of the sphere in momentum space, within which any (anti)nucleon will coalesce to produce an (anti)nucleus.
This parameter has to be determined by fitting the theoretical model predictions to the available experimental data on $\bar{d}$ production, as collected by ALEPH, CLEO, CERN ISR, ZEUS, ALICE and BABAR. However, no single value of $p_0$ could be determined that simultaneously fits all the data. 
This uncertainty has quite dramatic implications on the search for cosmic antideuterons, which is due to the strong dependence of the antideuteron yield on the coalescence momentum $N_{\bar{d}}$ $\propto$ $p_{0}^{3}$.
The antideuteron production cross section is not further discussed in this document in the following sections, but it will be included in the context of a more detailed feasibility study in a forthcoming proposal.

\subsubsection{Feasibility of the measurement}
The production cross section for antiprotons produced in $p+p$ and $p+{}^{4}He$ collisions is 
known only with uncertainties of the order of $20\%$ to $30\%$, depending on their energy. As this cross section cannot simply be constrained by a measurement of the other products
of these interactions, a direct measurement is required.
Here, we explore the possibility to use a \compass-like magnetic spectrometer to measure the products of the interactions of protons of different momenta delivered by the SPS M2 beam line to a target of liquid hydrogen and liquid helium.

We simulate $p+p$ and $p+{}^{4}He$ interactions to characterize the features of these events in term of multiplicity, energy and angular distributions of the produced  particles, in particular the antiprotons, and we study the performance of the \compass spectrometer  to detect such events.  
We also discuss the measurement of the differential antiproton production cross section and the possible sources of systematic uncertainties.





We performed the simulation with two beam/target configurations:
\begin{enumerate}
  \item $190\,$GeV/c protons on liquid $H_2$;
  \item $190\,$GeV/c protons on liquid ${}^{4}He$.
\end{enumerate}

At $190\,$GeV/c beam momentum events with an antiproton represent ~7\% of the total number of events, as summarized in table \ref{tableI}.
\begin{table}[bt]
\centering
  \begin{tabular}{|l|c|c|c|c|}
  \hline
  &     $p+p$  & $p+{}^{4}He$ \\
  	\hline
     Beam Mom& 190\,GeV/c & 190\,GeV/c \\
    \hline
    Mult ($Z\neq0$) &   9.9 & 11.07 \\\
    $\bar{p}$ event fraction &   7.1\% & 7.7\% \\
    $\langle p \rangle \,\mathrm{of}\,\bar{p}$ (GeV/c) & 15.3 & 14.5\\
    \hline 
  \end{tabular}
  \caption{Simulation results: the average charged track multiplicity, the fraction of the antiproton events and the average momentum of the antiprotons are shown.}
  	\label{tableI}
\end{table}
We are interested to understand the characteristics of these events in terms of the number of charged tracks, their forward angle, the number of produced antiprotons and their momentum distribution. 

\begin{figure}[hbt]
\centering
  \includegraphics[width=0.6\columnwidth]{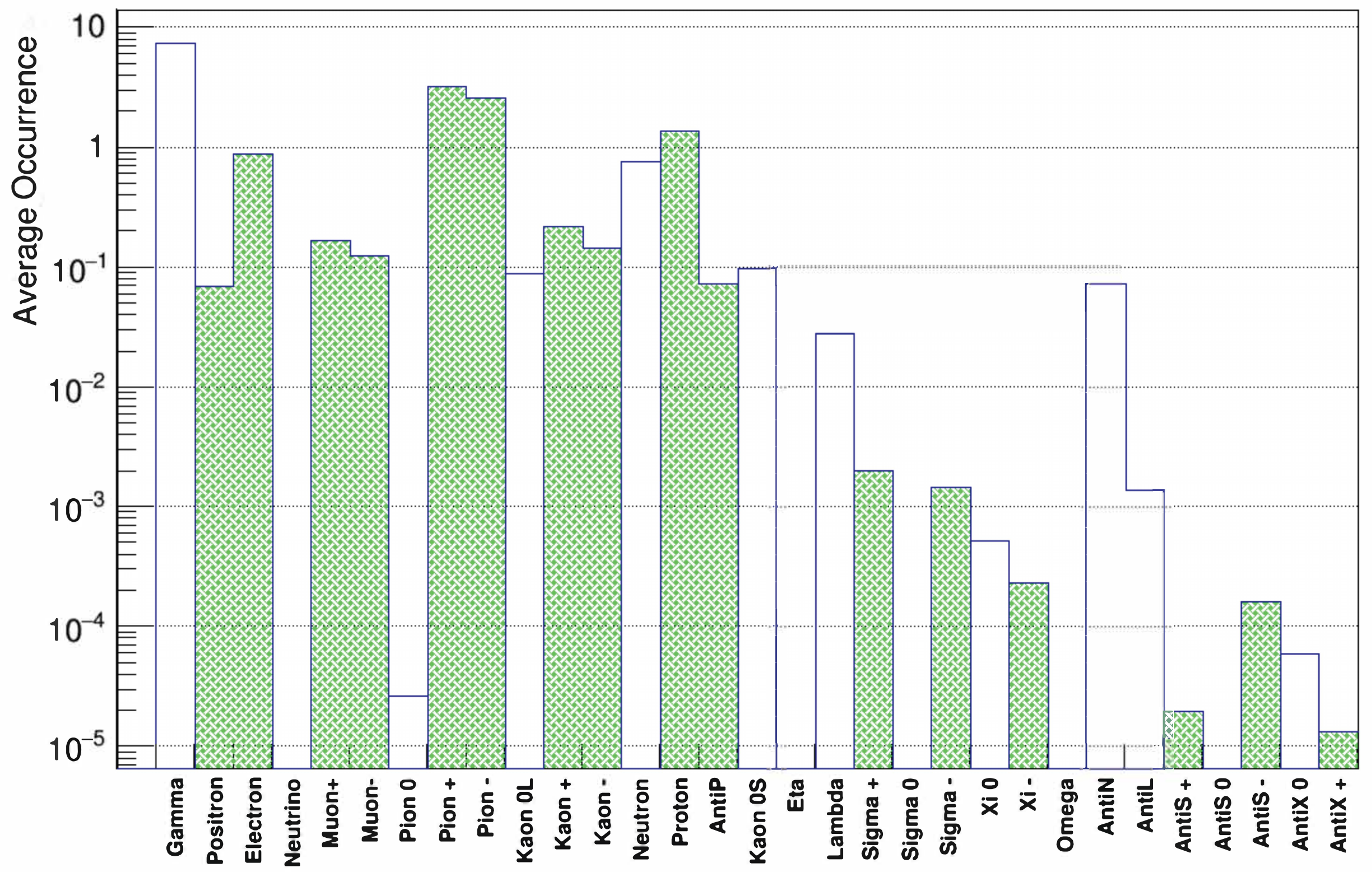}
  \caption{Particle type abundance in $p+p$ interactions at 190 GeV/c.}
  \label{fig:pkind}
\end{figure}

\begin{figure}
\centering
	\includegraphics[width=0.6\columnwidth]{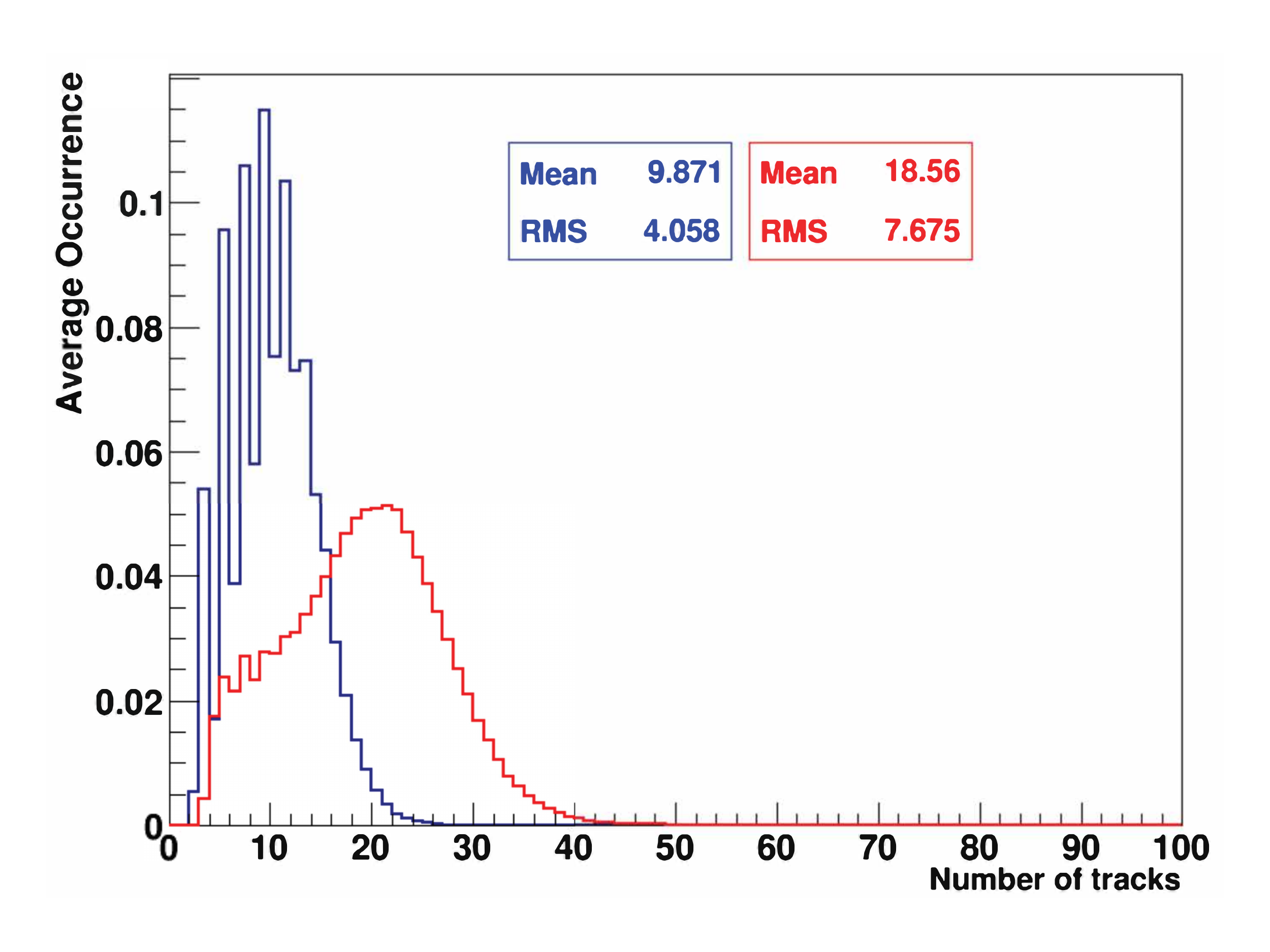}
  \caption{Track multiplicity in $p+p$ interactions at 190\,GeV/c: in blue the charged tracks, in red all tracks.}
  \label{fig:pmult}
\end{figure}

\begin{figure}[hbt]
\centering
  \includegraphics[width=0.6\columnwidth]{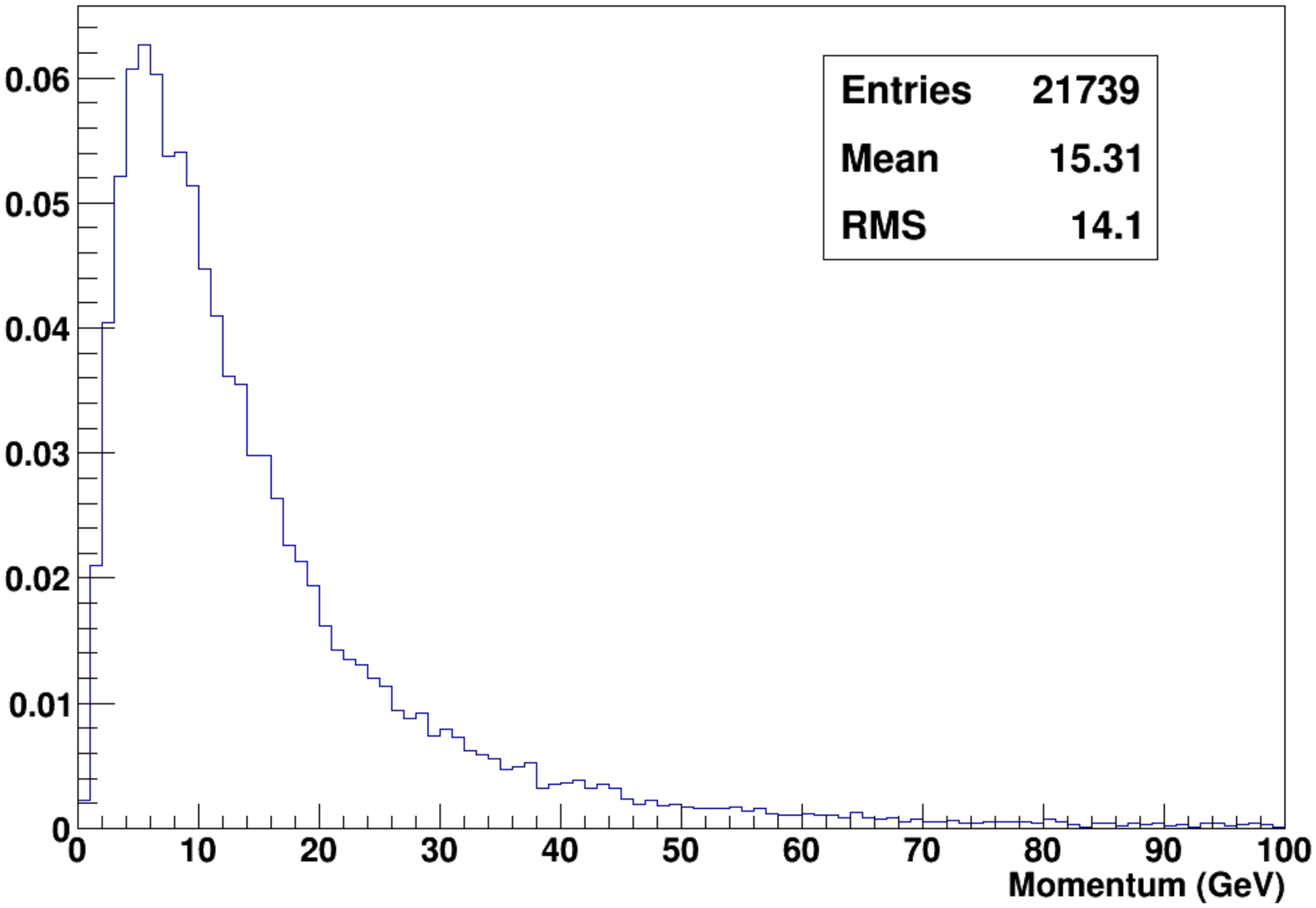}
  \caption{Momentum spectrum of $\overline{p}$  produced in $p+p$ interactions at 190\,GeV/c}
\label{fig:spectrum}
\end{figure}

Below we present some results for $p+p$ interactions at $190\,$GeV/c. The average occurrence of particles, the distribution of the final-state multiplicity and the energy spectrum of the produced antiprotons are shown in Figs. \ref{fig:pkind}, \ref{fig:pmult} and \ref{fig:spectrum} respectively.


\paragraph{Spectrometer performance}

{\makeatletter
\@beginparpenalty=10000 
\makeatother
We studied the performance of a \compass-like spectrometer in terms of the
\begin{itemize}
  \item ability to reconstruct the tracks within its geometric acceptance;
  \item resolution of the momentum measurement for a single track;
  \item vertex reconstruction and position resolution;
  \item particle identification (RICH).
\end{itemize}
}

\begin{figure}[hbt]
\centering
  \includegraphics[width=0.5\columnwidth]{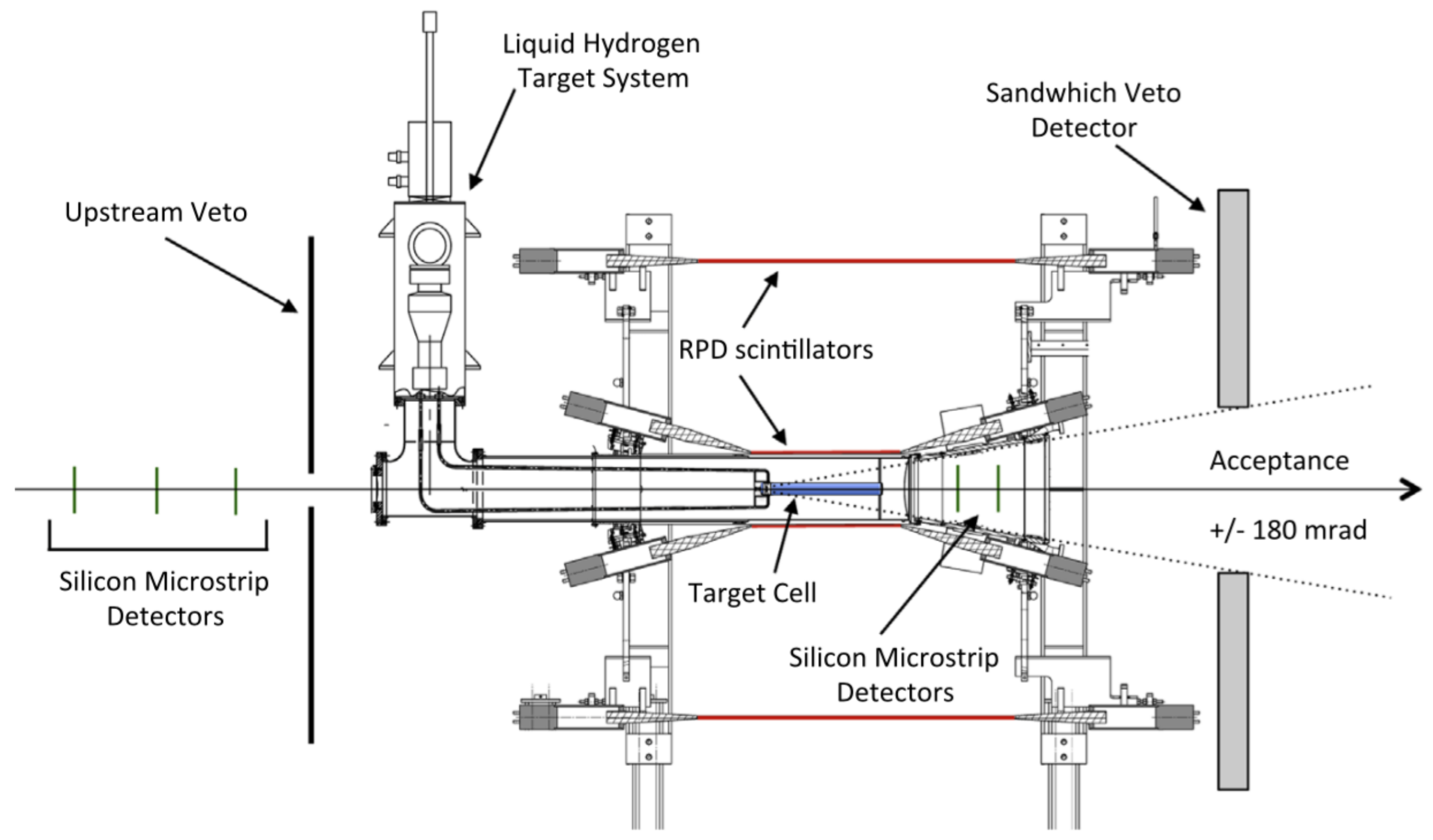}
  \caption{Longitudinal section of the \compass liquid $H2$ target.}
  \label{fig:target}
\end{figure}

The geometry of the \compass liquid hydrogen target allows for accepting particles with a polar angle smaller than $180$ mrad (or about 10 deg or pseudo-rapidity $\eta > 2.4$), see figure \ref {fig:target}.

\begin{figure}[hbt]
\centering
  \includegraphics[width=0.6\columnwidth]{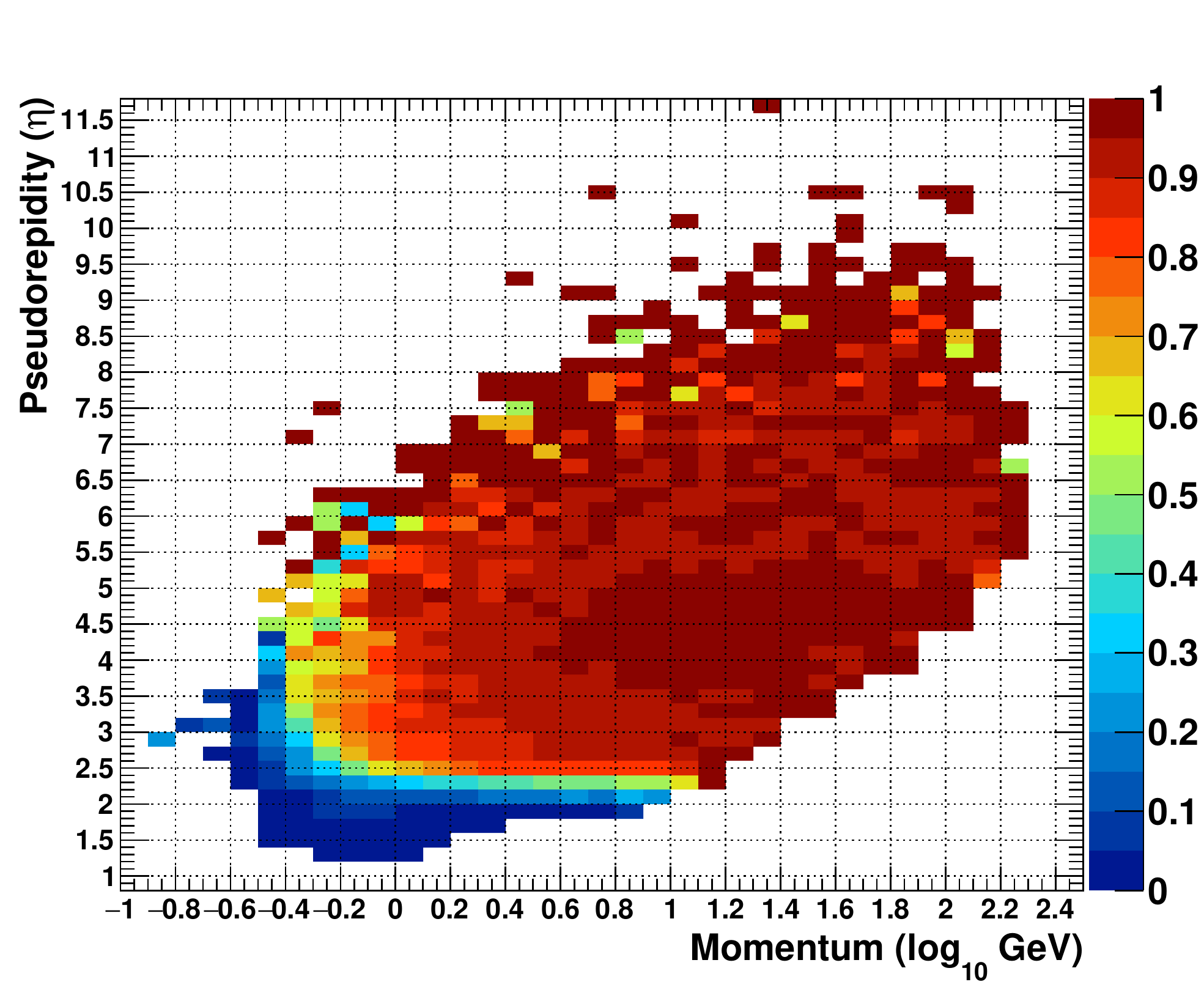}
  \caption{Double differential ($\eta$,$log_{10}(p)$) reconstruction efficiency for negative pions produced in 190\,GeV/c $p+p$ interactions.}
  \label{fig:effpim}
\end{figure}

Figure \ref{fig:effpim} shows the $\pi^-$ track-reconstruction efficiency as a function of momentum and pseudo-rapidity. For momenta above $1\,$GeV/c and pseudo-rapidity above 2.4, the tracking efficiency is larger than 90\% and mildly dependent on magnitude and direction of the momentum .
\begin{figure}[hbt]
\centering
  \includegraphics[width=0.6\columnwidth]{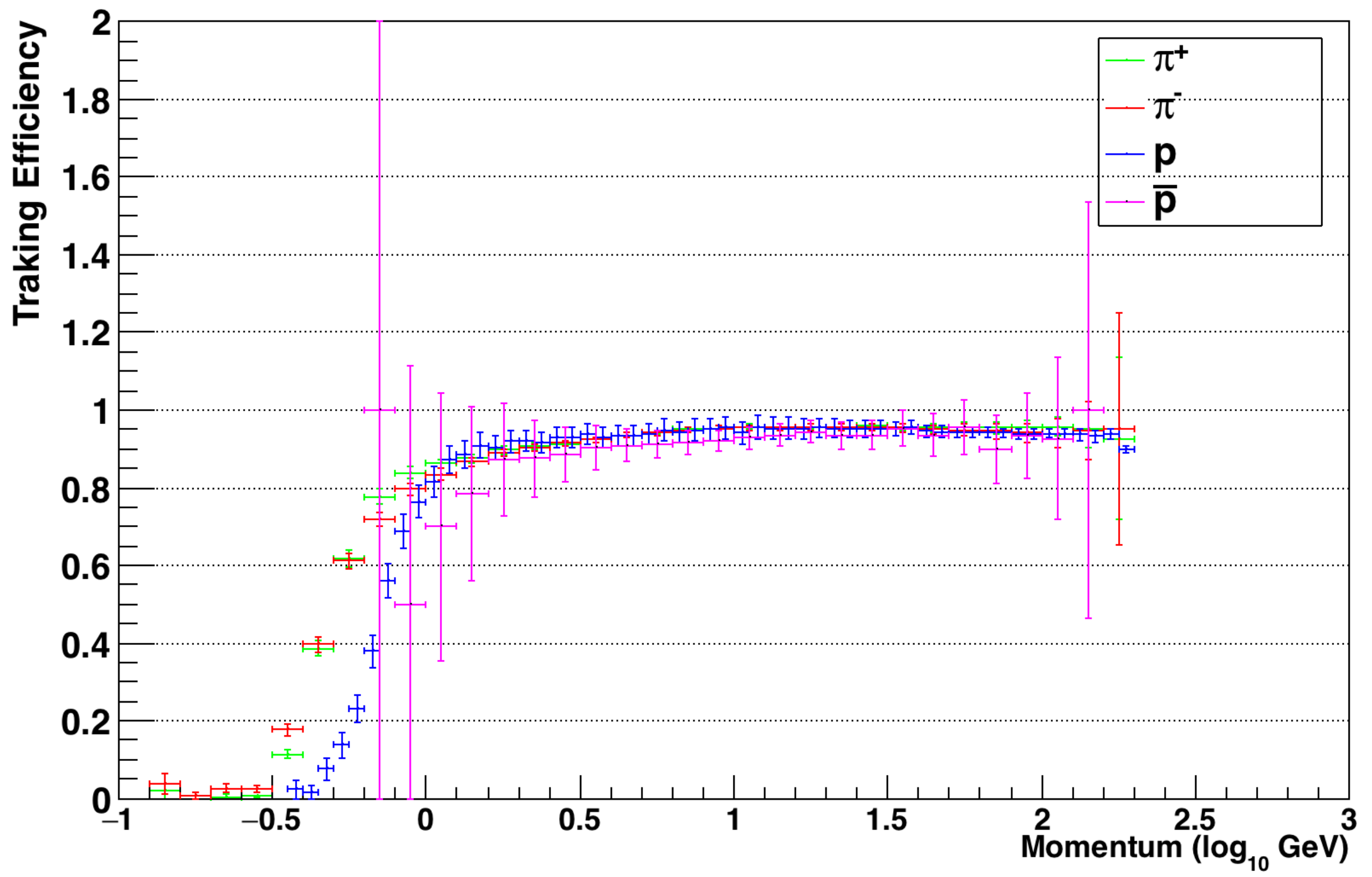}
  \caption{Tracking efficiency as function of the particle momentum, for $\pi^+$ (green), $\pi^-$ (red), $p$ (blue), $\overline{p}$ (purple)}
  \label{fig:eff2}
\end{figure}

Similar efficiencies were observed for $\pi^+$, $p$ and $\overline{p}$; Fig. \ref{fig:eff2} shows the momentum dependence of the tracking efficiency for these particles. The smaller $\overline{p}$ statistics leads to larger uncertainties in the corresponding efficiency curve.  The observed similarity between the $\pi^+$ and $\pi^-$ efficiencies suggests that the spectrometer behaves similarly for positive and negative tracks, hence it is safe to assume that $p$ and $\overline{p}$ reconstruction efficiencies are the same.  

The resolution in magnitude and direction of the momentum is also very good. The large-angle spectrometer yields $\sigma_p/p \approx 1\%$ for tracks with momentum smaller than $20\,$GeV/c,  while the small angle spectrometer yields $\sigma_p/p \approx 0.3\%$.
 The angular resolution has a typical value of $0.8\%$, while staying always better than $3\%$ in the pseudo-rapidity range $2.4 < \eta < 8$.  Track  association in vertices  is very efficient. Within the spectrometer acceptance ($\eta >2.4,  \;\;p\,>\,1\,$GeV/c), the ratio of the reconstructed track multiplicity to the MC multiplicity at the primary vertex is $0.98\pm0.05$.  The residual of the vertex position in the z direction has a width of about $0.7\,$mm.

Altogether, a \compass-like spectrometer  performs very well in reconstructing the event topology as well as the sign and the momentum of the tracks. The tracking efficiency is close to unity for momenta above about $1\,$GeV/c, see Fig. \ref{fig:eff2}.

Signals from the Ring-Imaging Cherenkov (RICH) detector allow us to measure the speed of the particle. An estimation  of the particle mass is then obtained from the velocity and the rigidity measurement, thereby providing the means for particle identification.
Considering a RICH position about $5\,$m downstream the target,  we expect to observe the following  particles: $e^\pm$, $\mu^\pm$, $\pi^\pm$, $K^\pm$, $p$ and $\overline{p}$ \footnote{Also a few hyperon tracks could trigger the RICH but their expected abundance is negligible.}. Muons are identified by their penetration capability and can  be neglected for the moment.

\begin{figure}[hbt]
\centering
  \includegraphics[width=0.6\columnwidth]{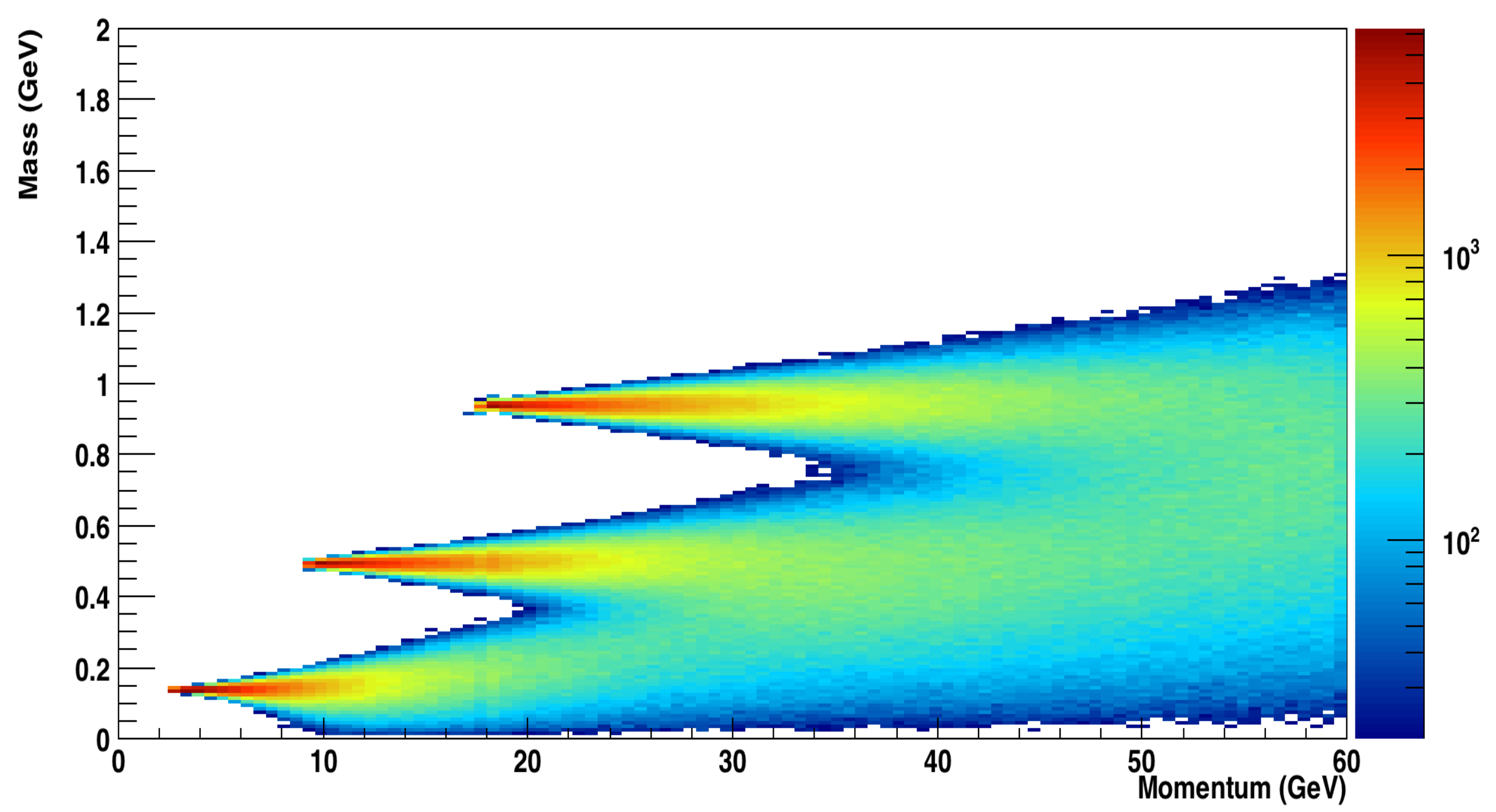}
  \caption{Particle mass reconstructed using $\beta$ and $p$ from the RICH, shown vs. reconstructed momentum. The selected sample has an equal abundance of $\pi$, $K$ and $p$. }
  \label{fig:rich_mass}
\end{figure}

The RICH radiator is a volume of $C_4F_{10}$ gas with refraction index n $\sim$ 1.0014; this corresponds to a threshold in velocity of $\beta=0.9986$. The corresponding momentum threshold depends on the particle mass: $p_{min}=2.6\,$GeV/c for pions, $p_{min}=9.3\,$GeV/c for kaons, $p_{min}= 17.7\,$GeV/c for protons.
The RICH beta resolution of $\sigma_{\beta}/\beta=0.6\%$ allows for an efficient separation of $\pi$, $k$ and $p$ by a measurement of the mass  $m=p/(\beta\gamma)$. 
Figure \ref{fig:rich_mass} shows the reconstructed mass vs. the reconstructed momentum for an equal population of $\pi$, $K$ and $p$. The very good separation of the proton signal from the $\pi$ and $K$ ones allows for an unambiguous identification of protons, and hence antiprotons in the momentum range 18 to 45\,GeV/c.
	

	Below the proton momentum threshold of about $18\,$GeV/c and above the Kaon threshold of about $10 \,$GeV/c,  the absence of a RICH signal can be interpreted as an identification of the particle as a proton (or antiproton). This identification must be corrected by the RICH efficiency, and a thorough study should be implemented to keep this efficiency under control. The use of this additional method would extend the (anti)proton identification range to momenta in the range  10 to 45\,GeV/c.

\paragraph{Measuring the antiproton cross section}
We  want to  measure the antiproton production cross section double differential in angle and momentum for the $p+p$ and $p+{}^{4}He$ process.
In each event the antiproton track is identified and counted as a function of momentum and pseudo-rapidity.
	
Events will be counted separately for momenta in the range 18-45\,GeV\,/c, where we can use the RICH to identify the antiprotons by their mass, and in the range 10-18\,GeV\,/c, where we use the absence of the RICH signal, i.e. the veto mode, to identify the particle as not $\pi$ or not $K$. In both cases the counting must be corrected for several effects including the track efficiency, the RICH detector efficiency and the particle identification efficiency.
These efficiencies can be estimated with the MC simulation and possibly directly from the data.

The antiproton cross section is obtained as ratio of the (corrected) number of events with an antiproton to the total number of interactions, where the latter is taken from the number of triggers including corrections for several effects.

The experiment will be operated with a minimum-bias trigger, which includes the:
	 \begin{itemize}
	 	\item beam trigger + hodoscope veto, which ensures that the particle reaches the experiment within the geometrical target acceptance, and it also includes a preselection of protons from the CEDAR beam line Cherenkov detectors;
	 	\item sandwich veto, which excludes events with signals outside the target acceptance;
	 	\item beam killer, which is located $32\,$m behind the target, to remove non-interacting beam particles.
	 \end{itemize} 
	 The beam intensity delivered to the M2 beam line will be adjusted to provide  about $10^{5}\,p/s$ at the target.
	 Then the above sketched trigger configuration yields a trigger rate of about $25\,$kHz, which is  well within the performances of the \compass DAQ system. In this configuration, and assuming a total of $10\,$s of SPS beam per minute, about $25\,$x$\,10^4$  events per minute will be collected and identified.
Considering the estimated production cross section for antiprotons, a conservative antiproton identification efficiency of $70\%$, and a measurement of the double-differential cross section with 20 bins in momentum and pseudo-rapidity each, a  statistical error of $1\%$ per data point will be reached after about 4~hours of beam time.
Including a contingency factor, 6~hours of beam time are required for each combination of target and beam settings. 
	  
Several corrections  to the event and trigger counts are needed to obtain an accurate measurement. Each of these correction contribute to the systematic uncertainty.
The trigger count rate must be corrected for trigger efficiency, DAQ dead time and the purity in selecting  protons in the secondary beam.
The antiproton event count needs to be corrected for the overall event reconstruction efficiency and the antiproton tracking  and  particle identification efficiencies, which possibly depend on momentum and pseudo-rapidity.
Altogether a total systematic uncertainty of about $5\%$ is expected.
		
It is planned to take data with proton momenta of 50, 100, 190\,GeV/c and at the maximum momentum achievable at the SPS M2 beam line.

\paragraph{Antihyperons and antineutrons}
In order to calculate the total amount of antiprotons produced in our galaxy, the contribution from antineutrons and antihyperons decaying into antiprotons has also to be taken into account. The total cross section is then obtained re-scaling from the prompt production, i.e. 
\begin{equation*}
\sigma_{tot} = \sigma_{prompt} (2 + \Delta_{IS} + 2\Delta_{\Lambda}) , 
\label{eq:totxsect}
\end{equation*}
where $\Delta_{IS}$ is the enhancement factor of antineutron-over-antiproton production due to isospin effects, and $\Delta_{\Lambda}$ is the hyperon factor, which assumes that antiproton and antineutron production from hyperons are equally abundant. The combined uncertainty arising from antineutron and hyperon-induced production was evaluated to not exceed $5\%$ and to be energy dependent~\cite{Winkler:2017xor}.
Moreover, with a \compass-like spectrometer secondary vertices can be reconstructed and distinguished from primary vertices, thus hyperon-induced antiproton production can be well separated from prompt production.

\subsubsection{Competitiveness and complementarity}

\begin{figure}[bt]
        \begin{center}
                \includegraphics[width=0.5\columnwidth]{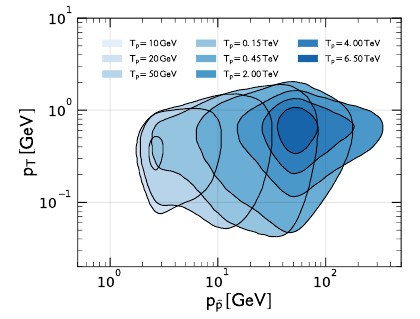}
        \end{center}
        \caption{Parameter space for the $p + {}^{4}He$ channel corresponding to an exemplary fixed-target experiment. The different
shaded areas correspond to different proton beam energies: a 3 $\%$ accuracy is required on the cross-section determination inside the blue shaded region, and a 30 $\%$ accuracy outside the contours.}
        \label{fig:dm.par_space}
\end{figure}  

As already mentioned in section \ref{sec:stdhad.xsect.phys}, the exceptional experimental accuracy of the order of a few $\%$ achieved by AMS-02 on the CR $\bar{p}$ flux and the $\bar{p}$/p flux ratio poses the challenge of achieving a similar precision in phenomenological models that describe the CR $\bar{p}$ flux as produced by the interaction of the CR primary components with the ISM. Such a phenomenological prediction is currently spoiled by the large uncertainty on the antiproton  production cross section.
In order to cover all the AMS-02  $\bar{p}$ energy range, precise $p + p \rightarrow \bar{p} + X $ and $p + {}^{4}He \rightarrow \bar{p} + X $ cross-section data are needed with a proton-beam kinetic energy $T_p$ ranging from 10\,GeV to 6\,TeV and a pseudorapidity $\eta$ ranging from 2 to almost 8.
Figure~\ref{fig:dm.par_space} shows the parameter space that has to be covered as a function of momentum $p_{\bar{p}}$ and transverse momentum $p_T$  at different kinetic energy $T_{\bar{p}}$ of the antiproton, for the case of the $p + {}^{4}He \rightarrow \bar{p} + X $ cross section. In this figure a 3\,$\%$ accuracy is required on the cross-section determination inside the blue shaded region, and a 30 $\%$ accuracy outside the contours, in order to guarantee the AMS-02 precision level on the $\bar{p}$ source term~\cite{Donato:2017ywo}.

The only data available so far from high-energy proton scattering on helium nuclei are those collected in May 2016 by LHCb 
operated in fixed-target mode with the SMOG device~\cite{Korsmeier:2018gcy} 
at 6.5 TeV ($\sqrt{S_{NN}}= 110$\,GeV, pseudo-rapidity range $ 2 < \eta < 5 $ and detected antiproton momentum range $12 < T_{\bar{p}}< 110\,\GeV/c$). A not-yet-published second dataset was collected by LHCb at $\sqrt{S_{NN}}= 86.6$\,GeV in November 2016.\\
A new fixed-target experiment at the CERN SPS M2 beam line could perform measurements of antiproton production in $p + {}^{4}He$ collisions at different momenta of the proton beam, from a few tens of\,GeV/c up to about $250\,$GeV/c and for pseudo-rapidities $\eta>2.4$.
Combined with the LHCb measurements at very high energy, the new data could yield the necessary kinematic coverage. This would contribute to a significant reduction of the uncertainty on the expected amount of secondary antiprotons produced by spallation of primary cosmic rays on the interstellar medium, which is currently one of the most limiting factors when using the AMS-02 data on the $\bar{p}$ flux and the $\bar{p}$/$p$ flux ratio as input for the indirect Dark Matter search.

\svnid{$Id: rfsep.tex 4376 2018-09-11 09:50:37Z avauth $} 
\section{Hadron physics with RF-separated beams}
\label{sec:rfsep}
\svnid{$Id: rfsep_beams.tex 4376 2018-09-11 09:50:37Z avauth $} 
\subsection{Beam line}
\label{sec:rfsep.beams}

In view of several proposals to perform new experiments with high-energy hadron beams at CERN, a study of a possible enrichment of desired particle species in the M2 beam has been launched by EN-EA in the context of the Physics Beyond Colliders Initiative.
Contrary to the case of lower energies as described in Sec.\,\ref{sec:lepbar.beam}, at higher energies an enrichment of antiprotons is not naturally given by decays of other particles over the length of a beam line, which is due to higher lifetimes of particles in the laboratory frame. In addition, several proposals prefer a higher content of kaons and positive pions in the beam.\\ 
Starting from studying limitations in terms of production of particles, there are several possibilities to enrich the content of a wanted particle species in the beam, usually by suppression of unwanted particles. Due to the $1/p^3$ dependence of electro-static separators, it is not reasonable to use such a method at beam energies higher than a few GeV. While in principle an enrichment by differential absorption would be feasible, the very low efficiency, high losses, and small suppression factors for unwanted particles leave only the possibility of radio-frequency (RF) separated beams.\\
The method of RF-separation was first employed at CERN in the 1960s based on ideas of Panofski and Schnell as for instance described in Ref.\,\cite{Bernard:275757}. The main idea is based on the different velocities of particle species in a beam with defined momentum. 

\begin{figure}[tbp]
  \begin{center}
    \includegraphics[width=0.8\columnwidth]{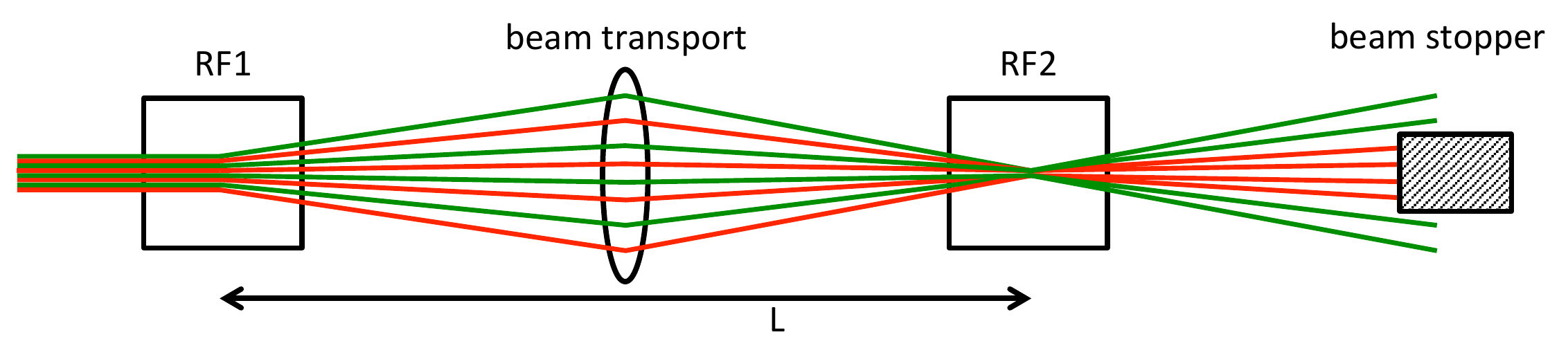}
    \caption{Panofsky-Schnell method for RF-separated beams. The unwanted particles (red) are stopped by a beam stopper while the wanted particles (green) receive a net deflection by the combination of the RF1 and RF2 dipole RF cavities out of the central axis, which is sufficient to go around the stopper.} 
    \label{fig:panofsky-schnell}
  \end{center}
\end{figure}

\begin{figure}[tbp]
  \begin{center}
    \includegraphics[width=0.8\columnwidth]{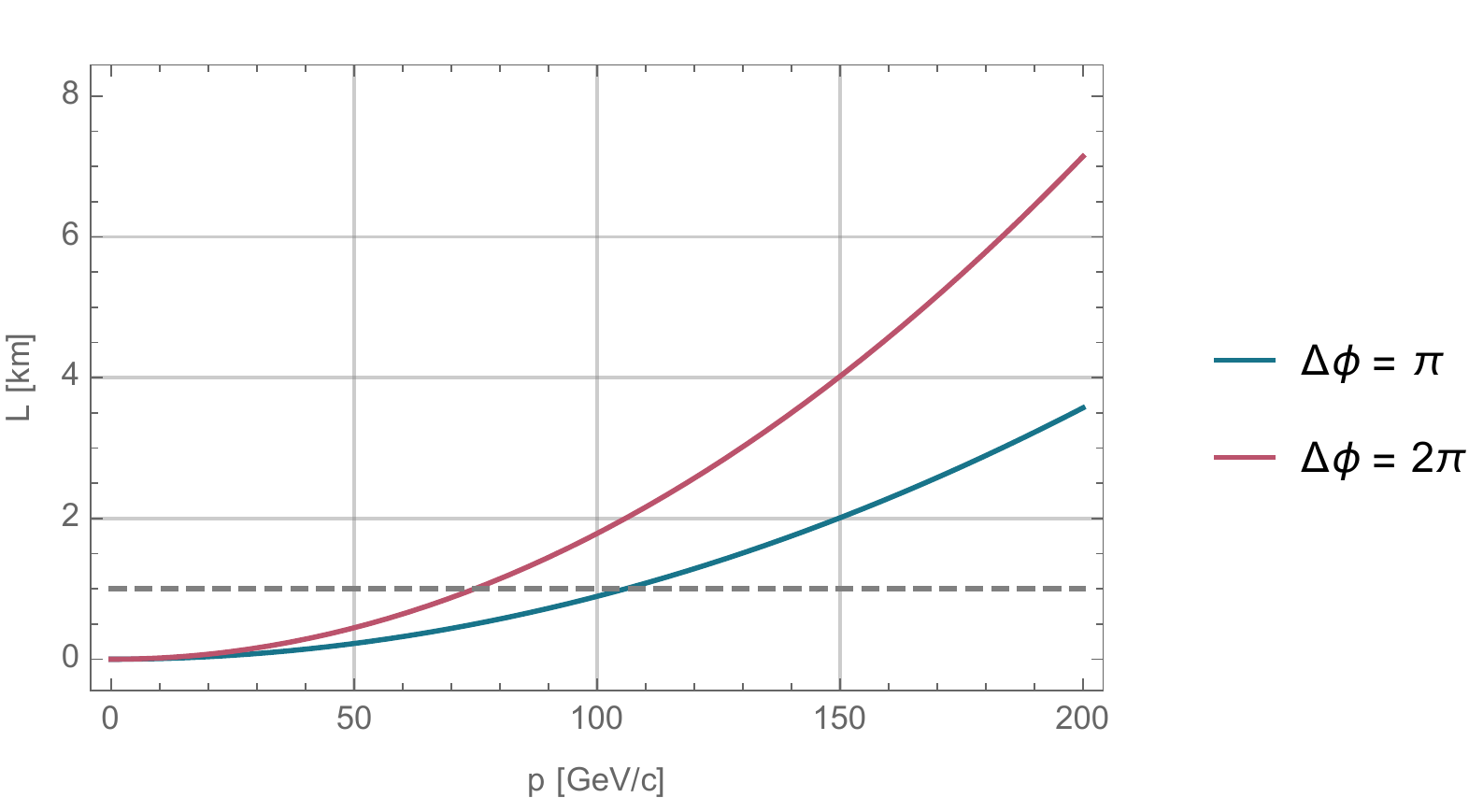}
    \caption{Dependence of the final beam momentum $p$ as a function of length $L$ between the RF cavities for two different phase differences. The case of $\Delta \Phi = 2 \pi$ (red) corresponds to kaons as wanted particles while $\Delta \Phi = \pi$ (blue) would be the choice for antiprotons. The dashed horizontal line indicates the maximum momentum possible for the given particle type at the M2 beam line.} 
    \label{fig:JPlotwholerange}
  \end{center}
\end{figure}

As displayed in Fig.\,\ref{fig:panofsky-schnell}, two dipole RF cavities (RF1 + RF2) with frequency $f$ are implemented at a given distance $L$. The transverse kick of RF1 is either amplified or compensated by RF2 depending on the phase difference between both. This phase difference is given by the difference of velocities of the various particle species. For two species $i$ $(i=1,2)$ with masses $m_i$ and velocities $\beta_i$, the phase difference reads $\Delta \Phi = 2 \pi (L f / c) (\beta_1^{-1} - \beta_2^{-1})$. In the limit of large momenta, the phase difference can be expressed as a mass difference between the two species at the beam momentum $p$: $$ \Delta \Phi = 2 \pi (L f / c) \frac{m_1^2-m_2^2}{2p^2}$$
For kaons as wanted particles, the phase difference could be chosen at $\Delta \Phi_{\pi p} = 2\pi$, which results in $\Delta \Phi_{\pi K} = 94^{\circ}$. This means that the kick for both protons and pions would be compensated by RF2 and they would be absorbed in the beam stopper. The kaons would receive a close-to-maximum transverse kick and mostly go around the stopper. For antiproton beams, the phase difference could be chosen at $\Delta \Phi_{\pi \bar{p}} = \pi$, which results in $\Delta \Phi_{\bar{p} K} = 133^{\circ}$ and $\Delta \Phi_{\bar{p} e} = 184^{\circ}$. In this case, the antiprotons would receive an acceptable deflection while electrons and pions are dumped effectively.
Based on a study by J.Doornbos at TRIUMF for CKM, we assume a similar input for frequency ($f=3.9$\,GHz) and kick strength of the RF cavities ($dp_T=15$\,MeV$/c$). Given the length of 1.1\,km of the M2 beam line, the length $L$ between cavities cannot be chosen larger. In such a study case, the upper momentum limitation for RF-separated kaon beams would be about 75\,GeV$/c$ and about 108\,GeV$/c$ for RF-separated antiproton beams, see Fig.\,\ref{fig:JPlotwholerange}. As the phase difference depends quadratically on the chosen momentum, such beams would deliver acceptable separation only in a small momentum band. In addition, the dispersion of the beam $\Delta p / p$ needs to be limited to about 1\,\% in order to prevent a phase shift of $\Delta \Phi_f = \Delta \Phi_i (1-2\Delta p / p)$ and thus a lower separation efficiency.

With the given acceptance values and target efficiency as explained in Sec.\,\ref{sec:lepbar.beam}, an exemplary calculation was performed for the case of a 100\,GeV$/c$ antiproton beam. Assuming that 80\,\% of the antiprotons would pass the beam stopper and an optimisation of the solid angle to $10$ $\pi$ $\upmu$sterad, one would expect about $8\cdot10^{7}$ antiprotons in the experimental hall EHN2 for $10^{13}$ incident protons at the T6 target. Due to the current radiation protection restrictions for EHN2 of $10^8$ particles per $4.8$\,s spill, the limit would be given only by the to-be-achieved purity of the beam. Assuming 50\,\% purity, this would be about $5\cdot10^{7}$ antiprotons per spill.

\svnid{$Id: rfsep_spect.tex 4400 2018-10-05 08:46:26Z avauth $} 
\subsection{Spectroscopy of kaons}
\label{sec:rfsep.spect}
\newcommand*{\twoPi}{\ensuremath{{\pi^-\pi^+}}\xspace}
\newcommand*{\threePi}{\ensuremath{{\pi^-\pi^-\pi^+}}\xspace}
\newcommand*{\reaction}{\ensuremath{\pi^- + p \to \threePi + p_\text{recoil}}\xspace}
\newcommand*{\Kpi}{\ensuremath{{K^-\pi^+}}\xspace}
\newcommand*{\Kpipi}{\ensuremath{{K^-\pi^-\pi^+}}\xspace}
\newcommand*{\reactionK}{\ensuremath{K^- + p \to \Kpipi + p_\text{recoil}}\xspace}
\newcommand*{\tpr}{\ensuremath{{t'}}\xspace}

\subsubsection{Physics case}
\begin{figure}[tbp]
\centering
\includegraphics[width=0.78\linewidth]{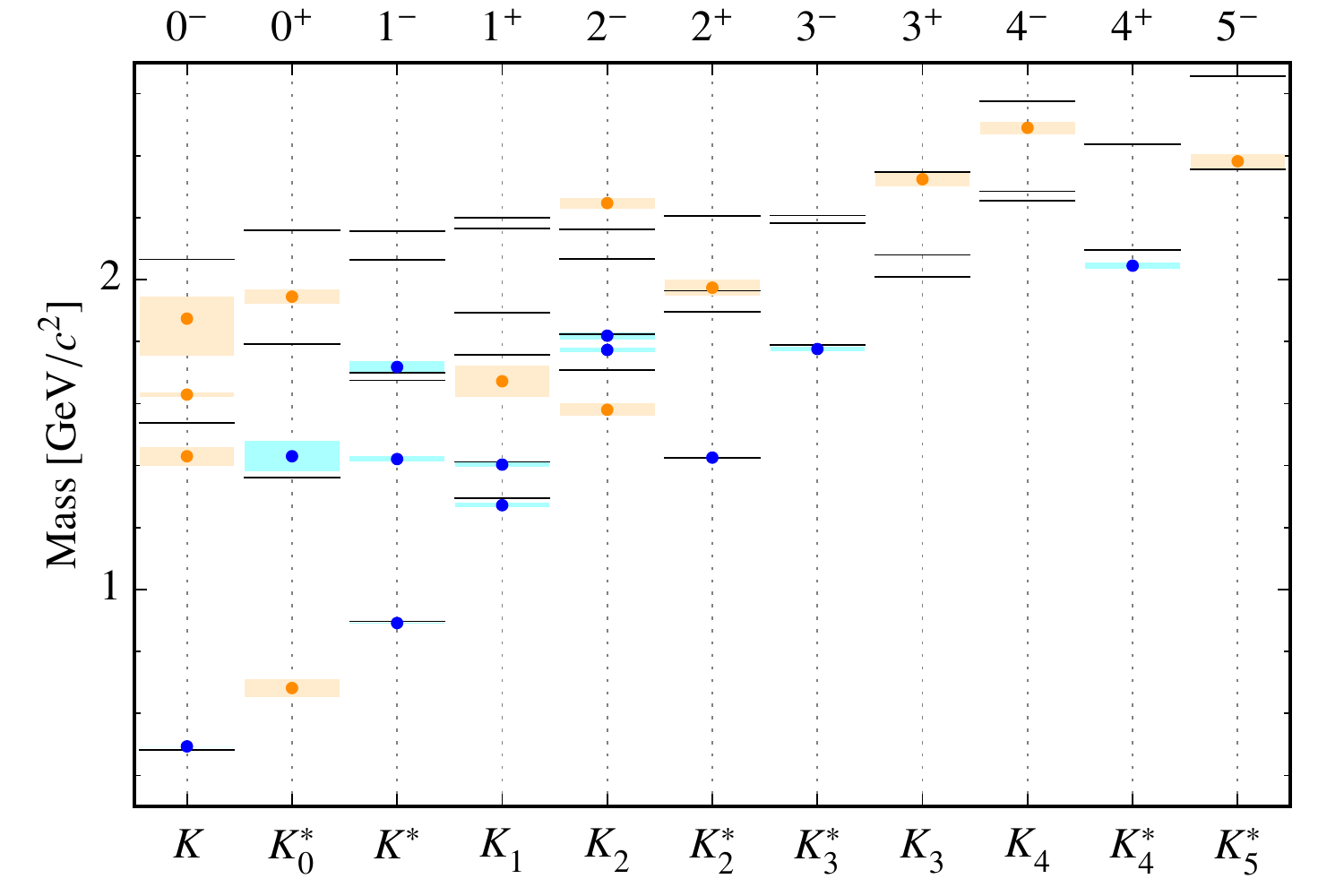}
\caption[Excitation spectrum of strange mesons from PDG.]{Excitation spectrum of strange mesons from PDG~\cite{Olive:2016} (points and shaded boxes representing the central value and uncertainty of the measurements, respectively) compared to a relativistic quark-model prediction from Ref.~\cite{Ebert2009} (black lines). States included in the PDG summary table are shown in blue, the remaining states are shown in orange. The states are grouped by their \JP quantum numbers.}
\label{fig:rfsep.spec:KaonSpectrum}
\end{figure}
The Particle Data Group lists 25 strange mesons that have been measured in the mass range from $0.5$ to $3.1\,\GeV/c^2$~\cite{Olive:2016}. Only 12 of them are included in the summary tables. The remaining 13 states still need further clarification. For two of them, even their spin-parity quantum numbers \JP are not yet determined.
Figure \ref{fig:rfsep.spec:KaonSpectrum} shows the masses of the observed strange mesons and compares them to a relativistic quark-model calculation from Ref.~\cite{Ebert2009}. For some well-known states, like e.g. the $K$ ground state, the $K^*(892)$, the $K_1(1270)$, and the $K_1(1400)$, the quark-model prediction agrees well with the experimental observations. However, many predicted states have not yet been observed and some of the observed states do not fit into the quark-model picture. While the PDG lists e.g. three excited $K$ states with $J^{PC}=0^{-+}$ in the region below $2.5\,\GeV/c^2$, the quark model predicts only two states, neither of those matching with the observed states.
Another example are the $K^*_0$ states, among which the $K^*_0(1430)$ is the best established one. 
There is also some experimental evidence for an excited $K^*_0(1950)$, but the observed mass is between the masses of two $K^*_0$ states predicted in Ref.~\cite{Ebert2009}. However, another quark-model calculation in Ref.~\cite{Godfrey1985} predicts only one excited state in better agreement with the experimental observations.
The most disputed state is the $K^*_0(800)$ or $\kappa$. The quark-model calculations in Refs.~\cite{Godfrey1985,Ebert2009} predict no $K^*_0$ state below $1\,\GeV/c^2$. Also the experimental situation is not clear. In many experiments, significant intensity is observed below $1\,\GeV/c^2$, which is typically parameterised as an ``effective-range non-resonant'' component with a phase shift~\cite{Bevan2014}. However, more advanced analyses using a $K$-matrix approach~\cite{Palano2017} or Roy-Steiner equations~\cite{Descotes-Genon2006}, find a pole below $1\,\GeV/c^2$ associated with the $K^*_0(800)$.
This situation is similar to the challenges in understanding the $S$-wave in the $\pi\pi$ system. There, only high-precision data~\cite{Batley2010} in combination with advanced models~\cite{GarciaMartin:2011jx} allowed one to establish the $f_0(500)$ state and to determine its parameters.
These examples show that in the strange-meson sector there are still many missing states and open questions, which need to be addressed. The final goal is to identify all strange and light non-strange mesons in the quark-model multiplets. This would allows us to single out supernumerous states and identify multiplets beyond the quark model, including e.g. gluonic excitations. 

Most of the experimental data on strange mesons are based on experiments that were performed more than 30 years ago. Since PDG listings from 1990~\cite{Bertsch1990}, only four additional kaon states were included in the PDG listings and only one state entered the summary tables.
Nevertheless, strange mesons appear in many processes in modern hadron and particle physics. An example are searches for CP violation in multi-body heavy-meson decays, e.g. in $B^\pm\to D^0 K^{\pm}$ with $D^0\to K^{0}_\text{S}\pi^+\pi^-$, which are currently under study at $B$-meson factories like BaBar~\cite{Aubert2008}, Belle~\cite{TheBelleCollaboration2010}, and LHCb~\cite{Aaij2014} and will remain an interesting topic with the upcoming high-precision measurements at Belle II and LHCb. These CP-violation searches are usually Dalitz-plot amplitude analyses~\cite{Bevan2014}. Typically, the isobar model is used, where the decay amplitudes are parameterised by intermediate resonances appearing in the various subsystems of the final-state particles.
In order to keep pace with the high statistical precision of the data, the models require as input an accurate knowledge of these appearing intermediate states, e.g. the strange mesons appearing in the $K^{0}_\text{S}\pi^\pm$ subsystems in the example above.
The large datasets allow to directly study strange mesons in heavy meson decay, as done e.g. in Refs.~\cite{Aubert2008,TheBelleCollaboration2010}. However, even with the biggest datasets, in the employed isobar models typically the masses and widths of only a few selected states can be fitted to the data while the parameter values of most of the kaon states included in the fit need to be taken from other measurements.

\subsubsection{Previous measurements}
Complementary to directly studying strange mesons is the investigation of peripheral production, i.e.\ diffractive production or charge-exchange reactions, in scattering a high-energy kaon beam off a fixed target. In the past, this reaction was measured by experiments at BNL~\cite{Etkin1980}, CERN~\cite{Otter1976,Vergeest1979,Daum1981a,Baubillier1982,Armstrong1983}, and SLAC~\cite{Brandenburg1976,Aston1987,Aston1988,Aston1993} to study strange mesons decaying into various final states.
For example, the LASS experiment measured $10^5\,\text{events}$ of the $K^-\omega$ final state using an $11\,\GeV/c$ $K^-$ beam~\cite{Aston1993}. The analysis of these data contributed significantly to establishing the $K_2(1820)$ state.
One of the largest datasets was acquired by the ACCMOR collaboration using a $63\,\GeV/c$ $K^-$ beam~\cite{Daum1981a}. They analyzed $2\times 10^5\,\text{events}$ of the reaction \reactionK. From this dataset, they extracted the parameters of five strange mesons and studied the excitation spectrum of the $J^P=0^-$ $K$ states.
However, even these large datasets are not sufficient to resolve the details of the kaon spectrum, especially for kaon states at higher masses, e.g. the excited $J^P=0^-$ $K$ states.

Also, COMPASS has measured strange mesons in peripheral production in the years 2008 and 2009 using the about $2.5\,\%$ $K^-$ fraction in their secondary hadron beam. In a first analysis of $2.7\times10^5\,\text{events}$ of the reaction \reactionK~\cite{Jasinik:2012}, the results were found to be consistent with previous measurements~\cite{Daum1981a,Aston1993}. Currently, the analysis is being improved. By major enhancements of the applied PID methods, a dataset of $8\times10^5\,\text{events}$ is envisaged. This would increase the statistical precision by a factor of two as compared to the ACCMOR analysis, thereby allowing one to measure masses and widths with improved precision and possibly give access to some higher-lying kaon states.

\subsubsection{Novel analysis techniques}
The main goal of the COMPASS physics programme in the years 2008 and 2009 was the measurement of the spectrum of light non-strange mesons. In analogy to the \Kpipi channel for strange-meson spectroscopy, the flagship channel for light non-strange meson spectroscopy is \reaction, for which $50\times10^6$ exclusive events were acquired~\cite{Adolph:2015tqa}. This huge dataset allows one to apply novel analysis methods, i.e. to perform the partial-wave analysis independently in narrow bins of the squared four-momentum transfer \tpr between the beam pion and the target proton. The additional information from this \tpr-resolved analysis helps to better separate the resonant and non-resonant contributions. For the first time, also the \tpr dependence of individual signals can be extracted~\cite{Paul2017,Krinner2017}. Especially for the spectroscopy of strange mesons, a dataset  large enough to perform a \tpr-resolved analysis would be helpful to separate the many overlapping states with the same \JP quantum numbers, e.g. the $K_1(1270)$ and $K_1(1400)$.
In the \threePi analysis, the \twoPi subsystem of the \threePi final state is studied by using the so-called ``freed-isobar'' approach~\cite{Adolph:2015tqa}. Applied to strange-meson spectroscopy, this approach would allow one the study of e.g. the $K^*_0$ states in the \Kpi subsystem of the \Kpipi final state. However, large datasets are mandatory in order to apply this method.

The high statistical precision of such measurements allows the study of much weaker signals, like e.g. the $a_1(1420)$ signal discovered in the above mentioned  $50\times10^6$ exclusive $\pi^- \pi- \pi+$ final state events~\cite{Adolph:2015pws}. Although this signal contributes only $0.3\,\%$ to the total intensity, a clear $a_1(1420)$ signal is observed and its parameters are extracted with high precision. We note that some of the missing strange-meson states, which are predicted by the quark-model, could have such small signals.
These examples clearly show that large datasets would not only improve the statistical precision of the measurements, but first and foremost would open a whole new field of novel analysis methods and thus would lead to new insights into the strange-meson sector. In order to apply the methods discussed above also to strange-meson spectroscopy, a dataset of at least $10\times10^6$ to $20\times10^6\,\text{events}$ of the flagship channel \reactionK needs to be acquired, which is a factor 15 to 25 more than what was measured so far.

\subsubsection{Future measurements at the SPS M2 beam line}
In order to obtain such a unique dataset for strange-meson spectroscopy, the $K^-$ fraction in the beam has to be considerably increased. One possibility is the use of an RF-separated beam.
With a kaon-beam intensity of $5\times10^7\,\text{per spill}$ at the target position of the experiment, a \Kpipi sample of about $20\times10^6\,\text{events}$ can be recorded within one year of data taking.\footnote{In total, $50\times10^6$ \threePi events were acquired within one year of data taking with a $\pi^-$ beam intensity of $5\times 10^7\,\text{per spill}$~\cite{Abbon:2014aex}. Assuming that due to the final-state PID the detection efficiency for \Kpipi is approximately $50\,\%$ of the one for \threePi, $20\times10^6$ \Kpipi events are expected for one year of data taking.}
Diffractive production does not depend strongly on the beam energy. With a beam momentum of at least $50\,\GeV/c$, diffractive production will be the dominant process and beam excitations can be well separated from target excitations. This is very important in order to obtain a clean sample of exclusive events and to keep systematic uncertainties from contributions of other processes small.
The most important requirement for the experimental setup is a uniform detection efficiency over a broad kinematic range.
Apart from precise tracking and vertex reconstruction, a good particle identification is mandatory. As the RF separation does not lead to a pure kaon beam, an efficient beam-particle identification with a low misidentification probability by the CEDAR detectors is required. This requires a small beam divergence at the position of the CEDARs.
Additionally, kaons have to be distinguished from pions in the final-state, e.g. for the \Kpipi final state. This requires a good final-state particle identification over most of the momentum range from about $1\,\GeV/c$ up to the beam momentum, with high efficiency above $50\,\%$. In order to study also final states with neutral particles, like $K^-\pi^0\pi^0$, the detection of photons over a broad kinematic range by electromagnetic calorimeters is important.

\subsubsection{Planned or proposed measurements at other facilities}
There are also proposals and plans for future measurements of strange mesons at other facilities.
The investigation of $\tau$ decays, in which strange mesons can appear in subsystems, as e.g. in $\tau^-\to K^-\pi^+\pi^-\nu_\tau$, will be pursued at Belle II, BES III and LHCb to study strange mesons. However, the largest possible mass of the strange subsystem is limited by the rather low $\tau$ mass of $1.8\,\GeV/c^2$, so that many of the observed or predicted kaon states are out of reach (see figure \ref{fig:rfsep.spec:KaonSpectrum}). Furthermore, the event samples are typically an order of magnitude smaller than those of measurements using peripheral production~\cite{Asner2000,Epifanov2007}. Nevertheless, the low-mass tails of the higher-lying kaon states may still play a role in the mass range of the $\tau$ decays. This means that the analysis of $\tau$ decays would benefit from a high-precision measurement of those states at the CERN SPS M2 beam.
The situation is similar for heavy-meson decays. In $D$ decays like e.g. $D\to K\pi\pi$~\cite{Link2007,Aaij2016}, the mass range is limited by the $D$ mass of $1.86\,\GeV/c^2$. In $B$ decays, the limited size of the dataset precludes studying strange mesons with high precision.

Another process, by which strange mesons can be studied, is photo-production. For example, GlueX proposed a measurement of the $K K \pi \pi$ final state, for which they expect a dataset of $100\times10^6\,\text{events}$~\cite{Dugger}. Using an approach similar to the ``freed-isobar'' method, they could study strange mesons in e.g. the $K\pi$ and $K\pi\pi$ subsystems.
However, it might be challenging to obtain accurate insight into the strange subsystems from four-body final states, as compared to direct strange-meson production discussed above.
Recently, measurements with a secondary $K_L$ beam were proposed at GlueX~\cite{Amaryan2017}. In their proposal, they focus on hyperon spectroscopy. For a strange-meson spectroscopy program, they mention only the charged and neutral $K\pi$ final state, which gives them access only to $K^*_J$ states.

At J-PARC, a new beam line with a separated kaon beam will be built in the near future~\cite{Fujioka2017}. They aim for a $K^-$ intensity of $10^7\,\text{per spill}$, similar to the above proposal; however, with a much lower beam momentum of $2$ to $10\,\GeV/c$. At such low momenta, the separation between beam and target excitations will become difficult and might lead to larger systematic uncertainties. To our knowledge, no strange-meson spectroscopy programme was proposed at J-PARC so far and there are no plans for a general-purpose detector with high-precision tracking and calorimetry, which is needed for spectroscopy as discussed above.

Altogether, most of the elsewhere planned or proposed measurements in the strange-meson sector can either not compete with the measurements proposed above or are complementary to them. Therefore, a spectroscopy programme at CERN using an RF-separated kaon beam from the M2 beam line would be a unique opportunity to study the excitation spectrum of strange mesons in great detail using the advanced methods that were developed by COMPASS for the study of the \threePi final state. This would significantly improve the precision of known states, allow one to search for new states that complete the light-meson multiplets, and it would clarify some of the open questions.

\svnid{$Id: rfsep_dryan.tex 4405 2018-10-09 12:15:17Z vandrieu $} 
\subsection{Drell-Yan physics with high-intensity kaon and antiproton beams}
\label{sec:rfsep.dryan}

In the conventional quark model, the properties of hadrons are mainly determined by their valence-quark structure. An exchange of a $u$ quark by a $d$ quark makes the neutron different  from the proton. Similarly, a replacement of the $d$ quark by an $s$ quark makes the kaon different from the pion. The heavier quark in the kaon leads to a significantly heavier hadron mass, much larger than the difference between the $s$ and $d$ quark masses. The mass scale in each hadron,  which is generated through dynamical chiral-symmetry breaking, is associated with gluon propagation;  the massless gluons acquire running mass, which is then transmitted to the quark sector. Exploring the hadron structure, and particularly the quark and gluon distributions in the lightest mesons,  provides a glimpse of the appearance of the hadron mass and its connection with colour confinement,  as explained in Sec. \ref{ssec:stdhad.dryan.intro}. At present, the kaon valence distributions are nearly  unknown and no information exists neither on the kaon sea, nor on the kaon gluon distribution. On the  theoretical side, the situation is rapidly evolving: a number of calculations based on various approaches are now addressing the kaon PDFs, usually as an extension of pion PDF studies.  

The availability of a kaon beam, such as the one envisaged by radio-frequency separation of charged hadrons at the SPS M2 beam line, would provide a unique opportunity for performing  extensive studies of the parton structure of the kaon. A high-intensity kaon beam would allow for  Drell-Yan measurements with unprecedented statistics. A detailed comparison between the quark structure  of the two lightest hadrons becomes possible. The Drell-Yan kaon data should be complemented with J/$\psi$ production  and prompt-photon measurements, thereby also paving the way for a determination of the gluon structure of the kaon. 

The availability of an RF-separated antiproton beam would make measurements of nucleon single-spin asymmetries with reduced systematic uncertainties possible. Because of charge symmetry, the antiproton-induced Drell-Yan process  is only sensitive to convolutions of valence-quark TMD PDFs of the nucleon. The M2 beam line, with an RF-separated  beam tuned for high-intensity antiprotons with energies of about 100\,GeV, would provide the only presently  foreseeable possibility for such measurements in the world.

\subsubsection{Studies of the spin structure of the nucleon with an antiproton beam}
\label{ssec:rfsep.dryan.antiproton}

Studying the Drell-Yan process with an antiproton beam and a transversely polarised proton target shall provide an ideal opportunity to study the transverse-momentum-dependent (TMD) PDFs of the nucleon.  Compared to the pion-induced Drell-Yan studies presently being performed at \compass, the uncertainties  related to the limited knowledge of the pion structure shall be eliminated. Additionally, an  extended $x_N$ region shall be explored.

The antiproton-induced Drell-Yan process using a transversely polarised proton target is the most promising way to access the Boer-Mulders function of the nucleon. In the Drell-Yan cross section,  two transverse spin-dependent modulations can be measured, which result from convolutions of the $\bar u$-valence Boer-Mulders function in the antiproton with the $u$-valence transversity function in the proton or with the $u$-valence pretzelosity function in the  proton. The former can be accessed through the $\cos(2\phi_{CS}-\phi_s)$ modulation and the latter through the  $\cos(2\phi_{CS}+\phi_s)$ modulation. Given the present knowledge of the $u$-transversity distribution  in the nucleon, which was extracted from the SIDIS results of the \compass\ and HERMES experiments, one can aim at  accessing the $u$-Boer-Mulders distribution of the nucleon. 

\begin{figure}[tb]
\centering
\includegraphics[width=0.6\textwidth]{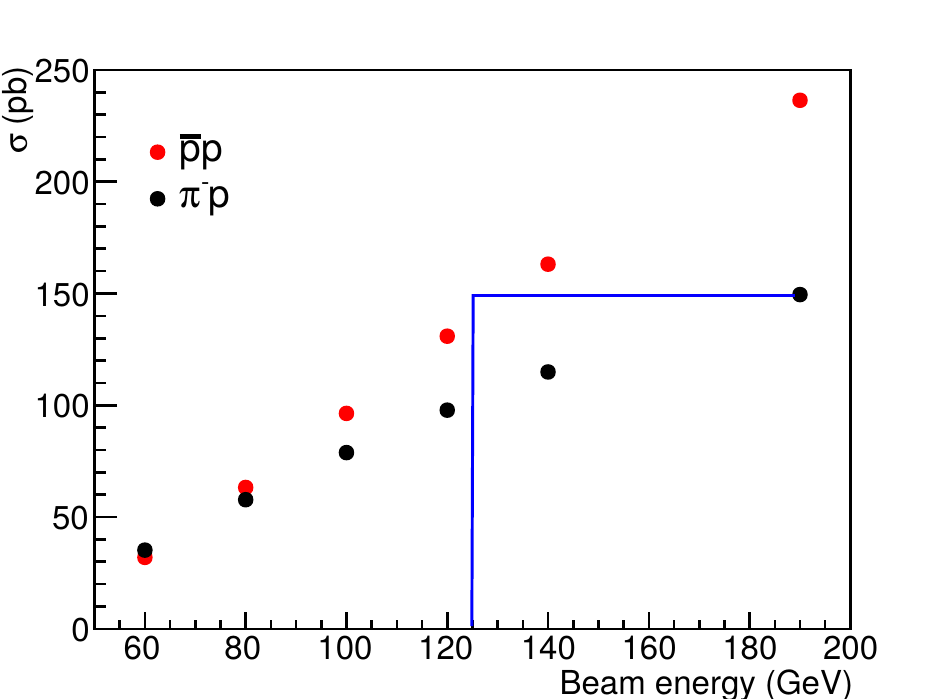}
\caption{Drell-Yan cross-section dependence on the beam energy for $\pi^-$-induced and antiproton-induced processes.}
\label{fig:antiproton_xsection}
\end{figure}

As compared to the pion-induced Drell-Yan process, the antiproton-induced case has a larger cross section. In spite of the expected limitations of the RF-separation discussed in Sec.~\ref{sec:rfsep.beams},  we also explore the case of an increased beam energy, which is an option that would require additional R$\&$D work. Figure~\ref{fig:antiproton_xsection} compares the cross-section dependence on the beam energy for $\pi^-$-induced and antiproton-induced Drell-Yan processes, which clearly demonstrates the advantage of larger beam energies.

\begin{figure}[tb]
\centering
\includegraphics[width=.48\textwidth]{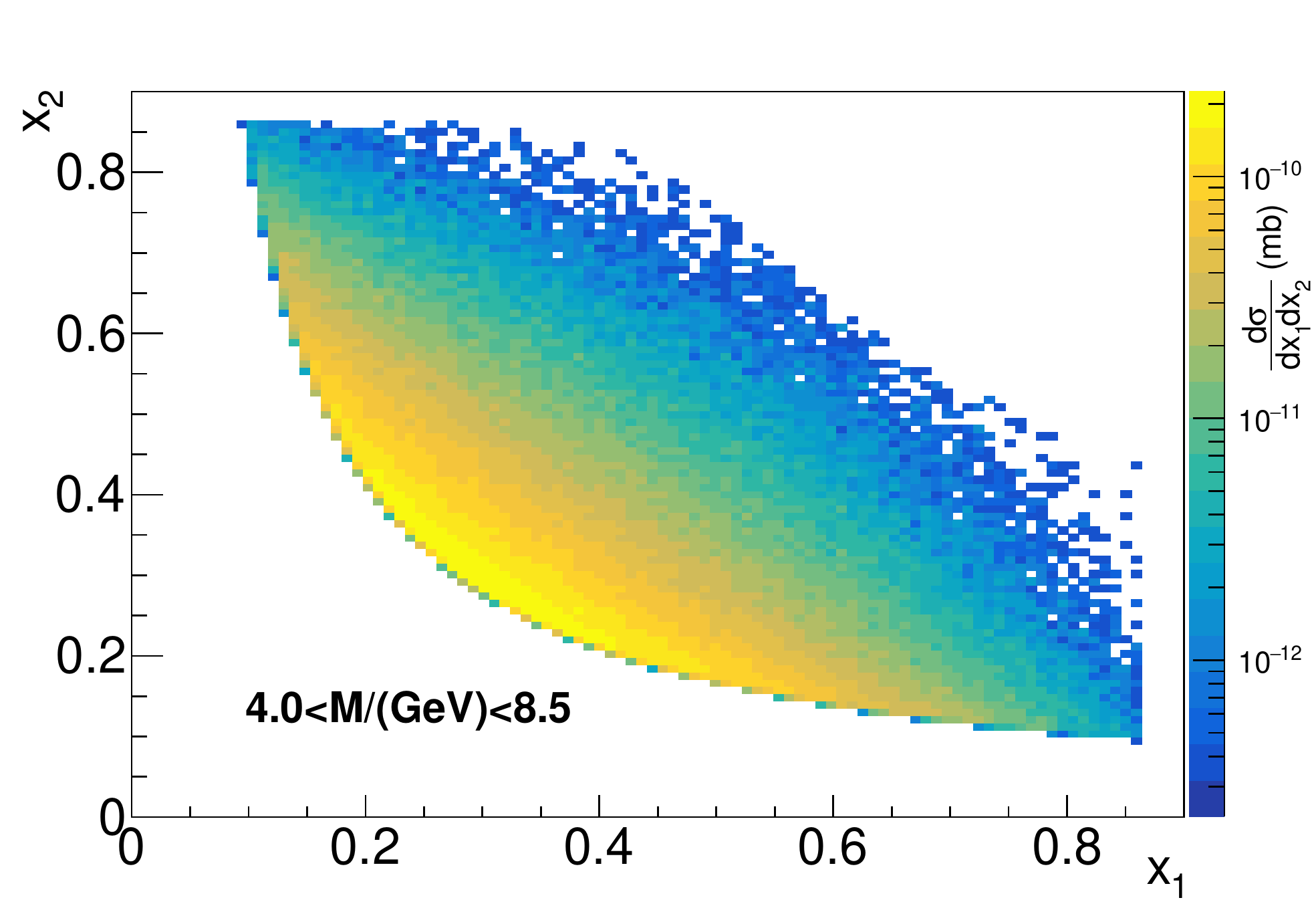}
\includegraphics[width=.45\textwidth]{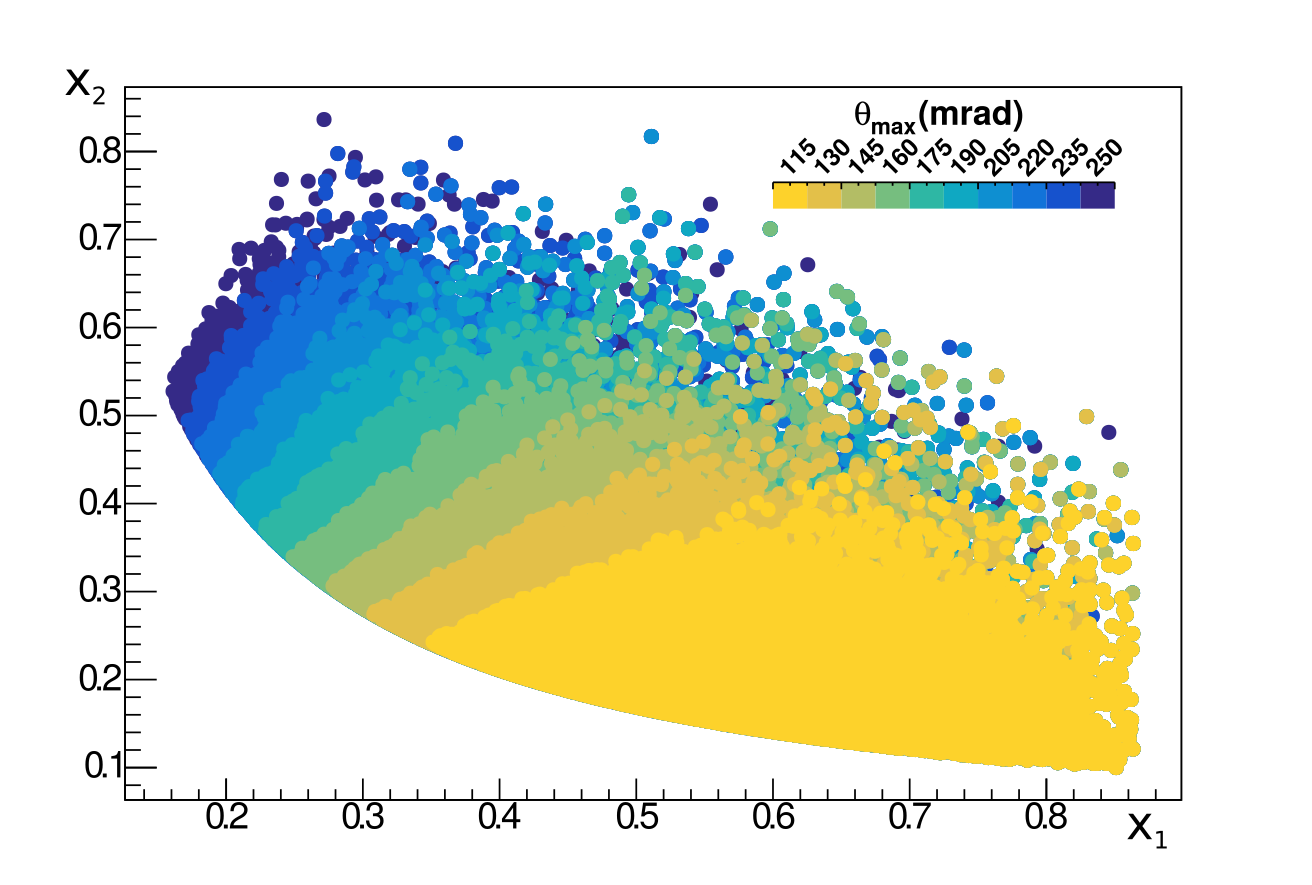}
\caption{Drell-Yan dimuon cross section for a 100\,GeV antiproton beam and a NH$_3$ target of 110\,cm length (left), where dimuons with $4.0 < M < 8.5$\,GeV are considered. The angular acceptance for muons in the proposed experiment is also shown (right).}
\label{fig:antiproton_accept}
\end{figure}

For a 100\,GeV antiproton beam, the requirement that the overall acceptance of the setup is above 40\% implies to cover polar angles of up to 250\,mrad.  This is illustrated in the right panel of Fig.~\ref{fig:antiproton_accept}.

These simple studies illustrate that a change of paradigm with respect to earlier Drell-Yan experiments  is needed to achieve the large statistics mandatory for studies of azimuthal asymmetries. Only a compressed setup allows to reach a coverage of $\pm$250\,mrad. While in earlier experiments this was achieved by using  a hadron absorber at the cost of dramatically reducing their mass and vertex-position resolution, there are  now technical solutions that may be explored in an innovative way to serve this purpose. As will be explained in Sec.~\ref{sec:inst-dyrfsep}, a highly-segmented active absorber with embedded magnetic field may be the ideal  device for the detection of large-angle lepton pairs, providing dielectron tracking, dimuon vertex-pointing capabilities,  dilepton auto-trigger and muon-momentum measurement. Layers of magnetised iron with tungsten-silicon detectors  sandwiched in between seem to be a viable option, the feasibility of which will be further explored. Simple calculations show that a detector with transverse dimensions of $1.5\times1.5$\,m$^2$ and 250\,cm length  could be located 75\,cm downstream of the polarised target, and would still provide $\pm$250\,mrad coverage. 
Table~\ref{table:antiproton.dy} gives the achievable statistics for 140 days of beam time using a polarised NH$_3$  target and the above sketched active absorber.
\begin{table}[h]
\centering
{\footnotesize
\begin{tabular}{lccccccc}\toprule
  \multirow{2}{*}{Experiment }          &    Target                    &   Beam                    & Beam intensity                   & Beam energy  & DY mass         & \multicolumn{2}{c}{DY events}  \\
                                        &     type                     &   type                    &   (part/sec)                     &      (GeV)  &  (GeV/c$^2$)    & $\mu^+\mu^-$ & $e^+e^-$ \\\midrule\midrule
  \multirow{3}{*}{\color{red}This exp.} &\multirow{3}{*}{110\,cm NH$_3$} &\multirow{3}{*}{$\bar{p}$} &\multirow{3}{*}{$3.5 \times 10^7$} & 100        & 4.0 -- 8.5      & 28,000      & 21,000  \\
                                        &                              &                           &                                   & 120        & 4.0 -- 8.5      & 40,000      & 27,300  \\
                                        &                              &                           &                                   & 140        & 4.0 -- 8.5      & 52,000      & 32,500  \\\bottomrule
\end{tabular}}
\caption{Achievable statistics of the new experiment with an active absorber and 140 days of beam time.}
\label{table:antiproton.dy}
\end{table}

\subsubsection{Valence-quark distributions in the kaon}
\label{ssec:rfsep.dryan.kaon-val}

\begin{figure}[tb]
\centering
\includegraphics[width=0.49\textwidth]{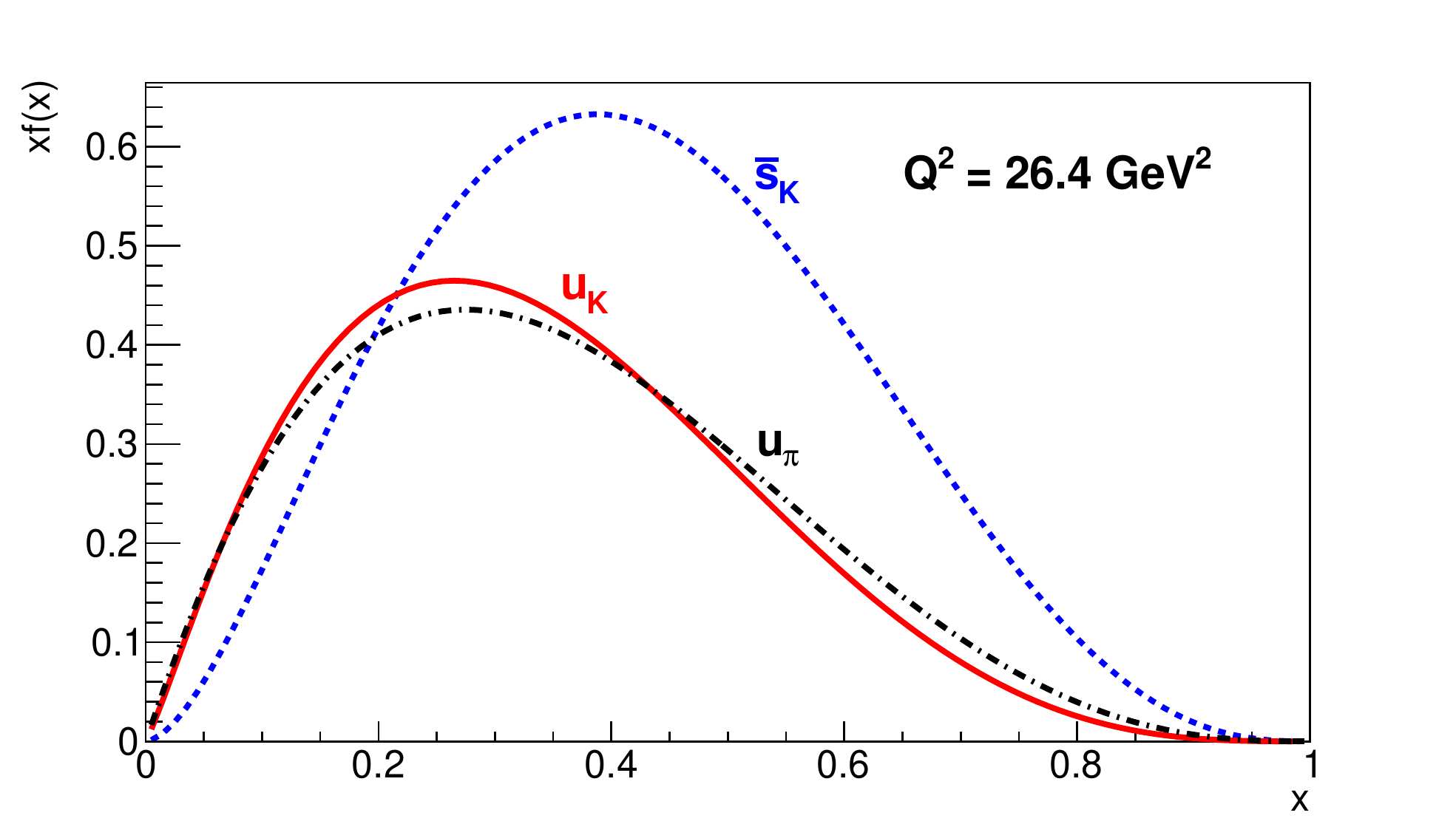}
\includegraphics[width=0.49\textwidth]{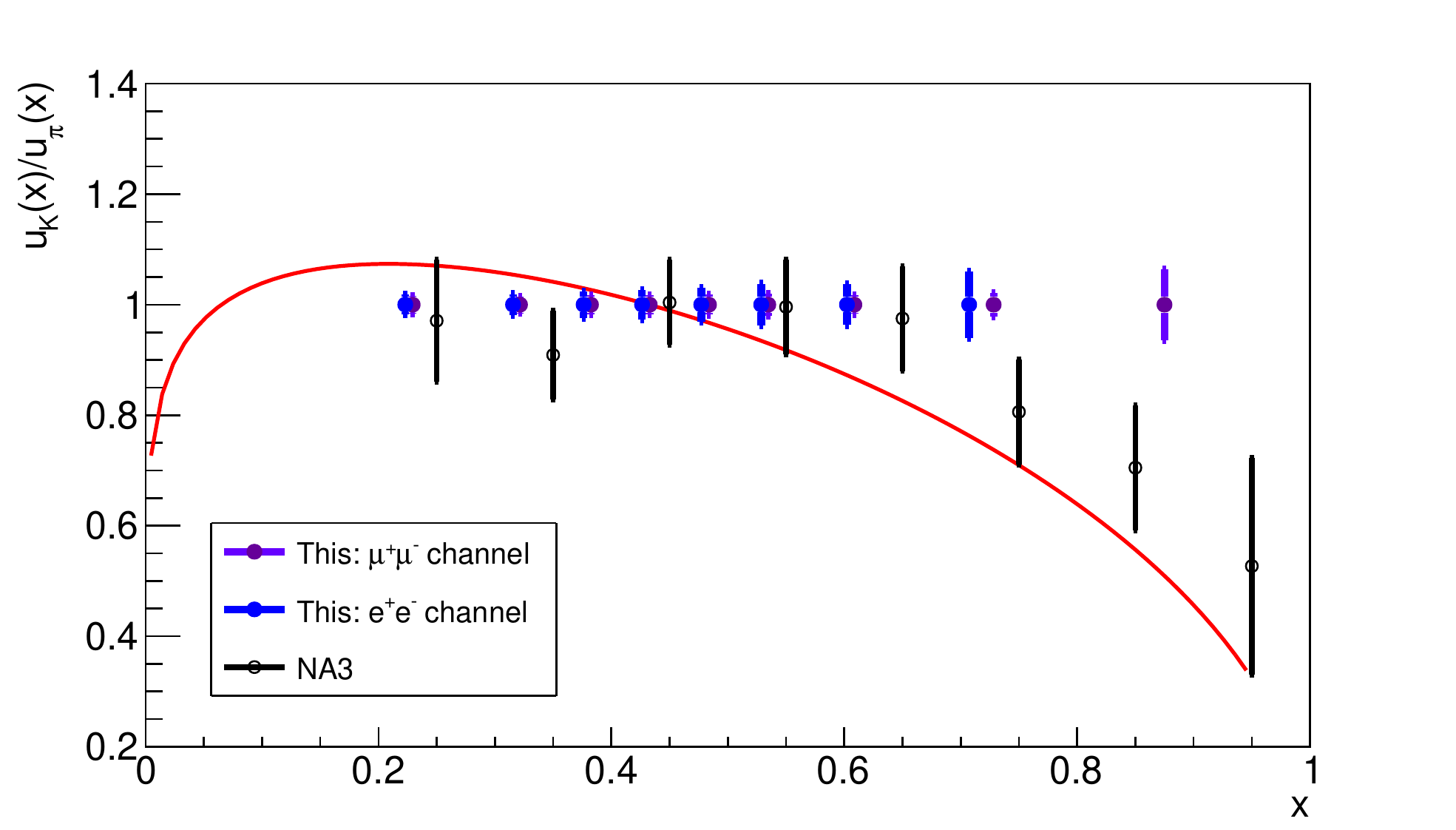}
\caption{Left: Valence PDFs for the $u$-quark in the pion and for the $u$ and $s$-quarks in the kaon, following the framework described in Ref.~\cite{Roberts:2017prc}.  Right: Projected statistical uncertainties on the kaon-to-pion  Drell-Yan yield ratio, assuming a 100\,GeV kaon beam and 140 days of data taking with a carbon target. The projections are given for both dimuon and dielectron channels. The results are compared to those of the NA3 measurement and to a calculation using the  DSE functional forms from~\cite{Chen:2016sno}. }
\label{DSE-val}
\end{figure}

The presence of the valence strange-quark significantly alters the properties of the  kaon in comparison to those of the pion. Being much heavier than the light quarks, it carries  a larger fraction of the kaon momentum. Accordingly, the valence contents of the kaon is expected to be significantly different from that of the pion. At the same $Q^2$ scale, the $\bar{s}_K(x)$ and $u_K(x)$ valence-quark distributions of the kaon are expected to peak at values  that are larger and smaller, respectively, than in the case of the pion. These distributions,  calculated in the framework of the Dyson-Schwinger Equations~\cite{Chen:2016sno},  are compared to the $u_\pi(x)$ valence-quark distribution of the pion in the left panel  of Fig.~\ref{DSE-val}. All three PDFs are evaluated at a small, non-perturbative QCD scale and then evolved to 5.2\,GeV, which is a scale typical for fixed-target Drell-Yan experiments.

Since the $u$-quark valence distribution in the kaon carries a momentum fraction smaller than that in the pion, it should show a somewhat faster decrease for large $x$ values.  This  behaviour is qualitatively confirmed by the first and only available experimental comparison between K$^-$ and $\pi^-$-induced Drell-Yan measurements~\cite{Badier:1980jq} by the NA3 collaboration, as shown in the right-hand side of Fig.~\ref{DSE-val}. The presented NA3 result is based on 700 Drell-Yan events produced with kaons as compared to 21,000 events  produced with pions. The ratio is consistent with unity up to $x_{\pi}=0.6$ and starts  dropping beyond $x_{\pi}=0.7$~.

The kaon valence distribution $u_K(x)$ can be determined with much improved accuracy in a dedicated measurement using a RF-separated kaon beam, as explained in Sec.~\ref{sec:rfsep.beams}. A 100\,cm long carbon target ($4 \times 25$\,cm) could be used in conjunction with a new large-acceptance active absorber downstream of it. The active absorber which is needed to  achieve an extended spectrometer acceptance is considered to be 250\,cm thick with an inner  radius of 9\,cm and an outer one of 135\,cm. A reconstruction efficiency of 80\% is assumed,  similar for dimuons and dielectrons. For electron-positron pairs, a solid estimate requires  further studies that are not available at the moment. Using a 100\,GeV hadron beam with an intensity of  $7 \times 10^7$ p/second, a kaon purity of about 30\%, and assuming $2 \times 140$ days of data-taking, about 65,000  kaon-induced Drell-Yan events should be collected in total.

Table~\ref{table:kpluskminus} presents a first estimate of the achievable statistics for kaon-induced Drell-Yan  measurements under the assumption of equal time sharing between the two beam charges, which is chosen for a good determination of the kaon-valence distributions. 

\begin{table}[h]
\centering
{\footnotesize
\begin{tabular}{lccccccc}\toprule
  \multirow{2}{*}{Experiment }          & Target                    &   Beam                   & Beam intensity                   & Beam energy  & DY mass         & \multicolumn{2}{c}{DY events}  \\
                                        &  type                     &   type                   &   (part/sec)                     &      (GeV)  &  (GeV/c$^2$)    & $\mu^+\mu^-$ & $e^+e^-$ \\\midrule\midrule
  NA3                                   &  6\,cm Pt                   & K$^{-}$                   &                                  & 200        & 4.2 -- 8.5      & 700         &  0 \\\midrule
  \multirow{6}{*}{\color{red}This exp.} &  \multirow{6}{*}{100\,cm C} & \multirow{3}{*}{K$^{-}$}  &\multirow{3}{*}{$2.1 \times 10^7$} & 80        & 4.0 -- 8.5      & 25,000      & 13,700  \\
                                        &                           &                          &                                   & 100       & 4.0 -- 8.5      & 40,000      & 17,700  \\
                                        &                           &                          &                                   & 120      & 4.0 -- 8.5      &  54,000      & 20,700  \\\cmidrule{3-8}
                                        &                           & \multirow{3}{*}{K$^+$}   &\multirow{3}{*}{$2.1 \times 10^7$} & 80       & 4.0 -- 8.5      &  2,800       & 1,300   \\
                                        &                           &                          &                                   & 100      & 4.0 -- 8.5      &  5,200       & 2,000  \\
                                        &                           &                          &                                   & 120      & 4.0 -- 8.5      &  8,000       & 2,400 \\\midrule
  \multirow{3}{*}{\color{red}This exp.} &  \multirow{3}{*}{100\,cm C} & \multirow{3}{*}{$\pi^-$} &\multirow{3}{*}{$4.8 \times 10^7$} & 80        & 4.0 -- 8.5      &  65,500      & 29,700  \\
                                        &                           &                          &                                   & 100       & 4.0 -- 8.5      &  95,500      &  36,000  \\
                                        &                           &                          &                                   & 120       & 4.0 -- 8.5      & 123,600      &  39,800  \\\bottomrule
\end{tabular}}
\caption{Achievable statistics of the new experiment, assuming $2 \times 140$ days of  data taking with equal time sharing between the two beam charges. For comparison, the collected statistics from NA3 is also shown.}

\label{table:kpluskminus}
\end{table}

Drell-Yan production by incoming $K^-$ and $\pi^-$ will be measured simultaneously, thus in the  ratio of kaon-to-pion yields some of the systematic uncertainties cancel. Within small corrections for  different sea-to-valence ratios in pion and kaon, the kaon-to-pion ratio is proportional to the one between their respective $u(x)$ distributions. The projected accuracy of this ratio is shown in the right panel of Fig.~\ref{DSE-val}.

\subsubsection{Separation of valence and sea-quark contributions in the kaon}
\label{ssec:rfsep.dryan.kaon-sea}

The kaon sea distribution and its composition are presently unknown. They can only be determined through a comparison between $K^+$ and $K^-$-induced Drell-Yan measurements. In such  measurements, the $K^+$ cross section is sensitive to valence-sea and sea-sea terms only, so that the difference between $K^-$ and $K^+$ beams is sensitive to $\bar{u}^{K^-} u^N$ valence-valence terms only. With a light isoscalar target one can define~\cite{Londergan:1996vh} the sea-to-valence ratio $R_{s/v}$ as:
\begin{equation}
R_{s/v}=\sigma^{K^{+}A}/\Sigma_{val}
\end{equation}
with
\begin{equation}
\Sigma_{val}=\sigma^{K^{-}A}-\sigma^{K^{+}A}~.
\end{equation}

\begin{figure}[tb] 
\centering 
\includegraphics[width=.98\textwidth]{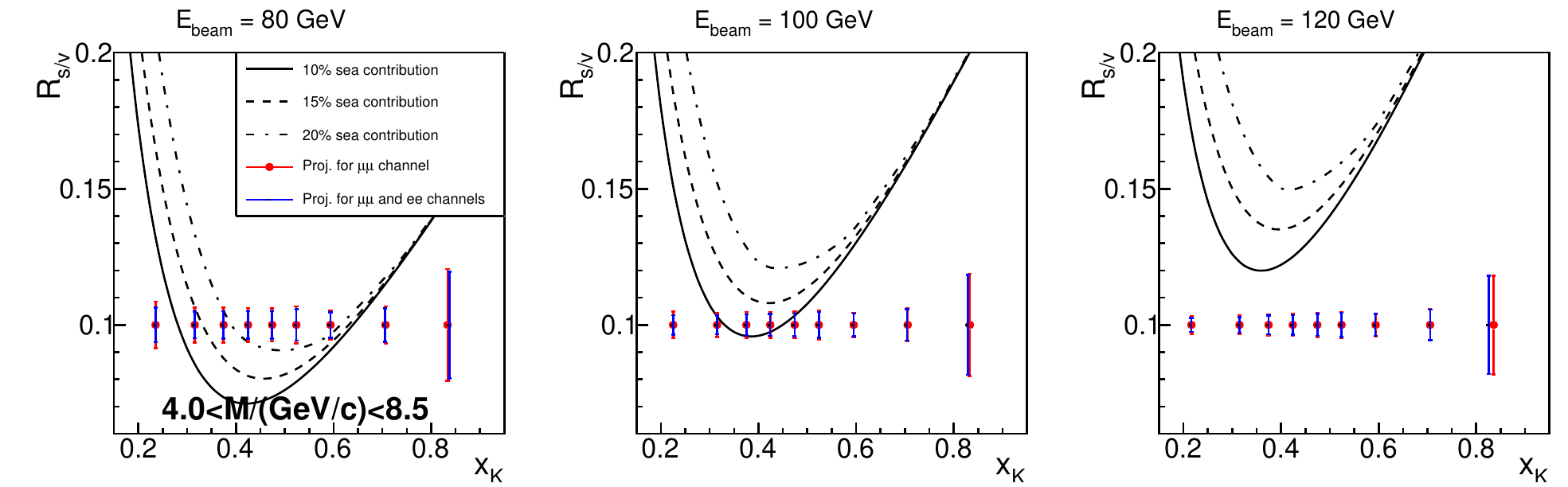}
\caption{The 3 panels show, for 3 different momenta of the incoming kaon beam, the ratio $R_{s/v}$ as a function of x$_{K}$ for three hypotheses on the kaon sea-quark content. The projected statistical uncertainties of the experiment are compared to  the sensitivity of $R_{s/v}$ to a variation of this content. The three curves representing 10\%, 15\%, and 20\% of kaon momentum carried by sea quarks are derived from SMRS pion PDFs by interchanging $d$-quarks with $s$-quarks.}
\label{kaon-sea}
\end{figure}

Figure~\ref{kaon-sea} shows the computed ratio $R_{s/v}(x)$ using three different assumptions for the amount of kaon sea, and for three different kaon-beam energies.  Since kaon-sea distributions are not available, the parameterisations of Ref.~\cite{Sutton:1991ay} for the pion were used replacing $d$ by $s$-quarks. The three distributions were obtained assuming sea-quark momentum contributions between 10\% and 20\%. For $x_K= 0.4$, the  difference between the two extreme values of the sea contribution reaches about 25\%. With decreasing $x_K$ the difference increases, uo to about a factor of 1.6 at $x_{K}= 0.2$~. Three different kaon-beam  momenta are shown. The largest is the most favourable one from the physics point of view, but it  requires additional R$\&$D studies to optimise the RF-separation of the kaon beam. 

Since the cross section for positive kaons is smaller than that for negative kaons, the statistical uncertainties of the sea-valence separation can be further minimised by a better sharing of the beam time for the two kaon charges. For a 280 days data-taking period, such an optimisation leads to 210 days with a K$^+$ beam and 70 days with a K$^-$ beam.

\subsubsection{The J/$\psi$ production mechanism and the gluon distribution in the kaon}
\label{ssec:rfsep.dryan.kaon-glu}

It is expected that the heavier quark in the kaon radiates less gluons than the lighter quarks in the pion. A natural consequence of this expectation is that the gluons in the kaon carry less momentum  than the gluons in the pion. Using the Dyson-Schwinger-Equation (DSE) approach, the authors of Ref.~\cite{Chen:2016sno} find that at the hadronic scale the gluons contribute to only 5\% of  the total momentum in the kaon, as compared to about one third in the pion. A stringent check of this  prediction requires the measurement of the presently unknown gluon distribution in the kaon.

The gluon distribution in the kaon can in principle be inferred through a measurement of  kaon-induced J/$\psi$ production. An important advantage of this process is its large cross section,  which reaches 100 nb/nucleon for small values of $x_F$, as compared to a fraction of nb per nucleon  for the high-mass Drell-Yan production at the fixed-target energies available at the CERN SPS. As discussed in Sec.~\ref{ssec:stdbeam.dryan.Jpsi}, J/$\psi$ production is not well understood. For fixed-target energies, the simple Color Evaporation Model (CEM) does not agree with the more thorough  NRQCD approach, and the relative contributions of the $gg$ fusion and $q\bar q$ annihilation terms depend on the model considered~\cite{Vogt:1999dw}. In both models, the $gg$ component is larger at small  $x_F$, whereas the $q\bar q$ term is dominant at large $x_F$, although with somewhat different intensities.

\begin{figure}[tb]
\centering
\includegraphics[width=0.7\textwidth]{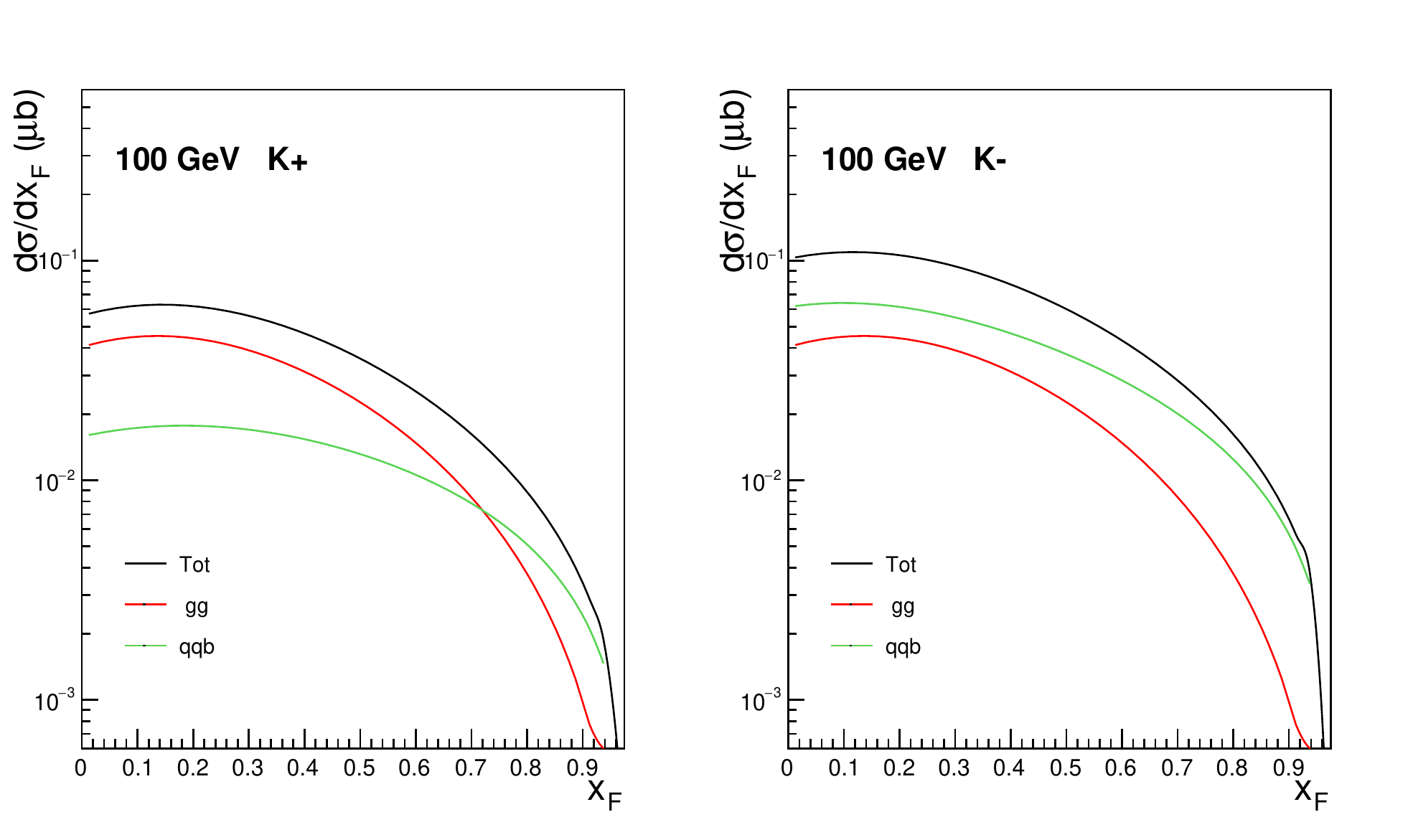}
\caption{Differential cross section as function of $x_F$, as obtained in the
Color Evaporation Model, on a $^{12}$C target for kaon-induced $J/\psi$ production with a 100\,GeV positive kaon beam (left) or negative kaon beam (right).}
\label{fig:kaon-glu}
\end{figure}

Here, the availability of the two signs of beam kaon charge can greatly help. A comparison between  cross sections measured with the two beam-charge signs can be used to both improve our understanding of the J/$\psi$ production mechanism and infer the gluon distribution in the kaon. Indeed, the J/$\psi$ cross section  for the positive kaon beam is different from the one for the negative kaon beam. The main difference comes from the valence $\bar u$ quark in the negative kaon, which annihilates the valence $u$ quark in the target. In contrast, there are no valence $\bar u$ quarks in the  positive kaon. Therefore, for $K^+$ the $q\bar q$ term is solely generated from the valence-sea and sea-sea contributions; those terms contribute an order of magnitude less to the $q\bar q$ annihilation term. The impact of the valence strange quark in the $K^+$ is also suppressed, as there are no valence strange  quarks in the target. A comparison between the $K^+$ and $K^-$-induced cross sections for J/$\psi$ production on a $^{12}$C target, as calculated in LO CEM, is shown in Fig.~\ref{fig:kaon-glu}. While the $gg$ term is identical for both kaon charges, the $q\bar q$ terms differ  by more than a factor of three.

Since the $gg$ contributions for both K$^-$ and K$^+$ beams are the same, the difference between the K$^-$ and K$^+$-induced J/$\psi$ cross sections is then equal to the $\bar{u}^{K^-}u^N$ valence-valence term:
\begin{equation}
\sigma_{J/\psi}^{K^{-}}-\sigma_{J/\psi}^{K^{+}}\propto \bar{u}^{K^-}u^N~.
\end{equation}

In this difference, the identical $gg$ contributions from $K^+$ and $K^-$ beams cancel. All other sea-valence, valence-sea, and sea-sea terms are also identical and cancel as well. The  difference from the two cross sections thus provides an alternative way for accessing the  $\bar{u}^{K^-}$-valence distribution in the $K^-$, after unfolding the well known $u(x)$  distribution in the nucleon target. This determination of the kaon valence density can then  be compared to the valence density determined using the Drell-Yan process. Both Drell-Yan and  $J/\psi$ production methods could be used simultaneously to minimise any model dependences in  the extraction.

An unambiguous determination of the $q\bar q$ annihilation term through a measurements of the K$^-$ vs. K$^+$ difference also gives access to the remaining $gg$ contribution,  within a given model. The $gg$ term is a convolution of the well-known gluon distribution of the nucleon and the unknown gluon distribution in the kaon, and will open the way to determine the  gluon distribution in the kaon.

\subsubsection{Other experiments}
\label{ssec:rfsep.dryan.competitors}

The interest for an improved understanding of the kaon structure is rapidly rising. Experiments dedicated to the measurement of the kaon valence structure with lepton beams  are planned~\cite{Park:2017pr,ANL-Workshop}, which rely on the validity of the meson-cloud model. The  JLab experiment plans to cover the $x_{\K}$ region between 0.4 and 0.95~. The kaon structure studies will be extended to lower $x_{\K}$ values at the forthcoming Electron-Ion-Collider. For both mentioned  experiments at JLab and EIC, the interpretation of their future kaon data strongly relies on a  model-dependent kaon flux determination. The proposed kaon-induced Drell-Yan  measurement at the CERN M2 beam line does not suffer from these limitations and is therefore a  much more direct way to access kaon structure.

Secondary kaon and antiproton beams are also under preparation at the J-PARC facility in Japan.  The J-PARC beams are expected to reach intensities of up to $10^6$ particles/second  for incident momenta of up to a maximum of 15\,GeV/$c$. Because of the lower momenta and intensity, no future experiment at J-PARC can compete with the above proposed studies of the meson and nucleon structure.


\subsubsection{Run plan: physics goals and required beam time}
\label{ssec.rfsep.dryan.timeline}

There are presently worldwide no planned facilities providing high-energy and high-intensity kaon and  antiproton beams. Although the RF-separation project at CERN is unique, the underlying technique is known and used since some time.

The availability of an RF-separated beam can be expected after the long CERN shutdown LS3. With the  technologies presently available, kaon and antiproton beams with momenta up to 80 and 100\,GeV, respectively,  could be envisaged. Further R\&D work might increase these limits, with obvious advantages to the physics cases presented above.

The long lead time before the start of this phase allows us to embark on an ambitious development  concerning the spectrometer. A new detector is envisaged, which comprises calorimetry and tracking embedded in a magnetic field  that provides a momentum measurement, and which in itself behaves as active absorber, is being considered. Such a detector would provide the largest geometrical acceptance ever achieved in a fixed-target Drell-Yan experiment.

With a nominal kaon-beam intensity of $2 \times 10^7$ kaons/second and a 100\,cm long carbon target, about 40,000 DY events could be collected in one year of data taking with a  $K^-$ beam, which is enough for a good determination of the $\bar{u}^{K^-}$-valence PDF. The corresponding number of J/$\psi$ events would be above 1 million. An additional  year with a $K^+$ beam, or an optimised sharing for a two-years data-taking period, would allow for a  sea-valence separation in the kaon. Additionally, the pion component in the hadron beam could be used for checks and kaon-to-pion ratio measurements.

The antiproton-induced Drell-Yan measurement requires one more year of data taking. Considering a 100\,GeV beam  with an intensity of $3.5\times 10^7$ antiprotons/second and a 110\,cm long polarised NH$_3$ target,  some 50,000 Drell-Yan events could be collected, which would allow for nucleon transverse-spin-asymmetry studies without requiring pion information as it was the case in the earlier COMPASS measurements.

\svnid{$Id: rfsep_prmpt.tex 4404 2018-10-08 18:17:46Z avauth $} 
\subsection{Study of the gluon distribution in the kaon via prompt-photon production }
\label{sec:rfsep.prmpt}
\subsubsection{Gluon PDFs for mesons}
Recent progress in theoretical calculations (see Sec.~\ref{ssec:stdhad.dryan.intro} and \ref{sec:rfsep.dryan}) makes the gluon distributions in pion and kaon especially important. Gluons not only significantly contribute to the internal structure of mesons; they also play a major role in the generation of their mass~\cite{doi:2542484}.  The available experimental information is however severely limited. In contrast to the rather well mapped gluon distribution in the nucleon, the gluon content of mesons is essentially unknown. The planned RF-separated beams facility at the M2 beam line will provide a unique opportunity for dedicated measurements of the pion and kaon gluon distributions.   

In order to measure the gluon PDF for the pion, the following hard processes can be used: i) $J/\psi$ and $\Upsilon$-states production; ii) di-jet production in $gg$ and $qg$ scattering; and iii) prompt-photon production in gluon Compton scattering. In the first approach, it is assumed that quarkonia production mainly proceeds through gluon fusion into quark-antiquark pairs. Here, uncertainties arise because contributions of 
other production mechanisms have to be taken into account. The second approach requires a high-energy meson beam, which has low sensitivity and its systematics is dominated by insufficient knowledge of fragmentation functions. In the case of the third approach, albeit the cross section of prompt-photon production is low, it is known at least up to NLO \cite{Aurenche:1987fs}. The systematics of this method is mainly given by experimental conditions and its dependence on the model assumptions is minimal. 

\subsubsection{Prompt photons}
Prompt photons are photons that are produced by hard scattering of partons. According to the factorisation theorem, the inclusive cross section for the production of a prompt photon in a collision of hadrons $h_A$ and $h_B$ can be written as follows:
\begin{equation}
d\sigma_{AB\to \gamma X}=d\sigma_{dir}+d\sigma_{frag}=
\sum_{a,b=q,\bar{q},g}\int{dx_a dx_b}f_a^A(x_a,Q^2)f_b^B(x_b,\mu^2)d\sigma_{ab\to \gamma x}(x_a, x_b, Q^2)+d\sigma_{frag}.
\end{equation}
Here, $d\sigma_{dir}$ is the contribution of photons emitted via direct coupling to a quark (direct photons) and $d\sigma_{frag}$ represents the contribution of photons produced from the fragmentation of a final parton (fragmentation photons). The function $f_a^A$ ($f_b^B$) is the parton density for hadron $h_A$ ($h_B$), $x_a$ ($x_b$) is the
fraction of the momentum of hadron $h_A$ ($h_B$) carried by parton $a$ ($b$), $Q^2$ is  the square of the 4-momentum transferred in the hard scattering process, and $\sigma_{ab\to \gamma x}(x_a, x_b, Q^2)$ represents the cross section for the hard scattering of partons $a$ and $b$. 
The contribution of fragmentation photons does not exceed 10 -- 20\% in the discussed kinematic range, even for much higher energies \cite{Hanks:2009fj} and can be taken into account.
There are two main hard processes for the production of direct photons: i) gluon Compton scattering, $gq(\bar{q})\rightarrow \gamma q(\bar{q})$, which dominates, and ii) quark-antiquark annihilation, $q \bar{q} \rightarrow \gamma g$. Measurements of the differential cross section of prompt-photon production 
in pion-nucleon collisions were already used  by the fixed-target experiments WA70 \cite{Bonesini:1987mq}, E706 \cite{Alverson:1993da} and others for the determination of the gluon structure of the pion.

\subsubsection{Prompt-photon production in COMPASS kinematics}
\begin{figure}[tb]
  \begin{minipage}{15pc}
 \includegraphics[width=20pc]{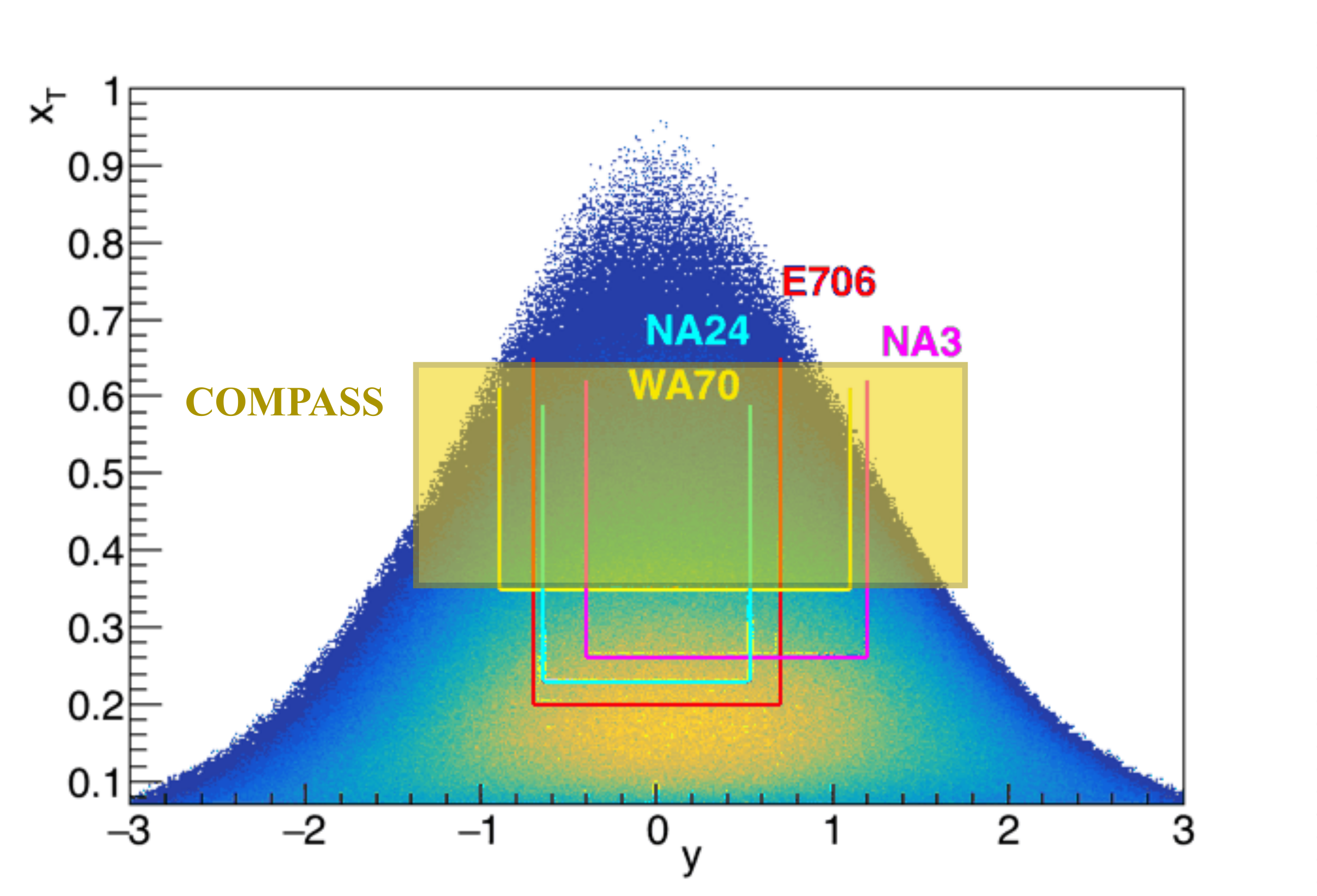}
 \end{minipage}\hspace{5pc}%
 \begin{minipage}{15pc}
  \includegraphics[width=20pc]{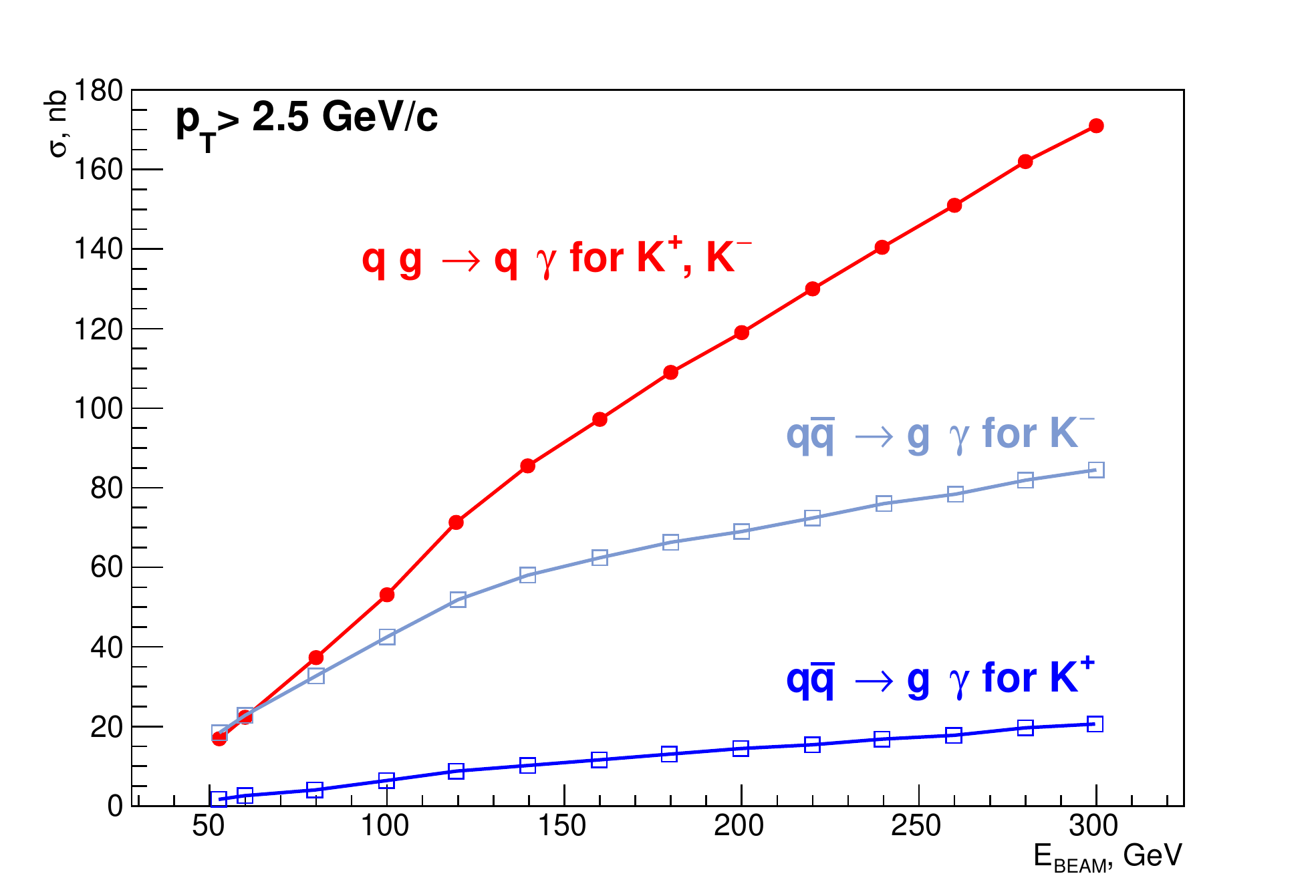}
  \end{minipage}
 \caption{\label{pic:PPsec} a) Kinematic distribution for $x_{T}=2p_{T}/\sqrt{s}$ vs. $y$ for the gluon Compton scattering process, for a 100\,GeV $K^{+}$ beam scattered off a proton target. The kinematic ranges covered by previous low-energy pion beam experiments and the possible coverage by COMPASS are also shown in different colors; b) energy dependence of the prompt-photon production cross section for $p_{T}>$2.5 GeV/$c$, for both production mechanisms and positive as well as negative kaon beams under the assumption $g_{\pi}(x, Q^{2})=g_{K}(x, Q^{2})$. }
  \end{figure}

In order to determine the gluon structure of charged kaons we propose to measure the differential cross section of prompt-photon production, $Ed^3\sigma/dp^{3}$,  in the kinematic range of transverse momentum $p_T$ larger than $p_{T0}=$2.5\,GeV/$c$ and center-of-mass rapidity $-1.4<y<1.8$, using a positive kaon beam of 100\,GeV/$c$ ($\sqrt{s}=13.7$\,GeV). This range corresponds to $x_{g}>
0.05$ for kaons and $Q^2\sim p_T^2$. The corresponding kinematic distribution for $x_{T}=2p_{T}/\sqrt{s}$ vs. $y$ for the gluon Compton scattering process, together with the kinematic ranges covered by previous low-energy pion-beam experiments and the possible 
kinematic region for COMPASS, are shown in Fig. \ref{pic:PPsec}a (according to \cite{Vogelsang:1997cq}).

A positively charged kaon beam is chosen in order to reduce the number of prompt photons produced via $q \bar{q}$ annihilation. A separate data sample should be collected with a negative kaon beam in order to separate gluon Compton scattering and quark-antiquark annihilation production. Before taking data  with a kaon beam, one year of data should be taken with a pion beam under similar conditions. If sufficient, the pion component of the RF-separated beam could be used instead. This pion data will be used 
to improve the knowledge on the gluon structure of the pion and to study systematic effects. 

Figure 
 \ref{pic:PPsec}b represents the energy dependence of the prompt-photon production cross section for $p_{T}$ larger than $p_{T0}$=2.5 GeV/$c$ for both production mechanisms, for positive and negative kaon beams under the  assumption $g_{\pi}(x,Q^{2})=g_{K}(x,Q^{2})$. 
 The contribution of gluon Compton scattering to the cross section $\sigma_{AB\to \gamma X}$ is calculated to be 53~nb in LO approximation (Pythia6) in the kinematic range that is mentioned above for a 100\,GeV/$c$ pion beam interacting with a proton target. The corresponding contribution of the quark-antiquark annihilation process is 6~nb and 42~nb for positive and negative beam, respectively. Similar magnitudes of  cross sections are also expected for a kaon beam of the same momentum. 

\begin{figure}[tb]
 \begin{center}
 \includegraphics[width=20pc]{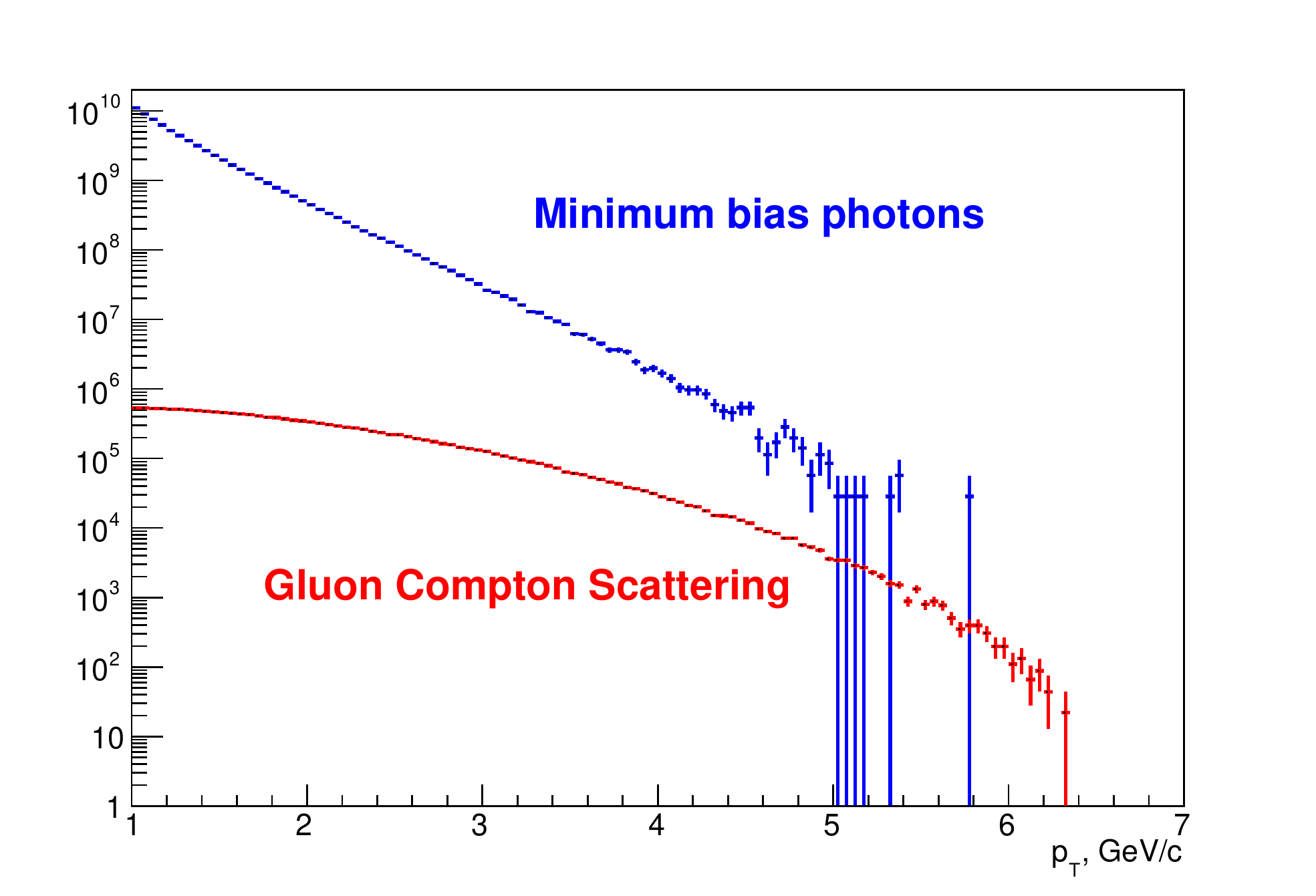}
 \caption{\label{pic:PtPP} The $p_{T}$ distributions for prompt photons from gluon Compton scattering (red) and for minimum-bias photons (blue), produced in the interaction of a 100\,GeV K$^+$ beam and a proton target (according to Pythia6 and assuming that $g_{\pi}(x, Q^{2})=g_{K}(x, Q^{2})$). The distributions are normalized to one year of data taking. For the prompt photons a K-factor of 1.4 is taken into account. }
 \end{center}
  \end{figure}

The main contribution to the systematic uncertainty, which will dominate over the statistical one, is expected to originate from the estimation of the number of photons produced from decays of secondary $\pi^{0}$ and $\eta$ mesons (minimum-bias photons). This kind of background is especially important at small $p_T$ and gives the lower limit  of the accessible $p_{T}$ range. The $p_{T}$ distributions for photons from  gluon Compton scattering  and for minimum-bias photons are shown in Fig. \ref{pic:PtPP}.  The value of $p_{T0}$ was assumed using the experience of previous experiments at similar values of $\sqrt{s}$. A limited spatial resolution of the electromagnetic calorimeters could lead to a misidentification of a cluster produced by both photons from the decay of a high-energy $\pi^0$ as a single-photon cluster. This effect becomes significant at high $p_T$. Background from 2$\gamma$ decays of $\pi^{0}$ and $\eta$ can be reduced by the reconstruction of such  decays. The subtraction of such background is based on a precise Monte-Carlo simulation of the setup. The detection of a photon produced far upstream  the target and the mis-association of such a photon with the interaction in the target may also lead to a significant overestimation of its transverse momentum. 

\begin{figure}[tb]
  \begin{minipage}{15pc}
 \includegraphics[width=20pc]{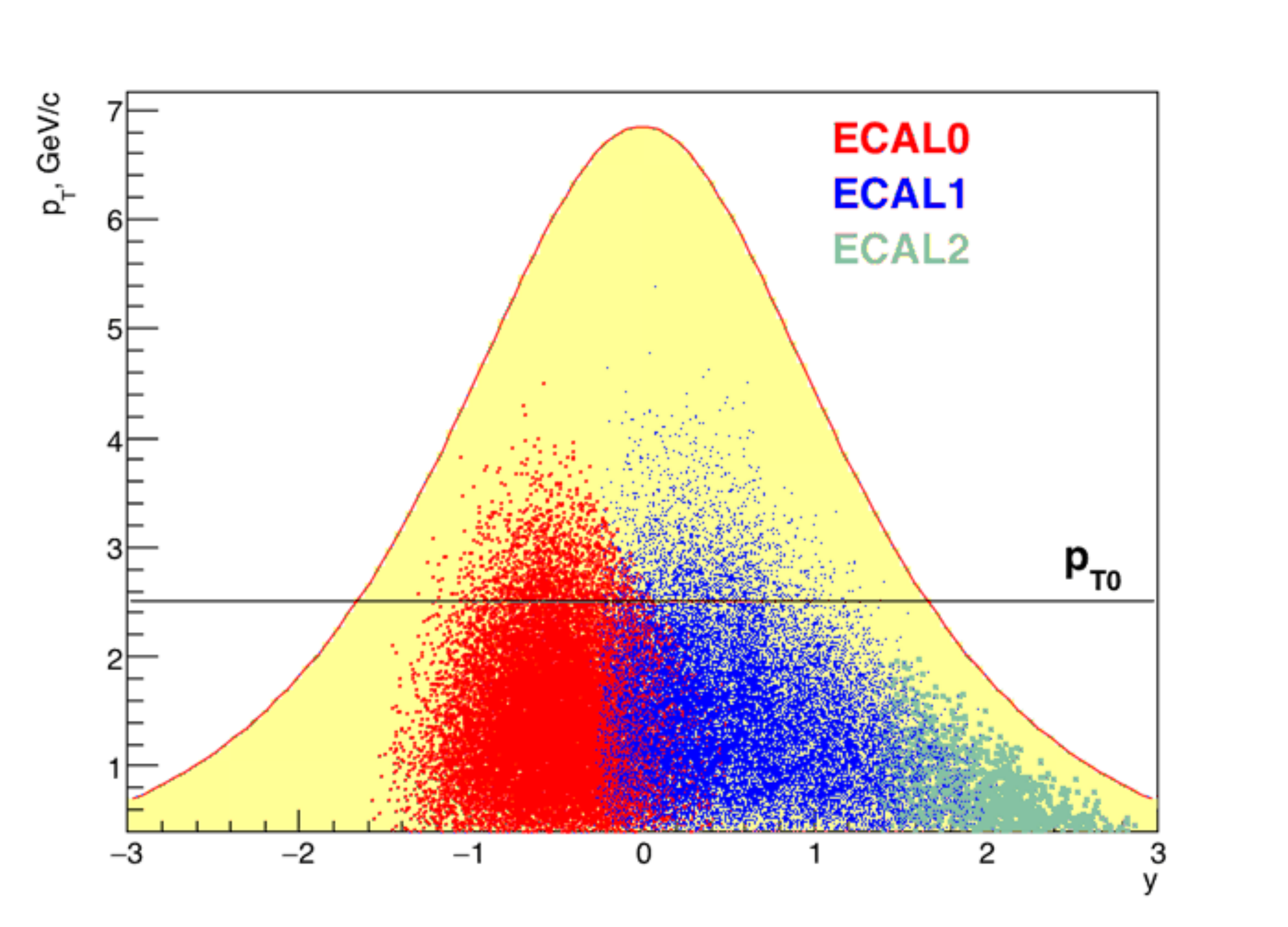}
 \end{minipage}\hspace{5pc}%
 \begin{minipage}{15pc}
  \includegraphics[width=20pc]{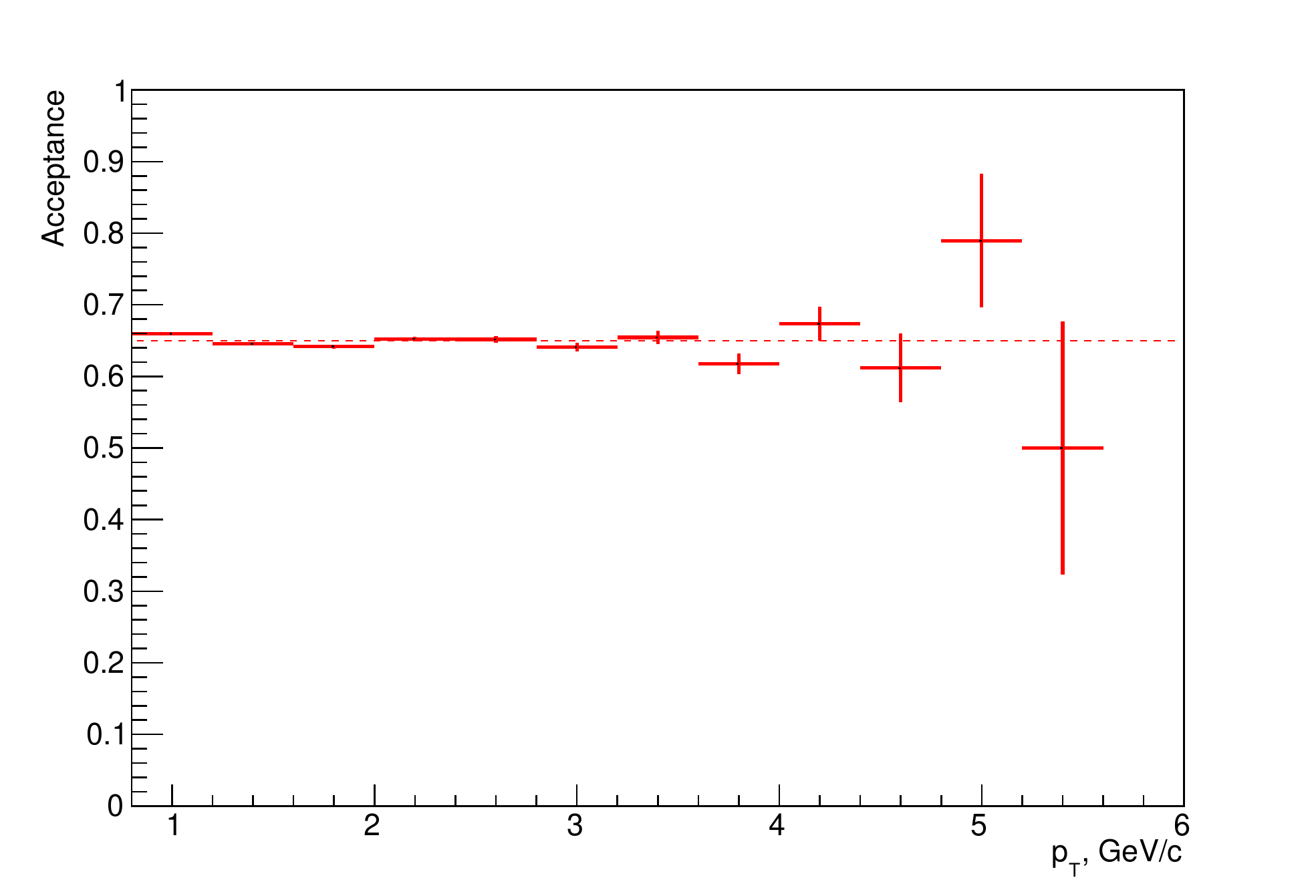}
  \end{minipage}
 \caption{\label{pic:yPtrange} a) kinematic range in rapidity $y$ and transverse momentum for prompt photons produced in gluon Compton scattering. Regions covered by the electromagnetic calorimeters ECAL0, ECAL1 and ECAL2 are shown in red, blue and green, respectively. The COMPASS setup for the GPD run in 2017 is used; b) acceptance of the COMPASS setup used for the GPD run in 2017 for prompt photons, as a function of their transverse momentum $p_{T}$.}
  \end{figure}

When studying prompt photon production, the following requirements should be fulfilled:
\begin{itemize}
\item A positive kaon beam of 100\,GeV/$c$ or higher momentum and intensity of $5\times10^6$ kaons per second should be delivered to the experimental area.
\item  CEDAR detectors should be used for the rejection of events produced by beam particles different from the kaon.
\item A two-meter long liquid hydrogen target ($\sim$0.2 $X_{0}$) should be used, which is transparent for any produced photon. A solid target of low-Z material could also be considered. 
\item The existing electromagnetic calorimeters ECAL0 and ECAL1 can provide sufficient capability for the detection of prompt photons in the rapidity range $-1.4<y<0.4$ and $-0.2<y<1.8$, respectively (see Fig. \ref{pic:yPtrange}a). They have to be included into dedicated triggers. The ECAL2 calorimeter should play an important role in the $\pi^0$ background subtraction. 
\item A stainless-steel shielding is required to be installed upstream the target to prevent illumination of the calorimeters by photons produced in the interaction of beam kaons with beam elements of the setup. 
\item A tracking detector (X and Y planes) with an aperture of about 2.3$\times$2.3 m$^{2}$ and a beam hole of 0.5$\times$0.5 m$^{2}$ should be installed in front of ECAL0 in order to provide the capability to identify "charged" clusters in ECAL0 and reject charged particles with high $p_T$. The spatial resolution of the detector is defined by the ECAL0 cell size (3.8 cm) and should be about 1 cm.
\item The transparency of the setup should be increased in order to reduce the number of secondary photons. Nevertheless we should be able to track charged particles in order to distinguish clusters produced by photons from "charged" clusters.
\end{itemize}

The acceptance of the COMPASS setup used for the GPD run in 2017 is shown in Fig. \ref{pic:yPtrange}b for prompt photons as a function of transverse momentum $p_{T}$.  The detector geometry, the material map and minimal thresholds for a cluster energy on the level of 0.5\,GeV, 1\,GeV and 2\,GeV  in ECAL0, ECAL1 and ECAL2  respectively, are taken into account. The acceptance is rather flat up to very high $p_T$ and is about 0.65.

In parallel, the production of high-$p_T$ charged and neutral pions and kaons could be studied, which can also be used to access the partonic structure of the kaon.

{\bf Rate estimation for prompt-photon production:} 
The following assumptions are made: i) 140 days of data taking with an accelerator efficiency of 0.8, which corresponds to an integrated flux of $5\times10^{12}$ kaons delivered to the 2 m long liquid-hydrogen target; 
ii) the LO gluon-Compton scattering cross section with  the K-factor 1.4 is 74 nb (for $p_T>2.5$ GeV/$c$); iii) the duty factor of the detector is 0.9; iv) the general acceptance (including geometry, photon conversion and selection criteria) is 30\%.
This leads to an expected statistics of gluon-Compton scattering events in the kinematic range $p_T>2.5$ GeV/$c$ and $-1.4<y<1.8$ of $0.85\times10^6$ events. 


We are not aware of any other plans to study the gluon structure of charged kaons. 

\svnid{$Id: rfsep_prima.tex 4406 2018-10-09 14:11:23Z aguskov $} 
\subsection{Primakoff Reactions}
\label{sec:rfsep.prima}

\subsubsection{Kaon polarisability}

The electric ($\alpha$) and magnetic ($\beta$) polarisabilities characterise a meson as a complex QCD system in terms of its interaction  with an external electromagnetic field. They are fundamental parameters in meson physics that can be probed in Compton scattering and hence provide a possibility to compare experimental results with theoretical predictions. The polarisabilities of the charged pion predicted by chiral perturbation theory ($\chi$PT) in the two-loop approximation are  $\alpha_{\pi}-\beta_{\pi} = (5.7\pm1.0)\times10^{-4}\,\text{fm}^{3}$ and $\alpha_{\pi}+\beta_{\pi} = 0.16\times10^{-4}\,\text{fm}^{3}$ \cite{Gasser:2006qa}. The currently most precise measurement of the electric pion polarisability is $\alpha_\pi = (2.0\pm0.6_{stat}\pm0.7_{syst})\times10^{-4}\,\text{fm}^3$, which is in agreement with the predictions of $\chi$PT \cite{Gasser:2006qa} and dispersion relations approach \cite{Pasquini:2009xa, PhysRevC.77.065211}. This result was obtained by the COMPASS experiment in the so-called Primakoff reaction $\pi^- Z \rightarrow \pi^- \gamma Z$ using a $190\,\text{GeV}/c$ negative pion beam and the assumption $\alpha_{\pi}+\beta_{\pi} = 0$ \cite{Adolph:2014kgj}. 

As the kaon is a more compact and more rigid object than the pion, the naive expectation is to observe smaller values for the polarisabilities. Indeed, the $\chi$PT prediction for the charged kaon polarisability in one-loop approximation is $\alpha_K = (0.64\pm0.10)\times10^{-4}\,\text{fm}^3$ under the assumption that $\alpha_{K}+\beta_{K} = 0$ \cite{Guerrero:1997rd}. The quark confinement model predicts values of $\alpha_{K} = 2.3\times10^{-4}\,\text{fm}^3$ and $\alpha_{K}+\beta_{K} = 1.0\times10^{-4}\,\text{fm}^3$ \cite{Ivanov:1991kw}. Experimentally, only an upper limit $\alpha_{K} < 200\times10^{-4}\,\text{fm}^3$ (CL=90\%) was established from the analysis of X-ray spectra of kaonic atoms \cite{Backenstoss:1973jx}.

A measurement of the kaon polarisability via the reaction $K^- Z \rightarrow K^- \gamma Z$, which is similar to the measurement of the pion polarisability performed by COMPASS, is challenging to prepare. The kaon component in a conventionally produced hadron beam is too small at high beam energies to collect the required amount of data on a reasonable timescale. Also, it is difficult to identify beam particles with high enough purity. To this end, a RF-separated hadron beam, in which kaons are enriched, would provide a unique opportunity to perform the first measurement of the kaon polarisability.
Additional difficulties for the kaon polarisability measurement are the small kinematic gap between the threshold in the invariant mass $M_{K^- \gamma}$ and the first resonance $\text{K}^*(892)$, when compared to  the pion case with the $\rho(770)$ resonance, and the one order of magnitude smaller Primakoff cross section compared to the case of the pion. 

For the kaon polarisability measurement with a $100\,\text{GeV}/c$ RF-separated kaon beam ($5\times10^{6}$ s$^{-1}$) we assume the following conditions: 
\begin{itemize}
\item a spectrometer configuration as it was used in 2009 and 2012 for the analogous measurements with the pion beam: the CEDAR detector in the beam line, a 0.3 $X_{0}$ thick nickel target, silicon-based telescopes upstream and downstream of the target, similar dead time of trigger and DAQ;
\item trigger on the high-energy deposit in the ECAL1 and ECAL2 calorimeters;
\item the new DAQ system with the capability to accept trigger rates up to 100 kHz.
\end{itemize}
\begin{figure}
\begin{minipage}{15pc}
  \includegraphics[width=20pc]{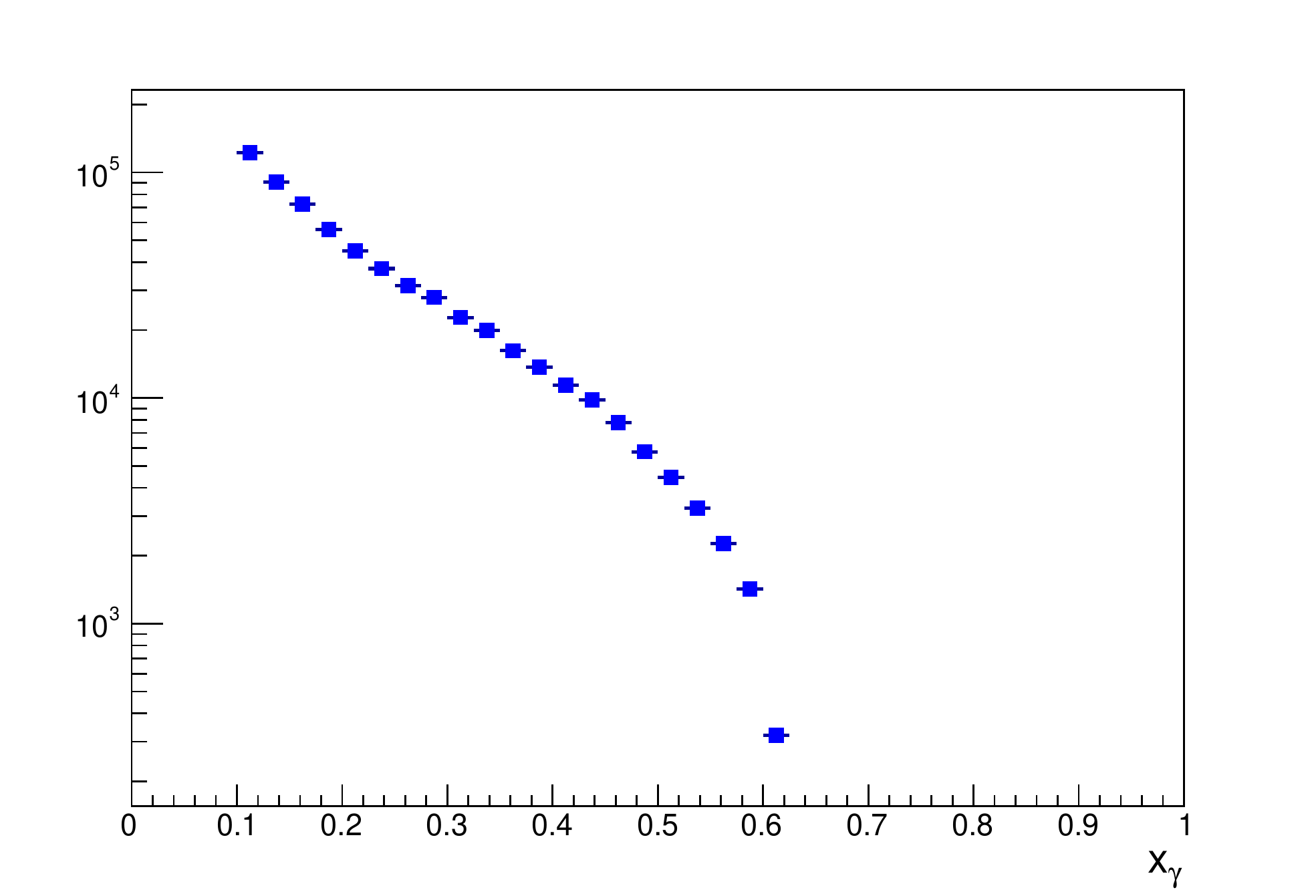}
\end{minipage}\hspace{5pc}%
\begin{minipage}{15pc}
  \includegraphics[width=20pc]{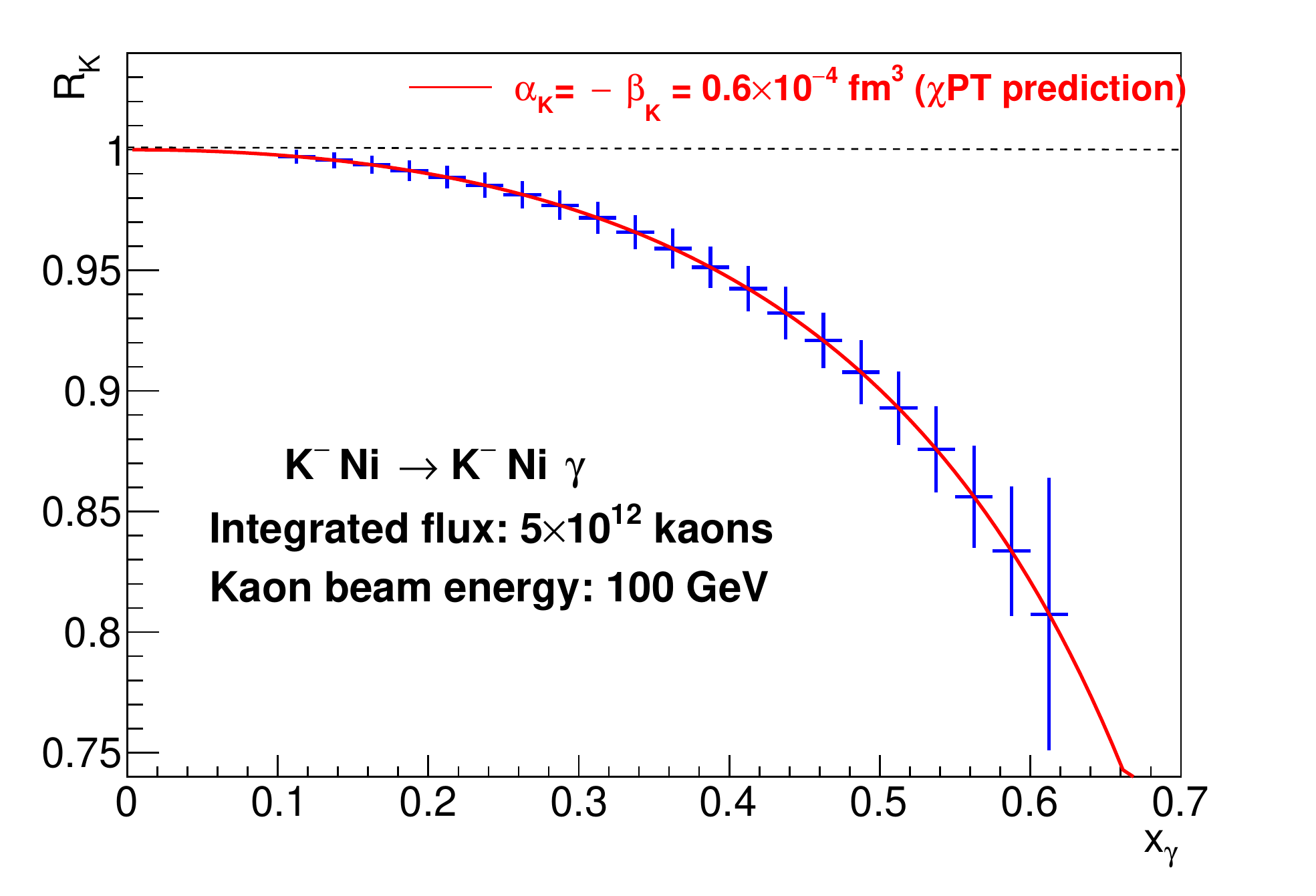}
\end{minipage}
\caption{\label{pic:Kpol} a) the expected $x_{\gamma}$ spectrum for $K^- \gamma$ events; b) the statistical accuracy for the measurement of the ratio $R_{K}$ of the differential cross section for the real kaon over the expected cross section for a hypothetical point-like kaon as a function of $x_{\gamma}$. The ratio $R_{K}$ corresponding to the $\chi$PT prediction is shown in red.}
\end{figure}
Assuming an integrated flux of $5\times10^{12}$ kaons after one year of data taking, we estimate the achievable statistics to be about $6\times10^5$ $K^- \gamma$ events in the kinematic range $0.1<x_{\gamma}<0.6$ and $M_{K^- \gamma}<0.8\,\text{GeV}/c^2$. Here, $x_{\gamma}$ is the energy of the produced photon normalized to the beam energy. The trigger efficiency is supposed to be close to 100\% in the whole range of $x_{\gamma}$. The expected $x_{\gamma}$ spectrum of $K^- \gamma$ events is shown in Fig.~\ref{pic:Kpol}a. The ratio $R_{K}$ of the differential cross section for the kaon over the expected cross section for a hypothetical point-like kaon can be approximately expressed as a function of $x_{\gamma}$ under the assumption $\alpha_{K}+\beta_{K}=0$ as
\begin{equation}
R_{K}=1-\frac{3}{2}\cdot \frac{x_{\gamma}^2}{1-x_{\gamma}}\cdot \frac{m_{K}^3}{\alpha}\cdot \alpha_{K}^{3},
\end{equation}
where $\alpha$ is the fine structure constant. It is important to emphasize that polarisation effects in case of the kaon are amplified by the factor $(m_{K}/m_{\pi})^3\approx 44$ when compared to the pion.
The statistical accuracy for the measurement of the ratio $R_{K}$ of the differential cross section for the kaon over the expected cross section for a hypothetical point-like kaon as a function of $x_{\gamma}$ is presented in Fig. \ref{pic:Kpol}b. Also shown is the ratio $R_{K}$ as expected from the $\chi$PT prediction. The statistical accuracy of the $\alpha_K$ extraction under the assumption $\alpha_{K}+\beta_{K}=0$ is $0.03\times10^{-4}\,\text{fm}^{3}$.
The main contributions to the systematic uncertainty are expected from (i) the uncertainty of the determination of the tracking detector efficiency from the Monte Carlo simulation; (ii) the statistical uncertainty of the $\pi^0$ background subtraction; (iii) the effect of the uncertainty on the estimate of the strong interaction background and its interference with the Coulomb contribution; (iv) the uncertainty in the subtraction of $\pi\gamma$ events due to the  pion contamination in the beam. The total systematic uncertainty is expected to be smaller than the statistical one. 

We are not aware of any other plans to measure the charged kaon polarisability.

\subsubsection{Direct measurement of the lifetime of the neutral pion}
The lifetime of the neutral pion is a quantity that is relevant in low-energy
QCD, as it is related to the chiral anomaly hypothesis. To-date  a 1\% accuracy has been reached in the theoretical lifetime prediction, 
while the experimental value $(8.52\pm0.18)\times10^{-17}$ s has an uncertainty of 2.1\% \cite{Patrignani:2016xqp}, 
which is not enough for a precise test of theoretical calculations.
The most precise experimental results for the lifetime of the neutral pion are $(8.32\pm0.15_{stat}\pm0.18_{sys})\times10^{-17}$ s, being obtained by PrimEx via the Primakoff effect \cite{Larin:2010kq}, and $(8.97\pm0.22_{stat}\pm0.17_{sys})\times10^{-17}$ s, which was measured directly \cite{Atherton:1985av}. 

The idea of the direct measurement is based on the estimation of the length of the decay path. This length
lies in the range of a few 100 micrometers for pions produced in the forward
direction with energies above 10\,GeV. As a consequence, in a thin foil made of 
a dense target material like tungsten, the conversion probability of the 
decay photons into $e^+/e^-$ pairs depends on the thickness and distribution of
the target material. Varying the distance between two tungsten foils, each with 
a thickness of about 30 $\mu m$, leads to a variation in the number of $e^+/e^-$ pairs arising from
$\pi^0$ decays. The first and only measurement of this kind was performed  at CERN by Atherton et al. 1982, with only observing the positron
modulation amplitude depending on the variation of two tungsten foils \cite{Atherton:1985av}. 
The spectrum of produced $\pi^0$ was not measured in the experiment, instead certain assumptions concerning its shape were used,
which turned out to be  the main contribution to the systematic uncertainty.

\begin{figure}
\begin{minipage}{15pc}
  \includegraphics[width=20pc]{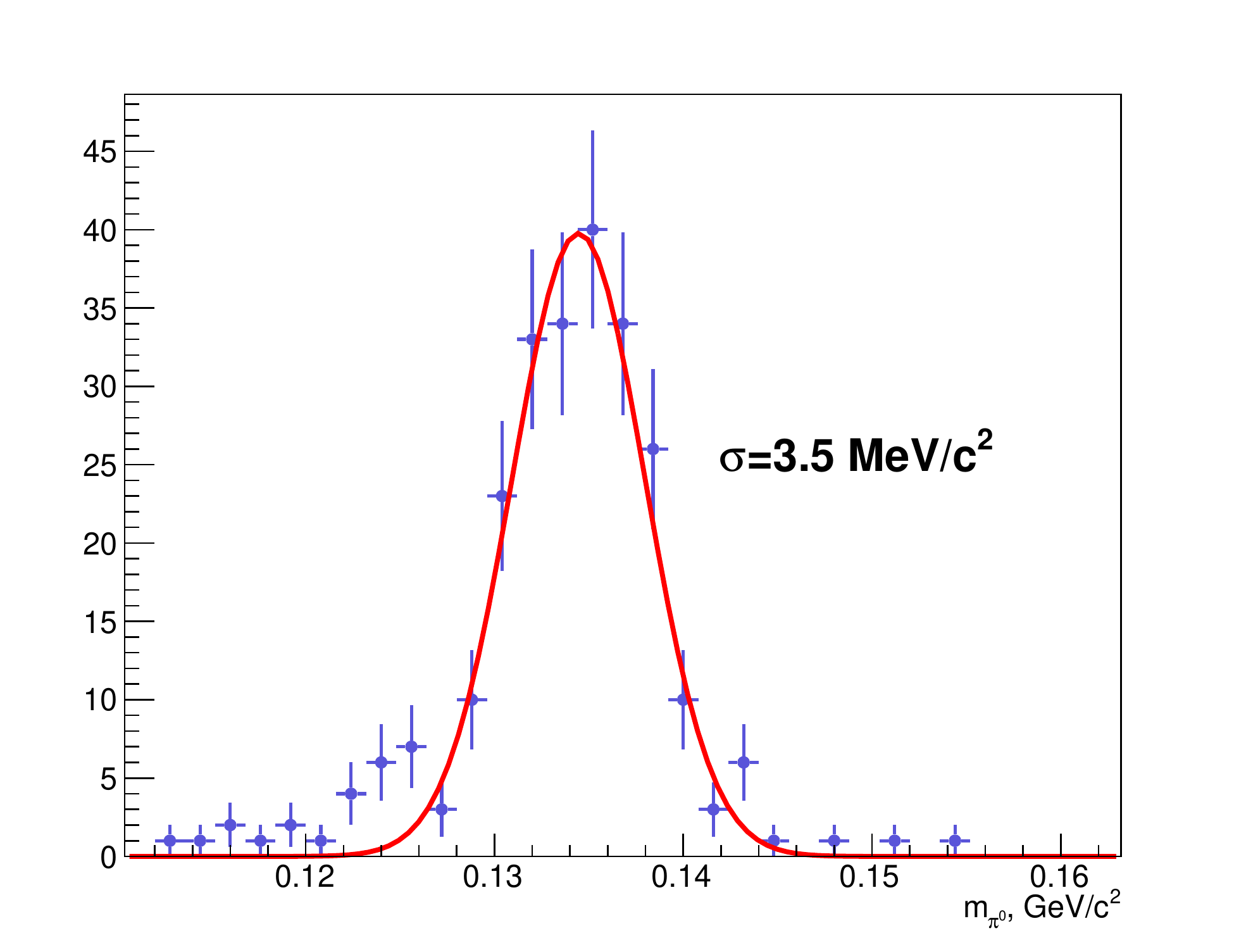}
\end{minipage}\hspace{5pc}%
\begin{minipage}{15pc}
  \includegraphics[width=20pc]{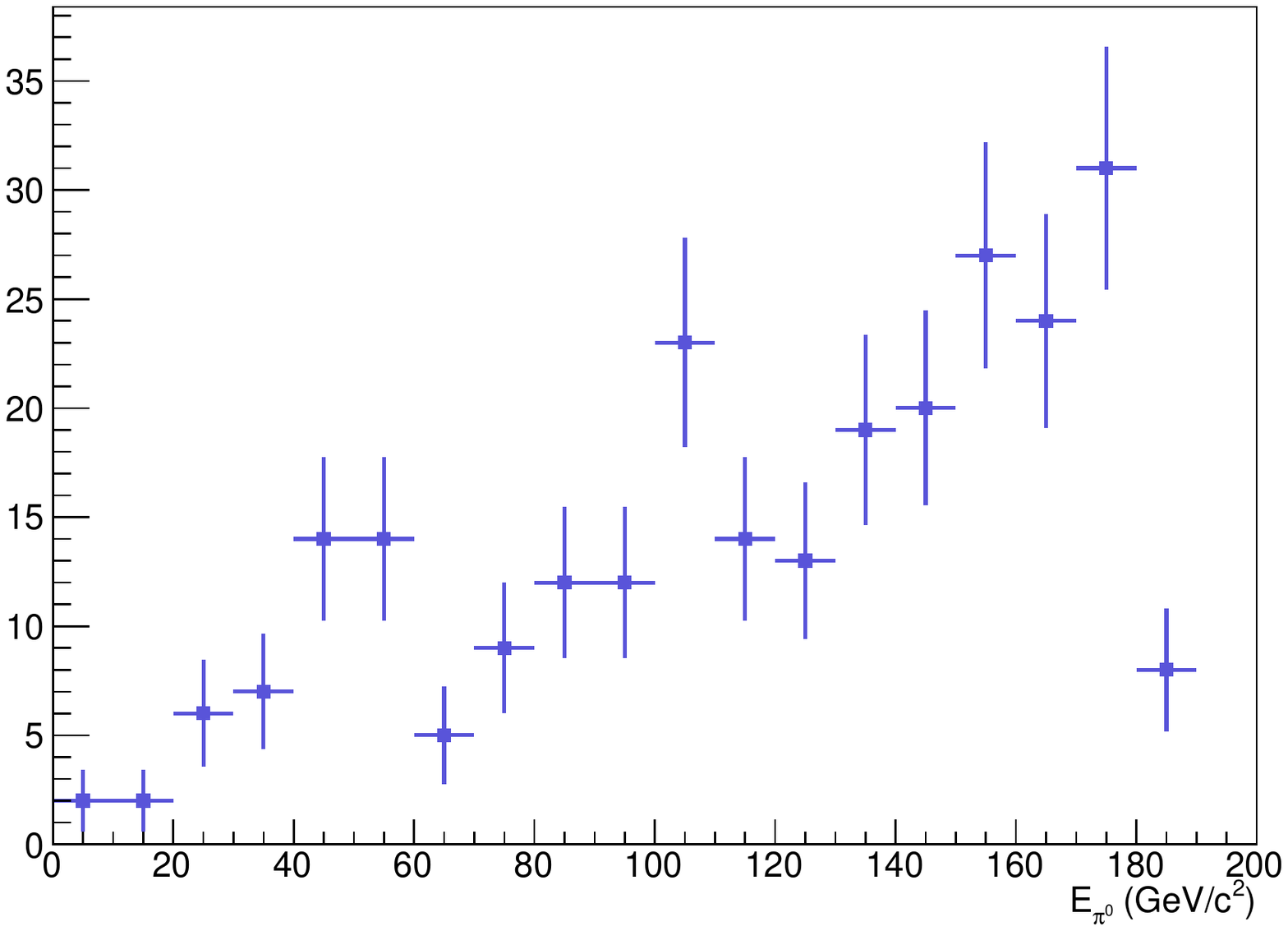}
\end{minipage}
\caption{\label{pic:pi0spectra} The $2\gamma$ invariant mass distribution (a) and  the observed $\pi^0$ energy spectrum (b) for  photons produced at COMPASS by a 190 GeV/$c$ negative hadron beam (2009)
in the exclusive reaction $\pi^- Ni\to \pi^-Ni\,\pi^0$}
\end{figure}

The measurement proposed here makes use of the full spectrometer in order to
reconstruct $\pi^0$ decays, both the main decay into two photons and the decay into
$e^+e^-\gamma$, which has to
be taken into account when interpreting the fraction with converted photons.
Such a complete reconstruction of $\pi^0$ decays 
is the main advantage of the proposed measurement when compared  to the previous one \cite{Denisenko:2017ofo}.
The ability of the COMPASS-like setup to reconstruct $\pi^0\to2\gamma$ decay is illustrated in 
Fig. \ref{pic:pi0spectra}. The $2\gamma$ invariant mass distribution  for photons produced by a 190 GeV/$c$ negative hadron beam 
in the exclusive reaction $\pi^- Ni\to \pi^-Ni\,\pi^0$ is shown in the left panel and the observed $\pi^0$ energy spectrum in the right one.
As the measurement requires  only a very thin target, corresponding to 0.25-0.5 \%
radiation lengths, it  can be combined with e.g. the Primakoff running,
which has similar requirements concerning the spectrometer setup. 
The kaon component of the beam produces a smooth, well-defined background of neutral and charged pions via the decays $K^-\to\pi^-\pi^0$, and $K^-\to\pi^-\pi^0\pi^0$, which allows one to control the systematics related to the conversion probability in the tungsten foils.

For a
 parasitic running of one year along with the kaon Primakoff run, it can be 
estimated that the statistical precision on the $\pi^0$ lifetime can be better than
1\%, thus clarifying the question whether the direct measurement of this
important quantity really deviates significantly from that found in
the dedicated Primakoff measurements at JLab/PrimEx \cite{Atherton:1985av}.

\svnid{$Id: rfsep_vector.tex 4400 2018-10-05 08:46:26Z avauth $} 
\subsection{Vector-meson production off nuclei by pion and kaon beams}
\label{sec:rfsep.vector}
\subsubsection{Physics case}
 The production of unstable particles off different nuclei provides the possibility  to  determine  the total cross section of the interaction of vector mesons with nucleons~\cite{RevModPhys.50.261}.  This interaction is  defined  by a set of amplitudes that correspond to the transverse  (helicity $\lambda=\pm 1$)  or  longitudinal 
 ($\lambda=0$) polarisation of the vector meson. The total cross section for the interaction of transversely polarised vector mesons with nucleons, $\sigma_T\equiv\sigma(V_T~N)$, can be extracted  from coherent photoproduction of such vector mesons off nuclei, while the investigation  of vector-meson production in the charge-exchange exclusive reactions $ \pi^-+A\to V(\rho,\omega,\varphi)+A'$ and
 $ K^++A\to K^*(892)+A'$  provides the unique opportunity to obtain  the not-yet-measured  total cross section for longitudinally polarised vector mesons interacting with nucleons,  $\sigma_L\equiv\sigma(V_LN)$.
 
In the late  1960s and early 1970s, numerous experiments on  photoproduction of vector mesons $V(\rho,\omega,\phi)$  off nuclei were performed  at SLAC, DESY, and Cornell ~\cite{RevModPhys.50.261}. They tested  predictions of the vector-meson-dominance model  and the naive   quark model, where the latter predicts for instance
$\sigma(\rho N)=\sigma (\omega N)=\frac{\sigma(\pi^+N)+\sigma(\pi^-N)}{2}$. At that  time, it was not considered that the polarisation of the vector meson could have an impact on its interaction with nucleons as  the  naive quark model  predicts that  $\sigma_T(VN)=\sigma_L(VN)$.

The knowledge of the  $\sigma_L(VN)$  cross section is important for treating the color-screening effect in vector-meson leptoproduction~\cite{1742-6596-678-1-012033}.
The idea of color transparency (CT) is that a hadron produced in certain hard processes has a smaller probability to interact in nuclear matter due to its smaller size compared to the   physical hadron. As a result, color transparency increases the nuclear transparency  defined as $A^{eff} = \frac{\sigma_A}{A\; \sigma_N}$, where $\sigma_A$ and  $\sigma_N$ are the vector-meson  production cross sections off nucleus and nucleon, respectively. On the other hand, the nuclear transparency $A^{eff}$  depends on the values of $\sigma_L(VN)$ and $\sigma_T(VN)$. The fraction of longitudinally polarised vector mesons rises with increasing values of $Q^2$. If for instance  $\sigma_T(VN)\gg \sigma_L(VN)$,  the absorption effect at large  $Q^2$ cannot be entirely described by the CT model and the difference  between the interactions of longitudinally and transversely polarised vector mesons with the nucleon has to be taken into account.

\begin{figure}[tb]
\begin{center}
\includegraphics[width=0.8\linewidth,angle=0]{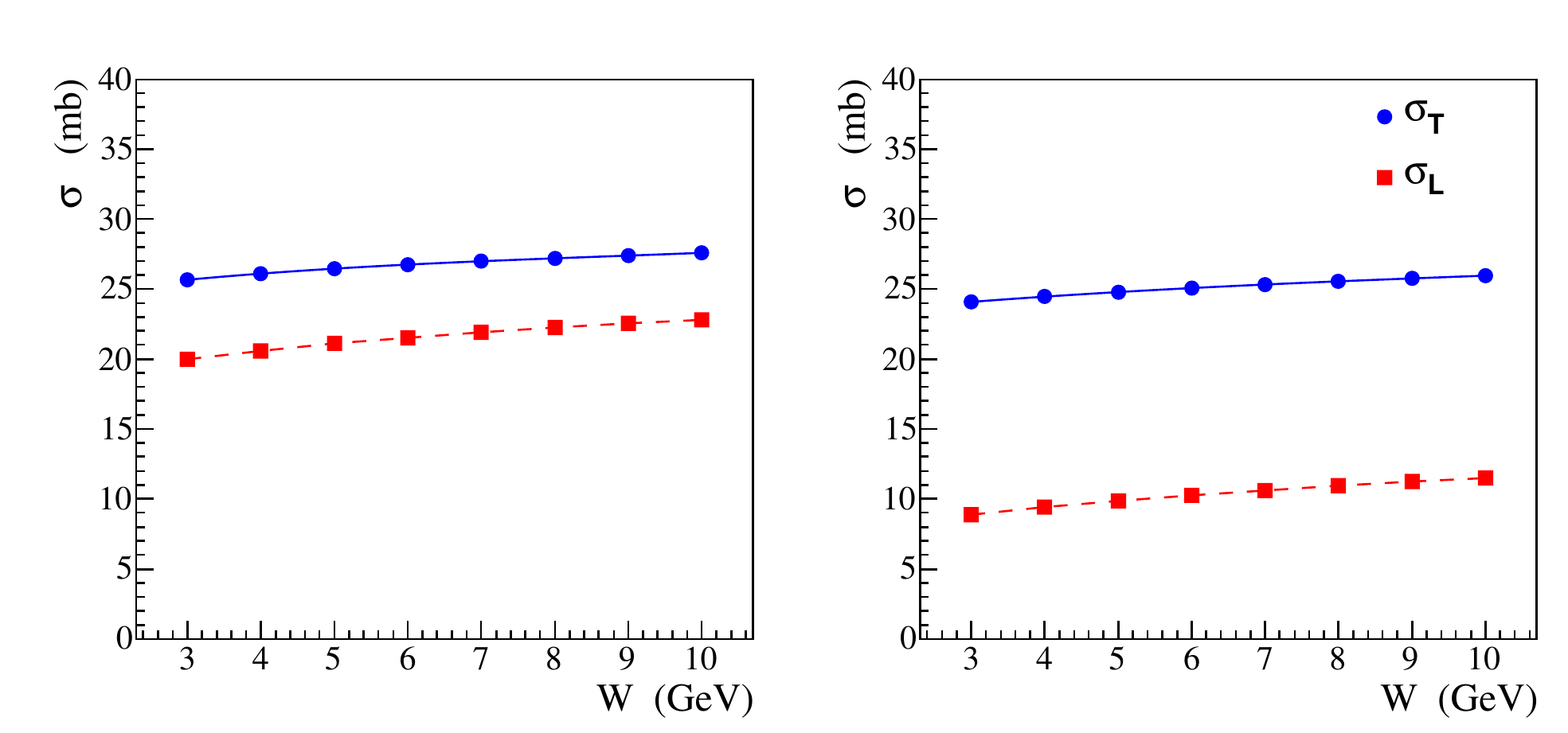}
\end{center}
\caption{  Energy dependence of longitudinal $\sigma_L$ and transverse $\sigma_T$  $\rho N$ cross sections of $\rho$N interactions for different parameterisation   of the $\rho$  meson wave function:
a)  boosted Gaussian model \cite{Kopeliovich:2000ra}; b) light-cone relativistic model \cite{Forshaw:2010py}. \label{vect1}}
\end{figure}

The energy dependence of the cross sections $\sigma_L(\rho N)$ and $\sigma_T(\rho N)$, calculated  in the  color-dipole model with different parameterisations of vector-meson wave functions~\cite{refId0}, is shown in Fig. \ref{vect1}. The impact of the polarisation on the total cross section of the vector-meson interaction with nucleons is huge.
Rather  different absorption strengths are predicted for a vector meson produced on nuclei depending on its polarisation.

An  attempt  to study the impact of the vector-meson polarisation on its  absorption strength in nuclear media  was made many years ago ~\cite{Arefev:1976uf} using the charge-exchange reaction  $\pi^- + A\to\rho^0+A^\prime$. The incoherent cross section and spin-density-matrix elements  of $\rho^0$  mesons were measured for C, Al, Cu, and Pb nuclei. Due to the dominance of  pion exchange in this process, most of the produced $\rho^0$ mesons are longitudinally polarised. At first glance,  the experimental data support  the assumption that  $\sigma_T(\rho N)\approx\sigma_L(\rho N)$.  However, there are strong arguments against  such a conclusion. It was  shown~\cite{Tarasov:1976} that, due to the relatively low energy of the primary  pion beam ($E_{\pi} = 3.7\;{\rm  GeV}$) and the large decay width of the $\rho$ meson, a significant fraction of  mesons decays inside the nucleus. This  complicates the  interpretation of the experimental data.
 
The total cross sections of  $\rho^0$ and $f^0(1270)$-meson interactions with a nucleon were measured at Argonne~\cite{Chaudhary:1974im} using the  charge-exchange process on a neon target:
 $\pi^++A\to \rho^0 [f^0(1270)]+A'$.  
 Taking into account the probability for a $\rho^0$ meson with a momentum of about 3.5 GeV/$c$ to decay inside the nucleus, the total cross section is required to
be $\sigma(\rho N)\approx 12$\,mb,  which contradicts the value of  $\sigma(\rho N)\approx 27$\,mb obtained from $\rho^0$ meson photoproduction off nuclei \cite{RevModPhys.50.261}.
This inconsistency may be due to the fact that
in the charge-exchange process  mainly  longitudinally
polarised $\rho^0$ mesons are produced,  whereas in  photoproduction  $\rho^0$ mesons  are transversely polarised due to s-channel helicity conservation. 

\begin{figure}[tb]
  \begin{minipage}{15pc}
\includegraphics[width=16pc]{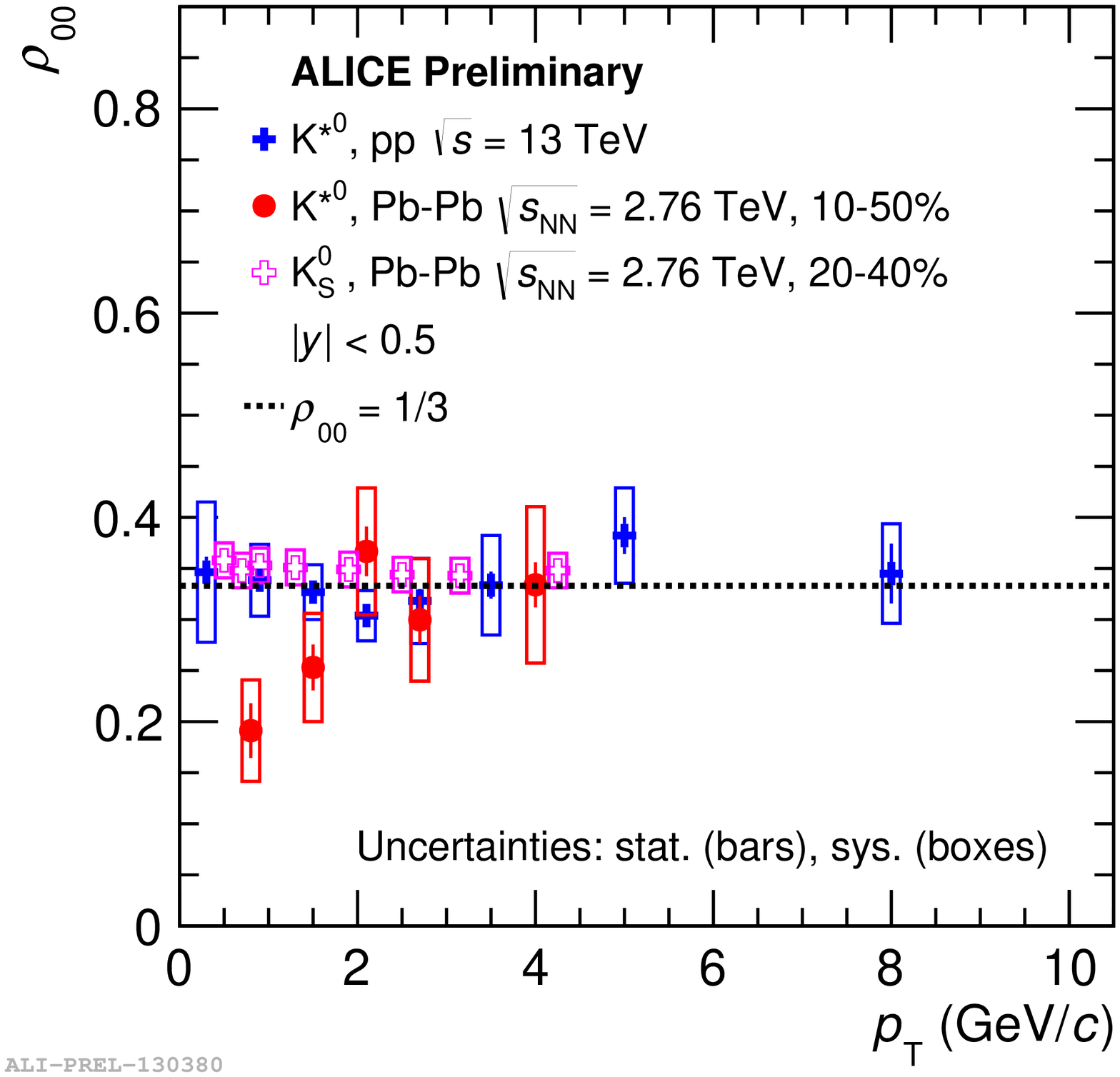}
\end{minipage}\hspace{1pc}
  \begin{minipage}{25pc}
\includegraphics[width=25pc]{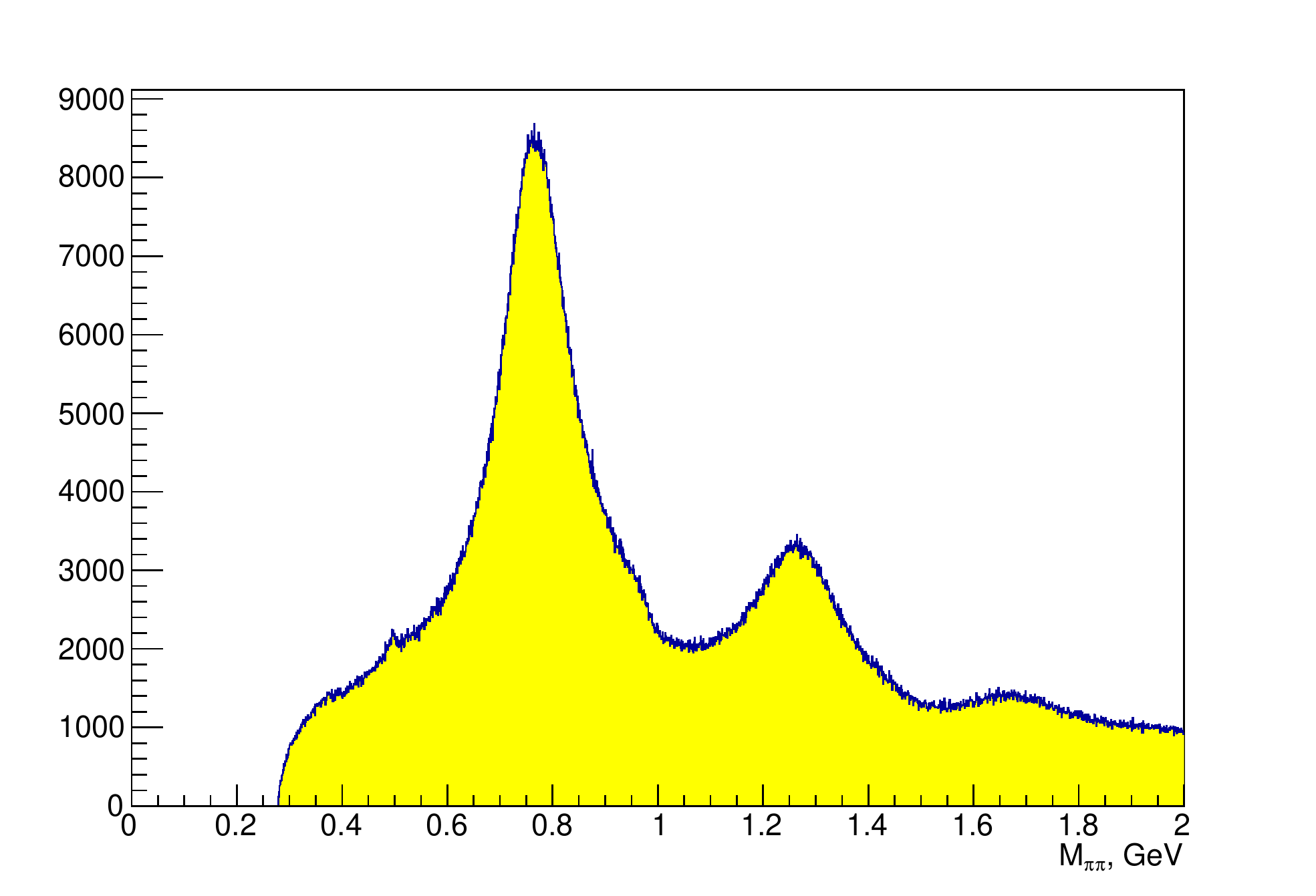}
\end{minipage}
\caption{ a) the  dependence of  $\rho_{00}$  on the  transverse momentum  of  $K^{*0}$ produced in p-p and Pb+Pb collisions; 
b) invariant mass of the $\pi^+\pi^-$ system produced exclusively by using the 190 GeV/$c$ pion beam and the COMPASS hydrogen target.\label{densmatrix} }
\end{figure}

Moreover, the absorption of polarised $K^*$ mesons in nuclear matter could be tested via the charge-exchange reaction $K^{\pm}+A\to K^*(892)+A'$. The ALICE collaboration has recently presented a preliminary result ~\cite{BedangadasMohantyfortheALICE:2017xgh}  on  $K^*$ production  in peripheral nucleus-nucleus collisions,  which shows that the vector-meson polarisation depends on transverse momenta and centrality, which is not the case in K* production in $p+p$ collisions.
 Fig.~\ref{densmatrix}a shows the spin density matrix element $\rho_{00}$  of $K^*$ produced in $p+p$ collisions and $Pb+Pb$ collisions.  The value $\rho_{00}=1/3$  corresponds to the production of unpolarised
vector mesons. One can see that in lead-lead collisions at moderate transverse momenta  the yield of longitudinally polarised $K^*$  is suppressed. 
Thus the investigation of $K^*$  production off nuclei is becoming  a topical issue, as well. 

\subsubsection{The proposed measurement}
We propose to measure the value of the spin density matrix element $\rho_{00}$ as a function of transverse momentum of the vector mesons produced off nuclear targets in order to extract the cross section $\sigma_L(V_L N)$. 
This can be accomplished using
 the exclusive reactions $\pi^- + A\to \rho^0(f^0)+A^\prime$ when using a pion beam and $K^- + A\to\K^0+A^\prime$ when using a kaon beam. A set of nuclear targets with different atomic numbers ranging from $H$ to $Pb$ is planned to be used. In order to avoid any effect related to meson decays inside the target nucleus, the beam momentum should be relatively large. However, it should not be too large since the cross section of the charge-exchange reaction decreases with increasing beam energy. A hadron beam momentum of about 100~GeV/$c$ appears to be a reasonable compromise. Basic tracking and photon calorimetry capabilities are required for this measurement. The expected cross section, 0.1-1~mb, is sufficiently large to collect a reasonable statistics within a short time or to run in parallel with other physics programmes.

 The feasibility of the proposed measurements with hadron beams could also be investigated using available COMPASS data since COMPASS already has some  data collected with a 190 GeV/$c$ negative hadron beam of high intensity and a set of nuclear targets. The invariant mass of the $\pi^+\pi^-$ final state produced exclusively off a hydrogen target by a 190 GeV/$c$ pion beam (COMPASS 2008 data) is shown in Fig. \ref{densmatrix}b.
 
 The proposed measurement is complementary  to the investigation of vector-meson  photoproduction  off nuclei, which is  planned to be performed at JLab with the GlueX setup \cite{Chudakov:loi,Chudakov:2015msa, GLUEX:loi17}.

\svnid{$Id: inst.tex 4582 2019-01-07 17:04:06Z friedric $} 
\section{Instrumentation}
\label{sec:inst}

\subsection{Summary table}
\label{sec:inst:summarytable}

Many programmes presented in this Letter of Intent are based on the concept of using the basic features of the present \compass\ setup \cite{ABBON2007455}, \cite{ABBON201569}: one or two large-gap dipole magnets with tracking stations around them, combined with particle identification detectors. The standard polarised target is described in Ref.~\cite{ABBON2007455} and the liquid hydrogen (LH2) target in Ref.~\cite{ABBON201569}. 


Most future programmes require additional specific detectors or other equipment, as will be explained in the text below (Sec.~\ref{sec:inst:general} for general upgrades and Sec.~\ref{sec:instmajor} for specific upgrades). The CEDARs, located at the beam line, are required in all hadron programmes for beam-particle identification. The RICH is necessary for the identification of produced hadrons in several programmes.

The phase-1 programmes will use the existing M2 muon or hadron beams. The phase-2 programmes are designed for using future RF-separated hadron beams in the M2 beam line with enhanced fractions of kaons and antiprotons (Sec.~\ref{sec:rfsep.beams}). 

The specific parameters and hardware upgrades for each programme are summarized in Table~\ref{tab:inst:programs}. 


\begin{table}[h]
  \begin{center}
  
  \end{center}
  \caption{Requirements for future programmes at the M2 beam line after 2021. {\color{blue}Muon beams} are in blue, {\color{ForestGreen}conventional hadron beams} in green, and {\color{red}RF-separated hadron beams} in red. }
  \label{tab:inst:programs}
\end{table}

\subsection{General upgrades}
\label{sec:inst:general}

The following general upgrades of the existing spectrometer in the EHN2 experimental hall are considered to form the experimental backbone of the ``New QCD facility at the M2 beam line of the CERN SPS'':
\begin{itemize}
\item New front-end electronics (FEE) and trigger logics that are compatible with triggerless readout, which include an FPGA-based TDC with time resolution down to 100\,ps and a digital trigger that is capable of rates up to 100-200\,kHz (Sec.~\ref{sec:inst:daqfeet}).
\item New large-size PixelGEMs as replacement and spares for existing large-area GEMs (Sec.~\ref{sec:inst:gems}). 
\item New large-area micro-pattern gaseous detectors (MPGD) based on GEMs or Micromegas technology to replace existing MWPCs (Sec.~\ref{sec:inst:mpgd}).
\item High-rate-capable CEDARs (Sec.~\ref{sec:inst:cedars}) for all hadron-beam programmes to identify the desired beam particle. 
\item The existing RICH-1 will be required by the spectroscopy  programmes (Secs.~\ref{sec:stdhad.spect} and \ref{sec:rfsep.spect}), the anti-matter cross section  measurement (Sec~\ref{sec:stdhad.xsect}), and the Primakoff programme (Sec.~\ref{sec:rfsep.prima}). A new, high-aperture RICH-0 would be desirable for  these programmes in order to identify hadrons at lower momenta (Sec.~\ref{sec:inst-rich}).
\end{itemize}

\subsubsection{Front-end electronics and DAQ} 
\label{sec:inst:daqfeet}

The purpose of the new front-end electronics and data-acquisition system is to read out detectors with best precision and minimum loss of efficiency. With particle rates on the target of up to $10^8$/sec, the optimum solution is the construction of triggered, pipelined front-end electronics with a maximum trigger-rate capability between 100-200\,kHz and a dead time of 2-3\,\%.
These requirements allow the usage of the well-performing APV25 ASIC for all MPGDs and silicon detectors, as well as for the RICH detector. Since many modern ASICs feature triggerless readout, the desired goal of a triggerless readout solution exists for every detector type. 
The newly developed FPGA-based TDC (iFTDC) has a time resolution down to 100\,ps. It is planned to equip all detectors with new modern FEE and to use the same kind wherever possible. This will allow a single expert to intervene on various equipments. 

\begin{figure}[tb]
  \centering
  \includegraphics[width=10cm]{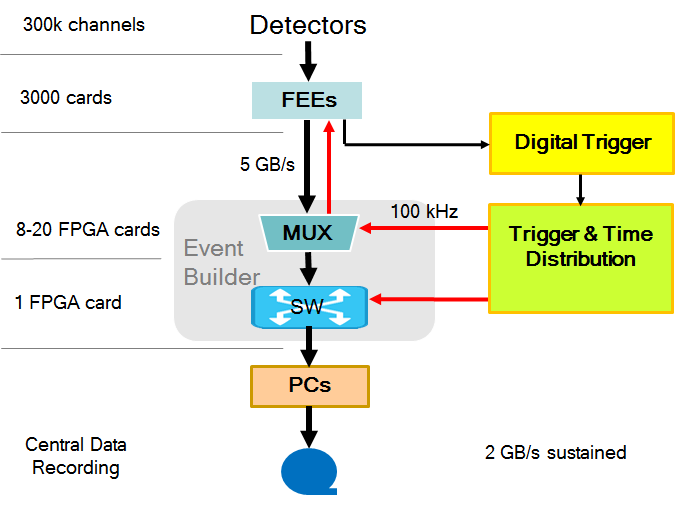}
  \caption{The new DAQ architecture.}
  \label{figure:DAQ_2020}
\end{figure}

The architecture of the new readout system is shown in Fig.~\ref{figure:DAQ_2020}.
The number of channels and the data rates are estimated using the performance of the existing \compass\ setup and the \compass\ DAQ \cite{1748-0221-11-02-C02025}. 
The front-end boards including digitisers will be placed near the detectors and will be equipped with two high-speed serial interfaces. One interface will transmit untriggered hit information and can be connected to the digital trigger module. The second link will transmit triggered information only to the DAQ. All high-speed serial interfaces within DAQ and digital trigger will employ the UCF protocol \cite{UCF:2016}, which transmits trigger and event identification information, slow control messages, and data via a single serial link.

The DAQ will consist of two stages of data processing.
At the first stage, the data will be buffered at the local SDRAM and then merged into sub-events.
At the second stage, complete events will be assembled and distributed between online PCs via 10\,Gb Ethernet.
The maximum data rate expected during a spill will be 5\,GB/s, while the sustained rate will be 2\,GB/s. The system is designed to handle sustained-data rates of 5\,GB/s.

\subsubsection{Large-area PixelGEM detectors} 
\label{sec:inst:gems}

New large-area PixelGEM detectors will be designed and ten such detectors will be built by 2021 as replacement and spares for the existing large-area GEMs \cite{KETZER2004314} in the \compass\ setup. Each detector will have 4,096 channels. The periphery will be read out with strip readout from both sides. The center will consist of hexagonal pads of 1.5\,mm outer radius and will be equipped with pixel readout. The active area of each detector will be between 30.7\,cm $\times$ 30.7\,cm and up to 40\,cm $\times$ 40\,cm, about a factor of 10 larger than the existing \compass\ PixelGEM detectors \cite{Ketzer2007}. The new PixelGEMs will be equipped with new Front-End Electronics allowing for higher rates and self-triggering.


\subsubsection{Large-area multi-pattern gaseous detectors (MPGD)} 
\label{sec:inst:mpgd}

New MPGDs will be designed and developed to replace the existing ageing 14 MWPC tracking stations in the \compass\ setup. 
The new detectors will be based on large-area GEM or Micromegas technology.
Each station will have an active area of about 1.5\,m$^2$ , with two or three coordinates planes and about 2\,mm pitch. The new detectors will be equipped with new front-end electronics with a rate capability of about 1\,MHz per channel. The total number of channels will be about 28,000. 

\subsubsection{CEDARs at high rates} 
\label{sec:inst:cedars}

The purpose of the CEDAR is the identification of the beam particle on an event-by-event basis. 
Presently, two 6\,m-long CEDAR stations are located in the M2 beam line 30\,m upstream of the \compass\ target. They are filled with helium gas at a pressure of approximately 10.5\,bar. The emerging Cherenkov photons are focused by a mirror and detected by eight PMTs arranged in a ring around the center. The pion, kaon, or proton ring is selected by tuning the diaphragm and the pressure. 

\begin{figure}[tb]
  \begin{center}
    \includegraphics[width=0.9\textwidth]{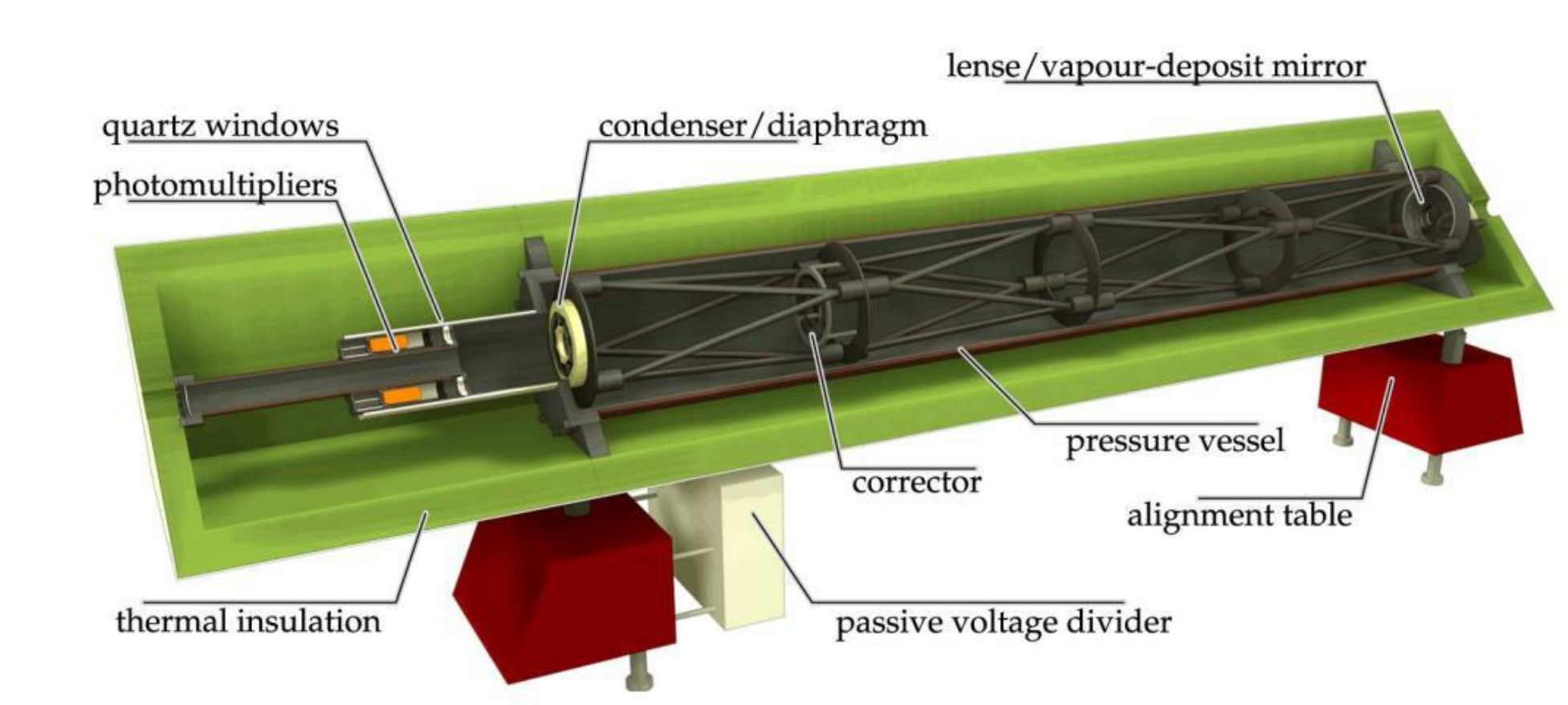}
  \end{center}
  \caption{CEDAR 2018 upgrade for better rate and thermal stability.} 
  \label{fig:inst:cedars}
\end{figure}

The existing CEDARs were upgraded in winter 2017/2018 for better rate and thermal stability in preparation of the 2018 \compass\ run with a pion beam. The project was carried out in collaboration of CERN and \compass\ representatives. New PMTs (fast Hamamatsu R11263-203 with pulse widths of 2-3\,ns), a new gain monitoring system, a new readout system, and a new thermalisation system are contained in the upgrade package. A conceptual sketch of the upgraded system is shown in Fig.~\ref{fig:inst:cedars}.

The decision whether further upgrades of the CEDAR system will be necessary during LS2 for the future hadron-beam programmes (as described for standard beams in Sec.~\ref{sec:stdhad} and for RF-separated beams in Sec.~\ref{sec:rfsep}) will be  based on the experience collected with the upgraded CEDARs during the 2018 \compass\ Drell-Yan run with the high-intensity pion beam (about $8\cdot 10^7$/sec).
We consider the possibility to install a tracking system for the CEDARs, which would correct the beam-track trajectories. One option could be the XBPF upgrade, a new SciFi-based instrumentation developed at the CERN North Area to measure beam profiles and momenta \cite{XBPF}.




\subsubsection{Hadron PID perspectives: RICH} 
\label{sec:inst-rich}

The RICH-1 Cherenkov-imaging detector  \cite{ALBRECHT2005215} \cite{ABBON2008371} \cite{Abbon:2010zza} \cite{Abbon:2011zza} is the backbone for hadron PID in the \compass\  setup.
It has a large acceptance, i.e.\ $\pm$200\,mrad in the vertical plane and $\pm$250\,mrad in the horizontal plane. It uses C$_4$F$_{10}$ as heavy and low-chromaticity radiator gas, where image focusing is provided by a wall of spherical UV mirrors. Presently, the photon detection system is formed by MAPMTS coupled to individual fused silica-lens telescopes in the central region, covering 25\% of the instrumented surface, where the rate is higher, and gaseous detectors in the peripheral region. Two types of gaseous detectors are in use, both equipped with CsI photo-converters: MWPC detectors and novel ones, which are based on a hybrid MPGD architecture with two THick GEM (THGEM) layers that are followed by a Micromegas multiplication stage, where the first THGEM also acts as photo-converter substrate. 
The RICH-1 detector provides hadron PID in the range from 3 to 60\,GeV/c, where 3\,GeV/c is the effective threshold for pion identification, and pions can be separated from kaons at 90\% confidence level at 60\,GeV/c \cite{Abbon:2011zza}.


For the future physics programme at the M2 beam line, RICH-1 can be complemented by new Cherenkov counters that enlarge the momentum range for the positive identification of hadrons both at lower and higher momenta. For low momenta (referred to as ``RICH-0'' here), a DIRC counter enriched with a focusing system \cite{Dey:2014fda} with horizontal radiator bars arranged in a planar configuration can be used in order to separate hadrons in the range 0.2\,GeV/c up to 5-6\,GeV/c. Fused silica bars are the default choice, while the use of Plexiglas bars \cite{ADOLPH2011185} is an alterative option to be analysed. The default readout sensors are MAPMTs, while other fast, pixelised photon detectors as MCP-PMTs can be considered. A relevant feature is the reduced physical length of such a detector that may require no more than a 20\,cm space-slot along the beam line.

\subsection{Specific upgrades} 
\label{sec:instmajor}

\subsubsection{Overview}

\begin{itemize}
\item muon-proton elastic scattering (more in Sec.~\ref{sec:inst-mup}): high-pressure active TPC target (similar to A2 at MAMI) or hydrogen tube surrounded by SciFis; SciFi trigger system on scattered muon; silicon trackers to veto on straight tracks (kink trigger). 
\item Hard exclusive reactions (more in Sec.~\ref{sec:inst-dvcs}): 3-layer silicon detector inside the existing but modified transversely polarised $\mathrm{NH}_3$ target, which operates at very low temperature, for tracking of the recoil proton produced in DVCS, as well as for PID via dE/dx. Alternatively: SciFis.
\item Input for DMS: liquid helium target.
\item $\overline{p}$-induced spectroscopy (more in Sec.~\ref{sec:inst-lowespec}): target spectrometer (tracking, barrel calorimeter) similar to WASA at COSY \cite{Adam:2004ch}; target: LH2, foil, wire.
\item Drell Yan: high-purity and high-efficiency dimuon trigger; dedicated precise luminosity measurement; dedicated vertex-detection system; beam trackers; targets: $^6\mathrm{LiD}\uparrow$, and C/W. 
\item Drell-Yan (RF) (see also Sec.~\ref{sec:inst-dyrfsep}): due to the lower beam energy, a wide aperture will be needed (up to $\pm300$\,mrad): a "magnetised spectrometer" (active absorber) is under consideration. It could possibly be similar to Baby MIND at JParc \cite{Antonova:2017thk} ("3-in-1" detector, spectrometer magnet, absorber). 
\item Prompt Photons (RF): 20-30\,cm steel absorber upstream of the target; new hodoscope upstream of the existing electromagnetic calorimeter ECAL0; transparent setup with as little material as necessary.
\item $K$-induced spectroscopy (RF): uniform acceptance; existing electromagnetic calorimeters; recoil TOF detector (see Fig.~\ref{fig:target}, called ``RPD'' there).
\end{itemize}


\subsubsection{High-pressure hydrogen TPC for proton-radius measurement } 
\label{sec:inst-mup}

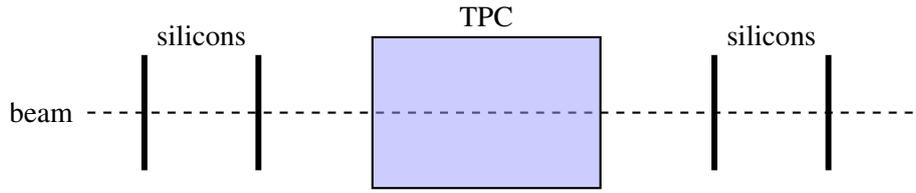
\begin{figure}
  \begin{center}
    \begin{tikzpicture}[thick,scale=0.5,
        axis/.style={postaction={decorate}, decoration={markings,mark=at position 1 with {\arrow{>}}}},
      ]
      \clip (-12.5,-2.8) rectangle (12.5,3.5);

      \draw[dashed,->] (-10.5,0) -- (11.5,0);
      \node[left] at (-10.6,0) {beam};

      \filldraw[fill=blue!40!white, fill opacity=0.5, draw=black] (-3,-2) rectangle (3,2);
      \node[above] at (0,2) {TPC};

      \filldraw[fill=black, draw=black] (-9.05,-1.5) rectangle (-8.95,1.5);
      \filldraw[fill=black, draw=black] (-6.05,-1.5) rectangle (-5.95,1.5);
      \node[above] at (-7.5,1.5) {silicons};
      \filldraw[fill=black, draw=black] (5.95,-1.5) rectangle (6.05,1.5);
      \filldraw[fill=black, draw=black] (8.95,-1.5) rectangle (9.05,1.5);
      \node[above] at (7.5,1.5) {silicons};
    \end{tikzpicture}%
    \caption{Sketch of the target for the proton-radius measurement.}
    \label{fig:stdmuo.mup.COMPASS_setup}
  \end{center}
\end{figure}

\begin{figure}
  \begin{center}
    \includegraphics[width=0.75\textwidth]{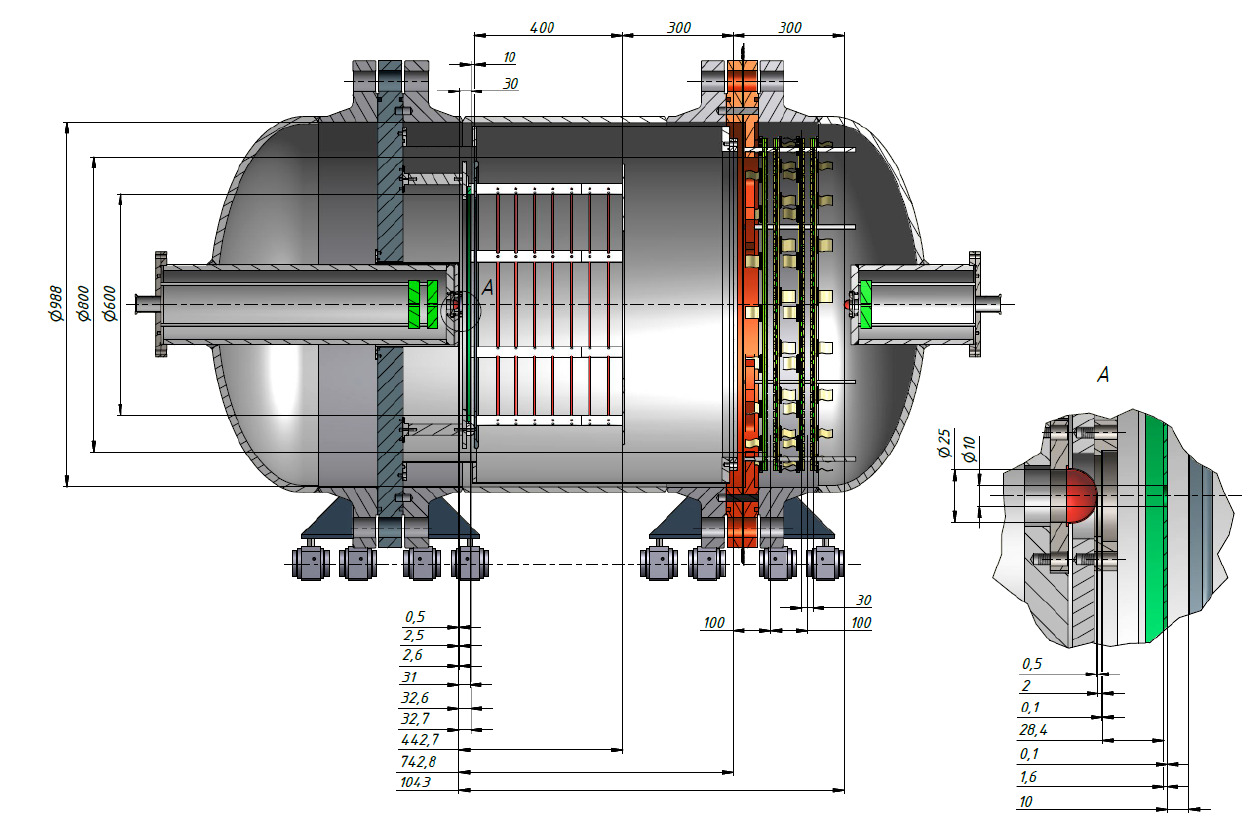}
    \caption{Sketch of the target TPC with pressure vessel as conceived for an elastic $e^{-}p$ scattering experiment at MAMI. The forward tracker system on the right side of the vessel will not be installed for the muon-proton measurement at the M2 beam line.} 
    \label{fig:stdmuo.mup.MZ_active_target}
  \end{center}
\end{figure}

The experimental setup for the proton radius measurement using elastic muon-proton scattering (Sec.~\ref{sec:stdmuo.mup}) is schematically shown in Fig.~\ref{fig:stdmuo.mup.COMPASS_setup}. The active hydrogen target (IKAR \cite{Ilieva:2012vhs}) is based on an existing setup used for an experiment at GSI (Germany), which is shown in Fig.~\ref{fig:stdmuo.mup.MZ_active_target}. Such a system was developed by the Gatchina group (PNPI, Russia) and has been employed for multiple radius measurements in the past. 

\paragraph{Proton recoil measurement}
\label{sec:stdmuo.mup.proton}

The proton recoil measurement can be achieved using a double-target scenario. For small values of $Q^{2}$ and proton kinetic energies up to a few MeV, a high-pressure hydrogen TPC, operated as ionisation chamber, can be used. The energy loss for incoming and outgoing muons is about 2\,keV/cm and thus small compared to the proton energy loss even for proton kinetic energies of 10\,MeV, as long as the path length traversed is smaller than 10\,cm. For $Q^{2}=10^{-4}\,($GeV$/c)^{2}$, the kinetic energy of recoil protons is 50$-$60\,keV. This value corresponds to the energy resolution obtained by \cite{Vorobyov:1974ds} in an experiment measuring $\pi p$ scattering in the Coulomb-nuclear interference region. This roughly determines the scale for the lowest value of $Q^2$ in the experiment. 

\begin{figure}[tb]
  \begin{center}
    \includegraphics[width=0.49\textwidth]{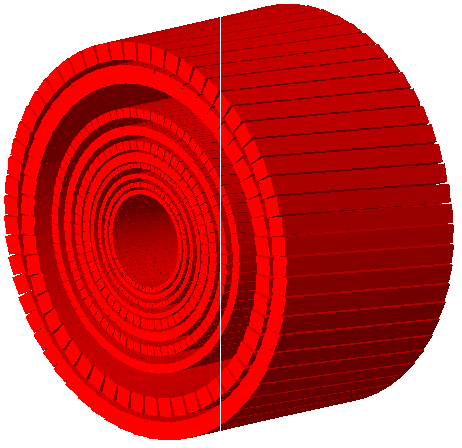}
    \caption{Layout of the recoil proton detector used for the high-$Q^2$ range.}
    \label{fig:stdmuo.mup.active_target_layout}
  \end{center}
\end{figure}

\begin{figure}[tb]
  \begin{center}
    \includegraphics[width=0.47\textwidth]{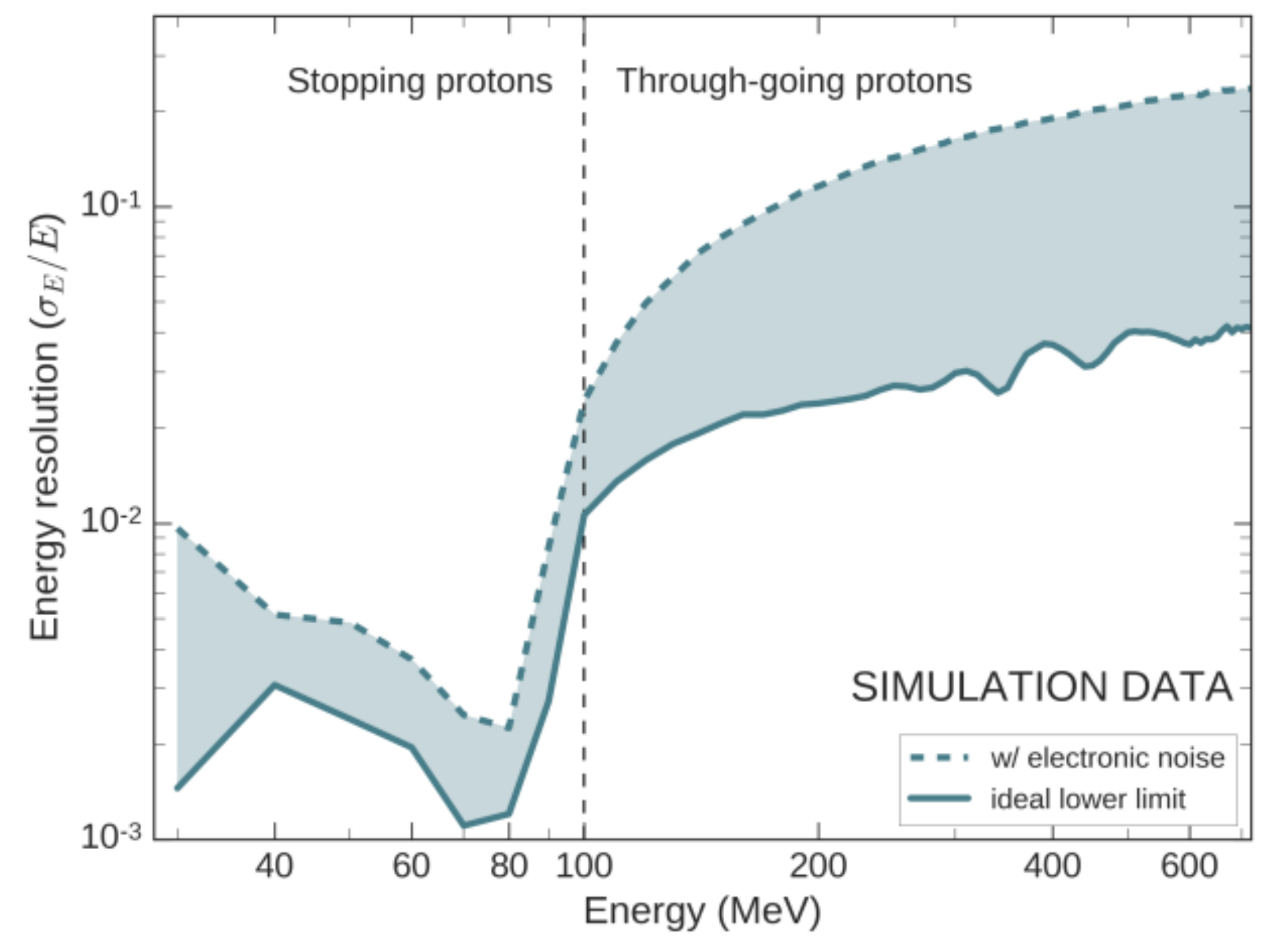} \hfill 
    \includegraphics[width=0.47\textwidth]{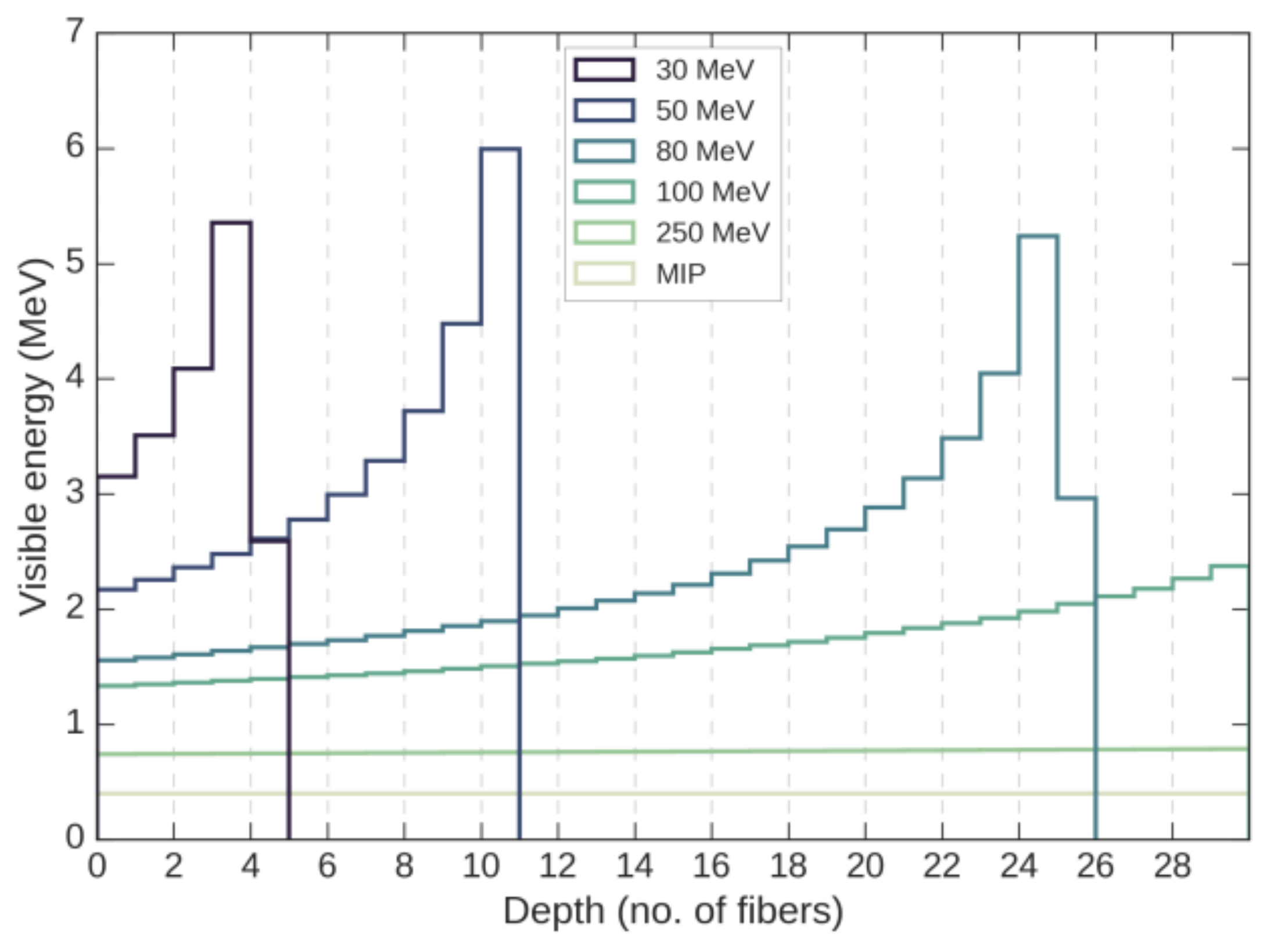}
    \caption{Left: Expected energy loss in individual fibres traversed by recoil protons for different proton energies. 
      Right: energy resolution obtained by Bragg-curve fitting using simulation data. Work in progress and data are still very preliminary.}
    \label{fig:stdmuo.mup.Bragg_curve_fitting}
  \end{center}
\end{figure}

At higher values of ${\textrm Q^{2}}$, when the recoiling protons are no longer stopped inside the hydrogen volume, one may envisage to surround the central part of the active target with a barrel made from scintillating fibres. Consecutive layers are arranged in a relative stereo angle of 6\textdegree. A possible setup is shown in Fig.~\ref{fig:stdmuo.mup.active_target_layout}. The scintillation light from the fibres is detected on one side by SiPM of high pixel density (Hamamatsu S13360-3025 or KETEK PM3325) to reduce saturation effects. The backend opposing the SiPM is aluminised. In the model used for simulation we assumed 10 layers of scintillating fibres, summing up to 2$-$3\,cm thickness. In order to perform a combined ($dE/dx,\;E$) analysis, we intend to surround the fibre tracker by 8 plates of scintillator, similar to the proton-recoil detector surrounding the liquid-hydrogen target of \compass\ in 2009. With such a detector, we should be able to stop protons up to 100\,MeV. By reconstructing the Bragg curve, we can obtain energy resolutions of the order of a few percent (Fig.~\ref{fig:stdmuo.mup.Bragg_curve_fitting}). We have performed test measurements on energy resolution up to energies of about 50\,MeV at PSI using various fibre materials and models of SiPM. Results from the analysis are expected soon.

As the range of low energy protons in the SciFi material is short, we need to keep the fibre thickness small in the inner layers ($2\times2\,$mm$^{2}$ or thinner). A requirement for the recoil proton of crossing at least 2 fibres to determine a 3D impact point imposes a lower limit for the kinetic energy of recoil protons of about 15-20\,MeV. This corresponds to a lower limit for $Q^{2}$ of $0.03-0.04\,($GeV$/c)^{2}$. The fibre cross section for the outer layers may grow to $4\times4\,$mm$^{2}$ and $8\times8\,$mm$^{2}$. 

The geometry of the scintillator barrel has not yet been optimised in terms of geometry, fibre cross sections and number of channels. However, the above sketched arrangement is feasible and has a reasonably flat acceptance across $Q^2$. Optimisation should allow us to further reduce an unwanted $Q^2$ dependence of the acceptance and obtain an effective threshold of $Q^2>0.3\,($GeV$/c)^{2}$. 

\paragraph{Muon measurement}
\label{sec:stdmuo.mup.muon}

The scattered muon can be identified using the \compass\ setup including the present muon identification system. 
The energy transfer in the reaction is very small and falls within the energy resolution of the spectrometer. However, \compass\ has demonstrated excellent angular resolution in the context of a measurement that used pions of 190\,GeV energy scattering off the electromagnetic field of heavy nuclei like Pb or Ni~\cite{Adolph:2014kgj}. Despite the presence of a solid target of thickness ${\textrm d=20\%\,X_{0}}$, \compass\ obtained a  $Q^{2}$-resolution of $\Delta Q^{2}=2\cdot 10^{-4}\,($GeV$/c)^{2}$. This was achieved by means of two silicon telescopes placed upstream and downstream of the solid target. The position resolution of each silicon station was about $\Delta x\approx 4\,\mu$m. In the future setup, it is intended to position the silicon stations within a telescope much further apart, i.e.\ larger than 1\,m, to provide a longer lever arm. 
It is assumed to improve the angular resolution such that a resolution of  $\Delta Q^{2}=1.4\cdot 10^{-4}\,($GeV$/c)^{2}$ is achieved, by: 
\begin{enumerate} 
\item replacing the thick solid target with a pressurised gaseous hydrogen target;
\item increasing the spacing of the layers of the silicon telescope in
  order to roughly match multiple scattering effects in the silicon itself.
\end{enumerate}

\paragraph{Trigger}
\label{sec:stdmuo.mup.trigger}

One of the challenges of this experiment will be the trigger. A trigger on the proton signal inside the TPC will require a trigger latency of the order of $20\,\mu$s owing to the drift time. This is not compatible with the current \compass\ readout scheme. We envisage two different approaches to circumvent this limitation. 

The approach compatible with current front-ends is the development of a trigger on a kink of the muon track. While two scintillating fibre detectors upstream of the target 
predict a straight track, a deviation from this straight track observed in a third scintillating fibre detector downstream of the target 
is a sign for an interaction inside the target. In order to suppress deviations from the straight line caused by multiple scattering, the distance between the predicted and measured position should be adjustable for this trigger. 

The more advanced approach, which is envisaged for the future, is based on the development of a triggerless readout scheme requiring the development or integration of new front-end electronics for the silicon strip detectors and the TPC. 

\subsubsection{Recoil detector with polarised target} 
\label{sec:inst-dvcs}


\begin{figure}[tb]
  \begin{center}
    \includegraphics[width=\textwidth]{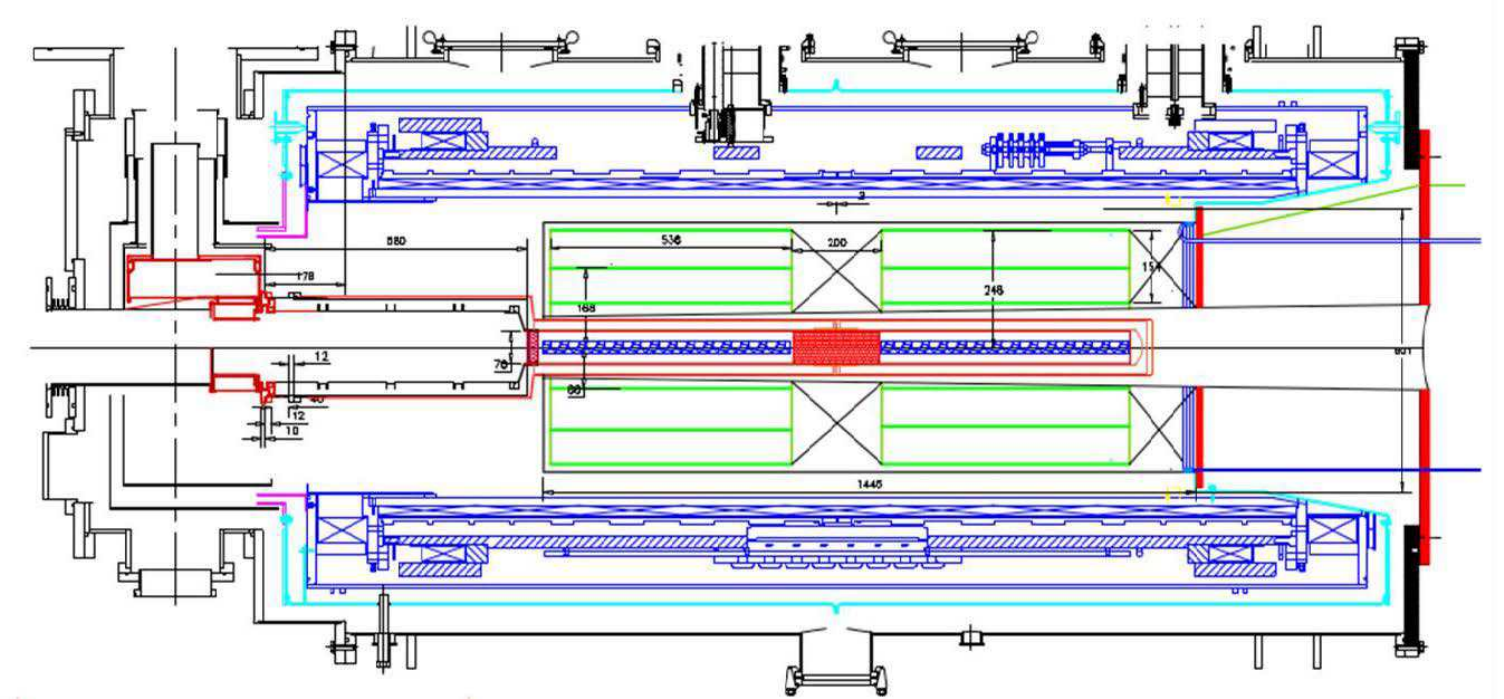}
  \end{center}
  \caption{Conceptual view of the \compass\ polarised target coupled with silicon detectors for tracking and identification of recoil particles.} 
  \label{fig:future.compass_dvcs_future_PT}
\end{figure}

The major technical challenge for the measurement of the $\mu p^\uparrow \to\ \mu \gamma p$ reaction (Sec.~\ref{sec:stdmuo.dvcs}) is the detection of the recoil particles produced in the solid-state transversely polarised target. The detection of the recoil particles, the momenta of which have to be determined with a precision of 10\% or better, is essential to ensure the exclusivity of the reaction. Missing-mass techniques cannot be employed here, as the best attainable experimental resolution would be about 2\,GeV. Two solutions can be envisaged to minimise the amount of material crossed by the recoil particles before being detected, and therefore to optimise the minimum accessible value of the momentum transfer $|t|$ to the recoil particle: 
\begin{enumerate}
\item The recoil particle detector is placed outside of a low-mass polarised target system with a thin super-conducting dipole located close to the target cells and enclosed inside a thin-walled cryostat. 
\item The recoil particle detector is inserted into the cryogenic vacuum volume surrounding the target cell and inside a large dipole magnet. 
\end{enumerate}
The first solution is technically very challenging, particularly from the point of view of the construction of a super-conducting dipole magnet with a low material budget. Moreover, it would require an additional external high-field solenoid magnet to re-polarise the target material every few days. 

\begin{figure}
  \begin{center}
    \includegraphics[width=0.43\textwidth]{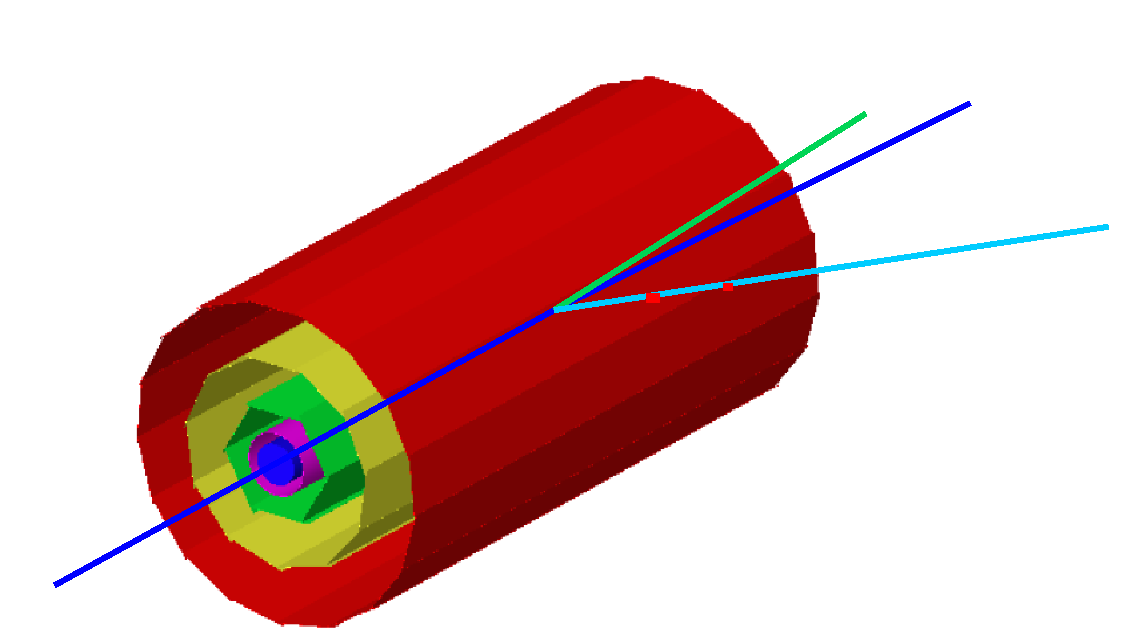}
    \includegraphics[width=0.25\textwidth]{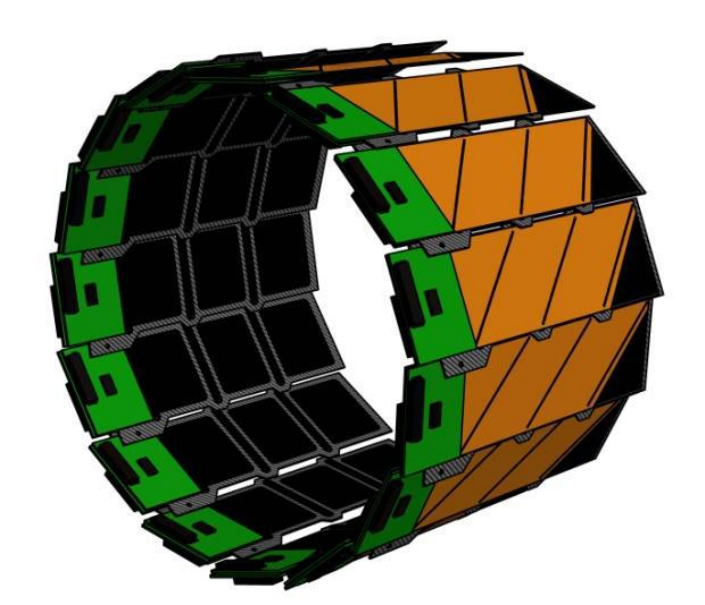}              
    \includegraphics[width=0.23\textwidth]{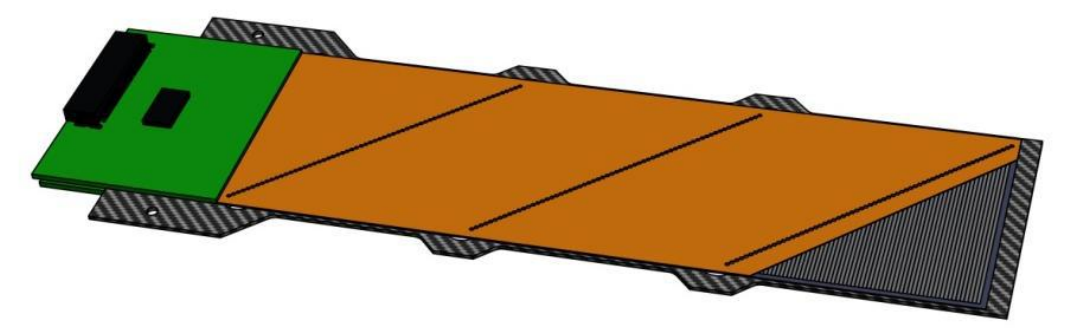}
  \end{center}
  \caption{Left: 3-layer silicon detector (SD) surrounding the polarised target, shown together with trajectories of particles emerging from an exclusive DVCS event: proton (light blue), photon (green), muon (blue). From inside to outside: target, MW cavity, inner SD, middle SD, outer SD. Each layer is 300\,$\mu$m thick.
    Middle: conceptual design of the outer SD layer with 300\,mm diameter. Right: SD ladder design.}
  \label{fig:future.compass_dvcs_silicon_detector}
\end{figure}

The feasibility of the second solution is currently under study, i.e.\ re-using the existing \compass\ polarised-target system, see Fig.~\ref{fig:future.compass_dvcs_future_PT}. In this scenario, the shape of the micro-wave (MW) cavity is modified and decoupled from the remainder of the inner target magnet volume, while sharing the same vacuum. The cylindrical part of the cavity consists of a 0.2-0.4\,mm thick copper foil to avoid distortion of the MW field by the presence of silicon detectors. Recoil-particle detection is based on two or three concentric barrels of silicon pixel detectors (Fig.~\ref{fig:future.compass_dvcs_silicon_detector}) in the empty space between the target cell and the superconducting magnets to measure particle trajectories and energy loss (dE/dx). Alternatively it is considered to use scintillating-fibre detectors instead of silicon detectors. SciFi detectors can be accommodated more easily in the target magnet with less challenging signal transport out of the magnet. 

\paragraph*{Performance of silicon detectors at low temperatures.}

Silicon detectors are capable of working (i) in a magnetic field (longitudinal or transverse) of about 0.5-2\,T, (ii) at low temperatures of about 5-10\,K \cite{Gusev2007}, (iii) in  the presence of a MW field and (iv) in a vacuum of about 10$^{-8}$\,bar.  Modifications of the inner volume of the existing target magnet are necessary in order to minimise the influence of the MW radiation on the silicon performance and to provide space for input/output connections. The MW cavity is cooled by circulation of liquid $^4$He. Part of this flow also cools a mesh surrounding the silicon-detector volume, keeping it at uniform temperature. This prevents decrepitation of the silicon wafers and dissipates the heat from their readout electronics. 

Double-sided Si-microstrip detectors developed at the Laboratory of High Energy Physics (LHEP) at the Joint Institute for Nuclear Research (JINR) meet the main requirements to serve as recoil detector inside a \compass-like polarised-target magnet. The LHEP JINR silicon detector is comparatively inexpensive. It has been tested in an experimental environment close to that of the present \compass\ polarised-target system. Tests with multi-layered flexible boards are under preparation with participation of LED Technologies of Ukraine (LTU / Kharkiv) \cite{ProtsenkoDAQFEET}.

\paragraph*{PID and momentum reconstruction of recoil particles.}

\begin{figure}[tb]
  \begin{center}
    \includegraphics[width=0.30\textwidth]{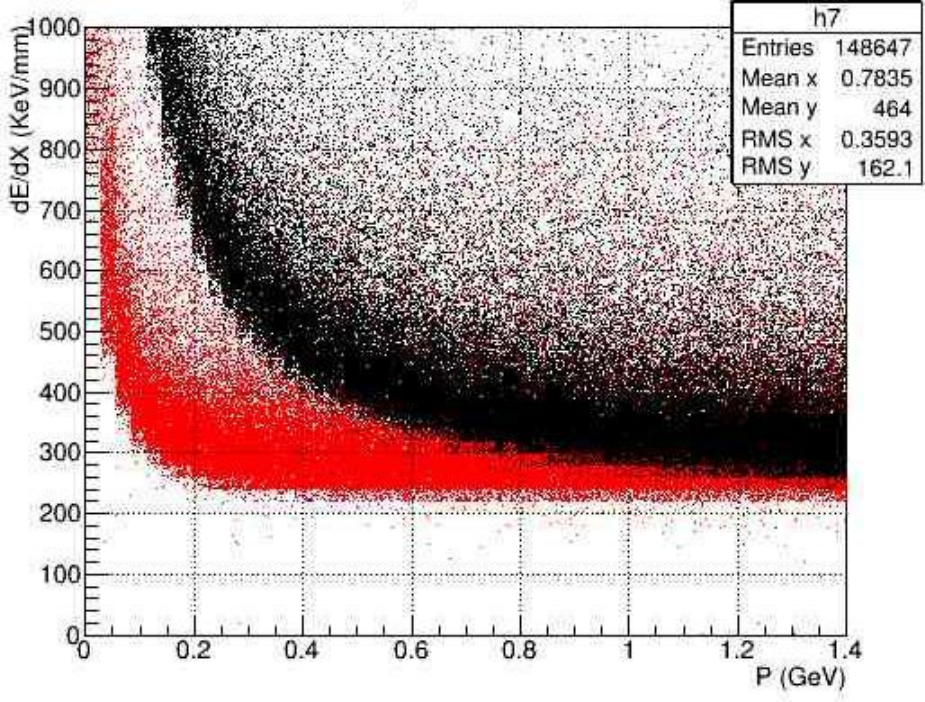} ~ 
    \includegraphics[width=0.33\textwidth]{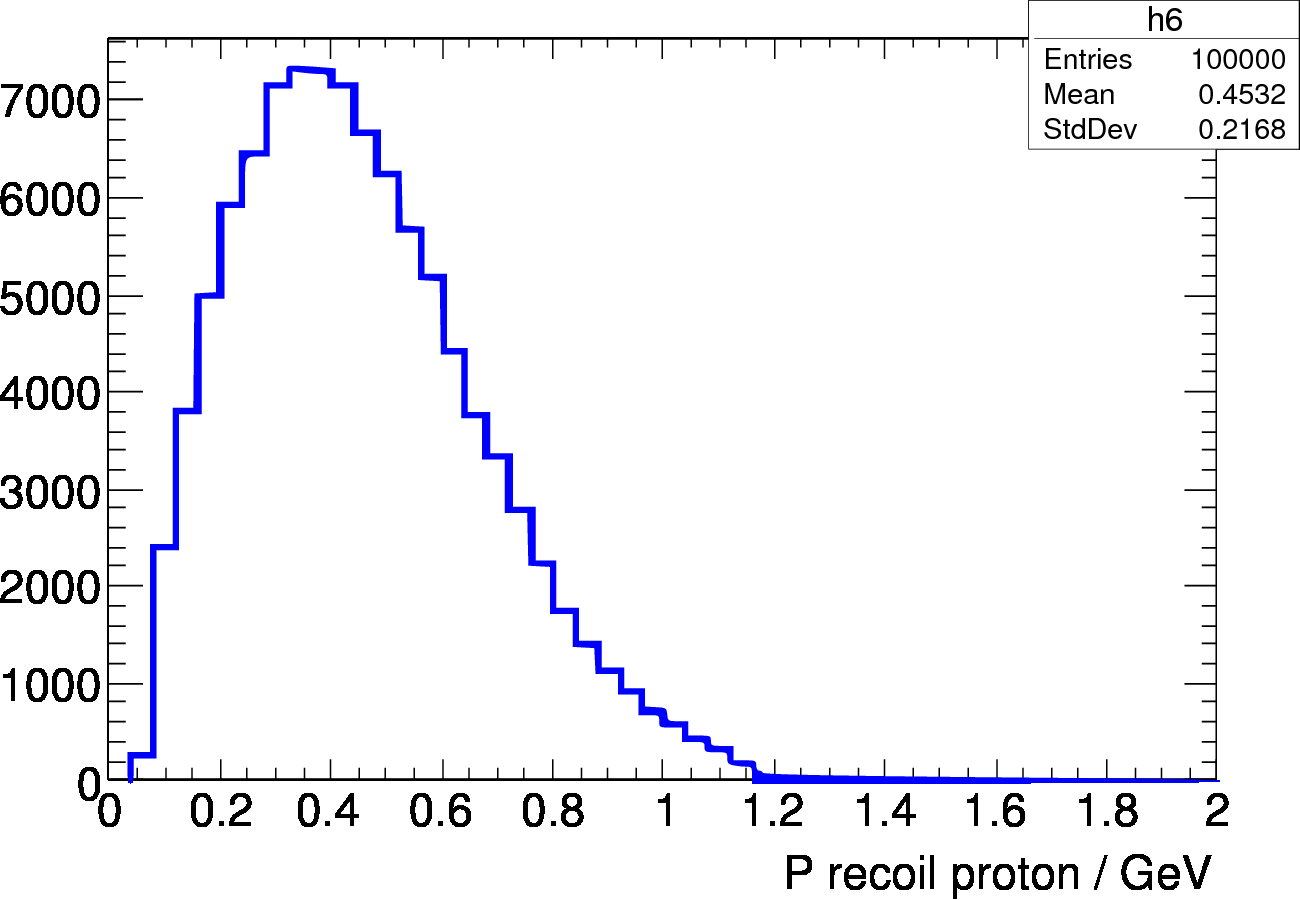} ~
    \includegraphics[width=0.27\textwidth]{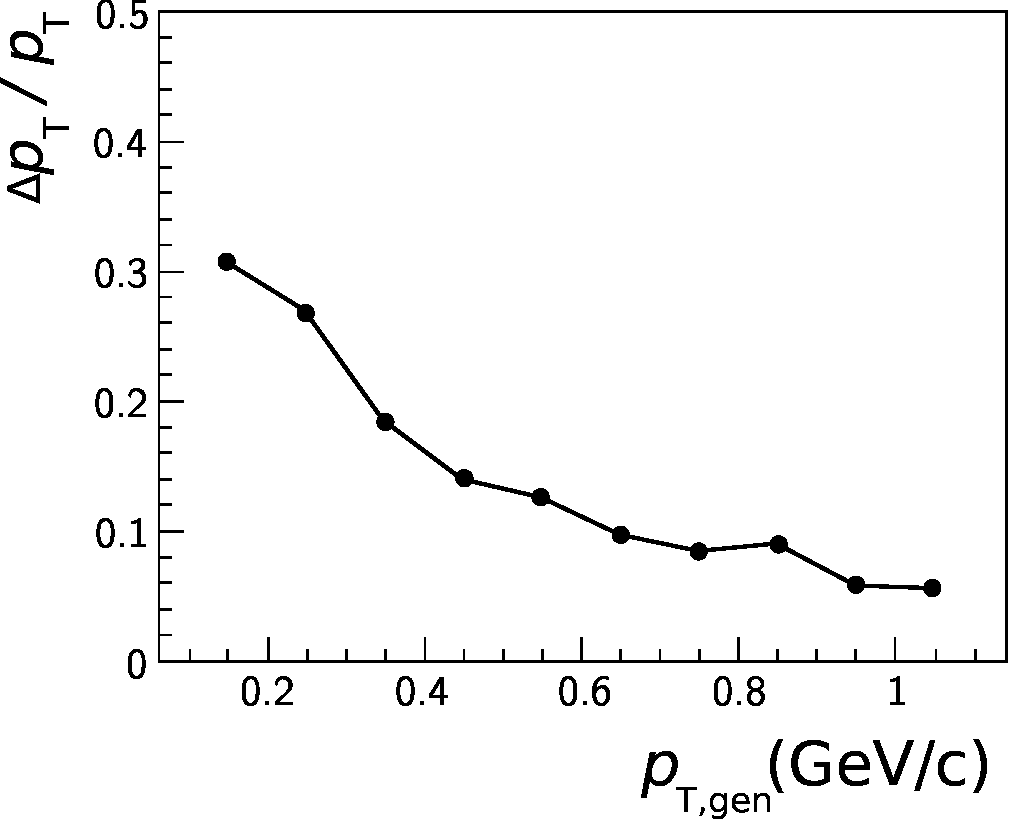}
  \end{center}
  \caption{Simulations with silicon detectors around the polarised target.
    Left: energy loss versus momentum in silicon for pions (red) and protons (black). Middle: momentum distribution for DVCS recoil protons. Right: momentum resolution for recoil protons.}
  \label{fig:future.compass_dvcs_silicon_momentum}
\end{figure}

Simulations based on the silicon geometry in Fig.~\ref{fig:future.compass_dvcs_silicon_detector} were carried out with the GEANT 4.6.10~\cite{GEANT42002} package using the HEPGEN~\cite{HEPGEN2012} generator for DVCS protons, and PYTHIA~\cite{Sjostrand:2006za} for SIDIS protons and pions. The dE/dx technique for Particle IDentification (PID) distinguishing protons, kaons and pions requires detectors that are capable of measuring: (i) space coordinates of the recoil particles with a precision of about 1\,mm in 3 space points at least, (ii) momentum reconstruction in the range between 100-1000\,MeV with a precision of about 5-10\% and (iii) dE/dx for each recoil particle with a precision of about 10\%. The particle momentum is determined from the reconstruction of its trajectory in the magnetic field, which requires at least 3 space points. Figure \ref{fig:future.compass_dvcs_silicon_momentum} shows the expected momentum distributions and resolutions of recoil protons and the PID performance.

\subsubsection{Target spectrometer for spectroscopy with low-energy antiprotons} 
\label{sec:inst-lowespec}

\begin{figure}[bt]
  \begin{center}
    \includegraphics[width=0.7\columnwidth]{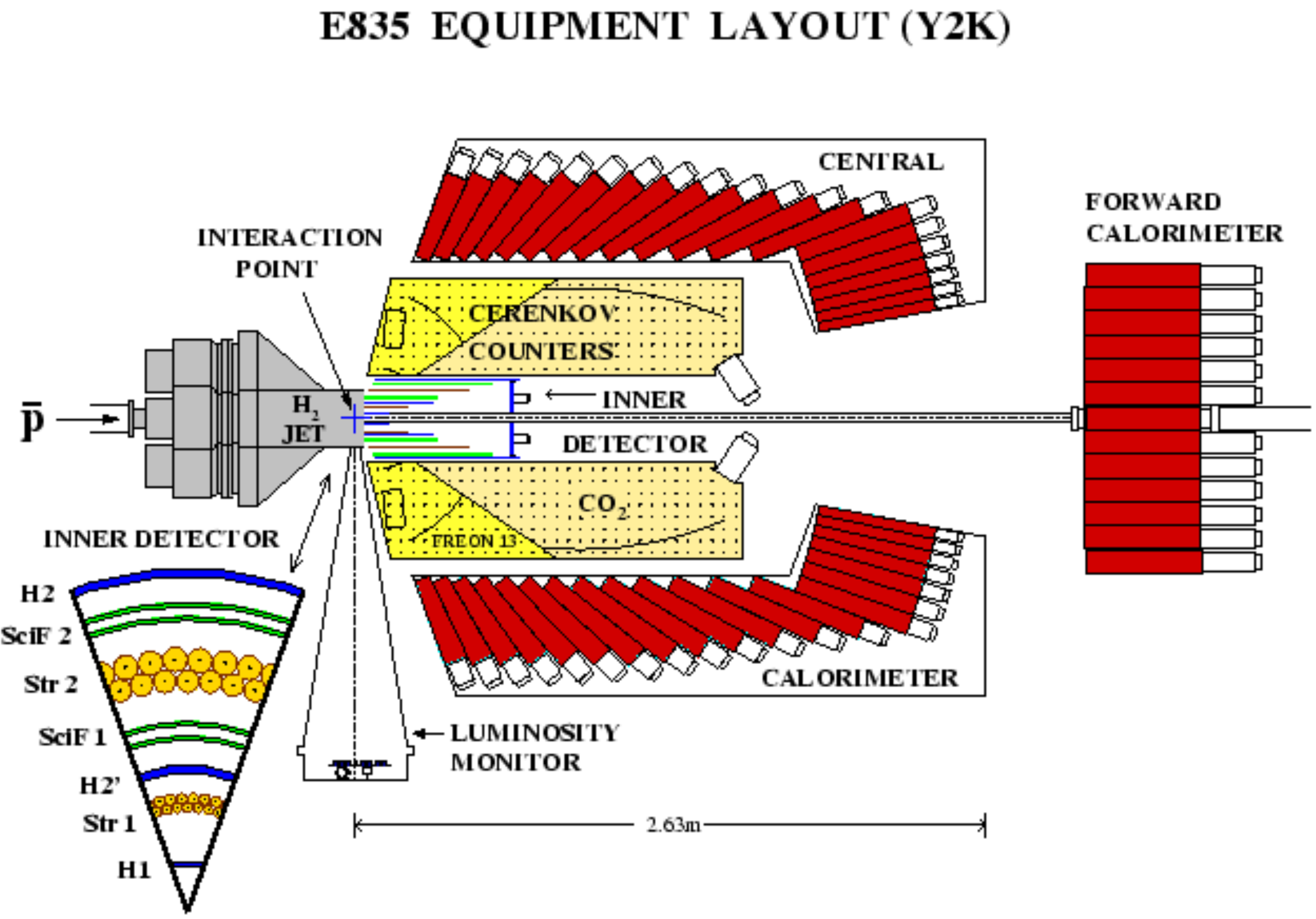}
    \caption{Schematic view of the E835 setup at Fermilab
      \cite{Garzoglio:2004kw}.} 
    \label{fig:E835}
  \end{center}
\end{figure}

\begin{figure}[tb]
  \begin{center}
    \hfill
    \includegraphics[width=0.5\columnwidth]{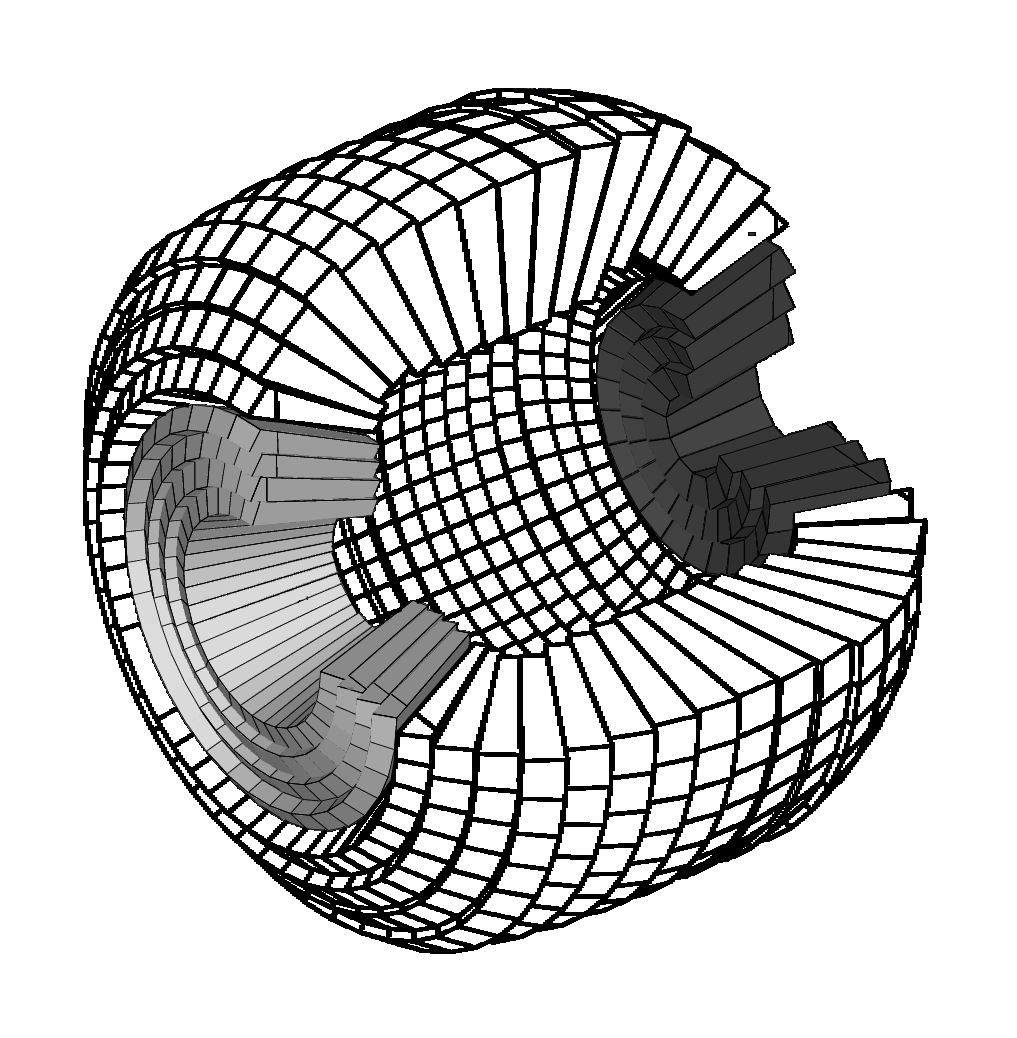} 
    \hfill
    \includegraphics[width=0.15\columnwidth]{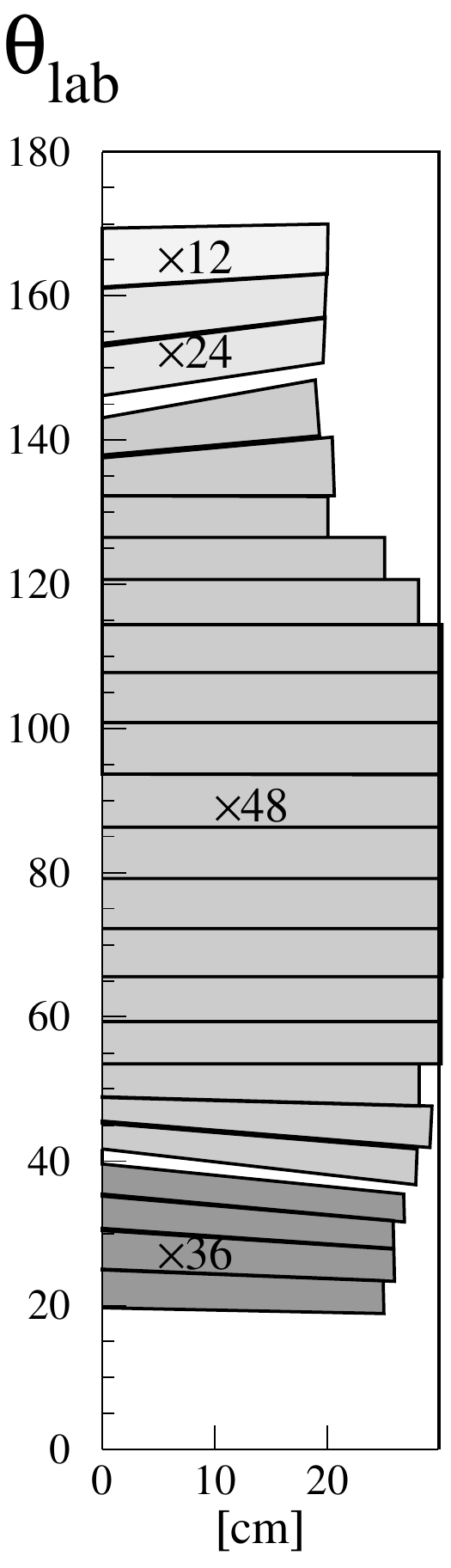}
    \hfill ~
    \caption{The scintillating electromagnetic calorimeter (SEC) of the WASA detector \cite{Adam:2004ch}. 
      Left: geometry of the crystals, with 4 layers with 36 elements each in the forward part, 17 layers with 48 elements in the central part, while the backward part consists of 2 layers with 24 elements and one layer closest to the beam pipe with 12 elements.
      Right: angular coverage in the laboratory system. } 
    \label{fig:wasa.ecal}
  \end{center}
\end{figure}

The exclusive measurements in the spectroscopy with low-energy antiprotons (Sec.~\ref{sec:stdhad.spect}) require the additional coverage of charged-particle tracking and calorimetry around the target. \Figref{fig:E835} shows as an example the setup of experiment E836 at Fermilab (USA) \cite{Garzoglio:2004kw}, which is split into a barrel part and a forward part. 
The PANDA experiment at FAIR/GSI was designed in a similar way, with improved calorimetry and charged-particle tracking also in the forward detector. With the components for PANDA not yet being fully available, a possible option that we are investigating at the moment is to re-use parts of the barrel spectrometer of the WASA detector \cite{Adam:2004ch}. It consists of an electromagnetic calorimeter made up of 1012 CsI(Na) scintillating crystals with a thickness corresponding to $16\,X_0$. It can measure photons, electrons, and positrons with energies up to 800\,MeV at a very low threshold of 2\,MeV. In its original shape, it covers scattering angles from $20^\circ$ up to $169^\circ$. 
\Figref{fig:wasa.ecal} shows the geometry and angular acceptance of the WASA calorimeter. 

In order to maximise the acceptance for the final-state particles emerging from antiproton annihilations in an experiment at the M2 beam line, it could be envisaged to rotate the WASA calorimeter by $180^\circ$, such that the coverage in forward direction increases and matches the acceptance of the first forward calorimeter ECAL0 (up to about $17^\circ$). Charged-particle tracking is performed in a $1.3\,\T$ solenoid magnetic field provided by a superconducting coil located inside the calorimeter. The originally used Straw tubes for charged-particle tracking will have to be replaced by a new tracking detector because of ageing. One option could be a continuously operating GEM-TPC as originally developed for PANDA and built and tested in FOPI \cite{Berger:2017tve}.  

The forward-going particles will be detected by the existing \compass\ detectors, including ECAL0. With this scenario, a high and uniform acceptance for charged and neutral particles will be achieved even at the low momenta foreseen for the antiproton beam. 

\subsubsection{Active absorber for Drell-Yan measurements with an RF-separated hadron beam} 
\label{sec:inst-dyrfsep}

For Drell-Yan physics with high-intensity kaon and antiproton beams (Sec.~\ref{sec:rfsep.dryan}), the tracking system needs to cover a wider aperture, i.e.\ up to $\pm300$\,mrad, to take the effect of the lower beam energy into account. For this purpose, we consider a "magnetised spectrometer", i.e.\ an active absorber. It could possibly be similar to Baby MIND at JParc \cite{Antonova:2017thk}, which is a "3-in-1" setup consisting of detector, spectrometer magnet, and absorber.


\bibliographystyle{./sty/elsart-num}
\bibliography{./bib/mpgd,./bib/compass,./bib/panda,./bib/3dstructure,./bib/prompt,./bib/prim,./bib/vector,./bib/drellyan,./bib/inst,./bib/hadron,./bib/stdmuo_mup,./bib/detectors,./bib/gpd,./bib/xsect}


\end{document}